\documentclass[a4paper,11pt,oneside]{report}

\usepackage{graphicx}
\usepackage{setspace}
\usepackage{array}
\usepackage{booktabs}
\usepackage{slashed}
\usepackage{subfigure}

\usepackage{amsmath}
\usepackage{amssymb}
\usepackage[british]{babel} 
\usepackage{hyperref}
\usepackage{mathrsfs}
\usepackage{tabulary}
\usepackage{mathdesign}
\usepackage{verbatim}
\usepackage{fancyhdr}
\def\doublesided{\evensidemargin=-15truemm}
\doublesided
\def\p{\mbox{\boldmath$\displaystyle\mathbf{p}$}}
\def\e{\mbox{\boldmath$\displaystyle\mathbf{\epsilon}$}}
\def\L{\mbox{\boldmath$\displaystyle\mathbf{\Lambda}$}}
\def\l{\mbox{\boldmath$\displaystyle\mathbf{\lambda}$}}
\def\th{\mbox{\boldmath$\displaystyle\mathbf{\theta}$}}
\def\k{\mbox{\boldmath$\displaystyle\mathbf{k}$}}

\def\q{\mbox{\boldmath$\displaystyle\mathbf{q}$}}
\def\bv{\mbox{\boldmath$\displaystyle\mathbf{\varphi}$}}

\def\0{\mbox{\boldmath$\displaystyle\mathbf{0}$}}
\def\O{\mbox{\boldmath$\displaystyle\mathbf{O}$}}
\def\s{\mbox{\boldmath$\displaystyle\mathbf{\sigma}$}}
\def\J{\mbox{\boldmath$\displaystyle\mathbf{J}$}}
\def\K{\mbox{\boldmath$\displaystyle\mathbf{K}$}}

\def\mJ{\mbox{\boldmath$\displaystyle\mathcal{J}$}}

\def\mK{\mbox{\boldmath$\displaystyle\mathcal{K}$}}
\def\x{\mbox{\boldmath$\displaystyle\mathbf{x}$}}
\def\y{\mbox{\boldmath$\displaystyle\mathbf{y}$}}

\def\lsim{\mathop{\hbox{${\lower3.8pt\hbox{$<$}}\atop{\raise0.2pt\hbox{$\sim$}}
$}}}
\def\gsim{\mathop{\hbox{${\lower3.8pt\hbox{$>$}}\atop{\raise0.2pt\hbox{$\sim$}}
$}}}
\def\sqr#1#2#3{{\vbox{\hrule height.#2pt \hbox{\vrule width.#2pt
height#1pt\kern#1pt\vrule width.#2pt}\hrule height.#2pt}\hbox{\hskip.#3em}}}

\def\scr{\scriptstyle}\def\scrscr{scriptscriptstyle}
\def\W#1{^{\raise2pt\hbox{$\scrscr#1$}}} \def\Y#1{^{\raise2pt\hbox{$\scr#1$}}}
\def\X#1{_{\lower2pt\hbox{$\scrscr#1$}}} \def\Z#1{_{\lower2pt\hbox{$\scr#1$}}}
 
\ifx\pdfximage\UnDeFiNeD
\ifx\pdfimage\UnDeFiNeD

\else

\fi
\else

\fi

\newcommand{\gdual}[1]{\overset{\:{}^{{}^{\boldsymbol{\neg}}}}{\smash[t]{#1}}} 
\newcommand{\dual}[1]{\overset{{}^{{}^{\boldsymbol{\neg}}}}{\smash[t]{#1}}} 

\begin{document}
\pagenumbering{roman}
\thispagestyle{empty}
\openup1\jot
\begin{center}{}
\end{center}
\openup-\jot
\vspace{3cm}

\centerline{\LARGE \bf Symmetries of  }
\vskip\baselineskip
\centerline{\LARGE \bf Elko and massive vector fields}
\vskip\baselineskip

\vskip 1.5\baselineskip
\centerline{\Large Cheng-Yang Lee}

    \begin{center}

      \vfill
      {\large A thesis submitted in partial fulfilment \par}
      {\large of the requirement for the degree of \par}
      {\large Doctor of Philosophy in Physics at the \par}
      {\large University of Canterbury \par}
    \end{center}

\vfill
\centerline{\includegraphics{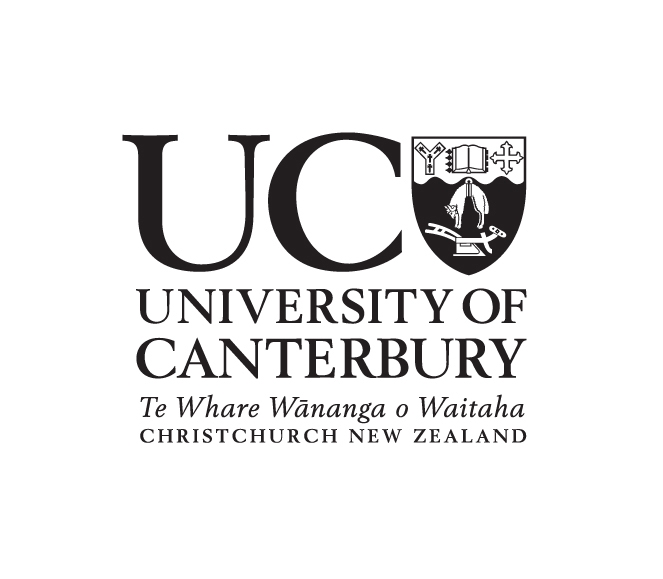}}
\vskip\baselineskip

\centerline{\large University of Canterbury 2012}

\vskip \baselineskip


\vskip \baselineskip
\vspace{2true cm}
\eject



\pagestyle{empty}  

\null\vfill
\setstretch{1.3}
\begin{center}
\textbf{\Large{Disclaimer}}
\end{center}
The research described in this thesis was carried out in the Department of Physics and Astronomy at the University of Canterbury under the supervision of Dr. D. V. Ahluwalia of the University of Canterbury. While much of the work contained in the thesis is my own, and is original, in writing this thesis I have felt free to incorporate the insights provided to me by other members of the group.

Chapter~\ref{Chapter2} is a review on quantum field theory mainly based on volume one of The Quantum Theory of Fields by S. Weinberg. The first section of chap.~\ref{Chapter3} is based on original research done in collaboration with D. V. Ahluwalia, S.~P. Horvath and D. Schritt and has yet to be published. The later section of chap.~\ref{Chapter3} is a review on Abelian gauge symmetry based on volume one of The Quantum Theory of Fields. The first section of chap.~\ref{Chapter4} is based on original research done in collaboration with D. V. Ahluwalia and D. Schritt and are published in Phys.Rev. \textbf{D83} (2011) 065017 and Phys.Lett. \textbf{B687} (2010) 248-252. The later research in chap.~\ref{Chapter4} has yet to be published.

\vfill\vfill\vfill\vfill\vfill\vfill\null
%
%
%
%
%
%
%
%
%

\clearpage  

\pagestyle{empty}  

\null\vfill

\setstretch{1.3}

\textit{If my history lesson has done nothing else, it should have reminded you that, during any given period in the evolving history of physics, the prevailing, main line, climate of opinion was likely as not to be wrong, as seen in the light of later developments. And yet, in those earlier times, with relatively few individuals involved, change did occur, but slowly... What is fundamentally different in the present day situation in high energy physics is that large numbers of workers are involved, with corresponding pressures to conformity and resistance to any deflection in direction of the main stream, and that the time scale of one scientific generation is much too long for the rapid pace of experimental discovery. I also have a secret fear that new generations may not necessarily have the opportunity to become familiar with dissident ideas. }
\begin{flushright}
"If you can't join 'em, beat 'em": Julian Schwinger's Conflicts in Physics. Directions in Cultural History, The UCLA Historical Journal, Volume 21, 2005-2006, p. 50
\end{flushright}

\vfill\vfill\vfill\vfill\vfill\vfill\null
\clearpage  

\abstract{
\addtocontents{toc}{\vspace{1em}}  

This thesis studies the symmetries and phenomenologies of the massive vector fields of indefinite spin with both scalar and spin-one degrees of freedom and Elko. The investigation is conducted by using and extending the quantum field theory formalism developed by Wigner and Weinberg. In particular, we explore the possibility that the $W^{\pm}$ and $Z$ bosons have an additional scalar degree of freedom and show that Elko is a fermionic dark matter candidate.

We show that the massive vector fields of indefinite spin are consistent with Poincar\'{e} symmetry and have physically desirable properties that are absent for their pure spin-one counterpart. Using the new vector fields, the decay of the $W^{\pm}$ and $Z$ bosons to leptons at tree-level are in agreement with the Standard Model (SM) predictions. For higher order scattering amplitudes, the theory has better convergent behaviour than the intermediate vector boson model and the Fermi theory. 

Elko has the unusual property that it satisfies the Klein-Gordon but not the Dirac equation and has mass dimension one instead of three-half. We show that the Elko fields are local only along a preferred axis and that they violate Lorentz symmetry. Motivated by the results obtained by Ahluwalia and Horvath that the Elko spin-sums are covariant under very special relativity (VSR) transformations, we derive the VSR particle states and quantum fields. We show that the VSR particles can only interact with the SM particles through gravity and massive scalar particles thus making them and hence Elko dark matter candidates. 

}

\clearpage  

\begin{center}
\textit{\Large{Acknowledgement}}
\end{center}
\addtocontents{toc}{\vspace{1em}}  

I would like express my gratitude to my supervisor Dr. Dharamvir Ahluwalia for his supervision. 
His hospitality, friendship and passion for physics has made our group 
a place of scholarship. I am grateful for the freedom and encouragement he 
has given me to pursue my own research and at times making sure I stayed on 
the right path.

I am grateful to Assoc. Prof. David Wiltshire for agreeing to be my co-supervisor.
I thank Adam Gillard, Assoc. Prof. Ben Martin and Prof. Sudhakar Panda for discussions. 

This work was partially funded by a University of Canterbury Doctoral Scholarship and
a departmental scholarship provided by the Department of Physics and Astronomy of the
University of Canterbury.

In our research group, I would like to thank Sebastian Horvath and Dimitri Schritt for their
friendship and constructive discussions. 

In the department, the friendship and companionship
of Ahsan Nazer, Ishwaree Neupane, Ewan Orr, Oscar Macias-Ramirez, Hamish Silverwood, Peter Smale, Jidi Sun
is very much appreciated.

In India, I am grateful to the warm hospitality offered by the family members
of Dr. Dharamvir Ahluwalia. At the Centre for Theoretical Physics in Jamia University 
and the Harish-Chandra Research Institute, I am thankful to the warm hospitality of 
Prof. M. Sami and Prof. Sudhakar Panda. 

At Lake Tekapo in New Zealand, I had the privilege of meeting Graeme Murray and enjoy his hospitality. The tranquil landscape at the Boatsman cottage provided the perfect atmosphere to conduct research.

Outside physics, I am grateful to my Tai-Chi Chuan teacher Yu-Shiun Tang and his wife Hui-Ying Chang. They have not only taught me
the art of Tai-Chi Chuan, but have also helped me to come
to appreciate the beauty and depth of Chinese culture. 

Lastly but not least, I would like to thank my brother and mother for their endless love and support during my thesis. They provided me with much needed care and understanding thus enabling me to focus on my research. The memory of the time we spent together will always be treasured and have a special place in my heart.

\clearpage  

%
%
%

\catcode`\@=11
\c@page=3
\catcode`\@=12
\tableofcontents
\vfil\eject
\pagestyle{plain}
\catcode`\@=11
\c@page=0
\catcode`\@=12
\pagenumbering{arabic}

\def\d{{\mathrm{d}}}
\def\e{{\mathrm{e}}}
\def\NN{{\cal N}}
\def\implies{\Rightarrow}
\font\sevenrm=cmr7
\def\etal{{\it et al.}}
\baselineskip=16pt plus.2pt minus.1pt

\setstretch{1.3}  

\pagestyle{empty}  
\null\vfill
\vspace{5cm}
\hspace{8cm}\includegraphics[scale=0.6]{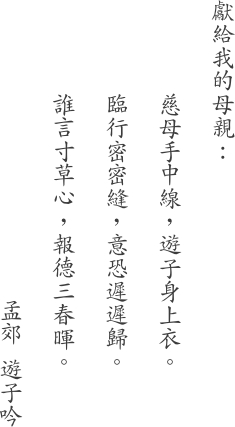}

\vfill\vfill\vfill\vfill\vfill\vfill\null
\clearpage

\clearpage


\chapter{Introduction} 
\label{Chapter1}
\lhead{Chapter 1. \emph{Introduction}}
The central theme of the thesis is on the connexions between space-time symmetries and quantum field theory and their application to particle physics. In this thesis we focus on the following two topics. The first topic involves the ab-initio construction of a massive vector particle with both scalar and spin-one degrees of freedom while the later studies a new fermionic quantum field called Elko constructed by Ahluwalia and Grumiller~\cite{Ahluwalia:2004ab,Ahluwalia:2004sz}. 

Although these two topics are not directly related, the results we have obtained suggest that beneath these topics are the existence of unexplored structures containing physics beyond the Standard Model (SM). Specifically, the massive vector fields offer deeper insights to the electroweak theory while Elko is a dark matter candidate.

Throughout the thesis, we closely follow the theme we have laid out. Our starting point of research is always based on a systematic approach to either explore the implications of the symmetry principles or to determine the symmetries of the proposed theories.
In this way, we can be confident that the theories and their phenomenologies are mathematically consistent and reliable.

\section{Possible generalisations of mass and spin}

Through the seminal work of Wigner, it is known that the simplest particle states in the Hilbert space can be obtained from the unitary irreducible representations of the Poincar\'{e} group and are uniquely characterised by their mass and spin~\cite{Wigner:1939cj}.\footnote{Strictly speaking, according to the unpublished notes of Bargmann, Wightman and Wigner, further classification of particles are possible when one includes discrete symmetries. See ref.~\cite{Ahluwalia:1993zt} for more details.} Quantum field theory, as shown by Weinberg is the only known way to describe the interactions of these particle states that is consistent with Poincar\'{e} symmetry~\cite{Weinberg:1964cn,Weinberg:1964ev,Weinberg:1995mt}.

Although the formalism developed by Wigner and Weinberg accommodates the SM particles and fields, from a theoretical perspective, it does not cover the most general possibilities. The axioms of quantum mechanics allow for the superposition of physical states, so in principle, we should be able to construct theories where particles have neither definite mass and spin. Therefore, the most general particle states that realise the principle of superposition would be a linear combination of eigenstates of different mass and spin~\cite{Ahluwalia:1997hc}. The corresponding quantum fields would then be the most general fields that satisfy Poincar\'{e} symmetry. In fact, we now know that nature utilises one of the possibilities through neutrino oscillation, where the neutrino states are linear superposition of mass eigenstates~\cite{Pontecorvo:1957cp,Bilenky:1978nj,Cleveland:1998nv}. 

In this thesis, we do not study the most general particle state nor neutrino oscillation. Instead, we realise one of the above mentioned possibilities by constructing a massive vector particle containing scalar and spin-one degrees of freedom. 

In the SM, the $W^{\pm}$ and $Z$ are taken to be massive spin-one vector bosons and their interactions are described by non-Abelian gauge theory. The theory requires scalar ghost fields and spontaneous symmetry breaking for it to be renormalisable and the particles to gain masses respectively.  Here, we explore the possibility that the $W^{\pm}$ and $Z$ bosons are described by massive vector particles of indefinite spin, containing scalar and spin-one degrees of freedom. In scattering processes, it is shown that the tree-level decay rates of $W^{\pm}$ and $Z$ to leptons are identical to the SM predictions up to an additive constant proportional to $m_{\ell}^{2}/m_{Z,W^{\pm}}^{2}$ where $\ell$ denotes the leptons. 
But more importantly, for physical processes considered in the thesis where the scattering amplitudes are divergent and unitary-violating in the intermediate vector boson model~\cite{Glashow:1961tr} and the Fermi theory~\cite{Fermi:1934hr}, they are shown to be finite and unitary for the proposed theory.

\section{Space-time symmetries and dark matter}

As the name suggest, dark matter is dark with respect to the SM particles so it has limited electromagnetic interactions with the SM particles. The term dark matter was first introduced in a  paper published in the Astronomical Institute of the Netherlands by Oort~\cite{Oort} in 1932 where he discovered irregularities in the velocity of stars (in direction perpendicular to the plane of the galaxies). A year later in 1933, Zwicky inferred missing masses when studying the motion of astrophysical objects in the Coma galaxy~\cite{Zwicky}. 

The findings of Zwicky and Oort now belong to a wide range of evidence for dark matter at various distance scales. On galactic and sub galactic scales, evidence include rotation curves~\cite{Borriello:2000rv} and gravitational lensing ~\cite{Hoekstra:2002nf,Hoekstra:2003pn,Metcalf:2003sz,Moustakas:2002iz}. At the scale of galaxy clusters, observations (radial velocities, weak lensing and X-ray emission) indicate a total matter density of $\Omega_{M}\approx0.2-0.3$ far greater than the density of ordinary baryonic matter~\cite{Bahcall:1998ur,Kashlinsky:1998fn,Tyson:1998vp,Dahle:2007wf}. The recent fitting of $\Lambda$CDM parameters with the latest WMAP observation yields a dark matter density that is five times greater than the baryonic matter density~\cite{Komatsu:2010fb}. 

The increasing observational evidence of non-baryonic dark matter and its dominance over baryonic matter makes it one of the most intriguing and important area of research in our quest to understand the fundamental laws of nature. Following the central theme of the thesis, a natural question to ask is what is the space-time symmetry of dark matter? We note, as currently there are no direct detections of dark matter, it is not possible to answer this question directly. But from a theoretical perspective, there are good reasons to assume dark matter satisfies Poincar\'{e} symmetry.

Poincar\'{e} symmetry is one of the cornerstone of modern physics, in particle physics it has given us quantum field theory, the language in which the SM is based on. On a broader picture, all the currently known laws of physics (including classical gravity) can be traced back to Poincar\'{e} symmetry.\footnote{The classical theory of gravity with non-vanishing torsion can be formulated as a gauge theory by gauging the Poincar\'{e} group~\cite{Kibble:1961ba,Hehl:1976kj,Blagojevic:2002du}.} Apart from the success of the SM, one of the most definitive evidence in favour of Lorentz symmetry is the validity of the dispersion relation for the SM particles. The most recent measurement made by the Fermi GBM/LAT Collaboration of photons in the vacuum emitted by distant astrophysical sources have shown that the dispersion relation is accurate up to $\sim 10^{19}$GeV~\cite{Ackermann:2009aa}. Moreover, Collins et al. have shown that a modification of the dispersion relation for massive particles due to the existence of Planck-scale preferred frame, with the inclusion of particle interactions, would give rise to Lorentz violation at the percent level unless the bare parameters are unnaturally fine-tuned~\cite{Collins:2004bp}.

Given the amount of evidence for Poincar\'{e} symmetry, it is natural and conservative for theorists to investigate the dark matter problem within the realm Poincar\'{e} symmetry. The inert interactions between dark matter and the SM particles can be achieved by demanding the dark matter particles to be electrically neutral. In this thesis, we argue that although this approach is pedagogic and well-motivated, it is not the only possibility. To see how this argument works, we need to review the connexions between Lorentz symmetry and local gauge symmetry.

The principle of local gauge invariance is a powerful formalism to determine the interactions between particles. In the SM, the gauge group $SU(3)_{C}\times SU(2)_{L}\times U(1)_{Y}$ where $C$, $L$ and $Y$ denote the colour, left-handed chirality and weak hypercharge respectively, successfully explain the interactions of the elementary particles. However, it also has its limitations as Weinberg has shown that the local Abelian gauge symmetry is a consequence of Lorentz symmetry~\cite{Weinberg:1964ev}. Specifically, the spin-one massless field $a^{\mu}(x)$ is not a four-vector. Instead, it transforms as a four-vector up to a local gauge transformation $a^{\mu}(x)\rightarrow \Lambda_{\nu}^{\,\,\mu}a^{\nu}(\Lambda x)+\partial^{\mu}\Omega(\Lambda,x)$.\footnote{See sec.~\ref{local_gauge_invariance} for details.}   Seen from this perspective, it is possible that the SM gauge symmetry may no longer be applicable to particles whose symmetry groups are not the Poincar\'{e} group. Here, it is important to realise that although there are stringent limits on Poincar\'{e} symmetry violation, all these experiments were conducted using equipments built from the SM particles~\cite{Ahluwalia:2009ia}. Consequently, there are no direct evidence suggesting that dark matter must satisfy Poincar\'{e} symmetry~\cite{Blas:2012vn}.

Our observation raise the possibility that the symmetry group for dark matter may not be the Poincar\'{e} group. Depending on the properties of the symmetry group, it may not admit the usual SM gauge symmetry thus naturally limits the interactions between dark matter and the SM. Therefore, a new direction of research for dark matter is to study particle states and quantum fields derived from symmetry groups that differ from the Poincar\'{e} group. Using the symmetry group, we may then proceed to determine the possible interactions and resulting gauge symmetries within the dark sector following the formalism developed by Weinberg~\cite{Weinberg:1965rz}.

In this thesis, Elko is shown to violate Lorentz symmetry but its spin-sums are shown to be covariant under very special relativity (VSR) transformations by Ahluwalia and Horvath~\cite{Ahluwalia:2010zn} where the VSR groups are subgroups of the Poincar\'{e} group. By studying the representations and symmetries of VSR, we show that the VSR particles can only interact with the SM particles through gravity and massive scalar particles, thus making them dark matter candidates.

\section{Outline of the thesis}

Chapter~\ref{Chapter2} reviews the foundations of quantum field theory with Poincar\'{e} symmetry based on the works of Wigner~\cite{Wigner:1939cj} and Weinberg~\cite{Weinberg:1964cn,Weinberg:1964ev,Weinberg:1995mt}. First, we derive the Poincar\'{e} algebra through the principles of special relativity and the axioms of quantum mechanics.

The massive and massless particles are obtained from the unitary irreducible representations of the Poincar\'{e} group and their transformations are derived. Subsequently the $S$-matrix formalism is introduced to compute the interacting observables between the particle states.  In the low energy limit (with respect to the Planck scale), quantum field theory as shown by Weinberg, is the inevitable consequence due to the demand of Poincar\'{e}-invariant $S$-matrix and the cluster decomposition principle~\cite[chap.~4]{Weinberg:1995mt}.

The formalism developed by Weinberg allows us to obtain the general constraints imposed by Poincar\'{e} symmetry on the quantum fields. Using these constraints, we can determine the expansion coefficients of the quantum fields and their field equations. To demonstrate the power of the formalism, we construct the massive and massless Dirac fields and generalise the former to arbitrary spin. 

Chapter~\ref{Chapter3} studies the massive and massless particles of the vector representation of the Poincar\'{e} group. Here, the massive vector particles have both scalar and spin-one degrees of freedom so they transform under a reducible rather than an irreducible representation of the Poincar\'{e} group. We show that by choosing the appropriate relative phases for the dual coefficients between the scalar and spin-one sector, the resulting propagator computed via the vacuum expectation value of the time-ordered product takes the form 
\begin{equation}
S^{(j=0,1)}_{\mu\nu}(p)=\frac{i}{(2\pi)^{4}}\frac{-\eta_{\mu\nu}}{p\cdot p-m^{2}+i\epsilon}
\end{equation}
in the momentum space. The numerator does not contain the usual $p^{\mu}p^{\nu}/m^{2}$ term associated with the massive spin-one vector field. As a result, it has better behaviour at large momentum than their spin-one counterpart. 

Using the massive vector fields, we then explore the possibility that the $W^{\pm}$ and $Z$ bosons have an extra scalar degrees of freedom. We show that at tree-level their decay rates to leptons are in agreement with the SM predictions up to an additive constant proportional to $m_{\ell}^{2}/m_{Z,W^{\pm}}^{2}$. Furthermore, the theory preserves unitarity and has better convergent behaviour due to the new propagator, unlike the intermediate vector boson model~\cite{Glashow:1961tr} and the Fermi theory~\cite{Fermi:1934hr}. 
Later, we show that the massless fields of the vector representation are four-vectors only up to a gauge-transformation. As a result, their interacting Lagrangian densities must not only be Lorentz invariant but also invariant under local gauge transformations. Seen from this perspective, local Abelian gauge symmetry is a consequence of Lorentz symmetry~\cite{Weinberg:1964ev}.

Chapter~\ref{Chapter4} studies Elko, a spin-half dark matter candidate proposed by Ahluwalia and Grumiller~\cite{Ahluwalia:2004ab,Ahluwalia:2004sz}.
We show that Elko satisfies the Klein-Gordon but not the Dirac equation. The Elko propagator obtained by computing the vacuum expectation value of the time-ordered product shows that it has mass-dimension one instead of three-half. As a result, Elko has a power-counting renormalisable self-interaction.

In the original papers published by Ahluwalia and Grumiller, the Elko fields are non-local with non-vanishing equal-time anti-commutators. The recent papers by Ahluwalia, Schritt and the author have shown that by choosing the appropriate phases for the Elko coefficients, the equal-time anti-commutators now enjoy locality along a preferred direction~\cite{Ahluwalia:2008xi,Ahluwalia:2009rh}. However, it was also noted that Elko violates Lorentz symmetry since its spin-sums contain a preferred plane and are not Lorentz-covariant.

Applying the Weinberg formalism presented in sec.~\ref{chap:quantum_fields}, we show that the Elko fields and their generalisation to arbitrary spin violate Lorentz symmetry. As a resolution, Ahluwalia and Horvath showed that Elko spin-sums are covariant under the $ISIM(2)$ transformations of VSR~\cite{Ahluwalia:2010zn}. Motivated by their results, we derive the particle states and quantum fields with $ISIM(2)$ symmetry. We show that the VSR and SM sector can only interact through gravity and massive scalar particles thus making VSR particles dark matter candidates.


\chapter{Foundations} 
\label{Chapter2}
\lhead{Chapter 2. \emph{Foundations}} 

In the low energy limit (with respect to Planck scale) where gravitational effects are negligible, all the known laws of physics originate from the Poincar\'{e} algebra
\begin{equation}	i[J^{\rho\sigma},J^{\mu\nu}]=\eta^{\rho\nu}J^{\mu\sigma}-\eta^{\rho\mu}J^{\nu\sigma}-\eta^{\sigma\mu}J^{\rho\nu}+\eta^{\sigma\nu}J^{\rho\mu},\label{eq:pa1}
\end{equation}
\begin{equation}
i[P^{\mu},J^{\rho\sigma}]=-\eta^{\rho\mu}P^{\sigma}+\eta^{\sigma\mu}P^{\rho}\label{eq:pa2},
\end{equation}
\begin{equation}
[P^{\mu},P^{\rho}]=0 \label{eq:pa3}
\end{equation}
where $J^{\mu\nu}$ and $P^{\mu}$ are the generators of the Lorentz group and space-time translation respectively and the metric in 3+1 dimensions is defined as
\begin{equation}
\eta_{\mu\nu}=\left(\begin{matrix}
1 & 0 & 0 & 0 \\
0 &-1 & 0 & 0 \\
0 & 0 &-1 & 0 \\
0 & 0 & 0 &-1 \end{matrix}\right).
\end{equation}
Historically, the Poincar\'{e} algebra was derived from the principles of special relativity~\cite{Einstein:1905ve}
\begin{enumerate}
\item The laws of nature are invariant in all inertial frames.
\item The speed of light in vacuum is invariant for all inertial observers.
\end{enumerate}
The principles of special relativity, at the start of the 20th century, marked a fundamental shift in our view on the structure of space-time. The hallmarks of special relativity are phenomenon such as time-dilation and length contraction as a result of the Lorentz transformations. 

Although the Poincar\'{e} algebra is traditionally obtained through Lorentz transformations and space-time translations, one should not consider the transformations to be more fundamental than the algebra. For alternatively, it is also possible for some intelligent alien life form to have discovered the Poincar\'{e} algebra by studying the properties of elementary particles. 

Here we take the latter view that the Lorentz transformations including space-time translations are reflections of the symmetries of the SM particles~\cite{Brown:2005,Ahluwalia:2007}. This view may seems meaningless at first, but is in fact motivated by the increasing evidence of dark matter. Cosmological data accumulated for the past decades indicated that the existence of non-baryonic dark matter is five times more than the baryonic matter described by SM~\cite{Komatsu:2010fb}. Since dark matter does not interact with the SM particles, it is possible that their underlying symmetries are not dictated by the Poincar\'{e} algebra. This topic is beyond the scope of this chapter, but is discussed in chapter~\ref{Chapter4}.
In this chapter, we focus on the Poincar\'{e} algebra and its implications on particle physics. 

In the first section, we follow the historical development and derive the Poincar\'{e} algebra from the Lorentz transformations. The Casimir invariants of the algebra naturally give us the physical interpretation of the simplest one-particle states, uniquely characterised by their mass and spin.

Sections \ref{one_particle_state} and \ref{PT_symmetries} reviews the formalism of Wigner~\cite{Wigner:1939cj}. The one-particle states are derived from the unitary irreducible representations of the Poincar\'{e} group and are classified into six subgroups of the Lorentz group called the little groups. Out of the six possibilities, we study the two little groups that correspond to the massive and massless particle states of positive-definite energy. Their continuous and discrete transformations are derived.

Sections \ref{chap:scattering_theory} and \ref{cluster_decomp} are devoted to the $S$-matrix. In these two sections, we study the symmetries of the $S$-matrix and derive the relevant formulae needed to compute observables for particle interactions. The important lesson to take here is that in order to construct a Poincar\'{e}-invariant $S$-matrix, the introduction of quantum field operators are inevitable. 

The subsequent section presents the Weinberg formalism for constructing quantum fields with Poincar\'{e} symmetry. In this formalism, the expansion coefficients of the quantum fields can be derived from symmetry consideration. The field equations and Lagrangians can then be obtained from the properties of the expansion coefficients. In secs.~\ref{massive_qf} and \ref{Massless QF}, we construct the massive and massless quantum fields. Their field equations and discrete symmetries are derived.

\section{The Poincar\'{e} algebra}\label{Poincare_algebra}

The Poincar\'{e} algebra in the low energy regime is the foundation of the known physical principles. The Casimir invariants of the algebra are
\begin{equation}
C_{1}=P^{\mu}P_{\mu}=m^{2},\hspace{0.5cm} C_{2}=W^{\mu}W_{\mu}=-m^{2}s(s+1)\label{eq:casimir_invariants}
\end{equation}
where $W_{\mu}=\frac{1}{2}\epsilon_{\mu\nu\rho\lambda}J^{\nu\rho}P^{\lambda}$ is the Pauli-Lubanski pseudo-vector where $\epsilon_{\mu\nu\rho\sigma}$ is the completely anti-symmetric Levi-Civita tensor which we define as
\begin{equation}
\epsilon_{0123}=-\epsilon^{0123}=1.
\end{equation}
Here $m^{2}$ and $-m^{2}s(s+1)$ are the eigenvalues of $C_{1}$ and $C_{2}$ respectively with $s$ taking integral and half-integral values. The two Casimir invariants naturally define the one-particle state  $|m,s\rangle$ as eigenstate of $C_{1}$ and $C_{2}$ where $m$ and spin $s$ are interpreted to be their respective mass and spin. Since $C_{1}$ and $C_{2}$ commute with all the generators of the Poincar\'{e} group, the mass and spin of the particle states are invariant in all inertial frames. However, this does not tell us how the particle states transform. For this purpose, we need to study the irreducible representations of the Poincar\'{e} algebra.

Following the traditional approach, we will first derive the Poincar\'{e} algebra via the principles of special relativity. Choosing an arbitrary coordinate system, $x=(t,\x)$, the principles of special relativity demand the following equality
\begin{equation}
\eta_{\mu\nu}dx^{\mu}dx^{\nu}=\eta_{\mu\nu}dx'^{\mu}dx'^{\nu} \label{eq:inv_scalar}
\end{equation}
to hold in all inertial frames with $x^{\mu}$ and $x'^{\mu}$ describing two different coordinate systems. 

Here we work with the natural unit where $c=\hbar=1$, so the speed of light is $|d\x/dt|=1$. Using eq.~(\ref{eq:inv_scalar}), one can verify that the speed of light is constant in all inertial frames. The two coordinate systems $x^{\mu}$ and $x'^{\mu}$ are related by a linear inhomogeneous Lorentz transformation
\begin{equation}
x'^{\mu}=\Lambda^{\mu}_{\,\,\,\nu}x^{\nu}+a^{\mu}
\end{equation}
where $a^{\mu}$ is space-time translation and $\Lambda$ represents boosts and rotations satisfying
\begin{equation}
\eta_{\mu\nu}\Lambda^{\mu}_{\,\,\,\rho}\Lambda^{\nu}_{\,\,\,\sigma}=\eta_{\rho\sigma} \label{eq:lg1}
\end{equation}
and
\begin{equation}
\Lambda^{\mu}_{\,\,\,\rho}\Lambda^{\nu}_{\sigma}\eta^{\rho\sigma}=\eta^{\mu\nu}. \label{eq:lg2}
\end{equation}
The Lorentz group, excluding space-time translations is then defined as
\begin{equation}
L=O(1,3)=\{\Lambda\in GL(4,R) | \Lambda^{T}\eta\Lambda=\eta\}
\end{equation}
and its proper orthochronous subgroup is
\begin{equation}
L_{+}^{\uparrow}=\{\Lambda\in O(1,3) | \det\Lambda=+1,\Lambda^{0}_{\,\,\,0}\geq +1\}.
\end{equation}

Any Lorentz transformations are either proper and orthochronous or can be written as a product between an element of $L_{+}^{\uparrow}$ with one of the discrete symmetry operators $\mathscr{P}$ or $\mathscr{T}$ where $\mathscr{P}$ and $\mathscr{T}$ are the parity and time-reversal operator respectively 
\begin{equation}
\mathscr{P}=\left(\begin{matrix}
1 & 0 & 0 & 0 \\
0 &-1 & 0 & 0 \\
0 & 0 &-1 & 0 \\
0 & 0 & 0 &-1 \end{matrix}\right),\hspace{0.5cm} 
\mathscr{T}=\left(\begin{matrix}
-1 & 0 & 0 & 0 \\
 0 & 1 & 0 & 0 \\
 0 & 0 & 1 & 0 \\
 0 & 0 & 0 &1 \end{matrix}\right).\label{eq:pt}
\end{equation}
The properties of the Lorentz group can then be determined from its proper orthochronous subgroup and discrete symmetries.

Since the Poincar\'{e} group contains the Lorentz transformations and space-time translations, to derive the Lie algebra, we need to consider the inhomogeneous Lorentz transformations. Given a coordinate system $x^{\mu}$, two successive transformations yield
\begin{eqnarray}
x''^{\mu}&=&\bar{\Lambda}^{\mu}_{\,\,\,\rho}(\Lambda^{\rho}_{\,\,\,\nu}x^{\nu}+a^{\rho})+\bar{a}^{\mu} \\ \nonumber
&=&(\bar{\Lambda}^{\mu}_{\,\,\,\rho}\Lambda^{\rho}_{\,\,\,\nu})x^{\nu}+(\bar{\Lambda}^{\mu}_{\,\,\,\rho}a^{\rho}+\bar{a}^{\mu}).
\end{eqnarray}
Therefore, an arbitrary coordinate transformation $T(\Lambda,a)$ of the Poincar\'{e} group satisfies the following product rule,
\begin{equation}
T(\bar{\Lambda},\bar{a})T(\Lambda,a)=T(\bar{\Lambda}\Lambda,\bar{\Lambda}a+\bar{a}). \label{eq:tt_product}
\end{equation}

The inhomogeneous Lorentz transformation on the coordinate $x^{\mu}$ induces the following transformation
\begin{equation}
|\Psi\rangle \rightarrow U(\Lambda,a)|\Psi\rangle. \label{eq:tran_psi}
\end{equation}
where $|\Psi\rangle$ is a physical state in the Hilbert space. However, the product rule for $U(\Lambda,a)$ is not the same as eq.~(\ref{eq:tt_product}). To see this, we consider two successive transformations $U(\bar{\Lambda},\bar{a})U(\Lambda,a)$ on state $|\Psi\rangle$, giving us $U(\bar{\Lambda},\bar{a})U(\Lambda,a)|\Psi\rangle$. The product rule tells us that this state is equivalent to  $U(\bar{\Lambda}\Lambda,\bar{\Lambda}a+\bar{a})|\Psi\rangle$. But in quantum mechanics, the equivalence of state does not imply equality, instead, it implies the two states are equal up to a phase
\begin{equation}
U(\bar{\Lambda},\bar{a})U(\Lambda,a)|\Psi\rangle=\exp[i\phi(\bar{\Lambda},\Lambda;\bar{a},a)]U(\bar{\Lambda}\Lambda,\bar{\Lambda}a+\bar{a})|\Psi\rangle.\label{eq:proj_rep}
\end{equation}
The phase has to be independent of the state $|\Psi\rangle$. To show this, we consider a state $|\Psi\rangle=|\Psi_{1}\rangle+|\Psi_{2}\rangle$ where $|\Psi_{1}\rangle$ and $|\Psi_{2}\rangle$ are linearly independent. Performing successive transformation on $|\Psi\rangle$ gives us
\begin{equation}
U(\bar{\Lambda}\Lambda,\bar{\Lambda}a+\bar{a})e^{i\phi_{\Psi}}|\Psi\rangle=
U(\bar{\Lambda}\Lambda,\bar{\Lambda}a+\bar{a})e^{i\phi_{\Psi_{1}}}|\Psi_{1}\rangle+
U(\bar{\Lambda}\Lambda,\bar{\Lambda}a+\bar{a})e^{i\phi_{\Psi_{2}}}|\Psi_{2}\rangle.
\end{equation}
where we have suppressed the $(\Lambda,a)$ dependence of the phases and assume they are state dependent. Multiply both side by $U^{-1}(\bar{\Lambda}\Lambda,\bar{\Lambda}a+\bar{a})$ we get
\begin{equation}
e^{i\phi_{\Psi}}|\Psi\rangle=e^{i\phi_{\Psi_{1}}}|\Psi_{1}\rangle+e^{i\phi_{\Psi_{2}}}|\Psi_{2}\rangle.
\end{equation}
Since $|\Psi_{1}\rangle$ and $|\Psi_{2}\rangle$ are linearly independent, the phase must be state independent.

Operators that satisfy eq.~(\ref{eq:proj_rep}) are the projective representation of the Poincar\'{e} group. Here we are only concerned with the case where $\phi(\bar{\Lambda},\Lambda;\bar{a},a)=0$ so that
\begin{equation}
U(\bar{\Lambda},\bar{a})U(\Lambda,a)=U(\bar{\Lambda}\Lambda,\bar{\Lambda}a+\bar{a}).\label{eq:u_rep}
\end{equation}
The operator $U(\Lambda,a)$ must then either be a linear and unitary or anti-linear and anti-unitary representation of the Poincar\'{e} group to ensure the inner-product is preserved under eq.~(\ref{eq:tran_psi})~\cite{Wigner:1931}. When $\Lambda\in L_{+}^{\uparrow}$, $U(\Lambda,a)$ is linear and unitary. The representations of parity and time reversal are studied in sec.~\ref{PT_symmetries}.

In the case of the Poincar\'{e} group, there is no loss of generality in studying the unitary and anti-unitary representations, for it is possible to enlarge the group such that the phase can be removed without changing the physical contents of the theory~\cite[sec.~2.7]{Weinberg:1995mt}. 

We start the derivation of the Poincar\'{e} algebra by considering the infinitesimal transformation of an element of the proper orthochronous Lorentz group,
\begin{equation}
\Lambda^{\mu}_{\,\,\,\nu}=\delta^{\mu}_{\,\,\,\nu}+\omega^{\mu}_{\,\,\,\nu},\hspace{0.5cm}a^{\mu}=\epsilon^{\mu}
\end{equation}
where the parameter $\omega_{\mu\nu}$ must be antisymmetric  $\omega_{\mu\nu}=-\omega_{\nu\mu}$ for $\Lambda$ to satisfy eqs.~(\ref{eq:lg1}) and (\ref{eq:lg2}). 
The corresponding unitary representation to the first order of $\omega$ and $\epsilon$ can be written as
\begin{equation}
U(I+\omega,\epsilon)=1-\frac{i}{2}\omega_{\mu\nu}J^{\mu\nu}+i\epsilon^{\mu}P_{\mu} \label{eq:u_expansion}
\end{equation}
where $J^{\mu\nu}=(J^{\mu\nu})^{\dag}$ and $P^{\mu}=(P^{\mu})^{\dag}$ are Hermitian operators. Since $\omega_{\mu\nu}$ is antisymmetric, we may choose $J^{\mu\nu}=-J^{\nu\mu}$. 

Now we consider the following product $U(\Lambda,a)U(I+\omega,\epsilon)U^{-1}(\Lambda,a)$. Expanding it using eq.~(\ref{eq:u_expansion}), we get
\begin{eqnarray}
&&U(\Lambda,a)U(I+\omega,\epsilon)U^{-1}(\Lambda,a) \label{eq:u_product} \\ \nonumber
&&\hspace{1cm}=1-\frac{i}{2}\omega_{\mu\nu}U(\Lambda,a)J^{\mu\nu}U^{-1}(\Lambda,a)+i\epsilon^{\mu}U(\Lambda,a)P_{\mu}U^{-1}(\Lambda,a).
\end{eqnarray}
Applying the product rule, the left-hand side of eq.~(\ref{eq:u_product}) becomes
\begin{eqnarray}
&&U(1+\Lambda\omega\Lambda^{-1},\Lambda\epsilon-\Lambda\omega\Lambda^{-1}a) \label{eq:u_product1} \\ \nonumber
&&\hspace{1cm}=1-\frac{i}{2}\omega_{\rho\sigma}\Lambda_{\mu}^{\,\,\,\rho}\Lambda_{\nu}^{\,\,\,\sigma}J^{\mu\nu}+i\epsilon^{\mu}\Lambda^{\rho}_{\,\,\,\mu}P_{\rho}-i\omega_{\rho\sigma}\Lambda_{\mu}^{\,\,\,\rho}\Lambda_{\nu}^{\,\,\,\sigma}a^{\nu}P^{\mu}.
\end{eqnarray}
Equating the right-hand side of eq.~(\ref{eq:u_product}) and (\ref{eq:u_product1}) with respect to $\omega_{\mu\nu}$ and $\epsilon^{\rho}$ gives us
\begin{equation}
U(\Lambda,a)J^{\rho\sigma}U^{-1}(\Lambda,a)=\Lambda_{\mu}^{\,\,\,\rho}\Lambda_{\nu}^{\,\,\,\sigma}(J^{\mu\nu}-a^{\mu}P^{\nu}+a^{\nu}P^{\mu}) \label{eq:trans_j}
\end{equation}
and
\begin{equation}
U(\Lambda,a)P_{\rho}U^{-1}(\Lambda,a)=\Lambda^{\sigma}_{\,\,\,\rho}P_{\sigma} \label{eq:trans_p}.
\end{equation}
This shows that $J^{\mu\nu}$ is a second-rank tensor under the Lorentz transformation $(a^{\mu}=0)$ and $P^{\mu}$ transforms as a four-vector.

Now let $\Lambda^{\mu}_{\,\,\,\nu}=\delta^{\mu}_{\,\,\,\nu}+\omega^{\mu}_{\,\,\,\nu}$ and $a^{\mu}=\epsilon^{\mu}$ where $\omega$ and $\epsilon$ are unrelated to the previously used parameters. Substitute them into the right-hand side of eq.~(\ref{eq:trans_j}), we obtain
\begin{eqnarray}
&&(\delta_{\mu}^{\,\,\,\rho}+\omega_{\mu}^{\,\,\,\rho})(\delta_{\nu}^{\,\,\,\sigma}+\omega_{\nu}^{\,\,\,\sigma})(J^{\mu\nu}-\epsilon^{\mu}P^{\nu}+\epsilon^{\nu}P^{\mu}) \\ \nonumber
&&\hspace{1cm}=J^{\rho\sigma}-\epsilon^{\rho}P^{\sigma}+\epsilon^{\sigma}P^{\rho}+\omega_{\nu}^{\,\,\,\sigma}J^{\rho\nu}+\omega_{\mu}^{\,\,\,\rho}J^{\mu\sigma}+O(\epsilon^{2},\omega^{2},\epsilon\omega).
\end{eqnarray}
Similarly, the left-hand side of eq.~(\ref{eq:trans_j}) reads
\begin{equation}
U(\Lambda,a)J^{\rho\sigma}U^{-1}(\Lambda,a)=J^{\rho\sigma}+\frac{i}{2}\omega_{\mu\nu}[J^{\rho\sigma},J^{\mu\nu}]+i\epsilon^{\mu}[P_{\mu},J^{\rho\sigma}].
\end{equation}
Equating the right and left-hand side with respect to the parameters $\omega$ and $\epsilon$, we obtain the following commutators,
\begin{equation}	i[J^{\rho\sigma},J^{\mu\nu}]=\eta^{\rho\nu}J^{\mu\sigma}-\eta^{\rho\mu}J^{\nu\sigma}-\eta^{\sigma\mu}J^{\rho\nu}+\eta^{\sigma\nu}J^{\rho\mu}
\end{equation}
and
\begin{equation}
i[P^{\mu},J^{\rho\sigma}]=-\eta^{\rho\mu}P^{\sigma}+\eta^{\sigma\mu}P^{\rho}.
\end{equation}
Performing a similar calculation with eq.~(\ref{eq:trans_p}), we get
\begin{equation}
[P^{\mu},P^{\nu}]=0.
\end{equation}
This completes our derivation of the Poincar\'{e} algebra. In 3+1 dimensions, we may now identify these generators as rotation, boost and momentum operators in quantum mechanics. The rotation and boost operators are identified as
\begin{equation}
\mathbf{J}=(J^{23},J^{31},J^{12})=(J^{1},J^{2},J^{3}),\label{eq:j123}
\end{equation}
and
\begin{equation} \mathbf{K}=(J^{01},J^{02},J^{03})=(K^{1},K^{2},K^{3})\label{eq:k123}
\end{equation}
respectively. The energy and momentum operators are
\begin{equation}
H=P^{0},\hspace{0.5cm}\mathbf{P}=(P^{1},P^{2},P^{3}).
\end{equation}
The commutation relations in terms of these generators are\footnote{In order to avoid possible confusions between upper and lower index generators, we write the Poincar\'{e} algebra using $\mathbf{J}=(J^{1},J^{2},J^{3})$, $\mathbf{K}=(K^{1},K^{2},K^{3})$ and $P=(P^{0},P^{1},P^{2},P^{3})$ with the placement of their indices corresponding to the actual tensors $J^{\mu\nu}$ and $P^{\mu}$. In this way, we can raise and lower indices of $K^{i}$ and $J^{i}$ without confusion.}
\begin{equation}
[J^{i},J^{j}]=i\epsilon^{ijk}J^{k}, \nonumber
\end{equation}
\begin{equation}
[J^{i},K^{j}]=i\epsilon^{ijk}K^{k},  \nonumber
\end{equation}
\begin{equation}
[K^{i},K^{j}]=-i\epsilon^{ijk}J^{k}, \nonumber
\end{equation}
\begin{equation}
[P^{i},J^{j}]=i\epsilon^{ijk}P^{k}, \nonumber
\end{equation}
\begin{equation}
[P^{i},K^{j}]=i\delta^{ij}H, \nonumber
\end{equation}
\begin{equation}
[H,K^{i}]=i P^{i}, \nonumber
\end{equation}
\begin{equation}
[H,P^{i}]=[H,J^{i}]=0. \label{eq:poincare}
\end{equation}
where $\epsilon^{ijk}$ is the Levi-Civita tensor defined as $\epsilon^{123}=1$. We note that $P^{0}=H$ is the Hamiltonian, so all the generators that commute with $H$ are conserved quantities. Therefore, $\mathbf{P}$ and $\mathbf{J}$ are conserved, but $\mathbf{K}$ is not. This is the reason why we label states with $\mathbf{J}$ and $\mathbf{P}$ but not $\mathbf{K}$.

\section{The one-particle state}\label{one_particle_state}

The identification of the Casimir invariants with mass and spin allow us to uniquely define the elementary particles with Poincar\'{e} symmetry, but it does not contain information on how the particle states transform between inertial reference frames. To determine this, we need to study the irreducible representations of the Poincar\'{e} group.

We note, apart from the mass and spin attributed to the one-particle state, we need to introduce additional labels which correspond to the eigenvalues of specific generators of the Poincar\'{e} group. These generators must satisfy the conservation of energy, that is, they must commute with $H$. By requiring the particle states to have definite energy and momentum, we define them to be eigenstates of $P^{\mu}$. Ignoring the mass and spin label, we get
\begin{equation}
P^{\mu}|p,\sigma\rangle= p^{\mu}|p,\sigma\rangle. \label{eq:particle_state}
\end{equation}
The extra index $\sigma$ represents all other possible labels that may be discrete or continuous. Under the action of an unitary Lorentz transformation $U(\Lambda)$ where $\Lambda$ is an element of the proper orthochronous Lorentz group, the generator $P^{\mu}$ transforms as
\begin{equation}
U(\Lambda)P^{\mu}U^{-1}(\Lambda)=\Lambda_{\nu}^{\,\,\,\mu}P^{\nu}
\end{equation}
so its action on a boosted one-particle state $U(\Lambda)|p,\sigma\rangle$ is
\begin{eqnarray}
P^{\mu}U(\Lambda)|p,\sigma\rangle &=&U(\Lambda)[\,U^{-1}(\Lambda)P^{\mu}U(\Lambda)\,]|p,\sigma\rangle \\ \nonumber
&=&\Lambda^{\mu}_{\,\,\,\nu}p^{\nu}U(\Lambda)|p,\sigma\rangle.
\end{eqnarray}
Therefore, the boosted one-particle state is a linear combination of $|\Lambda p,\sigma\rangle$,
\begin{equation}
U(\Lambda)|p,\sigma\rangle=\sum_{\bar{\sigma}}C_{\bar{\sigma}\sigma}(\Lambda,p)|\Lambda p,\bar{\sigma}\rangle.
\end{equation}
Generally, it is possible to choose a suitable linear combination of $|p,\sigma\rangle$ such that $C_{\sigma\bar{\sigma}}(\Lambda,p)$ is block-diagonal for each  $\sigma$. The components of $C(\Lambda,p)$ for a given $\sigma$ then furnishes a representation of the Poincar\'{e} group. Since $|p,\sigma\rangle$ is the defined one-particle state, it is natural to identify the corresponding component $C_{\sigma\bar{\sigma}}(\Lambda,p)$ to be the irreducible representation of the Poincar\'{e} group. 

To obtain the irreducible representations of the Poincar\'{e} group, we note that the only function of $p^{\mu}$ that is left invariant by all the proper orthochronous transformations $\Lambda^{\mu}_{\,\,\,\nu}$ is $p^{\mu}p_{\mu}$. We may therefore classify the invariant $p^{\mu}p_{\mu}$ by its signs and values of $p^{\mu}p_{\mu}$ and $p^{0}$ and assign a standard four-momentum $k^{\mu}$ to each case. The arbitrary momentum $p^{\mu}$, for each case of $k^{\mu}$ is
\begin{equation}
p^{\mu}=L^{\mu}_{\,\,\,\nu}(p)k^{\nu}
\end{equation}
where $L(p)$ is the standard Lorentz boost. We now define a state $|k,\sigma\rangle$ of standard four-momentum $k^{\mu}$ such that under a standard Lorentz boost $U(L(p))$ yields $|p,\sigma\rangle$,
\begin{equation}
|p,\sigma\rangle= N(p)U(L(p))|k,\sigma\rangle  \label{eq:little_u}
\end{equation}
where $N(p)$ is the normalisation factor. Operating $U(\Lambda)$ on $|p,\sigma\rangle$ gives us
\begin{eqnarray}
U(\Lambda)|p,\sigma\rangle&=&N(p)U(\Lambda L(p))|k,\sigma\rangle \\ \nonumber
&=&N(p)U(L(\Lambda p))[U(L^{-1}(\Lambda p)\Lambda L(p))]|k,\sigma\rangle.
\end{eqnarray}
The term $L^{-1}(\Lambda p)\Lambda L(p)$ leaves the momentum $k^{\mu}$ invariant, it first takes $k$ to $L(p)k$ then to $\Lambda L(p)k$ and finally back to $k$. Transformations that leave $k^{\mu}$ invariant form a group called the little group~\cite{Wigner:1939cj}. We define the element of the little group as 
\begin{equation}
W(\Lambda,p)= L^{-1}(\Lambda p)\Lambda L(p)\label{eq:little_group}.
\end{equation}
Therefore, the action of $U(W)$ on state $|k,\sigma\rangle$ is a linear combination of states $|k,\bar{\sigma}\rangle$,
\begin{equation}
U(W)|k,\sigma\rangle=\sum_{\bar{\sigma}}D_{\bar{\sigma}\sigma}(W(\Lambda,p))|k,\bar{\sigma}\rangle \label{eq:l}
\end{equation}
where $D(W(\Lambda,p))$ is the finite-dimensional representation of the little group. Substituting eq.~(\ref{eq:l}) into eq.~(\ref{eq:little_u}) we get
\begin{equation}
U(\Lambda)|p,\sigma\rangle=\frac{N(p)}{N(\Lambda p)}\sum_{\bar{\sigma}}D_{\bar{\sigma}\sigma}(W(\Lambda,p))|\Lambda p,\bar{\sigma}\rangle. \label{eq:boost}
\end{equation}
The task is then to find the finite-dimensional unitary irreducible representation of the little group $D(W(\Lambda,p))$. Using the classification schemes of particle states with $p^{\mu}p_{\mu}$ and $p^{0}$ described above, we find, there are six distinct little groups.

\begin{table}[!hbt]
\centering
\begin{tabular}{llc}
\toprule
											& Standard $k^{\mu}$ 			& Little group \\
\midrule
 (a) $p^{\mu}p_{\mu}=m^{2},p^{0}>0$ 		& $(m,0,0,0)$ 					& $SU(2$) \\
 (b) $p^{\mu}p_{\mu}=-m^{2},p^{0}<0$ 		& $(-m,0,0,0)$ 					& $SU(2)$ \\
 (c) $p^{\mu}p_{\mu}=0,p^{0}>0$ 			& $(\kappa,0,0,\kappa)$ 		& $ISO(2)$ \\
 (d) $p^{\mu}p_{\mu}=0,p^{0}<0$ 			& $(-\kappa,0,0,\kappa)$ 		& $ISO(2)$\\
 (e) $p^{\mu}p_{\mu}=-n^{2}<0$ 				& $(0,0,n,0)$					& $SO(2,1)$ \\
 (f) $p^{\mu}=0$ 							& $(0,0,0,0)$ 					& $SO(3,1)$ \\
\bottomrule
\end{tabular}
\caption{The little groups and their standard momentum.}\label{little_group_table}
\end{table}

Out of the six groups, only (a), (c) and (f) have known physical interpretations. Case (f) with $p^{\mu}=0$ is the vacuum, which is not of interest to us, we will focus on (a) and (c) which correspond to positive-mass and massless particle state respectively.

It is appropriate at this point to determine the normalisation factor $N(p)$. We normalise the one-particle state $|k,\sigma\rangle$ at rest to be 
\begin{equation}
\langle k',\sigma'|k,\sigma\rangle=\delta_{\sigma\sigma'}\delta^{3}(\mathbf{k}'-\mathbf{k}). \label{eq:norm}
\end{equation}
From eq.~(\ref{eq:l}), we see that the choice of the normalisation makes the representation of the little group $D(W(\Lambda,p))$ to be unitary
\begin{equation}
D^{\dag}(W(\Lambda,p))=D^{-1}(W(\Lambda,p)).
\end{equation}
As for the inner-product of arbitrary momenta $\langle p',\sigma'|p,\sigma\rangle$, using eq.~(\ref{eq:little_u}) we get
\begin{equation}
\langle p',\sigma'|p,\sigma\rangle=N(p)\langle p',\sigma'|U(L(p))|k,\sigma\rangle. \label{eq:norm1}
\end{equation}
Using eq.~(\ref{eq:boost}), we find
\begin{equation}
U^{-1}(L(p))|p',\sigma'\rangle=\frac{N(p')}{N(L^{-1}(p)p)}\sum_{\sigma''}D_{\sigma''\sigma'}(W(L^{-1}(p),p'))|L^{-1}(p)p',\sigma''\rangle
\end{equation}
where $k'= L^{-1}(p)p'$, so that $N(L^{-1}(p)p')=N(k')=1$  by eq.~(\ref{eq:norm}). Hence,
\begin{equation}
U^{-1}(L(p))|p',\sigma'\rangle=N(p')\sum_{\sigma''}D_{\sigma''\sigma'}(W(L^{-1}(p),p'))|k',\sigma''\rangle.\label{eq:u_inv_p}
\end{equation}
Substituting eq.~(\ref{eq:u_inv_p}) into eq.~(\ref{eq:norm1}), the inner-product becomes
\begin{eqnarray}
\langle p',\sigma'|p,\sigma\rangle&=&N^{*}(p')N(p)\sum_{\sigma''}\langle k',\sigma''|D^{*}_{\sigma''\sigma'}(W(L^{-1}(p),p'))|k,\sigma\rangle\\ \nonumber
&=&N^{*}(p')N(p)D^{*}_{\sigma\sigma'}(W(L^{-1}(p)),p')\delta^{3}(\mathbf{k}'-\mathbf{k}).
\end{eqnarray}
But since $k'= L^{-1}(p)p'$ and $k= L^{-1}(p)p$, the delta function $\delta^{3}(\mathbf{k}'-\mathbf{k})$ must be proportional to $\delta^{3}(\mathbf{p}'-\mathbf{p})$. Furthermore, the inner-product vanishes if $k'\neq k$, it follows that
\begin{equation}
\langle p',\sigma'|p,\sigma\rangle=|N(p)|^{2}\,\delta^{3}(\mathbf{k}'-\mathbf{k})\delta_{\sigma\sigma'}.
\end{equation}

To express the inner-product in terms of $\delta^{3}(\mathbf{p}-\mathbf{p}')$, we consider a Lorentz invariant integral for an arbitrary scalar function $f(p)$ over $p^{\mu}$ on the mass-shell where $p^{\mu}p_{\mu}=m^{2}\geq0$ and $p^{0}>0$,
\begin{eqnarray}
\hspace{-0.5cm}\int d^{4}p\,\delta(p^{\mu}p_{\mu}-m^{2})\theta(p^{0}) f(p)
&=&\int d^{3}p\, dp^{0} \delta[(p^{0})^{2}-|\mathbf{p}|^{2}-m^{2}]\theta(p^{0})f(p^{0},\p) \\ \nonumber
&=&\int d^{3}p \frac{f(\mathbf{p},\sqrt{|\p|^{2}+m^{2}})}{2\sqrt{|\p|^{2}+m^{2}}}.
\end{eqnarray}
But since $f(p)$ is a scalar $f(p)=f(\Lambda^{-1}p)$ and the integral is Lorentz invariant, it follows that when integrating on mass shell, the volume element $d^{3}p/\sqrt{|\p|^{2}+m^{2}}$ must be Lorentz-invariant. Similarly, by the definition of the delta-function any function $g(\mathbf{p})$ may be written as
\begin{eqnarray}
g(\mathbf{p})&=&\int d^{3}p'\, \delta^{3}(\mathbf{p}-\mathbf{p}')g(\mathbf{p}') \\ \nonumber
&=&\int\frac{d^{3}p'}{\sqrt{|\p|'^{2}+m^{2}}}g(\mathbf{p}')\Big[\sqrt{|\p|'^{2}+m^{2}}\delta^{3}(\mathbf{p}-\mathbf{p}')\Big].
\end{eqnarray}
Since $d^{3}p/\sqrt{|\p|^{2}+m^{2}}$ is Lorentz invariant, it follows that $\sqrt{|\p|^{2}+m^{2}}\delta^{3}(\mathbf{p}-\mathbf{p}')$ must also be Lorentz invariant. Since $p$ and $p'$ related to $k$ and $k'$ by the Lorentz transformation $L(p)$, we get
\begin{equation}
p^{0}\,\delta^{3}(\mathbf{p}'-\mathbf{p})=k^{0}\,\delta^{3}(\mathbf{k}'-\mathbf{k}).
\end{equation}
Therefore, the scalar product maybe written as
\begin{equation}
\langle p',\sigma'|p,\sigma\rangle=|N(p)|^{2}\left(\frac{p^{0}}{k^{0}}\right)\delta_{\sigma'\sigma}\delta^{3}(\mathbf{p}'-\mathbf{p}).
\end{equation}
We choose our normalisation factor $N(p)$ to be 
\begin{equation}
N(p)=\sqrt{\frac{k^{0}}{p^{0}}} \label{eq:normalisation}
\end{equation}
so that the one-particle states of arbitrary momenta are orthonormal 
\begin{equation}
\langle p',\sigma'|p,\sigma\rangle=\delta_{\sigma'\sigma}\delta^{3}(\mathbf{p}'-\mathbf{p}).
\end{equation}
Therefore, the Lorentz transformation on the one-particle state is given by
\begin{equation}
U(\Lambda)|p,\sigma\rangle=\sqrt{\frac{(\Lambda p)^{0}}{p^{0}}}\sum_{\bar{\sigma}}D_{\bar{\sigma}\sigma}^{(j)}(W(\Lambda,p))|\Lambda p,\bar{\sigma}\rangle \label{eq:Lorentz_particle_tran}
\end{equation}
We now have a well-defined scalar product of arbitrary momenta for the one-particle states of mass $m\geq0$ and energy $p^{0}>0$. In the following two sections we study the structure of the little groups for the massive and massless particle states.

\subsection{Massive particle state ($m>0,$ $p^{0}>0$)}

The little group for the one-particle states of positive-definite mass $m>0$ is the special unitary group $SU(2)$ a universal double cover of the rotation group $SO(3)$. The representation of $SU(2)$ can be decomposed into direct sums of its irreducible representations. The theory of group representation tells us, for a given $j=0,\frac{1}{2},1,\frac{3}{2},\cdots$ of half-integral and integral value, there exists an irreducible representation of dimension $(2j+1)$.\footnote{The rotation group $SO(3)$ only contains integral-representations.} They are generated by infinitesimal rotations, $R^{ij}=\delta^{ij}+i\Theta^{ij}$ where $\Theta^{ij}=-\Theta^{ji}$,\footnote{There is no summation on repeated indices.}
\begin{equation}
D_{\sigma\bar{\sigma}}^{(j)}(I+\Theta)=\delta_{\sigma\bar{\sigma}}+\frac{i}{2}\Theta^{ik}(J^{ik(j)})_{\sigma\bar{\sigma}},\label{eq:r1}
\end{equation}
\begin{eqnarray}
(J^{23(j)}\pm iJ^{31(j)})_{\sigma\bar{\sigma}}&=&(J^{1(j)}\pm iJ^{2(j)})_{\sigma\bar{\sigma}} \label{eq:r2} \\ \nonumber
&=&\delta_{\sigma,\bar{\sigma}\pm1}\sqrt{(j\mp\bar{\sigma})(j\pm\bar{\sigma}+1)},
\end{eqnarray}
\begin{equation}
(J^{12(j)})_{\sigma\bar{\sigma}}=(J^{3(j)})_{\sigma\bar{\sigma}}=\sigma\delta_{\sigma\bar{\sigma}}\label{eq:r3}
\end{equation}
where $\sigma=-j\cdots j$. In quantum mechanics, $\mathbf{J}=(J^{1},J^{2},J^{3})$ are simply the angular momentum matrices for a particle of spin-$j$. 

The particle state $|p,\sigma\rangle$ with mass $m>0$, under an arbitrary Lorentz transformation $U(\Lambda)$ is given by
\begin{equation}
U(\Lambda)|p,\sigma\rangle=\sqrt{\frac{(\Lambda p)^{0}}{p^{0}}}\sum_{\bar{\sigma}}D_{\bar{\sigma}\sigma}^{(j)}(W(\Lambda,p))|\Lambda p,\bar{\sigma}\rangle \label{eq:massive_tran}
\end{equation}
where $W(\Lambda,p)= L^{-1}(\Lambda p)\Lambda L(p)$ is the element of the little group. To completely determine the transformation of the particle state, we need to determine the little group when $\Lambda$ is a Lorentz boost and a rotation. 
Traditionally, the general Lorentz boost is given by
\begin{equation}
L^{i}_{\,\,\,k}=\eta^{i}_{\,\,\,k}-(\gamma-1)\hat{p}^{i}\hat{p}_{k}, \nonumber
\end{equation}
\begin{equation}
L^{i}_{\,\,\,0}=\hat{p}^{i}\sqrt{\gamma^{2}-1}, \nonumber
\end{equation}
\begin{equation}
L^{0}_{\,\,\,i}=-\hat{p}_{i}\sqrt{\gamma^{2}-1}, \nonumber
\end{equation}
\begin{equation}
L^{0}_{\,\,\,0}=\gamma \label{eq:lorentz_boost}
\end{equation}
where $\hat{p}^{\,i}= p^{i}/|\mathbf{p}|$ and $\gamma=\sqrt{|\mathbf{p}|^{2}+m^{2}}/m$. 
However, it is computationally intensive to explicitly evaluate the element of the little group for general boost and rotation. Fortunately, since the Lorentz group is a Lie group, the structure of the little group can be completely determined by considering the infinitesimal element about the identity. For this purpose, it is more convenient to express the boost and rotation in terms of the Lorentz generators.
\begin{equation}
L(p)=\exp(i\pmb{\mathscr{K}}\cdot\bv), 
\end{equation}
\begin{equation}
R(\theta)=\exp(i\pmb{\mathscr{J}}\cdot\th).
\end{equation}
where $\bv=\hat{\p}\varphi$ is the rapidity parameter with $\cosh\varphi=p^{0}/m$, $\sinh\varphi=|\p|/m$ and $\th=\theta\hat{\mathbf{n}}$ where $\hat{\mathbf{n}}$ is the unit vector denoting the axis of rotation. The rotation and boost generators in the four-vector representation are given by
\begin{equation}
\mathscr{J}^{1}=\frac{1}{2}
\left(\begin{array}{cccc}
0 & 0 & 0 & 0 \\
0 & 0 & 0 & 0 \\
0 & 0 & 0 &-i \\
0 & 0 & i & 0 \end{array}\right),
\mathscr{J}^{2}=\frac{1}{2}
\left(\begin{array}{cccc}
0 & 0 & 0 & 0 \\
0 & 0 & 0 & i \\
0 & 0 & 0 & 0 \\
0 &-i & 0 & 0 \end{array}\right),
\mathscr{J}^{3}=
\left(\begin{array}{cccc}
0 & 0 & 0 & 0 \\
0 & 0 &-i & 0 \\
0 & i & 0 & 0 \\
0 & 0 & 0 & 0 \end{array}\right),\label{eq:vector_j}
\end{equation}
\begin{equation}
\mathscr{K}^{1}=\frac{1}{2}
\left(\begin{array}{cccc}
 0 &-i & 0 & 0 \\
-i & 0 & 0 & 0 \\
 0 & 0 & 0 & 0 \\
 0 & 0 & 0 & 0 \end{array}\right),
\mathscr{K}^{2}=\frac{1}{2}
\left(\begin{array}{cccc}
 0 & 0 &-i & 0 \\
 0 & 0 & 0 & 0 \\
-i & 0 & 0 & 0 \\
 0 & 0 & 0 & 0 \end{array}\right),
\mathscr{K}^{3}=
\left(\begin{array}{cccc}
 0 & 0 & 0 &-i \\
 0 & 0 & 0 & 0 \\
 0 & 0 & 0 & 0 \\
-i & 0 & 0 & 0 \end{array}\right).\label{eq:vector_k}
\end{equation}
To the leading order of the parameters, the Lorentz transformations are linear, so it has the same structure as the Galilean transformations. For this reason, it is important for us to determine to what order must we expand $W(\Lambda,p)$ for it to exhibit relativistic effects. We note, for rotation $\Lambda=R(\theta)$, it is sufficient to consider the terms of the leading order since the commutator between boost and rotation generators are the same in the Galilean and Lorentz algebra. For boost $\Lambda=L(q)$, it is necessary to expand the terms to second order since the boost generators are non-commutating in the Lorentz algebra as opposed to the Galilean counterpart.

When $\Lambda=R(\theta)$ is an arbitrary rotation, we have 
\begin{equation}
W(R,p)=L^{-1}(Rp)R(\theta)L(p).
\end{equation}
The term $L(Rp)$ to the leading order is given by
\begin{eqnarray}
L(Rp)&=&I+i\pmb{\mathscr{K}}\cdot(R\hat{\p})\varphi\\
&=&I+i\pmb{\mathscr{K}}\cdot\hat{\p}\,\varphi+O(\theta\varphi).
\end{eqnarray}
Under rotation the rapidity parameter is unchanged since $(Rp)^{0}=p^{0}$ and $|R\p|=|\p|$. The little group element $W(R(\theta),p)$ is then
\begin{eqnarray}
W(R(\theta),p)&=&\left[I-i\pmb{\mathscr{K}}\cdot\hat{\p}\varphi\right][I+i\pmb{\mathscr{J}}\cdot\th][I+i\pmb{\mathscr{K}}\cdot\hat{\p}\varphi]\nonumber\\
&=&I+i\pmb{\mathscr{J}}\cdot\th+O(\varphi^{2},\theta\varphi).
\end{eqnarray}
Therefore, for an arbitrary rotation $\Lambda=R(\theta)$, the little group element remains a rotation
\begin{equation}
W(R(\theta),p)=R(\theta)
\end{equation}
This shows that a massive particle in relativistic quantum mechanics has the same transformation properties under rotation as massive particles in non-relativistic quantum mechanics. Therefore, the formalism developed for studying angular momentum in non-relativistic quantum mechanics may also be applied to relativistic particles.

Now we take $\Lambda=L(q)$ with $q\neq p$. The little group element is then 
\begin{equation}
W(L(q),p)=L^{-1}(L(q)p)L(q)L(p).
\end{equation}
To the second order, the product $L(q)L(p)$ is
\begin{eqnarray}
L(q)L(p)&=&\left[I+\pmb{\mathscr{K}}\cdot(\bv_{q}+\bv_{q})-\frac{1}{2}(\pmb{\mathscr{K}}\cdot\bv_{q})^{2}\right]
\left[I+\pmb{\mathscr{K}}\cdot(\bv_{p}+\bv_{p})-\frac{1}{2}(\pmb{\mathscr{K}}\cdot\bv_{p})^{2}\right]\nonumber\\
&=&I+i\pmb{\mathscr{K}}\cdot(\bv_{p}+\bv_{q})-\frac{1}{2}\left[\pmb{\mathscr{K}}\cdot(\bv_{p}+\bv_{q})\right]^{2}-\frac{1}{2}\left[\pmb{\mathscr{K}}\cdot\bv_{p},\pmb{\mathscr{K}}\cdot\bv_{q}\right].
\end{eqnarray}
Using the commutator $[\mathscr{K}^{i},\mathscr{K}^{j}]=-i\epsilon^{ijk}\mathscr{J}^{k}$, we get
\begin{equation}
L(q)L(p)=I+i\pmb{\mathscr{K}}\cdot(\bv_{p}+\bv_{q})-\frac{1}{2}\left[\pmb{\mathscr{K}}\cdot(\bv_{p}+\bv_{q})\right]^{2}+
\frac{1}{2}i\epsilon^{ijk}\mathscr{J}^{k}\varphi^{i}_{p}\varphi^{j}_{q}.
\end{equation}
Now we consider $L(L(q)p)$ so that
\begin{equation}
L(L(q)p)=I+i\pmb{\mathscr{K}}\cdot\hat{\p}'\varphi'-\frac{1}{2}(\pmb{\mathscr{K}}\cdot\hat{\p}'\varphi')^{2}
\end{equation}
where $p'=L(q)L(p)k$, $\cosh\varphi'=(L(q)p)^{0}/m$ and $\sinh\varphi'=|\mathbf{L(q)p}|/m$. To the second order, the explicit expression for $p'$ is
\begin{equation}
p'=m\left(\begin{matrix}
1+\frac{1}{2}|\bv_{p}+\bv_{q}|^{2} \\
\bv_{p}+\bv_{q}\end{matrix}\right).
\end{equation}
Therefore
\begin{equation}
\cosh\varphi'=1+\frac{1}{2}|\bv_{p}+\bv_{q}|^{2},\hspace{0.5cm}
\sinh\varphi'=|\bv_{p}+\bv_{q}|.
\end{equation}
Solving $\varphi'$ to the second order, we get
\begin{eqnarray}
\varphi'&=&\ln\left[1+|\bv_{p}+\bv_{q}|+\frac{1}{2}|\bv_{p}+\bv_{q}|^{2}\right]\\
&=&|\bv_{p}+\bv_{q}|+\mbox{higher-order terms}.
\end{eqnarray}
Here, the second order terms cancel thus leaving us with only the leading order term. Substituting the solutions of $p'$ and $\varphi'$ into $L(L(q)p)$ give us
\begin{equation}
L(L(q)p)=I+i\pmb{\mathscr{K}}\cdot(\bv_{p}+\bv_{q})-\frac{1}{2}\left[\pmb{\mathscr{K}}\cdot(\bv_{p}+\bv_{q})\right]^{2}.
\end{equation}
Finally, we substitute $L(L(q)p)$ and $L(q)L(p)$ into $W(L(q),p)$ to obtain
\begin{equation}
W(L(q),p)=I+\frac{1}{2}i\epsilon^{ijk}\mathscr{J}^{k}\varphi^{i}_{p}\varphi^{j}_{q}+\mbox{higher-order terms}
\end{equation}
where we have kept the second-order contribution. Therefore, under arbitrary boost, the element of the little group remains a rotation
\begin{equation}
W(L(q),p)=\exp\left(\frac{1}{2}i\epsilon^{ijk}\mathscr{J}^{k}\varphi^{i}_{p}\varphi^{j}_{q}\right).
\label{eq:Lorentz_little_group_B}
\end{equation}
But this time, the angle of rotation is $\theta=\varphi_{p}\varphi_{q}\sin\theta_{pq}$ where $\theta_{pq}$ is the angle between $\hat{\p}$ and $\hat{\mathbf{q}}$ and the axis of rotation is
\begin{equation}
\hat{n}^{k}=\frac{\epsilon^{ijk}\hat{p}^{i}\hat{q}^{j}}{\sin\theta_{pq}}
\end{equation}
so that $\hat{\mathbf{n}}$ remains a unit vector $|\hat{\mathbf{n}}|=1$.

\subsection{Massless particle state ($m=0$, $p^{0}>0$)}\label{Massless_particle_state}

We will start our study of the massless particle state by determining the structure of its little group. First, consider an arbitrary element of the little group $W^{\mu}_{\,\,\,\nu}$. By definition $W^{\mu}_{\,\,\,\nu}k^{\nu}=k^{\mu}$ where $k=(\kappa,0,0,\kappa)$ is the standard null four-momentum vector for a massless particle. Since $W$ as defined in eq.~(\ref{eq:little_group}) is a product of Lorentz transformation, it is also a Lorentz transformation. Therefore, given a time-like vector $t=(1,0,0,0)$, the action of $W$ on $t$ must preserve its norm as well as its scalar product with $k$,
\begin{equation}
(Wt)^{\mu}(Wt)_{\mu}=t^{\mu}t_{\mu}, \nonumber
\end{equation}
\begin{equation}
(Wt)^{\mu}k_{\mu}=t^{\mu}k_{\mu}. \label{eq:wtk}
\end{equation}
These two conditions can be satisfied with any four vector of the form
\begin{equation}
(Wt)^{\mu}=\left(\begin{array}{cccc}
1+\varrho \\
\alpha \\
\beta \\
\varrho \end{array}\right)
\end{equation}
where $\varrho=(\alpha^{2}+\beta^{2})/2$. 

We can find a $4\times4$ matrix $S(\alpha,\beta,x_{i})$ that satisfies eq.~(\ref{eq:wtk}),
\begin{equation}
S(\alpha,\beta,x_{i})=\left(\begin{array}{cccc}
1+\varrho & x_{1} & x_{2} & -\varrho \\
\alpha & x_{3} & x_{4} & -\alpha \\
\beta & x_{5} & x_{6} & -\beta \\
\varrho & x_{7} & x_{8} & 1-\varrho \end{array}\right)
\end{equation}
where $x_{i}$ with $i=1,2,\cdots,8$ are variables to be determined. They can be determined by demanding that $S(\alpha,\beta,x_{i})$ forms a group. To see this, we know that both $W$ and $S$ leave $k^{\mu}$ invariant and satisfy eq.~(\ref{eq:wtk}), but this does not mean $S=W$. Instead, it implies that $k^{\mu}$ is invariant under $S^{-1}W$, so it must be a pure rotation. In this case since $k=(\kappa,0,0,\kappa)$, then $S^{-1}W$ must be a rotation of angle $\phi$ about the $3$-axis,
\begin{equation}
S^{-1}W=R(\phi).
\end{equation}
Therefore, a general element of the little group takes the form,
\begin{equation}
W(\alpha,\beta,x_{i},\phi)=S(\alpha,\beta,x_{i})R(\phi)
\end{equation}
where
\begin{equation}
R(\phi)=\left(\begin{array}{cccc}
1 & 0 & 0 & 0 \\
0 & \cos\phi & -\sin\phi & 0 \\
0 & \sin\phi & \cos\phi & 0 \\
0 & 0 & 0 & 1
\end{array}\right). \label{eq:rphi}
\end{equation}
Since $W(\alpha,\beta,x_{i},\phi)$ forms the little group with rotation as its subgroup, then $S(\alpha,\beta,x_{i})$ must contain the identity and inverse. Therefore, we only have to show that $S(\alpha,\beta,x_{i})$ is closed under multiplication. This can be achieved by considering the product $S'W=S'SR$ , where $S'=S(\alpha',\beta',x_{i}')$. We see that most general solution in which $S'W$ remains an element of the group is when $S$ is closed under multiplication. Imposing the closure condition under multiplication, we find the following solutions
\begin{equation}
x_{1}=x_{7}=\alpha,\hspace{0.5cm}x_{2}=x_{8}=\beta, \nonumber
\end{equation}
\begin{equation}
x_{3}=x_{6}=1,\hspace{0.5cm} x_{4}=x_{5}=0 \nonumber
\end{equation}
thus giving us
\begin{equation}
S(\alpha,\beta)=\left(\begin{array}{cccc}
1+\varrho & \alpha & \beta & -\varrho \\
\alpha & 1 & 0 & -\alpha \\
\beta & 0 & 1 & -\beta \\
\varrho & \alpha & \beta & 1-\varrho \end{array}\right).\label{eq:sab}
\end{equation}
Explicit computation shows that $S(\alpha,\beta)$ is an Abelian group
\begin{equation}
S(\alpha+\bar{\alpha},\beta+\bar{\beta})=S(\alpha,\beta)S(\bar{\alpha},\bar{\beta}). 
\end{equation}

The structure of the little group for massless particle states of positive energy can now be summarised,
\begin{equation}
W(\alpha,\beta,\phi)=S(\alpha,\beta)R(\phi), \nonumber
\end{equation}
\begin{equation}
R(\phi)R(\bar{\phi})=R(\phi+\bar{\phi}), \nonumber
\end{equation}
\begin{equation}
S(\alpha+\bar{\alpha},\beta+\bar{\beta})=S(\alpha,\beta)S(\bar{\alpha},\bar{\beta}).
\end{equation}
These multiplication rules are the rules of group $ISO(2)$ consisting of translations (by vector $(\alpha,\beta)$) and rotation (by angle $\phi)$ in two dimensions.

Expanding $W(\alpha,\beta,\phi)$ about the identity to the first order of $\alpha$, $\beta$ and $\phi$ gives us
\begin{equation}
W^{\mu}_{\,\,\,\nu}(\alpha,\beta,\phi)=\delta^{\mu}_{\,\,\,\nu}+\omega^{\mu}_{\,\,\,\nu}
\end{equation}
where $\omega^{\mu}_{\,\,\,\nu}(\alpha,\beta)$ is defined as
\begin{equation}
\omega(\alpha,\beta)=\left(\begin{array}{cccc}
0 & \alpha & \beta & 0 \\
\alpha & 0 & \phi & -\alpha \\
\beta & -\phi & 0 & -\beta \\
0 & \alpha & \beta & 0 \end{array}\right).
\end{equation}
The infinitesimal expansion of the unitary representation for $W(\alpha,\beta,\phi)$ is
\begin{equation}
U(W(\alpha,\beta,\phi))=1-\frac{1}{2}i\omega_{\mu\nu}J^{\mu\nu} 
\end{equation}
where $\omega_{\mu\nu}=\eta_{\mu\lambda}\,\omega^{\lambda}_{\,\,\,\nu}$ and $J^{\mu\nu}$ is the generator of rotation and boost satisfying the Lorentz algebra. In terms of $\alpha$, $\beta$ and $\phi$, $U(W(\alpha,\beta,\phi))$ reads
\begin{equation} 
U(W(\alpha,\beta,\phi))=1+i\alpha A+i\beta B+i\phi J^{3} \label{eq:u_w}
\end{equation}
where $A$ and $B$ are defined as
\begin{equation}
A= K^{1}+J^{2},\hspace{0.5cm} B= K^{2}-J^{1}.
\end{equation}
Using the Poincar\'{e} algebra given in eq.~(\ref{eq:poincare}), we obtain
\begin{equation}
 [J^{3},A]=iB, \nonumber
\end{equation}
\begin{equation}
[J^{3},B]=-iA, \nonumber
\end{equation}
\begin{equation}
[A,B]=0. \label{eq:ab_j}
\end{equation}
Since $A$ and $B$ commute, they can be simultaneously diagonalised by state $|k,a,b\rangle$
\begin{equation}
A\,|k,a,b\rangle=a\,|k,a,b\rangle, \nonumber
\end{equation}
\begin{equation}
B\,|k,a,b\rangle=b\,|k,a,b\rangle.
\end{equation}

We now determine the spectrum of $A$ and $B$. First, we consider the identity
\begin{equation}
R(\phi)S(\alpha,\beta)R^{-1}(\phi)=S(\alpha\cos\phi+\beta\sin\phi,-\alpha\sin\phi+\beta\cos\phi).
\end{equation}
Its corresponding unitary representation in the Hilbert space is
\begin{equation}
U(R(\phi))U(S(\alpha,\beta))U^{-1}(R(\phi))=U(S(\alpha\cos\phi+\beta\sin\phi,-\alpha\sin\phi+\beta\cos\phi)) \label{eq:r_s}.
\end{equation}
Equation (\ref{eq:u_w}) with $\phi=0$ shows that $U(S(\alpha,\beta))=1+i\alpha A+i\beta B$. Substituting this into the left hand side of eq.~(\ref{eq:r_s}) and equating the coefficients with respect to $\alpha$ and $\beta$ gives us
\begin{equation}
U(R(\phi))AU^{-1}(R(\phi))=\cos\phi\, A-\sin\phi\, B, \nonumber
\end{equation}
\begin{equation}
U(R(\phi))BU^{-1}(R(\phi))=\sin\phi\, A+\cos\phi\, B. \label{eq:r_ab}
\end{equation}
Acting eq.~(\ref{eq:r_ab}) on $|k,a,b\rangle$ yields
\begin{equation}
A|k,a,b,\phi\rangle=(a\cos\phi-b\sin\phi)|k,a,b,\phi\rangle, \nonumber
\end{equation}
\begin{equation}
B|k,a,b,\phi\rangle=(a\sin\phi+b\cos\phi)|k,a,b,\phi\rangle \label{eq:ab_continuous_spectrum}
\end{equation}
where $|k,a,b,\phi\rangle= U^{-1}(R(\phi))|k,a,b\rangle$. Equation (\ref{eq:ab_continuous_spectrum}) shows that the spectrum of $A$ and $B$ to be continuous. However, experiments indicate massless particles do not carry any continuous degrees of freedom like $\phi$. Therefore, we must require our physical state $|k,a,b\rangle$ to be annihilated by $A$ and $B$,
\begin{equation}
A|k,\sigma\rangle=B|k,\sigma\rangle=0 \label{eq:a_b}
\end{equation}
where $\sigma$ represents other degrees of freedom. By eq.~(\ref{eq:ab_j}), we now have
\begin{equation}
[J^{3},A]|k,\sigma\rangle=[J^{3},B]|k,\sigma\rangle=0.
\end{equation}
Therefore, we may distinguish the states by their eigenvalues associated with $\mathbf{J}\cdot\hat{\mathbf{k}}$ where $\mathbf{k}=(0,0,\kappa)$.
This is possible because $\mathbf{J}$ is a conserved quantity, $[\mathbf{J},H]=0$. Since $\mathbf{k}$ is aligned to the $3$-axis, we have $\mathbf{J}\cdot\hat{\mathbf{k}}=J^{3}$ and the state is defined as
\begin{equation}
\mathbf{J}\cdot\hat{\mathbf{k}}|k,\sigma\rangle=\sigma|k,\sigma\rangle.
\end{equation}
The eigenvalue $\sigma$ is called the helicity and is always aligned to the direction of motion. In this case, it is aligned to the $3$-axis. Here we write $\mathbf{J}\cdot\hat{\mathbf{k}}$ in the expression to remind us that the helicity $\sigma$ has direction dependence.

We can now use the properties of the little group to determine the Lorentz transformations for the massless particle states. For finite values of $\alpha$ and $\beta$, $U(S(\alpha,\beta))$ reads
\begin{equation}
U(S(\alpha,\beta))=\exp[i(\alpha A+\beta B)]
\end{equation}
and similarly for finite $\phi$,
\begin{equation}
U(R(\phi))=\exp(iJ^{3}\phi).
\end{equation}
Hence, a general element of the little group is
\begin{equation}
U(W(\alpha,\beta,\phi))=U(S(\alpha,\beta))U(R(\phi))=\exp[i(\alpha A+\beta B)]\exp(i J^{3}\phi).
\end{equation}
Acting $U(W(\alpha,\beta,\phi))$ on $|k,\sigma\rangle$ yields,
\begin{equation}
U(W)|k,\sigma\rangle=\exp(i\sigma\phi)|k,\sigma\rangle
\end{equation}
where we have used eq.~(\ref{eq:a_b}). Therefore, by eq.~(\ref{eq:l}), the unitary representation of the little group for massless particle state is
\begin{equation}
D_{\sigma'\sigma}(W(\Lambda,p))=\exp[i\sigma\phi(\Lambda,p)]\delta_{\sigma'\sigma}.\label{eq:DW_massless}
\end{equation}
The Lorentz transformation of a massless particle state of arbitrary momentum is
\begin{equation}
U(\Lambda)|p,\sigma\rangle=\sqrt{\frac{(\Lambda p)^{0}}{p^{0}}}\exp(i\sigma\phi(\Lambda,p))|\Lambda p,\sigma\rangle.\label{eq:massless_trans}
\end{equation}
Unlike the Lorentz transformation for massive particle states, the helicity of the massless particle is a Lorentz invariant.

To completely determine the properties of the massless particle, we still need to construct an appropriate boost $L(p)$ that takes the standard vector $k=(\kappa,0,0,\kappa)$ to arbitrary momentum $p$. 
Here, we adopt the definition of~\cite{Weinberg:1964ev}, constructing $L(p)$ as a product of boost and rotation
\begin{equation}
L(p)=R(\hat{\mathbf{p}})B(|\mathbf{p}|)\label{eq:massless_boost}
\end{equation}
where $B(|\mathbf{p}|)$ is a boost along the $3$-axis taking $k=(\kappa,0,0,\kappa)$ to $Lk=(|\mathbf{p}|,0,0,|\mathbf{p}|)$,
\begin{equation}
B(|\mathbf{p}|)=\left(\begin{array}{cccc}
\cosh\varphi & 0 & 0 & \sinh\varphi \\
0 & 1 & 0 & 0 \\
0 & 0 & 1 & 0 \\
\sinh\varphi & 0 & 0 & \cosh\varphi\end{array}\right).
\end{equation}
with the rapidity parameter $\varphi$ defined as
\begin{equation}
\varphi(|\mathbf{p}|)= \ln(|\mathbf{p}|/\kappa).
\end{equation}
Subsequently, $R(\hat{\mathbf{p}})$  rotates $Lk$ to the direction of $\hat{\mathbf{p}}$. Taking the direction of motion to be $\hat{\mathbf{p}}= (\sin\theta\cos\phi,\sin\theta\sin\phi,\cos\theta)$, then $R(\hat{\mathbf{p}})$ can be constructed as two successive rotations by angle $-\theta$ and $-\phi$ about the $3$ and $2$-axis respectively,
\begin{eqnarray}
R(\hat{\mathbf{p}})&=& R_{3}(-\phi)R_{2}(-\theta) \label{eq:rot_y_z} \\
&=&
\left(\begin{array}{cccc}
1 & 0 & 0 & 0 \\
0 & \cos\phi & -\sin\phi & 0 \\
0 & \sin\phi & \cos\phi & 0 \\
0 & 0 & 0 & 1
\end{array}\right)
\left(\begin{array}{cccc}
1 & 0 & 0 & 0 \\
0 & \cos\theta & 0 &\sin\theta \\
0 & 0 & 1 & 0 \\
0 & -\sin\theta & 0 & \cos\theta \end{array}\right).  \nonumber
\end{eqnarray}
It is important to note that while $R(\hat{\mathbf{p}})$ remains unchanged when $\theta$ and $\phi$ are shifted by $2\pi$. Generally, this is not true for its corresponding operator in the Hilbert space,
\begin{equation}
U(R(\hat{\mathbf{p}}))=\exp(-i\phi J^{3})\exp(-i\theta J^{2})\label{eq:u_rot}
\end{equation}
where $J^{2}$ and $J^{3}$ are the generators of the rotation group for the little group. The rotation properties of $U(R(\hat{\mathbf{p}}))$ depends on both angles and the generators. It can then be shown that when acting on states of half-integer spin, a shift by $2\pi$ on $\theta$ or $\phi$ would give rise to a -1 factor. For this reason, we restrict the domain of the angles to $0\leq\theta\leq\pi$ and $0\leq\phi<2\pi$.

\section{Parity and time reversal}\label{PT_symmetries}

The elements of the Lorentz group are either proper and orthochronous ($\det\Lambda=1$, $\Lambda^{0}_{\,\,\,0}\geq1$), or are products of  proper orthochronous transformation and discrete symmetry operators in the form of parity $\mathscr{P}$ and time reversal $\mathscr{T}$ defined in eq.~(\ref{eq:pt}). 

In special relativity, given a coordinate system, $x=(t,\mathbf{x})$, we can always define a parity and time-reversal operator as given in eq.~(\ref{eq:pt}). As a result, it was believed that one can always define a parity and time-reversal operator
\begin{equation}
\mathsf{P}= U(\mathscr{P},0),\hspace{0.5cm} \mathsf{T}= U(\mathscr{T},0)
\end{equation}
such that they furnish a representation for $\mathscr{P}$ and $\mathscr{T}$ respectively and satisfy eq.~(\ref{eq:u_rep})
\begin{equation}
\mathsf{P} U(\Lambda,a)\mathsf{P}^{-1}=U(\mathscr{P}\Lambda\mathscr{P}^{-1},\mathscr{P}a), \label{eq:p_symm}
\end{equation}
\begin{equation}
\mathsf{T} U(\Lambda,a)\mathsf{T}^{-1}=U(\mathscr{T}\Lambda\mathscr{T}^{-1},\mathscr{T}a). \label{eq:t_symm}
\end{equation}

However, it is now established that eq.~(\ref{eq:p_symm}) only serves as an approximation~\cite{Lee:1956qn,Wu:1957my} while there are evidence suggesting that eq.~(\ref{eq:t_symm}) is also violated~\cite{Christenson:1964fg}. But in most cases such as quantum electrodynamics, it is still useful to consider eqs.~(\ref{eq:p_symm}) and (\ref{eq:t_symm}) as symmetry transformations. Under the infinitesimal transformations, 
\begin{equation}
\Lambda^{\mu}_{\,\,\,\nu}=\delta^{\mu}_{\,\,\,\nu}+\omega^{\mu}_{\,\,\,\nu},\hspace{0.5cm} a^{\mu}=\epsilon^{\mu}
\end{equation}
we obtain
\begin{equation}
\mathsf{P}iJ^{\rho\sigma}\mathsf{P}^{-1}=i\mathscr{P}_{\mu}^{\,\,\,\rho}\mathscr{P}_{\nu}^{\,\,\,\sigma}J^{\mu\nu}, \label{eq:pj}
\end{equation}
\begin{equation}
\mathsf{P}iP^{\rho}\mathsf{P}^{-1}=i\mathscr{P}_{\mu}^{\,\,\,\rho}P^{\mu}, \label{eq:pp}
\end{equation}
\begin{equation}
\mathsf{T}iJ^{\rho\sigma}\mathsf{T}^{-1}=i\mathscr{T}_{\mu}^{\,\,\,\rho}\mathscr{T}_{\nu}^{\,\,\,\sigma}J^{\mu\nu}, \label{eq:tj}
\end{equation}
\begin{equation}
\mathsf{T}iP^{\rho}\mathsf{T}^{-1}=i\mathscr{T}_{\mu}^{\,\,\,\rho}P^{\mu}. \label{eq:tp}
\end{equation}
The factor of $i$ remains in the equations because we need to determine whether $\mathsf{P}$ and $\mathsf{T}$ are linear and unitary or anti-linear and anti-unitary. Towards this end, consider eq.~(\ref{eq:pj}) with $\rho=0$
\begin{equation}
\mathsf{P}iH\mathsf{P}^{-1}=iH.
\end{equation}
If $\mathsf{P}$ is anti-unitary, then we would have $\mathsf{P}H\mathsf{P}^{-1}=-H$. But then for a state $|p,\sigma\rangle$ with positive-definite energy $H|p,\sigma\rangle=p^{0}|p,\sigma\rangle$, there would exist another state $\mathsf{P}^{-1}|p,\sigma\rangle$
\begin{equation}
H(\mathsf{P}^{-1}|p,\sigma\rangle)=-p^{0}(\mathsf{P}^{-1}|p,\sigma\rangle
\end{equation}
with negative-definite energy which is not physical. Therefore, the parity operator $\mathsf{P}$ must be linear and unitary. Following the same argument, we see that the time-reversal operator $\mathsf{T}$ must be anti-linear and anti-unitary. Rewriting eqs.~(\ref{eq:pj}-\ref{eq:tp}) in terms of $\mathbf{J}$ and $\mathbf{K}$ yields
\begin{equation}
\mathsf{P}\mathbf{J}\mathsf{P}^{-1}=\mathbf{J}, \label{eq:pjp}
\end{equation}
\begin{equation}
\mathsf{P}\mathbf{K}\mathsf{P}^{-1}=-\mathbf{K},
\end{equation}
\begin{equation}
\mathsf{P}\mathbf{P}\mathsf{P}^{-1}=-\mathbf{P}, \label{eq:ppp}
\end{equation}
\begin{equation}
\mathsf{T}\mathbf{J}\mathsf{T}^{-1}=-\mathbf{J}, \label{eq:tjt}
\end{equation}
\begin{equation}
\mathsf{T}\mathbf{K}\mathsf{T}^{-1}=\mathbf{K}, \label{eq:tkt}
\end{equation}
\begin{equation}
\mathsf{T}\mathbf{P}\mathsf{T}^{-1}=-\mathbf{P} \label{eq:tpt}
\end{equation}
and
\begin{equation}
\mathsf{P}H\mathsf{P}^{-1}=\mathsf{T}H\mathsf{T}^{-1}=H.\label{eq:tphtp}
\end{equation}
Equations (\ref{eq:pjp}-\ref{eq:tpt}) allow us to derive the discrete symmetry transformations of massive and and massless particle states.

\subsection{Parity: $m>0$}

The massive particle state $|k,\sigma\rangle$ with $k=(m,\0)$ are eigenstates of $H$, $\mathbf{P}$ and $J^{3}$ with eigenvalues $m$, $\mathbf{0}$ and
$\sigma=-j,\cdots j$ respectively. Equations (\ref{eq:pjp}) and (\ref{eq:ppp}) tell us that the states $\mathsf{P}|k,\sigma\rangle$ and $|k,\sigma\rangle$ have identical eigenvalues under $H$, $\mathbf{P}$ and $J^{3}$. Therefore, $\mathsf{P}|k,\sigma\rangle$ can only differ from $|k,\sigma\rangle$ up to a phase,
\begin{equation}
\mathsf{P}|k,\sigma\rangle=\eta_{\sigma}|k,\sigma\rangle. \label{eq:p_state}
\end{equation}
Acting $(J^{1}\pm iJ^{2})$ on $|k,\sigma\rangle$ with the help of eq.~(\ref{eq:l}), we get
\begin{equation}
(J^{1}\pm iJ^{2})|k,\sigma\rangle
=\sqrt{(j\mp\sigma)(j\pm\sigma+1)\,}|k,\sigma\pm1\rangle. \label{eq:j1_j2_k}
\end{equation}
Acting eq.~(\ref{eq:p_state}) from the left by $\mathsf{P}$, we obtain $\eta_{\sigma}=\eta_{\sigma+1}$, so it is $\sigma$-independent thus giving us
\begin{equation}
P|k,\sigma\rangle=\eta |k,\sigma\rangle.
\end{equation}
The phase $\eta$ is called the intrinsic parity, its value only depend on the species of the particles. 

We now determine the parity transformation of the massive one-particle state with arbitrary momentum. Such a state is given by eq.~(\ref{eq:little_u}) as
\begin{equation}
|p,\sigma\rangle=\sqrt{\frac{m}{p^{0}}}U(L(p))|k,\sigma\rangle\label{eq:boosted_massive_particle}
\end{equation}
so the action of parity on $|p,\sigma\rangle$ gives us
\begin{equation}
\mathsf{P}|p,\sigma\rangle=\sqrt{\frac{m}{p^{0}}}\mathsf{P}U(L(p))\mathsf{P}^{-1} \eta|k,\sigma\rangle.
\end{equation}
Using the definition of parity $\mathscr{P}$ and standard Lorentz boosts defined in eqs.~(\ref{eq:pt}) and (\ref{eq:lorentz_boost}) respectively, we find
\begin{equation}
\mathscr{P}L(p)\mathscr{P}^{-1}=L(\mathscr{P}p) \nonumber
\end{equation}
where $\mathscr{P}p=(p^{0},-\mathbf{p})$. Substituting this result into above equation for $\mathsf{P}|p,\sigma\rangle$ and use eq.~(\ref{eq:boosted_massive_particle}), we obtain
\begin{equation}
\mathsf{P}|p,\sigma\rangle=\eta|\mathscr{P}p,\sigma\rangle.
\end{equation}

\subsection{Time-reversal: $m>0$}

The effect of time reversal $\mathsf{T}$ on $|k,\sigma\rangle$ can be inferred from eqs.~(\ref{eq:tjt},\ref{eq:tpt},\ref{eq:tphtp}),
\begin{equation}
\mathsf{T}\mathbf{J}\mathsf{T}^{-1}=-\mathbf{J}, \nonumber
\end{equation}
\begin{equation}
\mathsf{T}\mathbf{P}\mathsf{T}^{-1}=-\mathbf{P}, \nonumber
\end{equation}
\begin{equation}
\mathsf{T}H\mathsf{T}^{-1}=H .\nonumber
\end{equation}
Applying the same argument as for parity, but with $\mathsf{T}\mathbf{J}\mathsf{T}^{-1}=-\mathbf{J}$, the spin-projection along the $3$-axis is reflected under time-reversal
\begin{equation}
\mathsf{T}|k,\sigma\rangle=\varrho_{\sigma}|k,-\sigma\rangle
\end{equation}
where $\varrho_{\sigma}$ is the time-reversal phase. Acting $\mathsf{T}$ from the left on eq.~(\ref{eq:j1_j2_k}), with some simplification we find
\begin{equation}\
\varrho_{\sigma}=-\varrho_{\sigma\pm1}
\end{equation}
so that $\varrho_{\sigma}$ depends on both the spin-projection $\sigma$ and the particle species. The phase $\varrho_{\sigma}$ may be written as
\begin{equation}
\varrho_{\sigma}=(-1)^{j-\sigma}\varrho
\end{equation}
where $\varrho$ only depends on the particle species. Therefore,
\begin{equation}
\mathsf{T}|k,\sigma\rangle=(-1)^{j-\sigma}\varrho|k,-\sigma\rangle.
\end{equation}
The phase $\varrho$ has no physical significance, for we can always redefine the state to eliminate the phase
\begin{equation}
|k,\sigma\rangle\rightarrow |k,\sigma\rangle'=\varrho^{1/2}|k,\sigma\rangle \nonumber
\end{equation}
so that
\begin{eqnarray}
\mathsf{T}|k,\sigma\rangle'&=&(\varrho^{1/2})^{*}\mathsf{T}|k,\sigma\rangle=(-1)^{j-\sigma}\varrho^{1/2}|k,-\sigma\rangle \\ \nonumber
&=&(-1)^{j-\sigma}|k,-\sigma\rangle'
\end{eqnarray}
where we have used the fact that $\mathsf{T}$ is an anti-unitary operator and $|\varrho|^{2}=1$. Nevertheless, we shall keep the phase and choose its value according to our convenience. 

The particle state of arbitrary momentum under time-reversal is then
\begin{equation}
\mathsf{T}|p,\sigma\rangle=\sqrt{\frac{m}{p^{0}}}U(\mathscr{T}L(p)\mathscr{T}^{-1})(-1)^{j-\sigma}|k,-\sigma\rangle.
\end{equation}
The term $\mathscr{T}L(p)\mathscr{T}^{-1}$ is identical to $\mathscr{P}L(p)\mathscr{P}^{-1}$
\begin{equation}
\mathscr{T}L(p)\mathscr{T}^{-1}=L(\mathscr{P}p).
\end{equation}
Therefore, we get
\begin{equation}
\mathsf{T}|p,\sigma\rangle=(-1)^{j-\sigma}\varrho|\mathscr{P}p,-\sigma\rangle.\label{eq:massive_time_reversal}
\end{equation}

\subsection{Parity: $m=0$}

The massless one-particle state $|k,\sigma\rangle$ has standard momentum four-vector $k=(\kappa,0,0,\kappa)$ and helicity $\sigma$ with respect to $\mathbf{J}\cdot\hat{\mathbf{k}}$. Since the massless particle state always have non-zero momentum where its helicity $\sigma$ is invariant under the continuous Lorentz transformations, its transformation under parity is more complicated.

The state $|k,\sigma\rangle$ is an eigenstate of $\mathbf{P}$ and $\mathbf{J}\cdot\hat{\mathbf{k}}$ with eigenvalue $\mathbf{k}$ and $\sigma$ respectively. Using eq.~(\ref{eq:ppp}), we see that the both the momentum and helicity are reflected under parity
\begin{equation}
\mathsf{P}|k,\sigma\rangle=\eta_{\sigma}|\mathscr{P}k,-\sigma\rangle
\end{equation}
where $\eta_{\sigma}$ is the intrinsic parity. Unlike massive particle states, due to the structure of the little group, the intrinsic parity $\eta_{\sigma}$ for massless particle states are in general $\sigma$-dependent. Exploiting this freedom, we  extract a factor of $(-1)^{j+\sigma}$ from the phase so that
\begin{equation}
\mathsf{P}|k,\sigma\rangle=(-1)^{j+\sigma}\eta_{\sigma}|\mathscr{P}k,-\sigma\rangle.
\end{equation}
Parity does not leave the four-momentum $k$ invariant so we perform a rotation by $180^{\circ}$ about the $2$-axis taking $\mathscr{P}k$ to $k$. Using eq.~(\ref{eq:massless_trans}) with $U(R_{2})=\exp(i\pi J^{2})$, we get
\begin{equation}
U(R_{2}^{-1})\mathsf{P}|k,\sigma\rangle=(-1)^{j+\sigma}\eta_{\sigma}|k,-\sigma\rangle.
\end{equation}

Now we consider parity on the massless particle state of arbitrary momentum. Using eq.~(\ref{eq:u}) and the normalisation given by eq.~(\ref{eq:normalisation}), we obtain
\begin{equation}
|p,\sigma\rangle=\sqrt{\frac{k^{0}}{p^{0}}}\,U(L(p))|k,\sigma\rangle
\end{equation}
where the boost $U(L(p))=U(R(\mathbf{\hat{\mathbf{p}}})B(|\mathbf{p}|))$ takes $k$ along the $3$-axis to magnitude $|\mathbf{p}|$ then rotate it to direction $\mathbf{\hat{p}}$. Acting $\mathsf{P}$ on the state, with some simplification, we obtain
\begin{equation}
\mathsf{P}|p,\sigma\rangle=(-1)^{j+\sigma}\eta_{\sigma}\sqrt{\frac{k^{0}}{p^{0}}}U(R(\mathbf{\hat{p}})R_{2}B(|\mathbf{p}|))|k,-\sigma\rangle \label{eq:parity_p}
\end{equation}
where we have used the identities that $\mathscr{P}R_{2}$ commutes with $B(|\mathbf{p}|)$ and $\mathscr{P}$ commutes with $R(\hat{\mathbf{p}})$. The rotation $R(\mathbf{\hat{p}})R_{2}$ takes the $3$-axis to $-\mathbf{\hat{p}}$, but its unitary representation $U(R(\mathbf{\hat{p}})R_{2})$ is not equal to $U(R(-\mathbf{\hat{p}}))$. To see this, we use eq.~(\ref{eq:u_rot}) to obtain
\begin{equation}
U(R(-\mathbf{\hat{p}}))=\exp[-i(\phi\pm\pi) J^{3}]\exp[-i(\pi-\theta)J^{2}]
\end{equation}
where $\phi+\pi$ and $\phi-\pi$ are for $0\leq\phi<\pi$ and $\pi\leq\phi<2\pi$ respectively. Now we compute the following product
\begin{eqnarray}
U^{-1}(R(-\mathbf{\hat{p}}))U(R(\mathbf{\hat{p}})R_{2})&=&\exp[i(\pi-\theta)J^{2}]\exp[i(\phi\pm\pi)J^{3}] \\ \nonumber
&&\times\exp[-i\phi J^{3}]\exp[-i\theta J^{2}]\exp[i\pi J^{2}] \\ \nonumber
&=&\exp[i(\pi-\theta)J^{2}]\exp[\pm i \pi J^{3}]\exp[i(\pi-\theta)J^{2}]\nonumber \\
&=& \exp[i(\pi-\theta)J^{2}]U[R_{3}(\pm\pi)R_{2}(\pi-\theta)].
\end{eqnarray}
Explicit computation of the product of the two rotation matrices shows that 
\begin{equation}
R_{3}(\pm\pi)R_{2}(\pi-\theta)=R_{2}(\theta-\pi)R_{3}(\pm\pi).
\end{equation}
Therefore,
\begin{equation}
U(R(\mathbf{\hat{p}})R_{2})=U(R(-\mathbf{\hat{p}}))\exp[\pm i\pi J^{3}]. \label{eq:u_rot_r2}
\end{equation}
Substituting eq.~(\ref{eq:u_rot_r2}) into eq.~(\ref{eq:parity_p}) and using with the fact that $J^{3}$ commutes with boost along the $3$-axis,  we obtain
\begin{equation}
\mathsf{P}|p,\sigma\rangle=(-1)^{j+\sigma}\eta_{\sigma}e^{\mp i\pi\sigma}|\mathscr{P}p,-\sigma\rangle.\label{eq:massless_particle_parity}
\end{equation}
where the top and bottom signs apply to states whose momenta along the $2$-axis are positive and negative respectively.

\subsection{Time-reversal: $m=0$}

The helicity of the massless one-particle state $|k,\sigma\rangle$ is invariant under time-reversal since $\mathsf{T}J^{3}\mathsf{T}^{-1}=-J^{3}$ and $\mathsf{T}\mathbf{P}\mathsf{T}^{-1}=-\mathbf{P}$. Therefore, we get
\begin{equation}
\mathsf{T}|k,\sigma\rangle=\varrho_{\sigma}|\mathscr{P}k,\sigma\rangle
\end{equation}
where $\varrho_{\sigma}$ is a $\sigma$-dependent phase. Following the same argument as before, we get
\begin{equation}
U(R_{2}^{-1})\mathsf{T}|k,\sigma\rangle=\varrho_{\sigma}|k,\sigma\rangle.
\end{equation}
Exploiting the fact that $\mathscr{T}R_{2}$ commutes with $B(|\mathbf{p}|)$ and $\mathscr{T}$ commutes with $R(\mathbf{\hat{p}})$, acting $\mathsf{T}$ on eq.~(\ref{eq:little_u}) yields
\begin{equation}
\mathsf{T}|p,\sigma\rangle=\varrho_{\sigma}\sqrt{\frac{k^{0}}{p^{0}}}U(R(\mathbf{\hat{p}})R_{2}B(|\mathbf{p}|))|k,\sigma\rangle.
\end{equation}
Using eq.~(\ref{eq:u_rot_r2}), we obtain
\begin{equation}
\mathsf{T}|p,\sigma\rangle=\varrho_{\sigma}e^{\pm i\pi\sigma}|\mathscr{P}p,\sigma\rangle
\label{eq:massless_particle_time_reversal}
\end{equation}
where the plus and minus sign apply to states whose momenta along the $2$-axis are positive and negative respectively. The difference of signs in the phases for eq.~(\ref{eq:massless_particle_parity}) and (\ref{eq:massless_particle_time_reversal}) is due to the fact that the helicity of massless particle states are reflected under parity while it remains unchanged under time-reversal.

\section{Scattering theory} \label{chap:scattering_theory}

All theories must be subjected to experimental tests. In particle physics these experiments often involve particle collisions at high energy. It is by studying the products of collisions that allow us to decipher the properties of various particles and discover new ones. In most experiments, the only measured quantities are the probability distribution and the cross-section. The purpose of this section is to outline the formalism for studying interactions between particles and how to calculate the quantities measured by experiments.

\subsection{In and out states}

In most particle physics experiments, particles travel long distances before interacting in a very small region then separate to a large distance. Therefore, we can effectively treat the particle states long before and after collision as non-interacting states. The free states involving more than one-particles can be described by taking symmetrised or anti-symmetrised tensor product of the one-particle state. The physical interpretation of the tensor products is discussed in sec.~\ref{cluster_decomp}.

Let us label the one-particle state with momentum $p^{\mu}$ with spin/helicity $\sigma$, we write the multi-particle state as
\begin{equation}
|p_{1}\sigma_{1}n_{1},p_{2}\sigma_{2}n_{2},\cdots\rangle
\end{equation}
where we have suppressed the tensor product sign between the particle states. The index $n_{i}$ represents particle of different species. The inhomogeneous Lorentz transformation for the multi-particle state is an extension from the one-particle case given by eq.~(\ref{eq:massive_tran}),\footnote{We adopt the notation $a\cdot b= a^{\mu}b_{\mu}$.}
\begin{eqnarray}
&&U(\Lambda,a)|p_{1}\sigma_{1}n_{1},p_{2}\sigma_{2}n_{2},\cdots\rangle \nonumber \\
&&=e^{ia\cdot(\Lambda p_{1}+\Lambda p_{2}+\cdots)}\sqrt{\frac{(\Lambda p_{1})^{0}(\Lambda p_{2})^{0}\cdots}{p^{0}_{1}p^{0}_{2}\cdots}}
\sum_{\sigma_{1}'\sigma_{2}'\cdots}D_{\sigma_{1}'\sigma_{1}}^{(j_{1})}(W(\Lambda,p_{1}))D_{\sigma_{2}'\sigma_{2}}^{(j_{2})}(W(\Lambda,p_{2}))\cdots \nonumber \\
&&\hspace{1cm}\times |\Lambda p_{1}\sigma_{1}'n_{1},\Lambda p_{2}\sigma_{2}'n_{2},\cdots\rangle. \label{eq:multi_state_trans}
\end{eqnarray}
Here we are concerned with the massive particle states where $D(W(\Lambda,p))$ are $(2j+1)\times(2j+1)$ matrices given by eqs.~(\ref{eq:r1}-\ref{eq:r3}). For massless particles, $D(W(\Lambda,p))$ is given by $\delta_{\sigma\sigma'}\exp[i\sigma\phi]$. The normalisation of the multi-particle state is a generalisation of the one-particle state
\begin{eqnarray}
&&\langle p_{1}'\sigma_{1}'n_{1}',p_{2}'\sigma_{2}'n_{2}',\cdots | p_{1}\sigma_{1}n_{1},p_{2}\sigma_{2}n_{2},\cdots\rangle \nonumber \\
&&\hspace{2cm}=\delta^{3}(\p_{1}'-\p_{1})\delta_{\sigma_{1}'\sigma_{1}}\delta_{n_{1}'n_{1}}\delta^{3}(\p_{2}'-\p_{2})\delta_{\sigma_{2}'\sigma_{2}}\delta_{n_{2}'n_{2}}\cdots \nonumber \\
&&\hspace{3.5cm} \pm \mbox{permutations}
\end{eqnarray}
where the top and bottom signs depends on whether the particle states are of integral or half-integral spin respectively, this will be discussed in more detail sec.~\ref{cluster_decomp}. For convenience, from now on we denote the multi-particle state as $|\alpha\rangle$ and write their inner-product as
\begin{equation}
\langle\alpha'|\alpha\rangle=\delta(\alpha'-\alpha).
\end{equation}
We define the following multi-integral as
\begin{equation}
\int d\alpha=\sum_{n_{1}\sigma_{1},n_{2}\sigma_{2}\cdots}\int d^{3}p_{1}\,d^{3}p_{2}\cdots
\end{equation}
where we sum over the spin and particle species and assume the multi-particle state satisfies the completeness relation
\begin{equation}
|\beta\rangle=\int d\alpha\,|\alpha\rangle \langle\alpha|\beta\rangle.
\end{equation}

The transformation given in eq.~(\ref{eq:multi_state_trans}) is only valid for free particles. Taking 
$\Lambda^{\mu}_{\,\,\,\nu}=\delta^{\mu}_{\,\,\,\nu}$ and $a^{\mu}=(\tau,0,0,0)$ we get $U(\Lambda,a)=e^{iH_{0}\tau}$, where $H_{0}$ is the free Hamiltonian. Equation (\ref{eq:multi_state_trans}) then tells us that the mutli-particle state must be an energy eigenstate
\begin{equation}
H_{0}|\alpha\rangle=p^{0}_{\alpha}|\alpha\rangle
\end{equation}
where $p^{0}_{\alpha}=p_{1}^{0}+p_{2}^{0}+\cdots$. Let us choose a frame such that the interaction takes place at $t=0$ so that $t\rightarrow-\infty$ and $t\rightarrow+\infty$ refer to particles long before and after interaction. In the limit $t\rightarrow\pm\infty$, the particle states are non-interacting so their transformation are governed by eq.~(\ref{eq:multi_state_trans}). Therefore, that there are two different sets of particle states before and after the interaction. The states which contain the free particle states in the limit $t\rightarrow-\infty$ and $t\rightarrow+\infty$ are called the 'in' state $|\alpha_{+}\rangle$ and 'out' state $|\alpha_{-}\rangle$ respectively.

Here, it is convenient to work in the Heisenberg picture where the particle states are time-independent as opposed to the Schr\"{o}dinger picture where the states are time-dependent. In the Heisenberg picture the in and out state do not change over time, so they are not the limit of the time dependent state $|\Psi(t)\rangle$ as $t\rightarrow\pm\infty$.

When we define a particle state, we have implicitly chosen an inertial reference frame. 
Suppose at time $t=0$ the observer in frame $\mathcal{O}$ sees the state $|\alpha\rangle$ takes part in some interaction. For the second observer in frame $\mathcal{O}'$ the interaction occurs at $t'=0$ which corresponds  $t=\tau$ in the frame $\mathcal{O}$. It follows that the two time coordinates are related by $t'=t-\tau$. If the observer in $\mathcal{O}$ sees state $|\Psi\rangle$, then the state in $\mathcal{O}'$ is obtained by time translation,  $U(1,-\tau)|\Psi\rangle=e^{-iH\tau}|\Psi\rangle$. Therefore as $\tau\rightarrow\pm\infty$ we obtain the state long before and after the interaction.

The free particle states we work with are defined as eigenstates of the free Hamiltonian $H_{0}$. These states do not constitute a physical picture since they cannot be localised in space-time. A proper treatment of requires the use of wave packet, a superposition of states $\int d\alpha\,g(\alpha)|\alpha\rangle$ where $g(\alpha)$ is a smooth varying function over some finite energy range $\Delta p^{0}$. Nevertheless, this does not obstruct our progress to formulate the scattering theory in terms of momentum eigenstate. The wave packet formalism, though more realistic, does not give us new physics.

Let us define the full Hamiltonian to be
\begin{equation}
H=H_{0}+V
\end{equation}
where $H_{0}$ and $V$ are the free and interacting Hamiltonian respectively. For collision processes, in the limit 
$t\rightarrow\pm\infty$, the interacting Hamiltonian $V$ vanishes. Therefore, in this case, it is reasonable assume that $H$ and $H_{0}$ have the same energy spectrum.\footnote{This assumption is justified for particle collision processes in absence of external fields by conservation of energy. The total energy of the in and out state must equal to the energy of the free state. However, this does not apply for system such as the Hydrogen atom where
$H=|\p|^{2}/{2m}-e^{2}/r$. The spectrum for $H$, to the leading order is discrete between $[-13.6\mbox{eV},0)$ and continuous from $(0,\infty)$ whereas the spectrum for $H_{0}$ is $(0,\infty)$.} The in and out states are defined as eigenstates of $H$
\begin{equation}
H|\alpha_{\pm}\rangle=p^{0}_{\alpha}|\alpha_{\pm}\rangle \label{eq:def_in_out_states}
\end{equation}
and they must satisfy the condition
\begin{equation}
\int d\alpha\, e^{-ip^{0}_{\alpha}\tau}g(\alpha)|\alpha_{\pm}\rangle\rightarrow
\int d\alpha\, e^{-ip^{0}_{\alpha}\tau}g(\alpha)|\alpha_{0}\rangle \label{eq:in_out_to_free_state}
\end{equation}
in the limit $\tau\rightarrow\mp\infty$ where $|\alpha_{0}\rangle$ is the free multi-particle state. This expression can be written in the operator form
\begin{equation}
e^{-iH\tau}\int d\alpha\, g(\alpha)|\alpha_{\pm}\rangle\rightarrow
e^{-iH_{0}\tau}\int d\alpha\, g(\alpha)|\alpha_{0}\rangle.
\end{equation}
This limit must hold for all $g(\alpha)$, therefore, the in and out state can be written in terms of the free particle state
\begin{equation}
|\alpha_{\pm}\rangle=\Omega(\mp\infty)|\alpha_{0}\rangle \label{eq:in_out_and_free_states}
\end{equation}
where
\begin{equation}
\Omega(\tau)= e^{iH\tau}e^{-iH_{0}\tau}.
\end{equation}
Since the inner product must be time-independent, eq.~(\ref{eq:in_out_to_free_state}) gives
\begin{equation}
\int d\alpha\,d\beta\,e^{-i(p^{0}_{\alpha}-p^{0}_{\beta})\tau}g(\alpha)g^{*}(\beta)\langle\beta_{\pm}|\alpha_{\pm}\rangle=
\int d\alpha\,d\beta\,e^{-i(p^{0}_{\alpha}-p^{0}_{\beta})\tau}g(\alpha)g^{*}(\beta)\langle\beta_{0}|\alpha_{0}\rangle.
\end{equation}
Therefore, the in and out states have the same inner product as their free particle states
\begin{equation}
\langle\beta_{\pm}|\alpha_{\pm}\rangle=\langle\beta_{0}|\alpha_{0}\rangle.
\end{equation}

Although we have defined the necessary states to describe particle interaction, it is useful and instructive to have an explicit mathematical relation between $|\alpha_{\pm}\rangle$ and $|\alpha_{0}\rangle$. 
This is given by the Lippmann-Schwinger equation
\begin{equation}
|\alpha_{\pm}\rangle=|\alpha_{0}\rangle+\int d\beta\frac{T_{\beta\alpha}^{\pm}\,|\beta_{0}\rangle}{p^{0}_{\alpha}-p^{0}_{\beta}\pm i\epsilon}
\label{eq:lippmann_schwinger}
\end{equation}
where $T_{\beta\alpha}^{\pm}=\langle\beta_{0}|V|\alpha_{\pm}\rangle$ and $\epsilon>0$ is a real infinitesimal number. For a detailed discussion and derivation see~\cite{PhysRev.79.469,Goldberger}. Here it is sufficient to see that $|\alpha_{\pm}\rangle\rightarrow|\alpha_{0}\rangle$ is satisfied in the limit $t\rightarrow\mp\infty$. To see this, let us define the following wave-packets
\begin{equation}
|g,t\rangle_{\pm}=\int d\alpha\,g(\alpha)e^{-ip^{0}_{\alpha}t}|\alpha_{\pm}\rangle.
\end{equation}
The Lippmann-Schwinger equation then gives us
\begin{equation}
|g,t\rangle_{\pm}=|g,t\rangle_{0}+\int d\alpha\,d\beta\frac{e^{-ip^{0}_{\alpha}t}g(\alpha)\,T_{\beta\alpha}^{\pm}\,|\beta_{0}\rangle}{p^{0}_{\alpha}-p^{0}_{\beta}\pm i\epsilon}.
\end{equation}
We now exchange the order of integration and examine the following complex integral
\begin{equation}
\mathscr{J}_{\beta}^{\pm}=\int d\alpha\
\frac{e^{-ip^{0}_{\alpha}t}g(\alpha)\,T_{\beta\alpha}^{\pm}\,|\beta_{0}\rangle}{p^{0}_{\alpha}-p^{0}_{\beta}\pm i\epsilon}.
\label{eq:jbeta}
\end{equation}
In the limit $t\rightarrow-\infty$ we have $\mathscr{J}_{\beta}^{+}$. We evaluate the integral by closing the contour in the upper-half plane. Since $\mbox{Im}(p^{0}_{\alpha})>0$, $\mathscr{J}_{\beta}^{+}$ is exponentially damped as $t\rightarrow-\infty$. The pole in the denominator ($p^{0}_{\alpha}=p^{0}_{\beta}-i\epsilon$) is not contained in the top-half plane so it does not contribute to the integral. The only other possible contributions to the integral come from the possible existence of poles in $T^{+}_{\beta\alpha}$ and $g(\alpha)$ in the region of positive $\mbox{Im}(p^{0}_{\alpha})$. But for large semi-circles, the contribution due to the poles of the two functions are also exponentially damped as $t\rightarrow-\infty$. Specifically, $|t|$ must be much larger than the time uncertainty of the wave packet and the period of interaction, since these two variables governs the position of the poles for $g(\alpha)$ and $T^{+}_{\beta\alpha}$ respectively. Therefore, in the limit $t\rightarrow-\infty$, we get
$|g,t\rangle_{+}\rightarrow |g,t\rangle_{0}$ as required. Similarly, in the case where $t\rightarrow\infty$, $|g,t\rangle_{-}\rightarrow |g,t\rangle_{0}$. It follows that the solution of the in and out state given by the Lippmann Schwinger equation satisfies eq.~(\ref{eq:in_out_to_free_state}).

\subsection{The $S$-matrix}

The $S$-matrix is defined as the probability transition amplitude from $|\alpha_{+}\rangle$ to $|\beta_{-}\rangle$
\begin{equation}
S_{\beta\alpha}=\langle\beta_{-}|\alpha_{+}\rangle. \label{eq:smatrix}
\end{equation}
In the absence of interaction, $S_{\beta\alpha}=\delta(\beta-\alpha)$. Since the in and out states are both orthonormal and complete, $S_{\beta\alpha}$ is unitary. This can be shown by computing the inner product,
\begin{eqnarray}
\int d\beta\,S^{\dag}_{\beta\gamma} S_{\beta\alpha}&=&\int d\beta\,\langle\gamma_{+}|\beta_{-}\rangle\langle\beta_{-}|\alpha_{+}\rangle \\ \nonumber
&=&\delta(\gamma-\alpha)
\end{eqnarray}
where we have used the completeness relation.

It is often more convenient to define the $S$-matrix in terms of free particle states by introducing the operator $S$ such that
\begin{equation}
S_{\beta\alpha}=\langle\beta_{0}|S|\alpha_{0}\rangle.
\end{equation}
Equation (\ref{eq:in_out_and_free_states}) tell us that $|\alpha_{+}\rangle=\Omega(-\infty)|\alpha_{0}\rangle$ and $|\beta_{-}\rangle=\Omega(+\infty)|\beta_{0}\rangle$, so the operator $S$ is given by
\begin{equation}
S=\Omega^{\dag}(+\infty)\Omega(-\infty)=U(+\infty,-\infty) \label{eq:smatrix1}
\end{equation}
where
\begin{equation}
U(\tau,\tau_{0})=\Omega^{\dag}(\tau)\Omega(\tau_{0})= e^{iH_{0}\tau}e^{-iH(\tau-\tau_{0})}e^{-iH_{0}\tau_{0}}.
\end{equation}
This definition of $S$ will be useful to us when we examine the Lorentz invariance of the $S$-matrix and deriving its perturbation expansion.

The $S$-matrix can also be written in an alternative form that yield a different but equivalent perturbation expansion derived using eq.~(\ref{eq:smatrix1}) and the Lippmann-Schwinger equation. The in state $|g,t\rangle_{+}$ using eq.~(\ref{eq:jbeta}) can be written as
\begin{equation}
|g,t\rangle_{+}=|g,t\rangle_{0}+\int d\beta \mathscr{J}_{\beta}^{+}.
\end{equation}
This time, we take the limit $t\rightarrow\infty$, so we choose the clockwise path and close the contour in the lower-half plane. Ignoring the poles of $g(\alpha)$ and $T_{\beta\alpha}^{+}$, the only contributing pole is $p^{0}_{\alpha}=p^{0}_{\beta}-i\epsilon$. Applying the residue theorem, in the limit $t\rightarrow\infty$ and $\epsilon\rightarrow0^{+}$, 
\begin{equation}
\mathscr{J}_{\beta}^{+}\rightarrow-2i\pi e^{-ip^{0}_{\beta}t}\int d\alpha\,\delta(p^{0}_{\alpha}-p^{0}_{\beta})g(\alpha)T_{\beta\alpha}^{+}.
\end{equation}
Therefore, in the limit $t\rightarrow\infty$,
\begin{equation}
|g,t\rangle_{+}\rightarrow\int d\beta\,e^{-ip^{0}_{\beta}t}|\beta_{0}\rangle\big[\,g(\beta)-2i\pi\int d\alpha\,\delta(p^{0}_{\alpha}-p^{0}_{\beta})g(\alpha)T_{\beta\alpha}^{+}\,\big]. \label{eq:gplus_limit1}
\end{equation}
The in state can also be written in terms of the $S$-matrix via the completeness relation
\begin{equation}
|g,t\rangle_{+}=\int d\alpha\,d\beta\,e^{-ip^{0}_{\alpha}t}g(\alpha)S_{\beta\alpha}|\beta_{-}\rangle.
\end{equation}
Energy conservation means that $S_{\beta\alpha}$ must contain a factor $\delta(p^{0}_{\beta}-p^{0}_{\alpha})$, so we replace $p^{0}_{\alpha}$ by $p^{0}_{\beta}$ in the exponential,
\begin{equation}
|g,t\rangle_{+}=\int d\beta e^{-ip^{0}_{\beta}t}|\beta_{-}\rangle\int d\alpha\,g(\alpha)S_{\beta\alpha}.
\end{equation}
In the limit $t\rightarrow\infty$, it becomes
\begin{equation}
|g,t\rangle_{+}\rightarrow\int d\beta\,e^{-ip^{0}_{\beta}t}|\beta_{0}\rangle\int d\alpha\,g(\alpha)S_{\beta\alpha}. \label{eq:gplus_limit2}
\end{equation}
Equating eq.~(\ref{eq:gplus_limit1}) and eq.~(\ref{eq:gplus_limit2}) yields a solution for $S_{\beta\alpha}$,
\begin{equation}
S_{\beta\alpha}=\delta(\beta-\alpha)-2i\pi \delta(p^{0}_{\alpha}-p^{0}_{\beta})T_{\beta\alpha}^{+}.
\end{equation}
where $T_{\beta\alpha}^{\pm}=\langle\beta_{0}|V|\alpha_{\pm}\rangle$. For weak interaction $V$, to the first-order, we may use the approximation $T_{\beta\alpha}^{\pm}\approx\langle\beta_{0}|V|\alpha_{0}\rangle$. The $S$-matrix is then
\begin{equation}
S_{\beta\alpha}\approx\delta(\beta-\alpha)-2i\pi\delta(p^{0}_{\alpha}-p^{0}_{\beta})\langle\beta_{0}|V|\alpha_{0}\rangle.
\end{equation}
This is known as the Born approximation which is useful to compute the first-order $S$-matrix. Later on we will derive the perturbation expansion to calculate high-order $S$-matrix elements.

\subsection{Symmetries of the $S$-matrix}

The $S$-matrix defined in the previous section will later be used to compute the probability distribution and cross-section of particle interactions. Here we establish the necessary conditions for which the $S$-matrix satisfies the Poincar\'{e} symmetry.

\subsubsection{Lorentz symmetry}\label{LsSmatrix}

We argued in the sec.~\ref{Poincare_algebra} that for any proper-orthochronous Lorentz transformation, it is possible to define a unitary operator $U(\Lambda,a)$. Given the operator $U(\Lambda,a)$, the in and out states are said to be Lorentz invariant when the same $U(\Lambda,a)$ acts on both states in the same way. Since $U(\Lambda,a)$ is unitary, we have
\begin{equation}
S_{\beta\alpha}=\langle\beta_{-}|\alpha_{+}\rangle=\langle\beta^{-}|U^{\dag}(\Lambda,a)U(\Lambda,a)|\alpha_{+}\rangle.
\end{equation}
The invariance of the inner product, with the help of eq.~(\ref{eq:multi_state_trans}) shows that the $S$-matrix is Poincar\'{e}-covariant
\begin{eqnarray}
&& S_{(p'_{1}\sigma_{1}'n_{1}',p'_{2}\sigma_{2}'n_{2}'\cdots)
(p_{1}\sigma_{1}n_{1},p_{2}\sigma_{2}n_{2}\cdots)}  \label{eq:smatrix_trans}\\  \nonumber
&&\hspace{1cm}=
e^{ia\cdot(\Lambda p_{1}+\Lambda p_{2}+\cdots-\Lambda p'_{1}-\Lambda p'_{2}\cdots)}
\sqrt{\frac{(\Lambda p_{1})^{0}(\Lambda p_{2})^{0}\cdots (\Lambda p'_{1})^{0}(\Lambda p'_{2})^{0}\cdots}{p^{0}_{1}p^{0}_{2}\cdots p'^{0}p'^{1}\cdots}} \\ \nonumber
&&\hspace{1.5cm}\times\sum_{\bar{\sigma}_{1}\bar{\sigma}_{2}\cdots}D_{\bar{\sigma}_{1}\sigma_{1}}^{(j_{1})}(W(\Lambda,p_{1}))D_{\bar{\sigma}_{2}\sigma_{2}}^{(j_{2})}(W(\Lambda,p_{2}))\cdots \\ \nonumber
&&\hspace{1.5cm}\times\sum_{\bar{\sigma}_{1}'\bar{\sigma}_{2}'\cdots}
D'^{(j'_{1})*}_{\bar{\sigma}'_{1}\sigma'_{1}}(W(\Lambda,p'_{1}))
D'^{(j'_{2})*}_{\bar{\sigma}'_{2}\sigma'_{2}}(W(\Lambda,p'_{2}))\cdots \\ \nonumber
&&\hspace{1.5cm}\times
 S_{(\Lambda p'_{1}\sigma_{1}'n_{1}',\Lambda p'_{2}\sigma_{2}'n_{2}'\cdots)
(\Lambda p_{1}\sigma_{1}n_{1},\Lambda p_{2}\sigma_{2}n_{2}\cdots)}. \nonumber
\end{eqnarray}
Since the left-hand side is energy-momentum independent, the same must hold for the right-hand side also. Therefore, the argument in the exponential on the right-hand side must vanish giving us energy-momentum conservation. Assuming that the interaction can be separated from free theory  $H=H_{0}+V$, we may write the $S$-matrix as
\begin{equation}
S_{\beta\alpha}-\delta(\beta-\alpha)=-2\pi i M_{\beta\alpha}\delta^{4}(p_{\beta}-p_{\alpha})\label{eq:master_s}
\end{equation}
where $M_{\beta\alpha}$ determines the structure of interaction and the delta function $\delta^{4}(p_{\beta}-p_{\alpha})$ ensures energy-momentum conservation at all times.

One should note, eq.~(\ref{eq:smatrix_trans}) is not a theorem. Instead, it should be interpreted as the definition of Lorentz invariance for the $S$-matrix. In fact, only a particular choice of Hamiltonian would give eq.~(\ref{eq:smatrix_trans}). To see this, it is convenient to use the definition
\begin{equation}
S_{\beta\alpha}=\langle\beta_{0}|S|\alpha_{0}\rangle. \nonumber
\end{equation}
The free particle states have well-defined Lorentz transformations. Let us denote the operator that acts on the free particle states as $U_{0}(\Lambda,a)$, then we have
\begin{eqnarray}
&&U_{0}(\Lambda,a)|p_{1}\sigma_{1}n_{1},p_{2}\sigma_{2}n_{2},\cdots\rangle \nonumber \\
&&=e^{ia\cdot(\Lambda p_{1}+\Lambda p_{2}+\cdots)}\sqrt{\frac{(\Lambda p_{1})^{0}(\Lambda p_{2})^{0}\cdots}{p^{0}_{1}p^{0}_{2}\cdots}}
\sum_{\sigma_{1}'\sigma_{2}'\cdots}D_{\sigma_{1}'\sigma_{1}}^{(j_{1})}(W(\Lambda,p_{1}))D_{\sigma_{2}'\sigma_{2}}^{(j_{2})}(W(\Lambda,p_{2}))\cdots \nonumber \\
&&\hspace{1cm}\times |\Lambda p_{1}\sigma_{1}'n_{1},\Lambda p_{2}\sigma_{2}'n_{2},\cdots\rangle. \label{eq:free_multi_particle_trans}
\end{eqnarray}
We see that the in and out states transform in the same way as the free states under $U(\Lambda,a)$ and $U_{0}(\Lambda,a)$ respectively. Substituting eq.~(\ref{eq:free_multi_particle_trans}) into eq.~(\ref{eq:smatrix_trans}), we see that $S$-matrix is Lorentz-invariant if $U_{0}(\Lambda,a)$ commutes with $S$,
\begin{equation}
U_{0}^{-1}(\Lambda,a)SU_{0}(\Lambda,a)=S. \label{eq:lorentz_invariant_smatrix}
\end{equation}
This equation translates to the following commutation relations 
\begin{equation}
[H_{0},S]=[\mathbf{P}_{0},S]=[\mathbf{K}_{0},S]=[\mathbf{J}_{0},S]=0
\end{equation}
where $H_{0}$, $\mathbf{P}_{0}$, $\mathbf{K}_{0}$ and $\mathbf{J}_{0}$ are the generators of the Poincar\'{e} group for the free particle states. Similarly, for $U(\Lambda,a)$ we can define $\mathbf{P}$, $\mathbf{J}$, $\mathbf{K}$ and the full Hamiltonian $H$ for the in and out states that satisfy the Poincar\'{e} algebra.

In almost all interacting theories, the full Hamiltonian $H$ can be decomposed into a sum of free and interacting term with $\mathbf{P}_{0}$ and $\mathbf{J}_{0}$ remain unchanged,
\begin{equation}
H=H_{0}+V,\hspace{0.5cm} \mathbf{P}=\mathbf{P}_{0},\hspace{0.5cm}\mathbf{J}=\mathbf{J}_{0}.
\end{equation}
These definitions imply
\begin{equation}
[J^{i},H]=[P^{i},H]=0,\hspace{0.5cm}
[J^{i},J^{j}]=i\epsilon^{ijk}J^{k}
\end{equation}
provided that the Hamiltonian satisfies
\begin{equation}
[H_{0},V]=[V,\mathbf{P}_{0}]=[V,\mathbf{J}_{0}]=0. \label{eq:smatrix_LI_condition}
\end{equation}
Assuming that eq.~(\ref{eq:smatrix_LI_condition}) is satisfied, one can see that $\mathbf{P}_{0}$ and $\mathbf{J}_{0}$ commute with $U(\tau)$ since it is a function of $H_{0}$ and $V$. Therefore, $\mathbf{P}_{0}$ and $\mathbf{J}_{0}$ commute with the $S$-matrix
\begin{equation}
[\mathbf{P}_{0},S]=[\mathbf{J}_{0},S]=0.
\end{equation}
This leaves us with the boost generator $\mathbf{K}$. It is not possible to take $\mathbf{K}=\mathbf{K}_{0}$ for this would lead to $[P_{i},K_{j}]=i\delta_{ij}H_{0}$ which is not possible for an interacting theory. Let us define $\mathbf{K}$ as
\begin{equation}
\mathbf{K}=\mathbf{K}_{0}+\mathbf{W}
\end{equation}
where $\mathbf{W}$ is an operator to be determined. We require $\mathbf{K}$ to satisfy the Poincar\'{e} algebra $[K_{i},H]=-iP_{i}$, this gives us
\begin{equation}
[\mathbf{K}_{0},V]=-[\mathbf{W},H]. \label{eq:k0v}
\end{equation}
This equation, by itself is meaningless for we can always define $\mathbf{W}$ with matrix element  
\begin{equation}
\langle\beta|\mathbf{W}|\alpha\rangle=\frac{-\langle\beta|[\mathbf{K}_{0},V]|\alpha\rangle}{(p^{0}_{\beta}-p^{0}_{\alpha})}
\end{equation}
that satisfies the above equation where $|\alpha\rangle$ and $\beta\rangle$ are energy eigenstates of $H$. The requirement of Lorentz invariance is that the in and out states are to transform under the same generators. The fact that $\mathbf{J}$, $\mathbf{K}$, $H$ and $\mathbf{P}$ satisfy the Poincar\'{e} algebra is a necessary but not sufficient condition.

Equation (\ref{eq:k0v}) becomes useful when we demand $\mathbf{W}$ to be a smooth function of energy with no singularities of the form $(p^{0}_{\alpha}-p^{0}_{\beta})^{-1}$. We now prove the commutativity between $\mathbf{K}_{0}$ and $S$. The commutator $[\mathbf{K}_{0},e^{iH_{0}t}]$ is 
\begin{equation}
[\mathbf{K}_{0},e^{iH_{0}t}]=t\mathbf{P}_{0} e^{iH_{0}t}
\end{equation}
where we have used the Poincar\'{e} algebra $[\mathbf{K}_{0},H_{0}]=-i\mathbf{P}$. Equivalently, we also have 
\begin{equation}
[\mathbf{K},e^{iHt}]=t\mathbf{P} e^{iHt}.
\end{equation}
The commutator between $\mathbf{K}_{0}$ and $U(\tau,\tau_{0})$ at finite time is then
\begin{equation}
[\mathbf{K}_{0},U(\tau,\tau_{0})]=-\mathbf{W}(\tau)U(\tau,\tau_{0})+U(\tau,\tau_{0})\mathbf{W}(\tau_{0}) \label{eq:k0u}
\end{equation}
where $\mathbf{W}(\tau)$ is defined as
\begin{equation}
\mathbf{W}(\tau)= e^{iH_{0}\tau}\mathbf{W}e^{-iH_{0}\tau}.
\end{equation}
In the limit $\tau\rightarrow\pm\infty$ and $\tau_{0}\rightarrow\mp\infty$, the in and out states $|\alpha_{\pm}\rangle$ become the free particle state $|\alpha_{0}\rangle$ and the interaction vanishes $V\rightarrow0$. It follows that the matrix elements of $\mathbf{W}(\tau)$ with smooth superposition of eigenstates of $H_{0}$ must vanish in this limit thus giving us
\begin{equation}
[\mathbf{K}_{0},U(\infty,-\infty)]=[\mathbf{K}_{0},S]=0.
\end{equation}
We see that by choosing an appropriate operator $\mathbf{W}$ with smooth energy functions, the boost generator $\mathbf{K}_{0}$ commutes with the $S$-matrix.

From eq.~(\ref{eq:k0u}), taking $\tau=0$ and $\tau_{0}\rightarrow\mp\infty$ we get
\begin{equation}
\mathbf{K}\Omega(\mp\infty)=\Omega(\mp\infty)\mathbf{K}_{0}.
\end{equation}
Since $\mathbf{P}=\mathbf{P}_{0}$, $\mathbf{J}=\mathbf{J}_{0}$ and they commute with $\Omega(\tau)$ for all $\tau$, we get
\begin{equation}
\mathbf{P}\Omega(\mp\infty)=\Omega(\mp\infty)\mathbf{P}_{0},\hspace{0.5cm}
\mathbf{J}\Omega(\mp\infty)=\Omega(\mp\infty)\mathbf{J}_{0}.
\end{equation}
Since $|\alpha\rangle$ and $|\alpha_{0}\rangle$ have the same energy spectrum with respect to $H$ and $H_{0}$, the same relation also hold for $H$ and $H_{0}$,
\begin{equation}
H\Omega(\mp\infty)=H_{0}\Omega(\mp\infty).
\end{equation}
Therefore, the generators $\mathbf{J}$, $\mathbf{K}$, $\mathbf{P}$ and $H$ are similarity transformations of the free-particle generators, they trivially satisfy the Poincar\'{e} algebra. More importantly, their action on the in and out state are the same thus satisfying the condition of Lorentz invariance.

\subsubsection{Discrete and internal symmetries}\label{discrete_symmetry_S_matrix}
In nature, besides continuous and discrete symmetries, there also exists internal symmetries. The internal symmetry is not directly related to Lorentz invariance. The symmetry between particle and anti-particle and the interchange of neutron and proton in nuclear reactions are examples of internal symmetry. Since the internal symmetries are not related to Lorentz invariance, they do not change the momentum and spin-projections of the particle states. Therefore, they must be the same in all inertial frames.

The internal symmetry $T$ acts on the particle state in the Hilbert space through its unitary representation $U(T)$, their action on the multi-particle state is defined as
\begin{equation}
U(T)|p_{1}\sigma_{1}n_{1},p_{2}\sigma_{2}n_{2},\cdots\rangle
=\sum_{\bar{n}_{1}\bar{n}_{2}\cdots}\mathscr{D}_{\bar{n}_{1}n_{1}}(T)\mathscr{D}_{\bar{n}_{2}n_{2}}(T)\cdots
|p_{1}\sigma_{1}\bar{n}_{1},p_{2}\sigma_{2}\bar{n}_{2},\cdots\rangle.\label{eq:int_sym}
\end{equation}
Since $U(T)$ furnishes a representation, it follows that the matrix $\mathscr{D}(T)$ satisfies
\begin{equation}
\mathscr{D}(\bar{T})\mathscr{D}(T)=\mathscr{D}(\bar{T}T).
\end{equation}
Additionally, Lorentz invariance imposes the requirement that $U(T)$ acts the same way on the in and out states. This will be the case if there exists a $U_{0}(T)$ that acts on the free multi-particle states according to eq.~(\ref{eq:int_sym}) and that it commutes with the free and interacting Hamiltonian,
\begin{equation}
U_{0}(T)H_{0}U_{0}(T)^{-1}=H_{0},\hspace{0.5cm} U_{0}(T)VU_{0}(T)^{-1}=V.
\end{equation}
Since the in and out states are given by $|\alpha_{\pm}\rangle=\Omega(\mp\infty)|\alpha_{0}\rangle$, we get 
\begin{equation}
U(T)=\Omega(\mp\infty)U_{0}(T)\Omega^{-1}(\mp\infty)\label{eq:int_symm_op}
\end{equation}
thus ensuring it acts on the in and out state in the same way and the unitarity of $\mathscr{D}(T)$
\begin{equation}
\mathscr{D}^{\dag}(T)=\mathscr{D}^{-1}(T).
\end{equation}
The $S$-matrix, under $U(T)$ transforms as
\begin{eqnarray}
S_{(p'_{1}\sigma_{1}'n_{1}',p'_{2}\sigma_{2}'n_{2}',\cdots)(p_{1}\sigma_{1}n_{1},p_{2}\sigma_{2}n_{2},\cdots)}&=&
\sum_{\bar{N}_{1}\bar{N}_{2}\cdots,\bar{N}'_{1}\bar{N}'_{2}\cdots}
\mathscr{D}_{\bar{N}_{1}n_{1}}(T)\mathscr{D}_{\bar{N}_{2}n_{2}}(T)\cdots
\mathscr{D}^{*}_{\bar{N}'_{1}n'_{1}}(T)\mathscr{D}^{*}_{\bar{N}'_{2}n'_{2}}(T)\cdots \nonumber \\ 
&&\hspace{1cm}\times
S_{(p'_{1}\sigma_{1}'\bar{N}'_{1},p'_{2}\sigma_{2}'\bar{N}'_{2},\cdots)(p_{1}\sigma_{1}\bar{N}_{1},p_{2}\sigma_{2}\bar{N}_{2},\cdots)}. \label{eq:int_sym_smatrix}
\end{eqnarray}
This shows that the matrix $\mathscr{D}(T)$ commutes with the $S$-matrix.

A special case of physical importance is a class of internal symmetry of the form
\begin{equation}
T(\bar{\theta})T(\theta)=T(\bar{\theta}+\theta).
\end{equation}
Its corresponding operator in the Hilbert space is
\begin{equation}
U(T(\theta))=e^{iQ\theta} 
\end{equation}
where $Q$ is a Hermitian operator. The matrix $\mathscr{D}(T)$ takes the form
\begin{equation}
\mathscr{D}_{n'n}(T(\theta))=\delta_{n'n}e^{iq_{n}\theta} \label{eq:int_sym_q}
\end{equation}
where $q_{n}$ is a set of real numbers that depend on the particle species. Substituting eq.~(\ref{eq:int_sym_q}) into eq.~(\ref{eq:int_sym_smatrix}), we see that since the left-hand side is independent of $q_{n}$ so it must also hold for the right-hand side. Therefore, for such a class of internal symmetry, the quantity $q_{n}$ is conserved
\begin{equation}
q_{n_{1}}+q_{n_{2}}+\cdots=q_{n_{1}'}+q_{n_{2}'}+\cdots.
\end{equation}
The conservation of electric charge is one example that belongs to this class of internal symmetry.

Discrete symmetries such as parity and time-reversal are not internal symmetry operators. However their invariances or violations involved in particle interactions are defined in a similar manner, in terms of the equality or inequality of intrinsic parity and time-reversal phases before and after the interactions.
\subsubsection{Parity}
If parity is a symmetry of a theory, then there must exist a parity operator $\mathsf{P}$ in the Hilbert space that acts in the same way on the in and out state
\begin{equation}
\mathsf{P}|p_{1}\sigma_{1}n_{1},p_{2}\sigma_{2}n_{2},\cdots_{\pm}\rangle=\eta_{n_{1}}\eta_{n_{2}}\cdots
|\mathscr{P}p_{1}\sigma_{1}n_{1},\mathscr{P}p_{2}\sigma_{2}n_{2},\cdots_{\pm}\rangle
\end{equation}
where $\eta_{n_{i}}$ is the intrinsic parity for particle of specie $n_{i}$. The $S$-matrix, under parity transforms as
\begin{eqnarray}
&& S_{(p_{1}'\sigma_{1}'n_{1}',p_{2}'\sigma_{2}'n_{2}',\cdots)(p_{1}\sigma_{1}n_{1},p_{2}\sigma_{2}n_{2},\cdots)}\\ \nonumber
&&\hspace{2cm}=\eta_{n_{1}}\eta_{n_{2}}\cdots \eta_{n_{1}}'^{*}\eta_{n_{2}}'^{*}\cdots
S_{(\mathscr{P}p_{1}'\sigma_{1}'n_{1}',\mathscr{P}p_{2}'\sigma_{2}'n_{2}',\cdots)(\mathscr{P}p_{1}\sigma_{1}n_{1},\mathscr{P}p_{2}\sigma_{2}n_{2},\cdots)}. \label{eq:p_smatrix}
\end{eqnarray}
Similar to the internal symmetry studied earlier, the parity operator $\mathsf{P}$ exists and satisfies Lorentz invariance as long as there is a $\mathsf{P}_{0}$ that acts on the free particle state and commutes with the free and interacting Hamiltonian.

The intrinsic parity $\eta_{n_{i}}$ can be measured through experiments, but they cannot be uniquely determined since $\mathsf{P}$ is not unique. Given a parity operator $\mathsf{P}$, one can always define a new parity operator $\mathsf{P}'$ as
\begin{equation}
\mathsf{P}'=\mathsf{P}\exp[i(\alpha B+\beta L+\gamma Q)]
\end{equation}
where $B$, $L$ and $Q$ are the baryon number, lepton number and electric charge respectively, the phases $\alpha$, $\beta$ and $\gamma$ are arbitrary real numbers. Since the phases are arbitrary, for a given particle, one can choose the phases such that the particle have an intrinsic parity of +1. Once this is chosen, the intrinsic parities of other particles involved in the interaction are fixed. For example, in neutron decay $n\rightarrow p+\pi^{-}$, if we choose $\eta_{n}=1$, then parity conservation imposes $\eta_{p}\eta_{\pi^{-}}=1$.

The parity operator maps the three-momentum $\mathbf{p}\rightarrow-\mathbf{p}$, so it is not an internal symmetry operator. However, since $\mathsf{P}^{2}$ leaves the three-momentum invariant, it is an internal symmetry operator
\begin{equation}
\mathsf{P}^{2}|p_{1}\sigma_{1}n_{1},p_{2}\sigma_{2}n_{2}\cdots_{\pm}\rangle=\eta^{2}_{n_{1}}\eta^{2}_{n_{2}}\cdots 
|p_{1}\sigma_{1}n_{1},p_{2}\sigma_{2}n_{2}\cdots_{\pm}\rangle.
\end{equation}

Independent of the choice of phases and intrinsic parities, from eq.~(\ref{eq:p_smatrix}) we can see that the $S$-matrix on the left-hand side is equal to the momentum-reflected $S$-matrix up to the product of intrinsic parities. Since the cross-section rate is proportional to the magnitude of the $S$-matrix squared, therefore, if parity is a symmetry of the theory, the cross-section of $\alpha\rightarrow \beta$ must then be equal to the cross-section of $\mathscr{P}\alpha\rightarrow\mathscr{P}\beta$. 

For example, parity is a symmetry in quantum electrodynamics. However, experiments studying the decay of $\mbox{Co}^{60}\rightarrow \mbox{Ni}^{60}+e^{-}+\bar{\nu}$ found an excess of electrons in the direction opposite to the spin of the nucleus~\cite{Wu:1957my}. This result would not be possible had parity been conserved. Therefore, parity is violated in the weak interaction. 

\subsubsection{Time reversal}

If there exists a time-reversal operator $\mathsf{T}_{0}$ for the free-particle state, then one can simply take $\mathsf{T}=\mathsf{T}_{0}$ to be the time-reversal operator for the in and out state provided that $\mathsf{T}_{0}$ commutes with the free and interacting Hamiltonian.

The action of $\mathsf{T}_{0}$ on the free multi-particle state is
\begin{eqnarray}
\mathsf{T}_{0}|p_{1}\sigma_{1}n_{1},p_{2}\sigma_{2}n_{2},\cdots_{\pm}\rangle&=&\varrho_{n_{1}}(-1)^{j_{n_{1}}-\sigma_{n_{1}}}\varrho_{n_{2}}(-1)^{j_{n_{2}}-\sigma_{n_{2}}}\cdots \\ \nonumber
&&\hspace{1cm}\times|\mathscr{P}p_{1}(-\sigma_{1})n_{1},\mathscr{P}p_{2}(-\sigma_{2})n_{2},\cdots_{\pm}\rangle.
\end{eqnarray}
For simplicity, we write this as $\mathsf{T}_{0}|\alpha_{0}\rangle=|\mathscr{T}\alpha\rangle$ where $\mathscr{T}$ denotes all the phases and reflects the momenta and spin-projections.
Recall that the time-reversal operator $\mathsf{T}$ is anti-unitary and $|\alpha_{\pm}\rangle=\Omega(\mp\infty)|\alpha_{0}\rangle$, it follows that the action of $\mathsf{T}$ interchanges the in and out states, 
\begin{equation}
\mathsf{T}|\alpha_{\pm}\rangle=|\mathscr{T}\alpha_{\mp}\rangle.
\end{equation}
As a result, the $S$-matrix under time-reversal becomes
\begin{eqnarray}
S_{(p_{1}'\sigma_{1}'n_{1}',p_{2}'\sigma_{2}'n_{2}',\cdots)(p_{1}\sigma_{1}n_{1},p_{2}\sigma_{2}n_{2},\cdots)}&=&
\varrho^{*}_{n_{1}'}(-1)^{j_{1}'-\sigma_{1}'}\varrho^{*}_{n_{2}'}(-1)^{j_{2}'-\sigma_{2}'}\cdots
\varrho_{n_{1}}(-1)^{j_{1}-\sigma_{1}}\varrho_{n_{2}}(-1)^{j_{2}-\sigma_{2}}\cdots  \nonumber \\
&&\hspace{-0.2cm}\times
S_{(\mathscr{P}p_{1}'(-\sigma_{1})'n_{1}',\mathscr{P}p_{2}'(-\sigma_{2})'n_{2}',\cdots)(\mathscr{P}p_{1}(-\sigma_{1})n_{1}\mathscr{P}p_{2}(-\sigma_{2})n_{2}\cdots)}
\end{eqnarray}
or simply as $S_{\beta\alpha}=S_{\mathscr{T}\alpha\mathscr{T}\beta}$.

Unlike parity, generally the conservation of time-reversal symmetry does not imply equal cross-section between $\alpha\rightarrow\beta$ and $\mathscr{T}\alpha\rightarrow\mathscr{T}\beta$ due to the $\sigma$-dependent phases. The case where this cross section is equal is when the $S$-matrix can be written in terms of a weak and dominant term
\begin{equation}
S_{\beta\alpha}=S^{(0)}_{\beta\alpha}+S^{(1)}_{\beta\alpha}\label{eq:s_split}
\end{equation}
where $S^{(0)}_{\beta\alpha}\gg S^{(1)}_{\beta\alpha}$. A detailed discussion can be found in Weinberg~\cite[sec.~3.3]{Weinberg:1995mt}.

\subsubsection{Charge-conjugation and product of discrete symmetries}

Charge-conjugation is the symmetry between particle and anti-particle. Let $\mathsf{C}$ be the unitary representation of charge-conjugation. The conjugation operator maps the one-particle state $|p,\sigma,n\rangle$ to its anti-particle state $|p,\sigma,n^{c}\rangle$
\begin{equation}
\mathsf{C}\,|p,\sigma,n\rangle=\varsigma_{n}|p,\sigma,n^{c}\rangle
\end{equation}
where $\varsigma_{n}$ is the charge-conjugation phase which may depend on the particle species. Since $\mathsf{C}$ is an internal symmetry operator, the existence of $\mathsf{C}_{0}$ for free-particle state ensures $\mathsf{C}$ is well-defined and act on the in and out state in the same way as given by eq.~(\ref{eq:int_symm_op}). The $S$-matrix under charge-conjugation transforms as
\begin{eqnarray}
S_{(p'_{1}\sigma_{1}'n_{1}',p'_{2}\sigma_{2}'n_{2}',\cdots)(p_{1}\sigma_{1}n_{1},p_{2}\sigma_{2}n_{2},\cdots)}&=&({\varsigma_{1}'}^{*}{\varsigma_{2}'}^{*}\cdots)(\varsigma_{1}\varsigma_{2}\cdots)\\ \nonumber
&&\hspace{0.5cm}
\times S_{(p'_{1}\sigma_{1}'{n_{1}^{c}}',p'_{2}\sigma_{2}'{n_{2}^{c}}',\cdots)(p_{1}\sigma_{1}n_{1}^{c},p_{2}\sigma_{2}n_{2}^{c},\cdots)}.
\end{eqnarray}
Noting that the charge-conjugation phases are $\sigma$-independent, for interactions that conserve charge-conjugation, the cross-section between $\alpha\rightarrow\beta$ equals the cross-section of $\mathscr{C}\alpha\rightarrow\mathscr{C}\beta$.

While charge-conjugation is not directly related to Lorentz symmetry, the product $\mathsf{CPT}$ plays an important role in space-time symmetry due to the $\mathsf{CPT}$ theorem~\cite{Pauli:1957,Jost:1957zz}, that any local interacting theory that satisfies Lorentz symmetry will also satisfy $\mathsf{CPT}$ symmetry. Although the theorem does not explain the origin of discrete symmetry violations, its conservation allows us to identify the conservation and violation of individual discrete symmetries. For example, in the study of long-lived neutral Kaon decay, it is found that $\mathsf{CP}$ is violated~\cite{Christenson:1964fg}. As a result, $\mathsf{T}$ must also be violated in order to conserve $\mathsf{CPT}$.

Like time reversal, the $\mathsf{CPT}$ operator is an anti-unitary operator, it maps the particle state to its anti-particle state with the spin-projection reversed,
\begin{equation}
\mathsf{CPT}|p_{1}\sigma_{1}n_{1},p_{2}\sigma_{2}n_{2}\cdots\rangle=\wp_{\sigma_{1}}\wp_{\sigma_{2}}\cdots
|p_{1}(-\sigma_{1})n_{1}^{c},p_{2}(-\sigma_{2})n_{2}^{c}\cdots\rangle
\end{equation}
where $\wp_{\sigma_{i}}$ is the combined $\mathsf{CPT}$ phases with $\sigma$-dependence. The $S$-matrix, under the action of $\mathsf{CPT}$ written in the compact form is
\begin{equation} 
S_{\alpha\beta}=S_{\mathscr{CPT}\alpha\mathscr{CPT}\beta}
\end{equation}
with the phases absorbed in $\mathscr{CPT}$. Since the phases $\wp_{\sigma_{i}}$ are $\sigma$-dependent, therefore, similar to the case of time-reversal, the cross section of $\alpha\rightarrow\beta$ will be the same as $\mathscr{CPT}\alpha\rightarrow\mathscr{CPT}\beta$ only when $S$-matrix can be separated into two terms according to eq.~(\ref{eq:s_split}). 

\subsection{Cross-sections}
In the previous sections, we have defined and studied the symmetries of the $S$-matrix. Here, we will use the $S$-matrix to calculate the cross-sections and particle decay rates measured in experiments. 

The observables such as cross-section and decay rates are proportional to the probability given by $|S_{\beta\alpha}|^{2}$ for the process $\alpha\rightarrow\beta$. The $S$-matrix encodes the information on the interaction between multi-particle states. In particular, the energy-momentum conservation for a given process $\alpha\rightarrow\beta$ is imposed through the delta function $\delta^{4}(p_{\alpha}-p_{\beta})$. Therefore, in calculating observables, we will get factors proportional to $[\delta^{4}(p_{\alpha}-p_{\beta})]^{2}$. In this section, we will show how to treat the square of the delta function.

We start by considering $N$ number of particles confined in a square box of finite volume of $L^{3}$. This quantises the momenta of the particles to
\begin{equation}
\p=(2\pi/L)(n_{1},n_{2},n_{3})\label{eq:quantised_momentum}
\end{equation}
where $n_{i}$ are integers specifying the states of momenta $\p$. The delta function is defined as
\begin{equation}
\delta^{3}_{V}(\mathbf{p-p'})=\frac{1}{(2\pi)^{3}}\int_{V}d^{3}x\,e^{i\mathbf{(p-p')}\cdot\mathbf{x}}=\frac{V}{(2\pi)^{3}}\delta_{\mathbf{pp'}}.
\label{eq:finite_v_delta}
\end{equation}
The inner product between $N$ multi-particle state in the box is then defined with respect to the finite volume delta function
\begin{eqnarray}
&&\langle p_{1}'\sigma_{1}'n_{1}',p_{2}'\sigma_{2}'n_{2}',\cdots | p_{1}\sigma_{1}n_{1},p_{2}\sigma_{2}n_{2},\cdots\rangle \nonumber \\
&&\hspace{2cm}=\left[\frac{V}{(2\pi)^{3}}\right]^{N}\big(\delta_{\mathbf{p_{1}'p_{1}}}\delta_{\sigma_{1}'\sigma_{1}}\delta_{n_{1}'n_{1}}\delta_{\mathbf{p_{2}'p_{2}}}
\delta_{\sigma_{2}'\sigma_{2}}\delta_{n_{2}'n_{2}}\cdots \nonumber \\
&&\hspace{7.5cm} \pm \mbox{permutations}\big).
\end{eqnarray}
To calculate the transition probability, we need to use states with unit norm, so we define the normalised state in the box as
\begin{equation}
|\alpha_{\mbox{\scriptsize{Box}}}\rangle=\left[\frac{(2\pi)^{3}}{V}\right]^{N_{\alpha}/2}|\alpha\rangle
\end{equation}
to obtain the orthonormal state $\langle \beta_{\scriptsize{\mbox{Box}}}|\alpha_{\scriptsize{\mbox{Box}}}\rangle=\delta_{\beta\alpha}$ where $\delta_{\beta\alpha}$ includes the products of all the delta functions and the permuting terms. Therefore, the $S$-matrix $S_{\alpha\beta}$ is
\begin{equation}
S_{\beta\alpha}=\left[\frac{V}{(2\pi)^{3}}\right]^{(N_{\alpha}+N_{\beta})/2}(S_{\mbox{\scriptsize{Box}}})_{\beta_\alpha}
\end{equation}
where $S_{\mbox{\scriptsize{Box}}}$ is the $S$-matrix in the box. When appropriate, we may take the infinite volume limit $V\rightarrow\infty$ to obtain the desired results.

The transition probability from $\alpha\rightarrow\beta$ in the box during the interaction is
\begin{equation}
P(\alpha\rightarrow\beta)=|(S_{\mbox{\scriptsize{Box}}})_{\beta_\alpha}|^{2}=\left[\frac{(2\pi)^{3}}{V}\right]^{N_{\alpha}+N_{\beta}}
|S_{\beta\alpha}|^{2}.
\end{equation}
Here we are interested in the differential probability $dP(\alpha\rightarrow\beta)$ in which the final state $\beta$ is found within the range of $d\beta=d^{3}p_{1}\cdots d^{3}p_{N_{\beta}}$. The number of particles within the momentum of $d^{3}p$ is given by $V d^{3}p/(2\pi)^{3}$ since this is the number of triplets $(n_{1},n_{2},n_{3})$ within $d^{3}p$.\footnote{Equation (\ref{eq:quantised_momentum}) shows that the number of triplets within $d^{3}p$ is $d^{3}n=dn_{1}dn_{2}dn_{3}=Vd^{3}p/(2\pi)^{3}$.} Therefore, the number of particles in the range of $d\beta$ is
\begin{equation}
dN_{\beta}=\left[\frac{V}{(2\pi)^{3}}\right]^{N_{\beta}}d\beta
\end{equation}
and the differential probability is
\begin{equation}
dP(\alpha\rightarrow\beta)=P(\alpha\rightarrow\beta)dN_{\beta}=\left[\frac{(2\pi)^{3}}{V}\right]^{N_{\alpha}}
|S_{\beta\alpha}|^{2}d\beta.
\end{equation}

In actual particle experiments, the interactions between particles only happens in a short period of time. We can model this using particle states in the box such that the interaction is switched on for a period of time $T$. This is achieved by defining the following delta function
\begin{equation}
\delta_{T}(p^{0}_{\alpha}-p^{0}_{\beta})=\frac{1}{2\pi}\int_{-T/2}^{T/2}dt\, e^{i(p^{0}_{\alpha}-p^{0}_{\beta})t}.
\end{equation}

Assuming that no states in $|\alpha\rangle$ and $|\beta\rangle$ have the same energies or momenta, we may write the $S$-matrix during the interaction as
\begin{equation}
S_{\beta\alpha}=-2i\pi\delta_{V}^{3}(\p_{\alpha}-\p_{\beta})\delta_{T}(p^{0}_{\beta}-p^{0}_{\alpha})M_{\beta\alpha}.\label{eq:smatrix_finite_v}
\end{equation}
The delta functions in the box of finite volume can now be written as
\begin{equation}
[\delta^{3}_{V}(\p_{\beta}-\p_{\alpha})]^{2}=\delta^{3}_{V}(\p_{\beta}-\p_{\alpha})\delta^{3}_{V}(\0)
=\delta^{3}_{V}(\p_{\beta}-\p_{\alpha})\frac{V}{(2\pi)^{3}},\label{eq:delta_v}
\end{equation}
\begin{equation}
[\delta_{T}(p^{0}_{\beta}-p^{0}_{\alpha})]^{2}=\delta_{T}(p^{0}_{\beta}-p^{0}_{\alpha})\delta_{T}(0)
=\delta_{T}(p^{0}_{\beta}-p^{0}_{\alpha}))\frac{T}{2\pi}.\label{eq:delta_t}
\end{equation}
Substituting eqs.~(\ref{eq:smatrix_finite_v}-\ref{eq:delta_t}) into $dP(\alpha\rightarrow\beta)$ we get
\begin{equation}
dP(\alpha\rightarrow\beta)=(2\pi)^{3N_{\alpha}-2}TV^{1-N_{\alpha}}\delta_{T}(p^{0}_{\beta}-p^{0}_{\alpha})
\delta^{3}_{V}(\p_{\beta}-\p_{\alpha})
|M_{\beta\alpha}|^{2}d\beta.
\end{equation}
The differential probability is proportional to the duration of the interaction $T$. Therefore, the differential rate of the interaction is
\begin{equation}
d\Gamma(\alpha\rightarrow\beta)=\frac{dP_{\alpha\rightarrow\beta}}{T}=(2\pi)^{3N_{\alpha}-2}V^{1-N_{\alpha}}
\delta^{4}(p_{\beta}-p_{\alpha})|M_{\beta\alpha}|^{2}d\beta.
\end{equation}

In the limit $T$ and $V$ become large, we obtain the Dirac-delta function so the $S$-matrix becomes
\begin{equation}
S_{\beta\alpha}=-2i\pi\delta^{4}(p_{\beta}-p_{\alpha})M_{\beta\alpha}.\label{eq:master_s_matrix}
\end{equation}
This formula for the $S$-matrix allows us to calculate the cross sections and decay rates measured by experiments. The interpretation of the square of the delta function is discussed below. We now consider some special cases that are important in particle interactions.

\subsubsection{One-particle decay: $\mathbf{N_{\alpha}=1}$}
Generally, this means the initial one-particle state $|\alpha\rangle$ decays into a multi-particle state $|\beta\rangle$. The differential
reaction rate is
\begin{equation}
d\Gamma(\alpha\rightarrow\beta)=(2\pi)\delta^{4}(p_{\beta}-p_{\alpha})|M_{\beta\alpha}|^{2}d\beta.
\end{equation}
This formula only makes physical sense if the period of interaction $T$ is much shorter than the average decay time $\tau_{\alpha}$. The decay rate is independent of the volume $V$.

\subsubsection{Two particle interactions: $\mathbf{N_{\alpha}=2}$}
This case describes the interaction between two initial particles and its subsequent products. The differential reaction rate is
\begin{equation}
d\Gamma(\alpha\rightarrow\beta)=(2\pi)^{4}V^{-1}\delta^{4}(p_{\beta}-p_{\alpha})|M_{\beta\alpha}|^{2},
\end{equation}
inversely proportional to the volume $V$. The cross-section of the interaction is defined as the transition rate per flux of the initial particle
\begin{equation}
d\sigma(\alpha\rightarrow\beta)=\frac{d\Gamma(\alpha\rightarrow\beta)}{\Phi_{\alpha}}
\end{equation}
where $\Phi_{\alpha}=u_{\alpha}/V$ and $u_{\alpha}$ is relative velocity between the initial particles which will be defined shortly. Therefore, the cross-section is given by
\begin{equation}
d\sigma(\alpha\rightarrow\beta)=\frac{d\Gamma(\alpha\rightarrow\beta)}{\Phi_{\alpha}}=(2\pi)^{4}u_{\alpha}^{-1}\delta^{4}(p_{\beta}-p_{\alpha})
|M_{\beta\alpha}|^{2}d\beta.
\end{equation}

We now consider how the transformation of the $S$-matrix relates to the transformation of the cross-section and decay rates. For this purpose, we need to determine the transformation properties of the matrix element $|M_{\beta\alpha}|^{2}$. Substituting eq.~(\ref{eq:master_s_matrix}) into the left and right-hand side of eq.~(\ref{eq:smatrix_trans}) and factoring out the delta-function, we get\footnote{The integral representation of the delta function is
\begin{equation}
\delta^{4}[(\Lambda p)_{\beta}-(\Lambda p)_{\alpha}]=\int\frac{d^{4}x}{(2\pi)^{4}}e^{i[(\Lambda p)_{\beta}-(\Lambda p)_{\alpha}]\cdot x}.\nonumber
\end{equation}
Perform a change of variable, $x=\Lambda x'$, we obtain
\begin{equation}
\delta^{4}[(\Lambda p)_{\beta}-(\Lambda p)_{\alpha}]=\det\Lambda\,\delta^{4}(p_{\beta}-p_{\alpha})=\delta^{4}(p_{\beta}-p_{\alpha}) \nonumber
\end{equation}
since $\Lambda$ is an element of the proper orthochronous Lorentz group.
}
\begin{eqnarray}
&& M_{(p'_{1}\sigma_{1}'n_{1}',p'_{2}\sigma_{2}'n_{2}'\cdots)
(p_{1}\sigma_{1}n_{1},p_{2}\sigma_{2}n_{2}\cdots)}  \\  \nonumber
&&\hspace{1cm}=
e^{ia\cdot(\Lambda p_{1}+\Lambda p_{2}+\cdots-\Lambda p'_{1}-\Lambda p'_{2}\cdots)}
\sqrt{\frac{(\Lambda p_{1})^{0}(\Lambda p_{2})^{0}\cdots (\Lambda p'_{1})^{0}(\Lambda p'_{2})^{0}\cdots}{p^{0}_{1}p^{0}_{2}\cdots p'^{0}p'^{1}\cdots}} \\ \nonumber
&&\hspace{1.5cm}\times\sum_{\bar{\sigma}_{1}\bar{\sigma}_{2}\cdots}D_{\bar{\sigma}_{1}\sigma_{1}}^{(j_{1})}(W(\Lambda,p_{1}))D_{\bar{\sigma}_{2}\sigma_{2}}^{(j_{2})}(W(\Lambda,p_{2}))\cdots \\ \nonumber
&&\hspace{1.5cm}\times\sum_{\bar{\sigma}_{1}'\bar{\sigma}_{2}'\cdots}
D'^{(j'_{1})*}_{\bar{\sigma}'_{1}\sigma'_{1}}(W(\Lambda,p'_{1}))
D'^{(j'_{2})*}_{\bar{\sigma}'_{2}\sigma'_{2}}(W(\Lambda,p'_{2}))\cdots \\ \nonumber
&&\hspace{1.5cm}\times
 M_{(\Lambda p'_{1}\sigma_{1}'n_{1}',\Lambda p'_{2}\sigma_{2}'n_{2}'\cdots)
(\Lambda p_{1}\sigma_{1}n_{1},\Lambda p_{2}\sigma_{2}n_{2}\cdots)}. \nonumber
\end{eqnarray}
Exploiting the unitarity of the $D(W(\Lambda,p))$ matrix, multiply both sides by its adjoint, sum over all the spin-projections and squaring them, we obtain
\begin{eqnarray}
&&\prod_{\alpha} p^{0}_{\alpha}\prod_{\beta}p^{0}_{\beta}\sum_{\sigma_{1}\sigma_{2}\cdots}\sum_{\sigma_{1}'\sigma_{2}'\cdots} 
|M_{(p'_{1}\sigma_{1}'n_{1}',p'_{2}\sigma_{2}'n_{2}'\cdots)(p_{1}\sigma_{1}n_{1},p_{2}\sigma_{2}n_{2}\cdots)}|^{2}\\ \nonumber
&&\hspace{1cm}=
\prod_{\alpha} (\Lambda p_{\alpha})^{0}\prod_{\beta}(\Lambda p_{\beta})^{0}
\sum_{\bar{\sigma}_{1}\bar{\sigma}_{2}\cdots}\sum_{\bar{\sigma}_{1}'\bar{\sigma}_{2}'\cdots}
|M_{(\Lambda p'_{1}\bar{\sigma}_{1}'n_{1}',\Lambda p'_{2}\bar{\sigma}_{2}'n_{2}'\cdots)(\Lambda p_{1}\bar{\sigma}_{1}n_{1},\Lambda p_{2}\bar{\sigma}_{2}n_{2}\cdots)}|^{2}.
\end{eqnarray}
In a more compact notation, this equation shows that the product
\begin{equation}
\sum_{\mbox{\scriptsize{spins}}}|M_{\beta\alpha}|^{2}\prod_{\beta}p^{0}_{\beta}\prod_{\alpha}p^{0}_{\alpha}=R_{\beta\alpha}
\end{equation}
is Lorentz invariant. 

This expression tells us how the spin-projection as well as the energies of the initial and final particles contribute to the cross-section. 
In the case of one-particle decay, the decay rate is
\begin{equation}
\sum_{\mbox{\scriptsize{spins}}}d\Gamma(\alpha\rightarrow\beta)=(2\pi)(p^{0}_{\alpha})^{-1}R_{\beta\alpha}\delta^{4}(p_{\beta}-p_{\alpha})d\beta/\prod_{\beta}p^{0}_{\beta}.\label{eq:total_spin_cs_1}
\end{equation}
For $N_{\alpha}=2$, the cross-section is
\begin{equation}
\sum_{\mbox{\scriptsize{spins}}}d\sigma(\alpha\rightarrow\beta)=(2\pi)^{4}u_{\alpha}^{-1}(p^{0}_{1}p^{0}_{2})^{-1}R_{\beta\alpha}
\delta^{4}(p_{\beta}-p_{\alpha})d\beta/\prod_{\beta}p^{0}_{\beta} \label{eq:total_spin_cs_2}
\end{equation}
where $p^{0}_{1}$ and $p^{0}_{2}$ represent the energy of the initial particles. 

On the right-hand side of eqs.~(\ref{eq:total_spin_cs_1}) and (\ref{eq:total_spin_cs_2}), all the terms $\delta^{4}(p_{\beta}-p_{\alpha})$, $R_{\beta\alpha}$ and $d\beta/\left(\prod_{\beta}p^{0}_{\beta}\right)$ are Lorentz invariant. Therefore, the cross-section for the $N_{\alpha}=1$, transforms as the inverse of energy.  This is in agreement with relativistic time-dilation where relativistic muons have longer life-time than their non-relativistic counterpart. 

Conventionally, the average cross-section summed over all spin-projection is taken to be a Lorentz-invariant function. So for $N_{\alpha}=2$, the product $u_{\alpha}p^{0}_{1}p^{0}_{2}$ must be Lorentz invariant.  The condition of Lorentz invariance and that $u_{\alpha}$ must be the physical velocity of the moving particle when measured in the rest frame of the other uniquely determines its expression to be
\begin{equation}
u_{\alpha}=\frac{\sqrt{(p_{1}^{\mu}\,p_{2\mu}^{})^{2}-(m_{1}m_{2})^{2}}}{p^{0}_{1}p^{0}_{2}}.
\end{equation}
Taking $p_{1}=(p^{0}_{1},\p_{1})$ and $p_{2}=(m,\0)$, we get
\begin{equation}
u_{\alpha}=\frac{\sqrt{(p^{0}_{1}m_{2})^{2}-(m_{1}m_{2})^{2}}}{p^{0}_{1}m_{2}}=\frac{|\p_{1}|}{p^{0}_{1}}
\end{equation}
giving us the physical velocity of particle 2 as measured by particle 1 in the rest frame. In the centre of mass frame where $p_{1}=(p^{0}_{1},\p)$ and $p_{2}=(p^{0}_{2},-\p)$, we get
\begin{equation}
u_{\alpha}=\frac{|\p|(p^{0}_{1}+p^{0}_{2})}{p^{0}_{1}p^{0}_{2}}=\left|\frac{\p_{1}}{p^{0}_{1}}-\frac{\p_{2}}{p^{0}_{2}}\right|.
\end{equation}
This justifies the term relative velocity for $u_{\alpha}$. However, this velocity is not physical, for ultra-relativistic particles where $|\p|\sim p^{0}$, it can get as large as 2.

We are now in the position to interpret the delta function. For simplicity, we choose the centre of mass frame for the initial particles where $\sum_{\alpha}\p_{\alpha}=\0$ so that $\delta^{4}(p_{\beta}-p_{\alpha})$ becomes
\begin{equation}
\delta^{4}(p_{\beta}-p_{\alpha})=\delta(p'^{0}_{1}+p'^{0}_{2}+\cdots - p^{0})\delta^{3}(\p_{1}'+\p_{2}'+\cdots)d^{3}\p_{1}'d^{3}\p_{2}'\cdots
\end{equation}
where $p^{0}$ is the total energy of the initial particles. Confining to the simple cases of interactions that produce two final particles $N_{\beta}=2$ with arbitrary number of initial particles in the centre of mass frame, the delta function becomes
\begin{equation}
\delta^{4}(p_{\beta}-p_{\alpha})=\delta(p'^{0}_{1}+p'^{0}_{2}- p^{0})\delta^{3}(\p_{1}'+\p_{2}')d^{3}\p_{1}'d^{3}\p_{2}'
\end{equation}
Integrating over the three-momentum $\p'_{2}$ by taking $\p_{2}'=-\p_{1}'$, we get
\begin{eqnarray}
\delta^{4}(p_{\beta}-p_{\alpha})&=&\delta(p'^{0}_{1}+p'^{0}_{2}-p^{0})|\p_{1}'|^{2}d
|\p_{1}'|d\Omega\nonumber\\
&=&\delta(\sqrt{|\p_{1}'|^{2}+m_{1}'}+\sqrt{|\p_{1}'|^{2}+m_{2}'}-p^{0})|\p_{1}'|^{2}d
|\p_{1}'|d\Omega.
\end{eqnarray}
where $d\Omega=\sin\theta\, d\theta d\phi$ is the solid angle of the particle with momentum $\p_{1}'$. 

The delta-function can be further simplified using the formula
\begin{equation}
\delta[f(x)]=\frac{\delta(x-x_{0})}{f'(x_{0})}
\end{equation}
where $f(x_{0})=0$ and $f'(x)=df/dx$. For the above delta function, the argument is a function of $|\p'_{1}|$, its root can be written in the following form
\begin{equation}
|\mathbf{k}'|=\frac{\sqrt{[(p^{0})^{2}-m_{1}'^{2}-m_{2}'^{2}]^{2}-4m_{1}'^{2}m_{2}'^{2}}}{2p^{0}}.\label{eq:EP_conserv1}
\end{equation}
The energy of the final particles of momentum in terms of $\mathbf{k}'$ are then
\begin{equation}
p'^{0}_{1}(\mathbf{k}')=\frac{(p^{0})^{2}+m_{1}'^{2}-m_{2}'^{2}}{2p^{0}},
\end{equation}
\begin{equation}
p'^{0}_{2}(\mathbf{k}')=\frac{(p^{0})^{2}+m_{2}'^{2}-m_{1}'^{2}}{2p^{0}}. \label{eq:EP_conserv2}
\end{equation}
Therefore, 
\begin{equation}
\frac{d}{d|\p_{1}'|}\left(\sqrt{|\p_{1}'|^{2}+m_{1}'}+\sqrt{|\p_{1}'|^{2}+m_{2}'}-p^{0}\right)\Big\vert_{|\mathbf{p}'_{1}|=|\mathbf{k}'|} =\frac{|\mathbf{k}'|p^{0}}{p'^{0}_{1}p'^{0}_{2}}
\end{equation}
Substituting this into the delta function, we get
\begin{equation}
\delta^{4}(p_{\beta}-p_{\alpha})\rightarrow\frac{|\mathbf{k}'|p'^{0}_{1}p'^{0}_{2}}{p^{0}}d\Omega\label{eq:delta_function_kinematic}
\end{equation}
where the final delta function $\delta(|\p_{1}'|-|\mathbf{k}'|)$ is removed by taking $|\p_{1}'|=|\mathbf{k}'|$. 


Replacing the delta function with eq.~(\ref{eq:delta_function_kinematic}), the decay rate for a single particle of energy $p^{0}$ into two particles becomes
\begin{equation}
\frac{d\Gamma}{d\Omega}(\alpha\rightarrow\beta)=2\pi |M_{\beta\alpha}|^{2}\frac{|\mathbf{k}'|p'^{0}_{1}p'^{0}_{2}}{p^{0}}\label{eq:decay_rate}
\end{equation}
and the differential cross-section for the $12\rightarrow1'2'$ process in the centre of mass frame reads
\begin{equation}
\frac{d\sigma}{d\Omega_{\mbox{\tiny{CM}}}}(\alpha\rightarrow\beta)=\frac{(2\pi)^{4} |\mathbf{k}'| p'^{0}_{1}p'^{0}_{2}}{p^{0}u_{\alpha}}|M_{\beta\alpha}|^{2}=
\frac{(2\pi)^{4} |\mathbf{k}'| p'^{0}_{1}p'^{0}_{2}p^{0}_{1}p^{0}_{2}}{(p^{0})^{2}|\mathbf{k}|}|M_{\beta\alpha}|^{2}\label{eq:diff_cross_section}
\end{equation}
where $|\mathbf{k}|=|\p_{1}|=|\p_{2}|$ and $p^{0}=p^{0}_{1}+p^{0}_{2}$.

\subsection{Perturbation theory}\label{perturbation_theory}

The definition of the $S$-matrix is simple, yet in most interacting theories, it is not possible to compute it exactly. This led to the development of perturbation theory, where the $S$-matrix is expanded in terms of the interacting Hamiltonian. 

In this section, we will show that the perturbation formalism of the $S$-matrix with the demand of Lorentz invariance put strong constraints on the type of interactions that are allowed. This result, along with cluster decomposition principle of the $S$-matrix will explain the reason behind the transition from relativistic quantum mechanics to quantum field theory.

Historically, there are two types of perturbations, the old-fashioned perturbation theory and the modern version based on the Dyson series~\cite{Dyson:1949bp}. The drawback of the former is that the Lorentz invariance is not manifest, which is the later is preferred. For the sake of completeness, both are derived. 

It should be noted that while the Dyson series is manifestly Lorentz invariant and unitary, there also exists alternative expansions that achieve the same goal. In particular, there are the Magnus and Fer series where the $S$-matrix is expanded as $S=\exp(i\eta)$ and $S=\prod_{k=1}^{\infty}\exp(F_{k})$ respectively which are manifestly unitary preserving to all orders of perturbation~\cite{Blanes2009151}. These expansions have not been as extensively explored as the Dyson series. Nevertheless, they may provide additional insights that are otherwise not obvious from the Dyson series. Therefore, this is a possible direction of future research.

\subsubsection{The old-fashioned perturbation theory}

We start with the Born approximation of the $S$-matrix
\begin{equation}
S_{\beta\alpha}=\delta(\beta-\alpha)-2i\pi\delta(p^{0}_{\beta}-p^{0}_{\alpha})T_{\beta\alpha}^{+}
\end{equation}
where $T_{\beta\alpha}^{+}=\langle\beta_{0}|V|\alpha_{+}\rangle$ and $|\alpha_{+}\rangle$ satisfies the Lippmann-Schwinger equation
\begin{equation}
|\alpha_{+}\rangle=|\alpha_{0}\rangle +\int d\gamma\frac{T_{\gamma\alpha}^{+} |\gamma_{0}\rangle}{p^{0}_{\alpha}-p^{0}_{\gamma}+i\epsilon}.
\end{equation}
Multiply from the left by $\langle\beta_{0}|V$ on both sides, we get
\begin{equation}
T_{\beta\alpha}^{+}= V_{\beta\alpha} + \int d\gamma\frac{T_{\gamma\alpha}^{+}V_{\beta\gamma}}{p^{0}_{\alpha}-p^{0}_{\gamma}+i\epsilon}.
\end{equation}
Substituting the entire expression into $T_{\gamma\alpha}^{+}$ on the right-hand side recursively gives us
\begin{equation}
T_{\beta\alpha}^{+}=V_{\beta\alpha}+\int d\gamma\frac{V_{\beta\gamma}V_{\gamma\alpha}}{p^{0}_{\alpha}-p^{0}_{\gamma}+i\epsilon}+
\int d\gamma d\gamma'\frac{V_{\beta\gamma}V_{\gamma\gamma'}V_{\gamma'\alpha}}{(p^{0}_{\alpha}-p^{0}_{\gamma}+i\epsilon)(p^{0}_{\alpha}-p^{0}_{\gamma'}+i\epsilon)}+\cdots\label{eq:t_beta_alpha}
\end{equation}
Substituting $T_{\beta\alpha}^{+}$ into the $S$-matrix given by the Born approximation yields the perturbative solution for the $S$-matrix. In eq.~(\ref{eq:t_beta_alpha}), the integration which sums over the all the momenta and spin-projections of the multi-particle states is not manifestly Lorentz invariant, it is therefore difficult to determine the types interactions that are allowed in the theory.

\subsubsection{The Dyson series}

The advantage of the Dyson series is the manifest Lorentz invariance. The derivation of the Dyon series uses the definition $S=U(\infty,-\infty)$ given in eq.~(\ref{eq:smatrix1}) where
\begin{equation}
U(\tau,\tau_{0})=e^{iH_{0}\tau}e^{-iH(\tau-\tau_{0})}e^{-iH_{0}\tau_{0}}.
\end{equation}
We will now derive a perturbative solution for $U(\tau,\tau_{0})$ and then take the limit $\tau\rightarrow\infty$ and $\tau_{0}\rightarrow-\infty$. Differentiating $U(\tau,\tau_{0})$ by $\tau$, after some manipulation, we obtain
\begin{equation}
i\frac{d}{d\tau}U(\tau,\tau_{0})=V(\tau)U(\tau,\tau_{0})\label{eq:ode_u}
\end{equation}
where $V(\tau)=e^{iH_{0}\tau}Ve^{-iH_{0}\tau}$ is the time evolution of $V$ in the interacting picture. Given the initial condition, $U(\tau_{0},\tau_{0})=1$, the solution for $U(\tau,\tau_{0})$ is
\begin{equation}
U(\tau,\tau_{0}) = 1 -i\int_{\tau_{0}}^{\tau}dt V(t)U(t,\tau_{0}).
\end{equation}
This is an iterative formula, so we can substitute the solution for $U(\tau,\tau_{0})$ to its right-hand side and perform further iterations to obtain higher-order terms
\begin{eqnarray}
U(\tau,\tau_{0})&=&1-i\int_{\tau_{0}}^{\tau} dt V(t)+
(-i)^{2}\int_{\tau_{0}}^{\tau}dt_{1}\int_{\tau_{0}}^{t_{1}}dt_{2}V(t_{1})V(t_{2})\nonumber \\
&&+(-i)^{3}\int_{\tau_{0}}^{\tau}dt_{1}\int_{\tau_{0}}^{t_{1}}dt_{2}\int_{\tau_{0}}^{t_{2}}dt_{3}V(t_{1})V(t_{2})V(t_{3})+\cdots.
\end{eqnarray}
Therefore, in the limit $\tau\rightarrow\infty$ and $\tau_{0}\rightarrow-\infty$ we obtain
\begin{eqnarray}
S&=&1-i\int_{-\infty}^{\infty}dt V(t)+
(-i)^{2}\int_{-\infty}^{\infty}dt_{1}\int_{-\infty}^{t_{1}}dt_{2}V(t_{1})V(t_{2})\nonumber \\
&&+(-i)^{3}\int_{-\infty}^{\infty}dt_{1}\int_{-\infty}^{t_{1}}dt_{2}\int_{-\infty}^{t_{2}}dt_{3}V(t_{1})V(t_{2})V(t_{3})+\cdots.
\end{eqnarray}

We note, for each term in the series, the integration of later time is always to the left of the earlier time. As a result, the $S$-matrix can be written in a more compact form by introducing the notion of a time-ordered product where functions of later time argument are arranged to the left of functions with earlier time argument,
\begin{equation}
T[V(t)]=V(t)
\end{equation}
\begin{equation}
T[V(t_{1})V(t_{2})]=\theta(t_{1}-t_{2})V(t_{1})V(t_{2})+\theta(t_{2}-t_{1})V(t_{2})V(t_{1})
\end{equation}
where $\theta(t)$ is the step function defined as $\theta(t)=1$ for $t>0$ and $\theta(t)=0$ when $t\leq0$. Each term in the time-ordered product contributes equally to the integral, so for a time-ordered product with $n$ terms, we need to divide the product by a factor of $n!$. Therefore, the $S$-matrix, in terms of the time-ordered product is given by
\begin{equation}
S=1+\sum_{n=1}^{\infty}\frac{(-i)^{n}}{n!}\int_{-\infty}^{\infty}dt_{1}\cdots dt_{n}T[V(t_{1})\cdots V(t_{n})].
\end{equation}
This is the Dyson series for the $S$-matrix.

Comparing the Dyson series to the $S$-matrix given by the old-fashioned perturbation theory, we note that the integral for the Dyson series integrates over the interaction $V(t)$ with respect to time. Manifest Lorentz invariance of the $S$-matrix is achieved by demanding the interaction $V(t)$ to be of the form
\begin{equation}
V(t)=\int d^{3}x\mathcal{V}(x)
\end{equation}
where $\mathcal{V}(x)$ is the interaction density which is a scalar quantity
\begin{equation}
U(\Lambda,a)\mathcal{V}(x)U^{-1}(\Lambda,a)=\mathcal{V}(\Lambda x+a).\label{eq:interaction_density_trans}
\end{equation}
The $S$-matrix can now be expressed in terms of manifestly Lorentz invariant integrals
\begin{equation}
S=1+\sum_{n=1}^{\infty}\frac{(-i)^{n}}{n!}\int_{-\infty}^{\infty}d^{4}x_{1}\cdots d^{4}x_{n}T[\mathcal{V}(x_{1})\cdots\mathcal{V}(x_{n})].\label{eq:dyson_series1}
\end{equation}
Up to this point, the Dyson series, in the manifest Lorentz invariant form could also be derived in non-relativistic quantum mechanics.
The question is then what is the difference between non-relativistic quantum mechanics and quantum field theory. A complete answer to this question would need to wait till the next section, but at this stage it is possible to identify the point of departure.

We note $S$-matrix in eq.~(\ref{eq:dyson_series1}) is not yet completely Lorentz invariant, an additional constraint on the interaction density $\mathcal{V}(x)$ must be imposed. The reason behind this is due to the time-ordered product and the fact that each interaction densities in the product have different time-arguments. The separation between space-time events can be classified into space-like, time like and null. For interaction densities whose separation between space-time arguments are time-like or null-like, the order of events is absolute for all inertial observers. Therefore, in the case where the separation between space-time events are time-like or null-like, the $S$-matrix is Lorentz-invariant. On the other hand, if the separation between space-time events are space-like, the order between events are no longer absolute. For example, given the product $\mathcal{V}(x_{1})\mathcal{V}(x_{2})$ where the separation between $x_{1}$ and $x_{2}$ are space-like, the order of product can change depending on the reference frame one chooses. Therefore, if the interaction density is non-commutative at space-like separation, the $S$-matrix would not be Lorentz invariant. Conversely, it follows that a sufficient condition, although not a necessary one for a Lorentz invariant $S$-matrix is to demand the interaction density to commute at space-like separation
\begin{equation}
[\mathcal{V}(x),\mathcal{V}(y)]=0.
\end{equation}
Under this condition, the $S$-matrix would always be Lorentz invariant. 

\section{The cluster decomposition principle}\label{cluster_decomp}

In the last section we have derived the perturbative $S$-matrix and the conditions for Poincar\'{e} invariance. These conditions are important since only a Poincar\'{e}-invariant $S$-matrix would give us physical particle cross-sections. In this section we present an additional physical condition due to the cluster decomposition principle~\cite{Eyvind:1963}. 

The cluster decomposition principle states that distant experiments must be uncorrelated. This is a fundamental principle that makes science possible. For the $S$-matrix, this condition must be imposed explicitly since it does not follow from the demand of unitarity and Poincar\'{e} invariance. 
But before we can consider its implication, it is necessary to introduce the creation and annihilation operators.


\subsection{Creation and annihilation operators}
The physical Hilbert space are spanned by $0,1,2,\cdots$ free particle states. Among these states, the vacuum state $|\,\,\rangle$, a state containing no particles, plays an important role. Since the vacuum state contains no particles, it has no energy and momentum so that
\begin{equation}
P^{\mu}|\,\,\rangle=0.
\end{equation}
This equation must hold in all inertial reference frame, otherwise Lorentz transformation to different inertial frames would result in particle creation thus violating the principle of energy-momentum conservation. Therefore, the vacuum state $|\,\,\rangle$, up to a global phase, is invariant under all inhomogeneous Lorentz transformations~\cite{Bogolyubov:1990kw}
\begin{equation}
U(\Lambda,a)|\,\,\rangle=|\,\,\rangle.
\end{equation}
While the vacuum state involved in particle scattering experiments are unique, there exists systems where the vacuum states are not unique and are unitarily inequivalent. One example is the coherent state where the associated vacuum state is in fact an eigenstate of the one-particle annihilation operator~\cite{Umezawa:1993yq}.

The existence of a unique vacuum state then allows us to construct the particle states spanning the physical Hilbert space by introducing the the creation and annihilation operators 
$a^{\dag}(\mathbf{p},\sigma,n)$ and $a(\mathbf{p},\sigma,n)$ respectively. We define these operators by their action on the vacuum state
\begin{equation}
a^{\dag}(\mathbf{p},\sigma,n)|\,\,\rangle=|p,\sigma,n\rangle,\hspace{0.5cm}a(\mathbf{p},\sigma,n)|\,\,\rangle=0.
\end{equation}
The successive application of creation operators creates a multi-particle state. Here we write these multi-particle state in the order which the creation operators on the vacuum
\begin{equation}
|p_{1}\sigma_{1}n_{1},p_{2}\sigma_{2}n_{2},\cdots\rangle=a^{\dag}(\mathbf{p}_{1},\sigma_{1},n_{1})a^{\dag}(\mathbf{p}_{2},\sigma_{2},n_{2})\cdots|\,\,\rangle. \label{eq:mstates1}
\end{equation}

However, this order is arbitrary. The physical properties of the multi-particle state is independent of the order of the one-particle states. Exchanging the two states $|p_{1}\sigma_{1}n_{1}\rangle$ and $|p_{2}\sigma_{2}n_{2}\rangle$ in eq.~(\ref{eq:mstates1}), the resulting state would must be proportional to the original states
\begin{equation}
|p_{1}\sigma_{1}n_{1},p_{2}\sigma_{2}n_{2},\cdots\rangle=\alpha|p_{2}\sigma_{2}n_{2},p_{1}\sigma_{1}n_{1},\cdots\rangle \label{eq:mstates2}
\end{equation}
where $\alpha$ is a constant global phase independent of $\sigma$. Exchanging the two states again, we find
\begin{equation}
\alpha^{2}=1.
\end{equation}
Therefore, the effect of exchanging particle states is either even or odd. 
More precisely, particle states of half-integral and integral spin anticommute and commute with itself respectively.
This result is known as the spin-statistics theorem~\cite{Weinberg:1995mt,Streater:1989vi} and give us the following commutation-anticommutation relations between the creation and annihilation operators\footnote{A detailed derivation is given in~\cite[sec.~4.2]{Weinberg:1995mt}.}
\begin{equation}
[a(\mathbf{p'},\sigma',n'),a^{\dag}(\mathbf{p},\sigma,n)]_{\pm}=\delta_{\sigma\sigma'}\delta_{nn'}\delta^{3}(\mathbf{p'-p}),\label{eq:aad}
\end{equation}
\begin{equation}
[a(\mathbf{p'},\sigma',n'),a(\mathbf{p},\sigma,n)]_{\pm}=[a^{\dag}(\mathbf{p'},\sigma',n'),a^{\dag}(\mathbf{p},\sigma,n)]_{\pm}=0.\label{eq:aadd}
\end{equation}

The inhomogeneous Lorentz transformation of the creation and annihilation operators are determined by the transformation of the one-particle state from eq.~(\ref{eq:boost})
\begin{equation}
U(\Lambda,b)a^{\dag}(\mathbf{p},\sigma,n)U^{-1}(\Lambda,b)=e^{i(\Lambda p)\cdot b}\sqrt{\frac{(\Lambda p)^{0}}{p^{0}}}\sum_{\bar{\sigma}}D^{(j)}_{\sigma\bar{\sigma}}(W^{-1}(\Lambda,p))a^{\dag}(\L\p,\bar{\sigma},n),\label{eq:ann}
\end{equation}
\begin{equation}
U(\Lambda,b)a(\mathbf{p},\sigma,n)U^{-1}(\Lambda,b)=e^{-i(\Lambda p)\cdot b}\sqrt{\frac{(\Lambda p)^{0}}{p^{0}}}\sum_{\bar{\sigma}}D_{\sigma\bar{\sigma}}^{(j)*}(W^{-1}(\Lambda,p))a(\L\p,\bar{\sigma},n).\label{eq:cre}
\end{equation}
Along with the introduction of the creation and annihilation operator comes a fundamental theorem, that any operator $\mathcal{O}$ may be expressed as a sum of products of creation and annihilation operators
\begin{eqnarray}
\mathcal{O}&=&\sum_{N=0}^{\infty}\sum_{M=0}^{\infty}\int d^{3}p_{1}'\cdots d^{3}p_{N}'d^{3}p_{1}\cdots d^{3}p_{M}\label{eq:basis} \\
&&\times a^{\dag}(\p_{1}')\cdots a^{\dag}(\p_{N}')a(\p_{1})\cdots a(\p_{M}) \nonumber \\
&&\times C_{NM}(\p_{1}',\cdots,\p_{N}',\p_{1},\cdots,\p_{M})\nonumber
\end{eqnarray}
where $C_{NM}(\p_{1}',\cdots,\p_{N}',\p_{1},\cdots,\p_{M})$ is momentum-dependent function. Intuitively, this result is expected since the basis of the Hilbert space is spanned by the particle states created and annihilated by these operators.

\subsection{Structure of the Hamiltonian}
The cluster decomposition principle states distant experiments are uncorrelated. In terms of the $S$-matrix it states that if $N$ multi-particle experiments $\alpha_{1}\rightarrow\beta_{1},\alpha_{2}\rightarrow\beta_{2},\cdots,\alpha_{N}\rightarrow\beta_{N}$ are each performed at distant locations, then $S$-matrix for the combined process factorises to
\begin{equation}
S_{\beta_{1}+\beta_{2}+\cdots+\beta_{N},\alpha_{1}+\alpha_{2}+\cdots+\alpha_{N}}\rightarrow
S_{\beta_{1}\alpha_{1}}S_{\beta_{2}\alpha_{2}}\cdots S_{\beta_{N}\alpha_{N}}.
\end{equation}
This assumes the interactions between particles are short-range and takes place in a confined region of space so that it is unaffected by distant particles. This property of the $S$-matrix does not follow from unitarity or Poincar\'{e} invariance, it is an additional physical requirement to make the interacting theories physical.

The full Hamiltonian $H=H_{0}+V$ by construction, is an operator in the Hilbert space, so it can be expanded in terms of the creation and annihilation operators,
\begin{eqnarray}
H&=&\sum_{N=0}^{\infty}\sum_{M=0}^{\infty}\int d^{3}p_{1}'\cdots d^{3}p_{N}'d^{3}p_{1}\cdots d^{3}p_{M} \label{eq:hamiltonian} \\ 
&&\times a^{\dag}(\p_{1}')\cdots a^{\dag}(\p_{N}')a(\p_{1})\cdots a(\p_{M})  \nonumber\\
&&\times h_{NM}(\p_{1}',\cdots,\p_{N}',\p_{1},\cdots,\p_{M})\nonumber
\end{eqnarray}
where $h_{NM}(\p_{1}',\cdots,\p_{N}',\p_{1},\cdots,\p_{M})$ is a multi-variable function of momentum. However, since any operators in the Hilbert space can be expressed in this form, eq.~(\ref{eq:hamiltonian}) without additional physical condition is meaningless. The additional constraint on the Hamiltonian comes from the demand of cluster decomposition principle on the $S$-matrix, 
that $h_{NM}(\p_{1}',\cdots,\p_{N}',\p_{1},\cdots,\p_{M})$ must only contain a single delta function
\begin{eqnarray}
h_{NM}(\mathbf{p}_{1}',\cdots \mathbf{p}_{N}',\mathbf{p}_{1},\cdots,\mathbf{p}_{M})&=&\delta^{3}(\mathbf{p}_{1}'+\cdots+\mathbf{p}_{N}'-\mathbf{p}_{1}-\cdots-\mathbf{p}_{M}) \\ \nonumber
&&\times\widetilde{h}_{NM}(\mathbf{p}_{1}',\cdots \mathbf{p}_{N}',\mathbf{p}_{1},\cdots,\mathbf{p}_{M}) \label{eq:h}
\end{eqnarray}
for it to satisfy the cluster decomposition principle~\cite[secs.~4.3-4.4]{Weinberg:1995mt}.

\section{Quantum field operators: The Weinberg formalism} \label{chap:quantum_fields}
In this chapter, so far we have derived the physical particle states from the irreducible representations of the Poincar\'{e} group. Ultimately, the objective is to compute observables such as cross-sections and decay rates and compare them with experiments. But before we can make the predictions, as a theorist, we must first make sure our theories are consistent with the principles of physics. For the $S$-matrix, these principles are Poincar\'{e} invariance and the cluster decomposition principle which amount to the following conditions on the full Hamiltonian $H$
\begin{enumerate}
\item The interacting Hamiltonian density $\mathcal{V}(x)$ must transform as a scalar and commute at space-like separation.
\item The full Hamiltonian expanded in terms of the creation and annihilation operators must only contain one momentum-conserving delta-function.
\end{enumerate}

Hamiltonians with these properties can be constructed using quantum field operators\footnote{The reason for choosing the normalisation factor $1/\sqrt{2p^{0}}$ will be apparent once we determine the coefficients.}
\begin{equation}
\psi^{+}(x)=(2\pi)^{-3/2}\int\frac{d^{3}p}{\sqrt{2p^{0}}}\sum_{\sigma n} u(\mathbf{x};\mathbf{p},\sigma,n)a(\mathbf{p},\sigma,n), \label{eq:f1}
\end{equation}
\begin{equation}
\psi^{-}(x)=(2\pi)^{-3/2}\int\frac{d^{3}p}{\sqrt{2p^{0}}}\sum_{\sigma n} v(\mathbf{x};\mathbf{p},\sigma,n)a^{\dag}(\mathbf{p},\sigma,n). \label{eq:f2}
\end{equation}
where $\psi^{+}(x)$ and $\psi^{-}(x)$ are the annihilation and creation field and $u(\mathbf{x};\p,\sigma,n)$ and $v(\mathbf{x};\p,\sigma,n)$ are their expansion coefficients to be determined. Here we will focus on quantum fields describing particle of a single specie so the summation over different species $n$ is dropped. 

%

Here, the quantum field operators are constructed from symmetry consideration. The field equation for the quantum fields and their solutions are derived as a consequence of the underlying symmetries. The Lagrangian can then be determined from the field equation. The demand of Poincar\'{e} symmetry places strict restrictions on the possible solutions on the quantum fields. One of the subtle, but important result derived from this formalism is that not all the solutions of the field equation satisfy Poincar\'{e} symmetry. For a quantum field corresponding to a given representation, its solution is uniquely determined by Poincar\'{e} symmetry up to a global phase. In this respect, the formalism presented here is more transparent than the canonical formalism since the later does not directly impose constraints on the solutions of the field equation for the quantum field.

The order of presentation in this thesis till this point, closely follows the monograph of Weinberg~\cite{Weinberg:1995mt}, which differs from the canonical formalism where the starting point is the Lagrangian.
Using the Lagrangians, one may derive the field equations, expand the solutions in terms of Fourier series and promote the general solutions to quantum field operators.  While the Lagrangians are useful for determining the symmetries of the theory, their origin in the canonical formalism are unclear. They are often taken as the starting point which can sometime lead to the wrong physics. It is trivial to construct manifestly Lorentz invariant Lagrangians but non-trivial to ensure the corresponding particle states have the correct degrees of freedom and are consistent with the Poincar\'{e} symmetry. The formalism developed by Weinberg emphasises the close connexion between quantum field operators and the underlying symmetries. The field equations and the Lagrangians are derived as a consequence of the symmetries and representations of the Poincar\'{e} group which ensure theories are consistency and unique.




The presence of the coefficients $u(\mathbf{x};\p,\sigma)$ and $v(\mathbf{x};\p,\sigma)$ means we cannot derive the transformation of the fields from the transformation of the creation and annihilation operators given by eqs.~(\ref{eq:ann}) and (\ref{eq:cre}). We must make an additional postulate that the fields are Poincar\'{e} covariant
\begin{equation}
U(\Lambda,b)\psi_{\ell}^{\pm}(x)U^{-1}(\Lambda,b)=\sum_{\bar{\ell}}\mathcal{D}_{\ell\bar{\ell}}(\Lambda^{-1})\psi_{\bar{\ell}}^{\pm}(\Lambda x+b) \label{eq:cov}
\end{equation}
where $\mathcal{D}(\Lambda)$ is a finite-dimensional matrix. Successive transformations on the field $\psi_{\ell}^{\pm}(x)$ shows that the matrix $\mathcal{D}(\Lambda)$ furnishes a representation of the Lorentz group.


The coefficients for the field, up to some proportionality constants can be determined by continuous Poincar\'{e} symmetry. The proportionality constants can subsequently be determined by the requirements of locality of the interaction density and discrete symmetry conservations without having to solve the field equations. 

We now derive the general constraints on the expansion coefficients. On the left-hand side of eq.~(\ref{eq:cov}), the operator $U(\Lambda,b)$ acts only on the creation and annihilation operators, therefore using eqs.~(\ref{eq:ann}) and (\ref{eq:cre}), we get
\begin{eqnarray}
U(\Lambda,b)\psi^{+}(x)U^{-1}(\Lambda,b)&=&(2\pi)^{-3/2}\int \frac{d^{3}(\Lambda p)}{\sqrt{2(\Lambda p)^{0}}} \sum_{\sigma\bar{\sigma}} u_{\ell}(\mathbf{x};\mathbf{p},\sigma)\label{eq:cov1}  \\ 
&&\times e^{-i(\Lambda p)\cdot b}D_{\sigma\bar{\sigma}}(W^{-1}(\Lambda,p))a(\L\p,\bar{\sigma}),\nonumber
\end{eqnarray}
\begin{eqnarray}
U(\Lambda,b)\psi^{-}(x)U^{-1}(\Lambda,b)&=&(2\pi)^{-3/2}\int \frac{d^{3}(\Lambda p)}{\sqrt{2(\Lambda p)^{0}}} \sum_{\sigma\bar{\sigma}} v_{\ell}(\mathbf{x};\mathbf{p},\sigma) \label{eq:cov2}\\ 
&&\times e^{i(\Lambda p)\cdot b}D_{\sigma\bar{\sigma}}^{*}(W^{-1}(\Lambda,p))a^{\dag}(\L\p,\bar{\sigma}). \nonumber
\end{eqnarray}
The above results are obtained by the Lorentz invariance of $d^{3}p/p^{0}$ and the unitarity of $D(W(\Lambda,p))$. Since the fields $\psi^{\pm}(x)$ are required to be Poincar\'{e} covariant, eqs.~(\ref{eq:cov1}) and (\ref{eq:cov2}) must be identical to the right-hand side of eq.~(\ref{eq:cov}) which reads
\begin{eqnarray}
&&\hspace{-1cm} U(\Lambda,b)\psi_{\ell}^{+}(x)U^{-1}(\Lambda,b)\nonumber\\
&&=\mathcal{D}_{\ell\bar{\ell}}(\Lambda^{-1})(2\pi)^{-3/2}\int\frac{d^{3}(\Lambda p)}{\sqrt{2(\Lambda p)^{0}}}\sum_{\sigma}
\left[u_{\bar{\ell}}(\Lambda x+b,\L\p,\sigma)a(\L\p,\sigma)\right],
\end{eqnarray}
\begin{eqnarray}
&&\hspace{-1cm}U(\Lambda,b)\psi_{\ell}^{-}(x)U^{-1}(\Lambda,b)\nonumber\\
&&=\mathcal{D}_{\ell\bar{\ell}}(\Lambda^{-1})(2\pi)^{-3/2}\int\frac{d^{3}(\Lambda p)}{\sqrt{2(\Lambda p)^{0}}}\sum_{\sigma}
\left[v_{\bar{\ell}}(\Lambda x+b,\L\p,\sigma)a^{\dag}(\L\p,\sigma)\right].
\end{eqnarray}
Equating both sides of the equations gives us
\begin{equation}
\sum_{\bar{\sigma}}u_{\bar{\ell}}(\Lambda x+b;\L\p,\bar{\sigma})D^{(j)}_{\bar{\sigma}\sigma}(W(\Lambda,p))=e^{-i(\Lambda p)\cdot b}\sum_{\ell}\mathcal{D}_{\bar{\ell}\ell}(\Lambda)u_{\ell}(\mathbf{x};\L\p,\sigma),\label{eq:con1}
\end{equation}
\begin{equation}
\sum_{\bar{\sigma}}v_{\bar{\ell}}(\Lambda x+b;\L\p,\bar{\sigma})D_{\bar{\sigma}\sigma}^{(j)*}(W(\Lambda,p))=e^{i(\Lambda p)\cdot b}\sum_{\ell}\mathcal{D}_{\bar{\ell}\ell}(\Lambda)v_{\ell}(\mathbf{x};\L\p,\sigma).\label{eq:con2}
\end{equation}
These two equations determine the form of the coefficients and how they transform under boost and space-time translation. 

For space-time translation, let $\Lambda=I$ and $b$ arbitrary, we get
\begin{equation}
u_{\bar{\ell}}(\mathbf{x};\mathbf{p},\sigma)=e^{-ip\cdot x}u_{\bar{\ell}}(\mathbf{p},\sigma), \label{eq:ti1}
\end{equation}
\begin{equation}
v_{\bar{\ell}}(\mathbf{x};\mathbf{p},\sigma)=e^{ip\cdot x}v_{\bar{\ell}}(\mathbf{p},\sigma). \label{eq:ti2}
\end{equation}
The creation and annihilation fields then take the form
\begin{equation}
\psi^{+}_{\ell}(x)=(2\pi)^{-3/2}\int \frac{d^{3}p}{\sqrt{2p^{0}}}\sum_{\sigma}e^{-ip\cdot x}u_{\ell}(\mathbf{p},\sigma)a(\mathbf{p},\sigma),\label{eq:t1}
\end{equation}
\begin{equation}
\psi^{-}_{\ell}(x)=(2\pi)^{-3/2}\int \frac{d^{3}p}{\sqrt{2p^{0}}}\sum_{\sigma}e^{ip\cdot x}v_{\ell}(\mathbf{p},\sigma)a^{\dag}(\mathbf{p},\sigma).\label{eq:t2}
\end{equation}
Substituting eqs.~(\ref{eq:ti1}) and (\ref{eq:ti2}) into eqs.~(\ref{eq:con1}) and (\ref{eq:con2}), they simplify to
\begin{equation}
\sum_{\bar{\sigma}}u_{\bar{\ell}}(\L\p,\bar{\sigma})D^{(j)}_{\bar{\sigma}\sigma}(W(\Lambda,p))=\sum_{\ell}\mathcal{D}_{\bar{\ell}\ell}(\Lambda)u_{\ell}(\mathbf{p},\sigma), \label{eq:con3}
\end{equation}
\begin{equation}
\sum_{\bar{\sigma}}v_{\bar{\ell}}(\L\p,\bar{\sigma})D^{(j)*}_{\bar{\sigma}\sigma}(W(\Lambda,p))=\sum_{\ell}\mathcal{D}_{\bar{\ell}\ell}(\Lambda)v_{\ell}(\mathbf{p},\sigma). \label{eq:con4}
\end{equation}
For boost we take $\mathbf{p}=\mathbf{0}$ and $\Lambda=L(p)$ where $L(p)$ takes particle at rest to momentum $p^{\mu}$, then we have $W(\Lambda,p)=I$ and $D(\Lambda,p)=I$. Equations (\ref{eq:con3}) and (\ref{eq:con4}) now give us the coefficients at arbitrary momentum\footnote{The normalisation factor $1/\sqrt{2p^{0}}$ for the quantum field is chosen specifically to obtain eqs.~(\ref{eq:b1}) and (\ref{eq:b2}).}
\begin{equation}
u_{\ell}(\mathbf{p},\sigma)=\sum_{\bar{\ell}}\mathcal{D}_{\ell\bar{\ell}}(L(p))u_{\bar{\ell}}(\mathbf{0},\sigma), \label{eq:b1}
\end{equation}
\begin{equation}
v_{\ell}(\mathbf{p},\sigma)=\sum_{\bar{\ell}}\mathcal{D}_{\ell\bar{\ell}}(L(p))v_{\bar{\ell}}(\mathbf{0},\sigma). \label{eq:b2}
\end{equation}
As for rotation, we take $\mathbf{p}=\0$ and $\Lambda=R$ to be a rotation. Therefore $W(R,p)=R$, so eqs.~(\ref{eq:con3}) and (\ref{eq:con4}) become
\begin{equation}
\sum_{\bar{\sigma}}u_{\bar{\ell}}(\mathbf{0},\bar{\sigma})D^{(j)}_{\bar{\sigma}\sigma}(R)=\sum_{\ell}\mathcal{D}_{\bar{\ell}\ell}(R)u_{\ell}(\mathbf{0},\sigma),
\end{equation}
\begin{equation}
\sum_{\bar{\sigma}}v_{\bar{\ell}}(\mathbf{0},\bar{\sigma})D^{(j)*}_{\bar{\sigma}\sigma}(R)=\sum_{\ell}\mathcal{D}_{\bar{\ell}\ell}(R)u_{\ell}(\mathbf{0},\sigma).
\end{equation}
Under an infinitesimal expansion about the identity,
\begin{equation}
\sum_{\bar{\sigma}}u_{\bar{\ell}}(\mathbf{0},\bar{\sigma})\J_{\bar{\sigma}\sigma}=\sum_{\ell}\mJ_{\bar{\ell}\ell}u_{\ell}(\mathbf{0},\sigma)\label{eq:rot1}
\end{equation}
\begin{equation}
\sum_{\bar{\sigma}}v_{\bar{\ell}}(\mathbf{0},\bar{\sigma})\J_{\bar{\sigma}\sigma}^{*}=-\sum_{\ell}\mJ_{\bar{\ell}\ell}v_{\ell}(\mathbf{0},\sigma)\label{eq:rot2}
\end{equation}
where $\J$ and $\mJ$ are the generators of $D(R)$ and $\mathcal{D}(R)$ respectively.

So far, we have derived the necessary conditions on the fields and their coefficients which ensure that the interaction density transforms as a scalar. However, they are not yet sufficient to guarantee the Lorentz invariance of the $S$-matrix. 
Generally, interaction densities constructed from $\psi^{\pm}(x)$ do not commute at space-like separation since the fields themselves do not commute or anti-commute at space-like separation,
%
%
\begin{equation}
[\psi_{\ell}^{+}(x),\psi_{\bar{\ell}}^{-}(y)]_{\pm}=\int d^{3}p\sum_{\sigma}\, u_{\ell}(\mathbf{p},\sigma)v_{\bar{\ell}}(\mathbf{p},\sigma)e^{-ip\cdot(x-y)}
\end{equation}
unless the coefficients vanish. The resolution is to consider a linear combination of creation and annihilation field,
\begin{equation}
\psi_{\ell}(x)=\kappa\psi_{\ell}^{+}(x)+\lambda\psi_{\ell}^{-}(x)\label{eq:def_psi}
\end{equation}
where $\kappa$ and $\lambda$ are constants chosen so that the field commutes or anti-commutes with itself and its adjoint at space-like separation
\begin{equation}
[\psi_{\ell}(x),\psi_{\bar{\ell}}(y)]_{\pm}=[\psi_{\ell}(x),\psi^{\dag}_{\bar{\ell}}(y)]_{\pm}=0. \label{eq:locality}
\end{equation}
The interaction density constructed from $\psi(x)$ would then commute at space-like separation. We can construct the interaction density using the covariant fields as
\begin{eqnarray}
\mathcal{V}(x)&=&\sum_{NM}\sum_{\ell'_{1}\cdots\ell'_{N}}\sum_{\ell_{1}\cdots\ell_{M}}g_{\ell'_{1}\cdots\ell'_{N},\ell_{1}\cdots\ell_{M}} \label{eq:coef}\\ 
&&\times\psi^{\dag}_{\ell'_{1}}(x)\cdots\psi^{\dag}_{\ell'_{N}}(x)\psi_{\ell_{1}}(x)\cdots\psi_{\ell_{M}}(x) \nonumber
\end{eqnarray}
where $N$ and $M$ denote particles of different species. Equation (\ref{eq:cov}) shows that the interaction density transforms as a scalar satisfying eq.~(\ref{eq:interaction_density_trans}) if the coefficient $g_{\ell'_{1}\cdots\ell'_{N},\ell_{1}\cdots\ell_{M}}$ satisfies
\begin{eqnarray}
&&\sum_{\ell'_{1}\cdots\ell'_{N}}\sum_{\ell_{1}\cdots\ell_{M}}\mathcal{D}^{\dag}_{\bar{\ell'}_{1}\ell'_{1}}(\Lambda^{-1})\cdots\mathcal{D}^{\dag}_{\bar{\ell'}_{N}\ell'_{N}}(\Lambda^{-1})
\mathcal{D}_{\ell_{1}\bar{\ell}_{1}}(\Lambda^{-1})\cdots\mathcal{D}_{\ell_{M}\bar{\ell}_{M}}(\Lambda^{-1}) \\ \nonumber
&&\hspace{1cm}\times g_{\ell'_{1}\cdots\ell'_{N}\ell_{1}\cdots\ell_{M}}=g_{\bar{\ell}'_{1}\cdots\bar{\ell}'_{N}\bar{\ell}_{1}\cdots\bar{\ell}_{M}}.
\end{eqnarray}

Interaction densities in most theories contain no more than three particle species and the coefficient $g$ can usually be decomposed into product of matrices
\begin{equation}
g_{\ell'_{1}\cdots\ell'_{N}\ell_{1}\cdots\ell_{M}}=g_{\ell'_{1}\ell_{1}}g_{\ell'_{2}\ell_{2}}\cdots g_{\ell'_{N}\ell_{M}}.
\end{equation}
The constraint on $g$ for these theories can then be simplified by taking $N=M=1$. In terms of matrices, we get
\begin{equation}
g\mathcal{D}(\Lambda^{-1})=\mathcal{D}^{\dag}(\Lambda)g.\label{eq:g_constraint}
\end{equation}
Since $\mathcal{D}(\Lambda)$ is a representation of the Lorentz group, we can expand it about the identity
\begin{equation}
\mathcal{D}(I+\omega)=I-\frac{i}{2}\omega_{\mu\nu}\mathcal{J}^{\mu\nu}
\end{equation}
to obtain
\begin{equation}
g\mathcal{J}^{\mu\nu}=\mathcal{J}^{\dag\mu\nu}g.\label{eq:g_constraint1}
\end{equation}

In an interacting theory, the particle states usually have some non-zero conserved charges $q(n)$ associated with a Hermitian charge operator $Q$ satisfying the commutation relations
\begin{equation}
[Q,a(\p,\sigma,n)]=-q(n)a(\p,\sigma,n),
\end{equation}
\begin{equation}
[Q,a^{\dag}(\p,\sigma,n)]=q(n)a^{\dag}(\p,\sigma,n)
\end{equation}
where the annihilation and creation operator act as the lowering and raising operator.  Since the commutators between $Q$ and the annihilation and creation operators do not vanish, we do not expect $[Q,\psi_{\ell}(x)]$ to vanish either. Instead, we must demand
\begin{equation}
[Q,\psi_{\ell}(x)]=-q(n)\psi_{\ell}(x)\label{eq:charged_qf}
\end{equation}
for $Q$ to commute with the interaction density to ensure charge conservation. Using eq.~(\ref{eq:charged_qf}), the commutator between $Q$ and $\mathcal{V}(x)$ is
\begin{equation}
[Q,\mathcal{V}(x)]=(-q'_{1}-\cdots -q'_{N}+q_{1}+\cdots +q_{M})\mathcal{V}(x).
\end{equation}
It vanishes if and only if
\begin{equation}
-q'_{1}-\cdots -q'_{N}+q_{1}+\cdots +q_{M}=0.
\end{equation}

Equation(\ref{eq:charged_qf}) cannot be satisfied if both  $\psi^{+}(x)$ and $\psi^{-}(x)$ describes the same particle of charge $q$. The only solution to eq.~(\ref{eq:charged_qf}) is when the two fields $\psi^{+}(x)$ and $psi^{-}(x)$ describe particles of opposite charges
\begin{equation}
[Q,\psi^{+}(x)]=-q\psi^{+}(x),
\end{equation}
\begin{equation}
[Q,\psi^{-}(x)]=-\bar{q}\psi^{-}(x)=q\psi^{-}(x).
\end{equation}
This implies a doubling in particle species. Lorentz invariance of the $S$-matrix demands that a particle of non-zero conserved charge $q$ to be always accompanied by a particle of charge $-q$ which is known as the anti-particle. The derivation given here is based on the quantum field theoretic consideration. An alternative argument using causality between space-time events can be found in~\cite[sec.~2.1.3]{Weinberg:1972}.

To put the formalism into practice, we must specify the representation $\mathcal{D}(\Lambda)$ of the Lorentz group which corresponds to a particular particle specie. While there are infinitely many representations, only the $j\leq1$ representations are relevant to the SM. We will explicitly construct both massive and massless quantum fields of $j\leq1$ representations and outline the generalisation to higher spin.

\section{Massive quantum fields}\label{massive_qf}

In this section, we provide explicit construction of the massive scalar and Dirac field and examine their discrete symmetries. Later we will generalise the results to higher-spin representations. The study of massive vector field is presented in the next chapter.

\subsection{Scalar field}

The scalar field $\phi(x)$, by definition is a single component field with trivial Lorentz representation $\mathcal{D}(\Lambda)=I$, $\mathcal{J}^{\mu\nu}=O$ so it transforms as
\begin{equation}
U(\Lambda,b)\phi(x)U^{-1}(\Lambda,b)=\phi(\Lambda x+b).
\end{equation}
The explicit form of the scalar field, in accordance with eq.~(\ref{eq:def_psi}) is
\begin{equation}
\phi(x)=\kappa\phi^{+}(x)+\lambda\phi^{-}(x)
\end{equation}
where $\phi^{+}(x)$ and $\phi^{-}(x)$ are the annihilation and creation fields
\begin{equation}
\phi^{+}(x)=(2\pi)^{-3/2}\int\frac{d^{3}p}{\sqrt{2p^{0}}}\sum_{\sigma}u(\p,\sigma)e^{-ip\cdot x}a(\p,\sigma),
\end{equation}
\begin{equation}
\phi^{-}(x)=(2\pi)^{-3/2}\int\frac{d^{3}p}{\sqrt{2p^{0}}}\sum_{\sigma}v(\p,\sigma)e^{-ip\cdot x}a^{\dag}(\p,\sigma).
\end{equation}
For now, we confined ourselves to the neutral scalar field where $[Q,\phi(x)]=0$, later on we will also consider charged scalar field. 

The rotation constraints on the expansion coefficients are
\begin{equation}
\sum_{\sigma}u(\0,\bar{\sigma})\mathbf{J}_{\bar{\sigma}\sigma}=0
\end{equation}
\begin{equation}
\sum_{\sigma}v(\0,\bar{\sigma})\mathbf{J}^{*}_{\bar{\sigma}\sigma}=0
\end{equation}
where $\mathbf{J}_{\bar{\sigma}\sigma}$ is given by eqs.~(\ref{eq:r2}) and (\ref{eq:r3}). The only solution where the coefficients are non-zero is $\J=\mathbf{O}$. Therefore, the coefficients are just momentum independent constants which can be absorbed into $\kappa$ and $\lambda$. The annihilation and creation fields now take the form
\begin{equation}
\phi^{+}(x)=(2\pi)^{-3/2}\int\frac{d^{3}p}{\sqrt{2p^{0}}}e^{-ip\cdot x}a(\p),
\end{equation}
\begin{equation}
\phi^{-}(x)=(2\pi)^{-3/2}\int\frac{d^{3}p}{\sqrt{2p^{0}}}e^{-ip\cdot x}a^{\dag}(\p).
\end{equation}

The constants $\kappa$ and $\lambda$ can be determined by evaluating the commutation/anti-commutation of the fields at space-like separation,
\begin{equation}
[\phi(x),\phi(y)]_{\pm}=|\kappa|^{2}\Delta_{+}(x-y)\pm |\lambda|^{2}\Delta_{+}(y-x)
\end{equation}
where $\Delta_{+}(x)$ is the standard function defined as
\begin{equation}
\Delta_{+}(x)=(2\pi)^{-3}\int\frac{d^{3}p}{2p^{0}}\,e^{-ip\cdot x}.
\end{equation}
Since $\Delta_{+}(x)$ is manifestly Lorentz invariant, for space-like separation, we can choose an equal-time frame where $t=t'$ and $x=(0,0,0,r)$ to simplify the calculation. Computing the integral in spherical polar coordinate, we get
\begin{equation}
\Delta_{+}(x)=\frac{1}{4\pi^{2}|\mathbf{x}|}\int_{0}^{\infty}dp\frac{p\sin(pr)}{\sqrt{p^{2}+m^{2}}}=\frac{m}{(2\pi\sqrt{-x^{\mu}x_{\mu}})^{2}}K_{1}(m\sqrt{-x^{\mu}x_{\mu}})
\end{equation}
where $K_{1}$ is the standard Hankel function and we have used the identity $|\mathbf{x}|=\sqrt{-x^{\mu}x_{\mu}}$. The second equality shows that when $x^{\mu}$ is a space-like vector, $\Delta_{+}(x)$ is an even function in $x^{\mu}$. Therefore, the commutator/anti-commutator simplifies to
\begin{equation}
[\phi(x),\phi(y)]_{\pm}=(|\kappa|^{2}\pm |\lambda|^{2})\Delta_{+}(\mathbf{x}-\mathbf{y}).
\end{equation}
This expression vanishes if the field satisfies bosonic statistics with $|\kappa|^{2}=|\lambda|^{2}$. Choosing $\kappa=\lambda=1$, then for arbitrary space-time separation, we get
\begin{equation}
[\phi(x),\phi(y)]=\Delta(x-y)
\end{equation}
where $\Delta(x-y)=\Delta_{+}(x-y)-\Delta_{+}(y-x)$. 

The charged scalar field takes the same coefficients as the neutral scalar field but with distinct particle and anti-particle,
\begin{equation}
\Phi(x)=(2\pi)^{-3/2}\int\frac{d^{3}p}{\sqrt{2p^{0}}}\left[e^{-ip\cdot x}a(\p)+e^{ip\cdot x}b^{\dag}(\p)\right]
\end{equation}
so that particle states created by $a^{\dag}(\p)$ and $b^{\dag}(\p)$ creates states of opposite charges
\begin{equation}
[Q,a^{\dag}(\p)]=qa^{\dag}(\p),
\end{equation}
\begin{equation}
[Q,b^{\dag}(\p)]=-qb^{\dag}(\p).
\end{equation}
We note, the statistics for $\Phi(x)$ cannot be determined by computing $[\Phi(x),\Phi(y)]_{\pm}$ since it identically vanishes for all space-time separation 
\begin{equation}
[\Phi(x),\Phi(y)]_{\pm}=0
\end{equation}
by virtue of eqs.~(\ref{eq:aad}) and (\ref{eq:aadd}) where particles and anti-particles are of different species. The statistics is determined by calculating $[\Phi(x),\Phi^{\dag}(y)]$ at space-like separation
\begin{equation}
[\Phi(x),\Phi^{\dag}(y)]_{\pm}=\Delta_{+}(x-y)\pm\Delta_{+}(y-x)
\end{equation}
which vanishes under bosonic statistics. 

The presence of non-zero charges realises the charge-conjugation symmetry. As discussed in sec.~\ref{discrete_symmetry_S_matrix}, charge-conjugation is an internal symmetry, so it cannot affect the statistics and transformations of the scalar fields.

\subsubsection{Parity}

Acting the parity operation on the scalar field, we get
\begin{equation}
\mathsf{P}\Phi(x)\mathsf{P}^{-1}=(2\pi)^{-3/2}\int\frac{d^{3}p}{\sqrt{2p^{0}}}\left[e^{-ip\cdot x}\eta^{*}a(-\p)+e^{ip\cdot x}\bar{\eta}b^{\dag}(-\p)\right]
\end{equation}
where $\eta$ and $\bar{\eta}$ are the intrinsic parities for particle and anti-particle respectively. Changing the integration variable from $\p\rightarrow-\p$, we see that parity is preserved if particle and anti-particle have even intrinsic parities
\begin{equation}
\eta^{*}=\bar{\eta}
\end{equation}
thus giving us
\begin{equation}
\mathsf{P}\Phi(x)\mathsf{P}^{-1}=\eta^{*}\Phi(\mathscr{P}x).
\end{equation}
Therefore, parity is conserved for the neutral scalar field if we take its intrinsic parity to be real $\eta=\eta^{*}$
\begin{equation}
\mathsf{P}\phi(x)\mathsf{P}^{-1}=\eta\phi(\mathscr{P}x).
\end{equation}

\subsubsection{Charge-conjugation}

Acting the charge-conjugation operator $\mathsf{C}$ on $\Phi(x)$ gives us
\begin{equation}
\mathsf{C}\Phi(x)\mathsf{C}^{-1}=(2\pi)^{-3/2}\left[e^{-ip\cdot x}\varsigma^{*}b(\p)+e^{ip\cdot x}\bar{\varsigma}a^{\dag}(\p)\right].
\end{equation}
The symmetry is conserved by taking the charge-conjugation phases to be even between particle and anti-particle
\begin{equation}
\varsigma^{*}=\bar{\varsigma}
\end{equation}
thus giving us
\begin{equation}
\mathsf{C}\Phi(x)\mathsf{C}^{-1}=\varsigma^{*}\Phi^{\dag}(x).
\end{equation}
As for the neutral scalar field, there is only one phase $\varsigma$ so charge-conjugation is conserved if $\varsigma=\varsigma^{*}$. Therefore, $\phi(x)$ is an eigen-function under charge-conjugation
\begin{equation}
\mathsf{C}\phi(x)\mathsf{C}^{-1}=\varsigma\phi(x).
\end{equation}

\subsubsection{Time-reversal}

Time-reversal is an anti-unitary operator so when acting on $\Phi(x)$, it is necessary to apply complex conjugation
\begin{equation}
\mathsf{T}\Phi(x)\mathsf{T}^{-1}=(2\pi)^{-3/2}\int\frac{d^{3}p}{\sqrt{2p^{0}}}\left[e^{ip\cdot x}\varrho^{*} a(-\p)+e^{-ip\cdot x}\bar{\varrho}b^{\dag}(-\p)\right].
\end{equation}
Taking $\p\rightarrow-\p$ in the integration, we see that time-reversal is preserved if 
\begin{equation}
\varrho=\bar{\varrho}^{*}
\end{equation}
thus giving us
\begin{equation}
\mathsf{T}\Phi(x)\mathsf{T}^{-1}=\varrho^{*}\Phi(\mathscr{T}x).
\end{equation}
Similar to charge-conjugation and parity, one obtains
\begin{equation}
\mathsf{T}\phi(x)\mathsf{T}^{-1}=\varrho\phi(\mathscr{T}x)
\end{equation}
by taking $\varrho=\varrho^{*}$ so that the time-reversal phase is real.

\subsection{Dirac field}\label{Dirac_field}
The Dirac field is constructed from one of the simplest non-trivial representations of the Lorentz group. The matrix $\mathcal{D}(\Lambda)$ with $\Lambda=I+\omega$ is taken to be
\begin{equation}
\mathcal{D}(I+\omega)=I-\frac{i}{2}\omega_{\mu\nu}\mathcal{J}^{\mu\nu}
\end{equation}
where $\mathcal{J}^{\mu\nu}$ is the finite-dimensional generator of the Lorentz group. The generators associated with the Dirac field and its generalisation to higher-spin representation takes the form of a direct sum of irreudible representations of the Lorentz group
\begin{equation}
\mathcal{J}^{ij}=\epsilon^{ijk}\left(\begin{array}{cc}
J^{k} & O \\
O & J^{k} \end{array}\right),\hspace{0.5cm}
\mathcal{J}^{0i}=i\left(\begin{array}{cc}
-J^{i} & O \\
O & J^{i} \end{array}\right).
\end{equation}
Generally, given a rotation generator $\J$, it is always possible to construct the boost generator as $\K=\pm i\J$ such that the Lorentz algebra is satisfied. In the literature, quantum fields which transform according to the $\K=i\J$ and $\K=-i\J$ representations are known as the right-handed and left-handed fields respectively. Therefore $\mathcal{J}^{\mu\nu}$, furnishes a reducible representation of the Lorentz group where $J^{k}$ is the irreducible representation of the rotation group of dimension $(2j+1)\times(2j+1)$ given by eqs.~(\ref{eq:r1}-\ref{eq:r3}) up to a similarity transformation. 
Using the identification given by eqs.~(\ref{eq:j123}) and (\ref{eq:k123}) where
\begin{equation}
\mJ=(\mathcal{J}^{23},\mathcal{J}^{31},\mathcal{J}^{12}),\hspace{0.5cm}
\mK=(\mathcal{J}^{01},\mathcal{J}^{02},\mathcal{J}^{03})
\end{equation}
the explicit expression of rotation and boost generators are
\begin{equation}
\mJ=\left(\begin{array}{cc}
\J &\O \\
\O & \J \end{array}\right),\hspace{0.5cm}
\mK=\left(\begin{array}{cc}
-i\J & \O \\
\O & i\J \end{array}\right).\label{eq:rb_generators}
\end{equation}
Defining the rotation and rapidity (boost) parameters of $\omega_{\mu\nu}$ as
\begin{equation}
\th=-(\omega^{23},\omega^{31},\omega^{12}),\hspace{0.5cm}
\bv=(\omega^{01},\omega^{02},\omega^{03}),\hspace{0.5cm}
\end{equation}
we obtain the following rotation 
\begin{equation}
\mathcal{D}(R(\theta))=\left(\begin{matrix}
\exp(i\J\cdot\th) & O \\
O & \exp(i\J\cdot\th) \end{matrix}\right)
\end{equation}
and boost
\begin{equation}
\mathcal{D}(L(p))=\left(\begin{matrix}
\exp(\J\cdot\bv) & O \\
O & \exp(-\J\cdot\bv) \end{matrix}\right).\label{eq:spin_half_boost}
\end{equation}
The rapidity parameter $\bv=\varphi\hat{\p}$ with $\hat{\p}=\p/|\p|$ is defined as
\begin{equation}
\cosh\varphi=\frac{p^{0}}{m},\hspace{0.5cm}
\sinh\varphi=\frac{|\p|}{m}.
\end{equation}

In the case of the Dirac field, we take
\begin{equation}
\J=\frac{\s}{2}
\end{equation}
where $\s=(\sigma^{1},\sigma^{2},\sigma^{3})$ are the Pauli matrices
\begin{equation}
\sigma^{1}=\left(\begin{array}{cc}
0 & 1 \\
1 & 0 \end{array}\right),\hspace{0.5cm}
\sigma^{2}=\left(\begin{array}{cc}
0 &-i \\
i & 0 \end{array}\right),\hspace{0.5cm}
\sigma^{3}=\left(\begin{array}{cc}
1 & 0 \\
0 &-1 \end{array}\right).
\end{equation}
The submatrices of $\mathcal{D}(L(p))$ are then given by
\begin{equation}
\exp\left(\pm\frac{\s}{2}\cdot\bv\right)=\sqrt{\frac{p^{0}+m}{2m}}\left(I\pm\frac{\s\cdot\p}{p^{0}+m}\right).\label{eq:left_right_boost}
\end{equation}

The expansion coefficients at rest for the Dirac field can now be solved by substituting $\mJ$ into eqs.~(\ref{eq:rot1}) and (\ref{eq:rot2}). Here, it is more convenient to  write the coefficients as
\begin{equation}
(U_{\pm})_{\ell\sigma}=u_{\ell}(\mathbf{0},\sigma),\hspace{0.5cm} (V_{\pm})_{\ell\sigma}=v_{\ell}(\mathbf{0},\sigma)
\end{equation}
where $U_{\pm}$ and $V_{\pm}$ are $2\times2$ matrices. The column index $\sigma$ ranges from $-\frac{1}{2}$ and $\frac{1}{2}$. The row index $\ell$ is defined so that for $U_{+}$, $\ell=1,2$ and for $U_{-}$, $\ell=3,4$. 
Equations (\ref{eq:rot1}) and (\ref{eq:rot2}) now become
\begin{equation}
\sum_{\bar{\sigma}}(U_{\pm})_{\ell\bar{\sigma}}\s_{\bar{\sigma}\sigma}=\sum_{\bar{\ell}}\s_{\ell\bar{\ell}}(U_{\pm})_{\bar{\ell}\sigma},
\end{equation}
\begin{equation}
\sum_{\bar{\sigma}}(V_{\pm}\sigma_{2})_{\ell\bar{\sigma}}\s_{\bar{\sigma}\sigma}=\sum_{\bar{\ell}}\s_{\ell\bar{\ell}}(V_{\pm}\sigma_{2})_{\bar{\ell}\sigma}
\end{equation}
where the second equation is obtained using the identity $\s^{*}=-\sigma^{2}\s\sigma^{2}$. Since the Pauli matrix are irreducible, by Schur's lemma, the matrices $U_{\pm}$ and $V_{\pm}\sigma^{2}$ must either vanish or be proportional to the identity matrix. Therefore,
\begin{equation}
U_{\pm}=c_{\pm} I,\hspace{0.5cm} V_{\pm}=-i d_{\pm}\sigma^{2}
\end{equation}
where $c_{\pm}$ and $d_{\pm}$ are proportionality constants. The Dirac coefficients, with the appropriate normalisation take the form
\begin{equation}
u(\mathbf{0},\textstyle{\frac{1}{2}})=\sqrt{m}\left(\begin{array}{cccc}
c_{+} \\
0 \\
c_{-} \\
0 \end{array}\right),\hspace{0.5cm}
u(\mathbf{0},-\textstyle{\frac{1}{2}})=\sqrt{m}\left(\begin{array}{cccc}
0 \\
c_{+} \\
0 \\
c_{-} \end{array}\right),
\end{equation}
\begin{equation}
v(\mathbf{0},\textstyle{\frac{1}{2}})=\sqrt{m}\left(\begin{array}{cccc}
0 \\
d_{+} \\
0 \\
d_{-} \end{array}\right),\hspace{0.5cm}
v(\mathbf{0},-\textstyle{\frac{1}{2}})=-\sqrt{m}\left(\begin{array}{cccc}
d_{+} \\
0 \\
d_{-} \\
0 \end{array}\right).
\end{equation}
Equations (\ref{eq:b1}) and (\ref{eq:b2}) give us the coefficients at arbitrary momentum
\begin{equation}
u(\p,\sigma)=\mathcal{D}(L(p))u(\0,\sigma),
\end{equation}
\begin{equation}
v(\p,\sigma)=\mathcal{D}(L(p))v(\0,\sigma).
\end{equation}

Field operators with these coefficients satisfy the continuous Poincar\'{e} symmetry, so the proportionality constants can take arbitrary values. These constants can be determined by demanding parity conservation. The action of parity operator on the creation and annihilation operator are
\begin{equation}
\mathsf{P}a(\mathbf{p},\sigma)\mathsf{P}^{-1}=\eta^{*}a(-\p,\sigma),\label{eq:p1}
\end{equation}
\begin{equation}
\mathsf{P}b^{\dag}(\mathbf{p},\sigma)U^{-1}\mathsf{P}^{-1}=\bar{\eta}b^{\dag}(-\p,\sigma)\label{eq:p2}
\end{equation}
where the phases $\eta$ and $\bar{\eta}$ are the intrinsic parities independent of $\sigma$. 

Parity is a symmetry transformation when $\mathsf{P}\Psi(x)\mathsf{P}^{-1}$ is proportional to $\Psi(\mathscr{P}x)$.
To find the relation between the two fields, we write the field operator as $\Psi(x)=\kappa\Psi^{+}(x)+\lambda\Psi^{-}(x)$, where
\begin{equation}
\Psi^{+}(x)=(2\pi)^{-3/2}\int\frac{d^{3}p}{\sqrt{2p^{0}}}\sum_{\sigma}e^{-ip\cdot x}u(\mathbf{p},\sigma)a(\p,\sigma),
\end{equation}
\begin{equation}
\Psi^{-}(x)=(2\pi)^{-3/2}\int\frac{d^{3}p}{\sqrt{2p^{0}}}\sum_{\sigma}e^{ip\cdot x}v(\mathbf{p},\sigma)b^{\dag}(\p,\sigma).
\end{equation}
Acting $\mathsf{P}$ on $\Psi^{\pm}(x)$ with the help of eqs.~(\ref{eq:p1}) and (\ref{eq:p2}), we get
\begin{equation}
\mathsf{P}\Psi^{+}(x)\mathsf{P}^{-1}=\eta^{*}(2\pi)^{-3/2}\int\frac{d^{3}p}{\sqrt{2p^{0}}}\sum_{\sigma}e^{-ip\cdot x}u(\mathbf{p},\sigma)a(-\mathbf{p},\sigma),\label{eq:pc1}
\end{equation}
\begin{equation}
\mathsf{P}\Psi^{-}(x)\mathsf{P}^{-1}=\bar{\eta}(2\pi)^{-3/2}\int\frac{d^{3}p}{\sqrt{2p^{0}}}\sum_{\sigma}e^{ip\cdot x}v(\mathbf{p},\sigma)b^{\dag}(-\mathbf{p},\sigma).\label{eq:pc2}
\end{equation}
For the fields $\Psi^{\pm}(x)$ to conserve parity, it is necessary to find a relation between the coefficients and their momentum-reflected counterpart. We use the following identity
\begin{equation}
u(\mathbf{-p},\sigma)=\Big[\Gamma\,\mathcal{D}(L(p))\,\Gamma\Big]u(\mathbf{0},\sigma)\label{eq:inversion}
\end{equation}
where
\begin{equation}
\Gamma=\left(\begin{array}{cc}
O & I \\
I & O \end{array}\right)\label{eq:Gamma}
\end{equation}
is acting as a parity operator on the boost matrix $\Gamma\mathcal{D}(L(p))\Gamma=\mathcal{D}(L(\mathscr{P}p))$. Therefore, a sufficient condition for parity conservation is
\begin{equation}
\Gamma u(\mathbf{0},\sigma)=b_{u}u(\mathbf{0},\sigma),\hspace{0.5cm}
\Gamma v(\mathbf{0},\sigma)=b_{v}v(\mathbf{0},\sigma) \label{eq:eigen}
\end{equation}
where $b_{u}$ and $b_{v}$ are sign factors with $b_{u}^{2}=b_{v}^{2}=1$. Multiply both sides of the equation from the left by $\mathcal{D}(L(p))$, we get
\begin{equation}
u(-\mathbf{p},\sigma)=b_{u}\Gamma\,u(\p,\sigma),\label{eq:u}
\end{equation}
\begin{equation}
v(-\mathbf{p},\sigma)=b_{v}\Gamma\,v(\mathbf{p},\sigma).\label{eq:v}
\end{equation}
Substituting eqs.~(\ref{eq:u}) and (\ref{eq:v}) into eqs.~(\ref{eq:pc1}) and (\ref{eq:pc2}), we obtain two parity-conserving fields
\begin{equation}
\mathsf{P}\Psi^{+}(x)\mathsf{P}^{-1}=\eta^{*}b_{u}\Gamma\Psi^{+}(\mathscr{P}x),
\end{equation}
\begin{equation}
\mathsf{P}\Psi^{-}(x)\mathsf{P}^{-1}=\bar{\eta}\,b_{v}\Gamma\Psi^{-}(\mathscr{P}x).
\end{equation}
Adjusting the overall scale of coefficients by setting $c_{+}=d_{+}=1$ and using eq.~(\ref{eq:eigen}), we get
\begin{equation}
u(\0,\textstyle{\frac{1}{2}})=\sqrt{m}\left(\begin{array}{cccc}
1 \\
0 \\
b_{u} \\
0 \end{array}\right),\hspace{0.2cm}
u(\0,-\textstyle{\frac{1}{2}})=\sqrt{m}\left(\begin{array}{cccc}
0 \\
1 \\
0 \\
b_{u} \end{array}\right),\hspace{0.2cm}
\end{equation}
\begin{equation}
v(\0,\textstyle{\frac{1}{2}})=\sqrt{m}\left(\begin{array}{cccc}
0 \\
1 \\
0 \\
b_{v} \end{array}\right),\hspace{0.2cm}
v(\0,-\textstyle{\frac{1}{2}})=-\sqrt{m}\left(\begin{array}{cccc}
1 \\
0 \\
b_{v} \\
0 \end{array}\right). \nonumber
\end{equation}

To determine the sign factors, we apply the locality condition, that $\Psi(x)$ must either commute or anti-commute with its adjoint $\Psi^{\dag}(x)$ at space-like separation
\begin{equation}
[\Psi(x),\Psi^{\dag}(y)]_{\pm}=0.
\end{equation}
Explicit computation gives us
\begin{equation}
[\Psi(x),\Psi^{\dag}(y)]_{\pm}=(2\pi)^{-3}\int\frac{d^{3}p}{2p_{0}}\Big[|\kappa|^{2}e^{-ip\cdot(x-y)}N(\mathbf{p})\pm|\lambda|^{2}e^{ip\cdot(x-y)}M(\mathbf{p})\Big]\label{eq:locality1}
\end{equation}
where $N(\mathbf{p})$ and $M(\mathbf{p})$ are the spin-sums
\begin{equation}
N(\mathbf{p})=\sum_{\sigma}u(\mathbf{p},\sigma)u^{\dag}(\mathbf{p},\sigma),\label{eq:sp1}
\end{equation}
\begin{equation}
M(\mathbf{p})=\sum_{\sigma}v(\mathbf{p},\sigma)v^{\dag}(\mathbf{p},\sigma).\label{eq:sp2}
\end{equation}
We can evaluate the spin-sums directly, but here it is more instructive to evaluate them at rest
\begin{equation}
N(\mathbf{0})=(b_{u}\Gamma+I)m,\hspace{0.5cm}
M(\mathbf{0})=(b_{v}\Gamma+I)m
\end{equation}
then boost them to arbitrary momentum. Later, this method is used obtain spin-sums of arbitrary momentum for higher-spin representations without explicit computation. The spin-sums at arbitrary momentum are obtained by applying the boost matrix $\mathcal{D}(L(p))$
\begin{eqnarray}
N(\mathbf{p})&=&\sum_{\sigma}\mathcal{D}(L(p))u(\mathbf{0},\sigma)u^{\dag}(\mathbf{0},\sigma)\mathcal{D}^{\dag}(L(p)) \\ \nonumber
&=& m\Big[\mathcal{D}(L(p))\mathcal{D}^{\dag}(L(p))+b_{u}\Gamma\Big]
\end{eqnarray}
and similarly for $M(\mathbf{p})$,
\begin{equation}
M(\mathbf{p})=m\Big[\mathcal{D}(L(p))\mathcal{D}^{\dag}(L(p))+b_{u}\Gamma\Big]
\end{equation}
where we have used the identity $\mathcal{D}(L(p))\Gamma\mathcal{D}^{\dag}(L(p))=\Gamma$. The term $\mathcal{D}(L(p))\mathcal{D}^{\dag}(L(p))$ is a quadratic polynomial in $p^{\mu}$. It can be written in a Lorentz covariant form,
\begin{equation}
\mathcal{D}(L(p))\mathcal{D}^{\dag}(L(p))=\frac{\gamma^{\mu}p_{\mu}}{m}\gamma^{0}
\end{equation}
where $\gamma^{\mu}$ is the Dirac matrix in the chiral representation
\begin{equation}
\gamma^{0}=\Gamma=\left(\begin{array}{cc}
O & I \\
I & O \end{array}\right),\hspace{0.5cm}
\gamma^{i}=\left(\begin{array}{cc}
O &-\sigma^{i} \\
\sigma^{i} & O \end{array}\right).
\end{equation}
They satisfy the Clifford algebra
\begin{equation}
\{\gamma^{\mu},\gamma^{\nu}\}=2\eta^{\mu\nu}I.
\end{equation}
 For later convenience, we define
\begin{equation}
\gamma^{5}=i\gamma^{0}\gamma^{1}\gamma^{2}\gamma^{3}=
\left(\begin{array}{cc}
I & O \\
O &-I \end{array}\right).
\end{equation}
The spin-sums in the covariant form are
\begin{equation}
N(\p)=(\gamma^{\mu}p_{\mu}+b_{u}m I)\gamma^{0},
\end{equation}
\begin{equation}
M(\p)=(\gamma^{\mu}p_{\mu}+b_{v}m I)\gamma^{0}.
\end{equation}
Substituting the spin-sums into eq.~(\ref{eq:locality1}) we get
\begin{eqnarray}
[\Psi(x),\Psi^{\dag}(y)]_{\pm}&=&(2\pi)^{-3}\int\frac{d^{3} p}{2p^{0}}\Big[\,|\kappa|^{2}e^{-ip\cdot(x-y)}(\gamma^{\mu}p_{\mu}+b_{u}m I)\label{eq:locality2}\\
&&\hspace{4cm}\pm |\lambda|^{2}e^{ip\cdot(x-y)}(\gamma^{\mu}p_{\mu}+b_{v}mI)\Big]\gamma^{0}\nonumber \\
&=&\Big[(|\kappa|^{2}\mp|\lambda|^{2})i\gamma^{\mu}\partial_{\mu}\Delta_{+}(x-y)+mI(|\kappa|^{2}b_{u}\pm|\lambda|^{2}b_{v})\Delta_{+}(x-y)\Big]\gamma^{0}.\nonumber
\end{eqnarray}
The third line uses the property that $\Delta_{+}(x)$ is an even function when $x$ is space-like. Equation (\ref{eq:locality2}) vanishes when the Dirac field furnishes fermionic statistics with $|\kappa|^{2}=|\lambda|^{2}=1$ and
\begin{equation}
b_{u}=-b_{v}=1\label{eq:bubv}
\end{equation}
thus completely determining the coefficients of the Dirac field
\begin{equation}
u(\0,\textstyle{\frac{1}{2}})=\sqrt{m}\left(\begin{array}{cccc}
1 \\
0 \\
1 \\
0 \end{array}\right),\hspace{0.2cm}
u(\0,-\textstyle{\frac{1}{2}})=\sqrt{m}\left(\begin{array}{cccc}
0 \\
1 \\
0 \\
1 \end{array}\right),\hspace{0.2cm}
\end{equation}
\begin{equation}
v(\0,\textstyle{\frac{1}{2}})=\sqrt{m}\left(\begin{array}{cccc}
0 \\
1 \\
0 \\
-1 \end{array}\right),\hspace{0.2cm}
v(\0,-\textstyle{\frac{1}{2}})=\sqrt{m}\left(\begin{array}{cccc}
-1 \\
0 \\
1 \\
0 \end{array}\right). \nonumber
\end{equation}

\subsubsection{Parity}

Re-examining the transformations of $\Psi^{+}(x)$ and $\Psi^{-}(x)$ under parity, we arrive at the conclusion that parity is preserved for $\Psi(x)$ if particles and anti-particles have opposite intrinsic parity
\begin{equation}
\eta^{*}=-\bar{\eta}
\end{equation}
thus giving us 
\begin{equation}
\mathsf{P}\Psi(x)\mathsf{P}^{-1}=\eta^{*}\gamma^{0}\Psi(\mathscr{P}x).
\end{equation}

We have so far made the distinction between particles and anti-particles. Poincar\'{e} symmetry also allows the possibility of neutral fermions known as the Majorana fermions described by the neutral Dirac field
\begin{equation}
\psi(x)=(2\pi)^{-3/2}\int\frac{d^{3}p}{\sqrt{2p^{0}}}\sum_{\sigma}\left[e^{-ip\cdot x}u(\p,\sigma)a(\p,\sigma)+e^{ip\cdot x}v(\p,\sigma)a^{\dag}(\p,\sigma)\right]
\label{eq:Majorana_fermion}
\end{equation}
so that particles are indistinguishable from anti-particles. For Majorana fermions, there is only one intrinsic parity phase. Therefore, parity is conserved for $\psi(x)$ if its intrinsic parity is imaginary $\eta=-\eta^{*}$
\begin{equation}
\mathsf{P}\psi(x)\mathsf{P}^{-1}=-\eta\gamma^{0}\psi(\mathscr{P}x).
\end{equation}
When particles and anti-particles are distinguishable, we have the freedom of choosing the intrinsic parity to be either real or imaginary (in most cases we take it to be real). However, for Majorana fermions parity conservation requires them to be imaginary.

The fact that fermions can have either real or imaginary parity can be understood as follow.\footnote{I am grateful to Matt Visser for this argument.} Fermions typically appear in the interaction an even number of times. Since the Hamiltonian is an observable, we must have $\mathsf{P}^{2}H\mathsf{P}^{-2}=H$. It follows that
\begin{equation}
\mathsf{P}^{2}\psi(x)\mathsf{P}^{-2}=\pm \psi(x)
\end{equation}
and so the intrinsic parity of fermions can be either real or imaginary 
\begin{equation}
\eta_{f}=\pm\eta_{f}^{*}.
\end{equation}
 By the same argument, since bosonic fields usually appear in the interaction an odd number of times, its intrinsic parity must be real
 \begin{equation}
 \eta_{b}=\eta_{b}^{*}.
 \end{equation}
\subsubsection{Charge-conjugation}

The charge-conjugated Dirac field is
\begin{equation}
\mathsf{C}\Psi(x)\mathsf{C}^{-1}=(2\pi)^{-3/2}\int\frac{d^{3}p}{\sqrt{2p^{0}}}\sum_{\sigma}\left[e^{-ip\cdot x}u(\p,\sigma)\varsigma^{*}b(\p,\sigma)+e^{ip\cdot x}v(\p,\sigma)\bar{\varsigma}a^{\dag}(\p,\sigma)\right].
\end{equation}
We find the coefficients $u(\p,\sigma)$ and $v(\p,\sigma)$ are related to their complex-conjugates by
\begin{equation}
u(\p,\sigma)=i\gamma^{2}v^{*}(\p,\sigma),
\end{equation}
\begin{equation}
v(\p,\sigma)=i\gamma^{2}u^{*}(\p,\sigma).
\end{equation}
Therefore, charge-conjugation is conserved if both particles and anti-particles have the same phase
\begin{equation}
\varsigma^{*}=\bar{\varsigma}
\end{equation}
so that the charge-conjugated field is related to its complex conjugate
\begin{equation}
\mathsf{C}\Psi(x)\mathsf{C}^{-1}=i\varsigma\gamma^{2}\Psi^{*}(x).
\end{equation}

In the case of neutral Dirac field, there exists only one phase. This phase has to be real to satisfy charge-conjugation symmetry $\varsigma^{*}=\varsigma$. Consequently, we obtain
\begin{equation}
\mathsf{C}\psi(x)\mathsf{C}^{-1}=\varsigma\psi(x)
\end{equation}
so the neutral Dirac field is invariant under charge-conjugation.

\subsubsection{Time-reversal}

The Dirac field describes particle states with non-trivial $\sigma$ label so its transformation under time-reversal involves $\sigma$-dependent phases in accordance with eq.~(\ref{eq:massive_time_reversal})
\begin{eqnarray}
\hspace{-1.5cm}\mathsf{T}\Psi(x)\mathsf{T}^{-1}&=&(2\pi)^{-3/2}\int\frac{d^{3}p}{\sqrt{2p^{0}}}\sum_{\sigma}(-1)^{1/2-\sigma}\Big[e^{ip\cdot x}u^{*}(\p,\sigma)\varrho^{*}a(-\p,-\sigma)\nonumber\\
&&\hspace{5.5cm}+e^{-ip\cdot x}v^{*}(\p,\sigma)\bar{\varrho}\,b^{\dag}(-\p,-\sigma)\Big].\label{eq:spin_half_time_reversal_dfield}
\end{eqnarray}
The complex-conjugated coefficients satisfy the following identities
\begin{equation}
u^{*}(\p,\sigma)=(-1)^{1/2-\sigma}i\gamma^{0}\gamma^{2}\gamma^{5}u(-\p,-\sigma),
\end{equation}
\begin{equation}
v^{*}(\p,\sigma)=(-1)^{1/2-\sigma}i\gamma^{0}\gamma^{2}\gamma^{5}v(-\p,-\sigma).
\end{equation}
Substituting these identities into eq.~(\ref{eq:spin_half_time_reversal_dfield}), the Dirac field satisfies the time-reversal symmetry under the condition
\begin{equation}
\varrho^{*}=\bar{\varrho}
\end{equation}
so that
\begin{equation}
\mathsf{T}\Psi(x)\mathsf{T}^{-1}=\varrho^{*} i\gamma^{0}\gamma^{2}\gamma^{5}\Psi(\mathscr{T}x).
\end{equation}
Similar to charge-conjugation and parity, time-reversal is conserved for $\psi(x)$ if the time-reversal phase is real $\varrho=\varrho^{*}$ thus giving us
\begin{equation}
\mathsf{T}\psi(x)\mathsf{T}^{-1}=\varrho\,i\gamma^{0}\gamma^{2}\gamma^{5}\psi(\mathscr{T}x).
\end{equation}

We have determined the Lorentz transformations of the Dirac field. We now shift our attention to derive the field equation of the Dirac field and how to construct Lorentz invariant interactions. Towards this end, we need to determine the matrix $g$  in eq.~(\ref{eq:g_constraint1}). This will allow us to construct the field adjoint and Lorentz invariant interactions. Since the rotation and boost generators in eq.~(\ref{eq:rb_generators})  are Hermitian and anti-Hermitian respectively, eq.~(\ref{eq:g_constraint1}) can be rewritten as
\begin{equation}
[g,\mJ]=\{g,\mK\}=0
\end{equation}
which gives us
\begin{equation}
g=
\left(\begin{array}{cc}
O & bI \\
aI & O \end{array}\right)
\end{equation}
where $a$ and $b$ are proportionality constants. The demand of parity conservation for scalar such as $\Psi^{\dag}(x)g\Psi(x)$ requires that $a=b$. We may take $a=1$ without losing generality to obtain
\begin{equation}
g=\gamma^{0}.
\end{equation}
The field-adjoint operator $\overline{\Psi}(x)$ defined for notational convenience is
\begin{equation}
\overline{\Psi}(x)=\Psi^{\dag}(x)\gamma^{0}
\end{equation}
up to a global phase, positivity of the free Hamiltonian for the Dirac field demands it to be unity. Using eqs.~(\ref{eq:cov}) and (\ref{eq:g_constraint1}), the field-adjoint transforms as
\begin{equation}
U(\Lambda,b)\overline{\Psi}(x)U^{-1}(\Lambda,b)=\overline{\Psi}(\Lambda x+b)\mathcal{D}^{-1}(\Lambda)
\end{equation}
so that general functions of $\overline{\Psi}(x)\Psi(x)$ transform as scalar and can therefore be used to construct Lorentz-invariant interactions. 

Expanding $\overline{\Psi}(x)$, we get
\begin{equation}
\overline{\Psi}(x)=(2\pi)^{-3/2}\int\frac{d^{3}p}{\sqrt{2p^{0}}}\sum_{\sigma}\left[e^{ip\cdot x}\bar{u}(\p,\sigma)a^{\dag}(\p,\sigma)+e^{-ip\cdot x}\bar{v}(\p,\sigma)b(\p,\sigma)\right]
\end{equation}
where $\bar{u}(\p,\sigma)$ and $\bar{v}(\p,\sigma)$ are the dual coefficients
\begin{equation}
\bar{u}(\p,\sigma)=u^{\dag}(\p,\sigma)\gamma^{0},\hspace{0.5cm}
\bar{v}(\p,\sigma)=v^{\dag}(\p,\sigma)\gamma^{0}.
\end{equation}
The norm of the coefficients are defined in terms of the dual and they are orthonormal
\begin{equation}
\bar{u}(\p,\sigma)u(\p,\sigma')=2m\delta_{\sigma\sigma'},\hspace{0.5cm}
\bar{v}(\p,\sigma)v(\p,\sigma')=-2m\delta_{\sigma\sigma'},
\end{equation}
\begin{equation}
\bar{u}(\p,\sigma)v(\p,\sigma')=\bar{v}(\p,\sigma)u(\p,\sigma')=0.
\end{equation}
The spin-sums are Lorentz covariant
\begin{equation}
\sum_{\sigma}u(\p,\sigma)\bar{u}(\p,\sigma)=\gamma^{\mu}p_{\mu}+mI, \label{eq:u_spin_sum}
\end{equation}
\begin{equation}
\sum_{\sigma}v(\p,\sigma)\bar{v}(\p,\sigma)=\gamma^{\mu}p_{\mu}-mI \label{eq:v_spin_sum}
\end{equation}
and they satisfy the completeness relation
\begin{equation}
\frac{1}{2m}\sum_{\sigma}\left[u(\p,\sigma)\bar{u}(\p,\sigma)-v(\p,\sigma)\bar{v}(\p,\sigma)\right]=I.
\end{equation}
Therefore, the coefficients and their dual form an orthonormal and complete basis. 

Multiply eqs.~(\ref{eq:u_spin_sum}) and (\ref{eq:v_spin_sum}) by $u(\p,\sigma')$ and $v(\p,\sigma')$ respectively and using the orthonormality relations, we obtain the Dirac equation in the momentum space
\begin{equation}
(\gamma^{\mu}p_{\mu}-mI)u(\p,\sigma)=0,
\end{equation}
\begin{equation}
(\gamma^{\mu}p_{\mu}+mI)v(\p,\sigma)=0.
\end{equation}
The difference in the signs for the mass term can be traced back to the fact that particles and anti-particles have opposite intrinsic parity phases given by eq.~(\ref{eq:bubv}). In the configuration space, the field equation for $\Psi(x)$ and $\psi(x)$ are
\begin{equation}
(i\gamma^{\mu}\partial_{\mu}-mI)\Psi(x)=0,
\end{equation}
\begin{equation}
(i\gamma^{\mu}\partial_{\mu}-mI)\psi(x)=0.
\end{equation}
Therefore, the Lagrangian densities for the charged and neutral Dirac fields are
\begin{equation}
\mathscr{L}_{\Psi}=\overline{\Psi}(i\gamma^{\mu}\partial_{\mu}-mI)\Psi+\mbox{h.c.}
\end{equation}
\begin{equation}
\mathscr{L}_{\psi}=\mathscr{L}_{\Psi\rightarrow\psi}
\end{equation}

As promised at the beginning of sec.~\ref{chap:quantum_fields}, we have derived the expansion coefficients, field equation and Lagrangian density by the demand of Poincar\'{e} symmetry. The derivation shows the uniqueness of the Dirac field. The only freedom we have are the choice of global phases for the expansion coefficients and their dual which do not affect of the physical content of the theory.  However, in the subsequent chapter on vector fields, we show that the phases for the field-adjoint are local and of physical significance. The difference between the Dirac and the vector field is that the former furnishes an irreducible representation of the Poincar\'{e} group while the later is reducible.

It should be emphasized that the derivations of the Dirac spinors and their field equations are far from trivial. Indeed, browsing through various textbooks on quantum field theory, we find that many have the incorrect label identification for $v(\0,\sigma)$ and neglected important phases~\cite{Itzykson:1980rh,Ryder:1985wq,Zee:2010}. For Majorana fermions, a straightforward computation shows that the fields are non-local $\{\psi(\mathbf{x},t),\psi(\mathbf{y},t)\}\neq 0$ as opposed to being proportional to a Dirac delta function had one used the correct spinor identification and phases. Consequently, the $S$-matrix for Majorana fermions would not be Poincar\'{e}-invariant unless we choose the correct spinors.

\subsection{Higher spin fields}

We now generalise the Dirac field to higher-spin representation. Although this does not include all the possible representations of the Lorentz group, it does reveal certain general structures of quantum field theory such as the spin-statistics and discrete symmetries of particles of different spin and species. The construction of massive quantum fields of general irreducible representations of the Lorentz group can be found in~\cite[sec.~5.6-5.8]{Weinberg:1995mt}.

The generators of higher-spin fields are of the form given in eq.~(\ref{eq:rb_generators}). The rotation constraints on the coefficients are
\begin{equation}
\sum_{\bar{\sigma}}(U_{\pm})_{\ell\bar{\sigma}}\J_{\bar{\sigma}\sigma}=\sum_{\bar{\ell}}\J_{\ell\bar{\ell}}(U_{\pm})_{\bar{\ell}\sigma},
\end{equation}
\begin{equation}
\sum_{\bar{\sigma}}(V_{\pm})_{\ell\bar{\sigma}}\J^{*}_{\bar{\sigma}\sigma}=-\sum_{\bar{\ell}}\J_{\ell\bar{\ell}}(V_{\pm})_{\bar{\ell}\sigma}
\end{equation}
where $U_{\pm}$ and $V_{\pm}$ are now $(2j+1)\times(2j+1)$ matrices and the components of $\J_{\bar{\sigma}\sigma}$ given by eqs.~(\ref{eq:r1}-\ref{eq:r3}). Choosing $\J_{\ell\bar{\ell}}$ to be equal to $\J_{\sigma\bar{\sigma}}$, according to Schur's Lemma, the only non-trivial solution of $U_{\pm}$ must be proportional to the identity matrix
\begin{equation}
(U_{\pm})_{\ell\sigma}=c_{\pm}\delta_{\ell\sigma}.
\end{equation}
The solution for $V_{\pm}$ can be found by using the relation
\begin{equation}
\J^{*}_{\bar{\sigma}\sigma}=-(-1)^{\bar{\sigma}-\sigma}\J_{(-\bar{\sigma})(-\sigma)}.\label{eq:js_j}
\end{equation}
Substituting eq.~(\ref{eq:js_j}) into the constraint for $V_{\pm}$ and rearrange the indices, one obtains
\begin{equation}
\sum_{\bar{\sigma}}(V_{\pm})_{\bar{\ell}(-\bar{\sigma})}(-1)^{-\bar{\sigma}}\J_{\bar{\sigma}\sigma}=\sum_{\ell}\J_{\bar{\ell}\ell}(V_{\pm})_{\ell(-\sigma)}(-1)^{-\sigma}.
\end{equation}
Therefore, $(V_{\pm})_{\bar{\ell}(-\bar{\sigma})}(-1)^{-\bar{\sigma}}$ either vanishes or is proportional to the identity matrix. Adjusting the proportionality constants appropriately, we obtain
\begin{equation}
(V_{\pm})_{\ell\sigma}=d_{\pm}(-1)^{j+\sigma}\delta_{\ell(-\sigma)}.
\end{equation}

The coefficients, with appropriate normalisation take the form 
\begin{equation}
u(\0,j)=m^{j}\left(\begin{matrix}
c_{+} \\
0 \\
\vdots \\
0 \\
c_{-} \\
0 \\
\vdots \\
0
\end{matrix}\right),\,
u(\0,j-1)=m^{j}\left(\begin{matrix}
0 \\
c_{+} \\
\vdots \\
0 \\
0 \\
c_{-} \\
\vdots \\
0
\end{matrix}\right),\,\cdots\,,
u(\0,-j)=m^{j}\left(\begin{matrix}
0 \\
\vdots\\
0 \\
c_{+} \\
0 \\
\vdots\\
0 \\
c_{-} \\
\end{matrix}\right)
\end{equation}
\begin{equation}
v(\0,j)=(-1)^{2j}m^{j}\left(\begin{matrix}
0 \\
\vdots\\
0\\
d_{+}\\
0 \\
\vdots\\
0\\
d_{-}
\end{matrix}\right)\,,
v(\0,j-1)=(-1)^{2j-1}m^{j}\left(\begin{matrix}
0 \\
\vdots\\
d_{+}\\
0 \\
0 \\
\vdots\\
d_{-}\\
0 \end{matrix}\right),\,\cdots\,,
v(\0,-j)=m^{j}\left(\begin{matrix}
d_{+}\\
0\\
\vdots\\
0\\
d_{-}\\
0\\
\vdots\\
0\end{matrix}\right)
\end{equation}
where the dots in the entries represent null elements. The proportionality constants $c_{\pm}$ and $d_{\pm}$ can be determined by parity conservation using methods presented in the previous section for the Dirac field. A straightforward calculation yields
\begin{equation}
u(\0,j)=m^{j}\left(\begin{matrix}
1 \\
0 \\
\vdots \\
0 \\
b_{u} \\
0 \\
\vdots \\
0
\end{matrix}\right),\,
u(\0,j-1)=m^{j}\left(\begin{matrix}
0 \\
1 \\
\vdots \\
0 \\
0 \\
b_{u} \\
\vdots \\
0
\end{matrix}\right),\,\cdots\,,
u(\0,-j)=m^{j}\left(\begin{matrix}
0 \\
\vdots\\
0 \\
1 \\
0 \\
\vdots\\
0 \\
b_{u} \\
\end{matrix}\right)
\end{equation}
\begin{equation}
v(\0,j)=(-1)^{2j+1}m^{j}\left(\begin{matrix}
0 \\
\vdots\\
0\\
1\\
0 \\
\vdots\\
0\\
b_{v}
\end{matrix}\right)\,,
v(\0,j-1)=(-1)^{2j}m^{j}\left(\begin{matrix}
0 \\
\vdots\\
1\\
0 \\
0 \\
\vdots\\
b_{v}\\
0 \end{matrix}\right),\,\cdots\,,
v(\0,-j)=-m^{j}\left(\begin{matrix}
1\\
0\\
\vdots\\
0\\
b_{v}\\
0\\
\vdots\\
0\end{matrix}\right).
\end{equation}

The sign factors $b_{u}$ and $b_{v}$ are determined by the locality condition
\begin{equation}
[\Psi(x),\Psi^{\dag}(y)]_{\pm}=(2\pi)^{-3}\int\frac{d^{3}p}{2p^{0}}
\left[e^{-ip\cdot(x-y)}N(\p)\pm e^{ip\cdot(x-y)}M(\p)\right]\label{eq:general_locality}
\end{equation}
where the spin-sums are given by
\begin{equation}
N(\p)=\sum_{\sigma}u(\p,\sigma)u^{\dag}(\p,\sigma)=m^{2j}\left[b_{u}\Gamma+\mathcal{D}(L(p))\mathcal{D}^{\dag}(L(p))\right],
\end{equation}
\begin{equation}
M(\p)=\sum_{\sigma}v(\p,\sigma)v^{\dag}(\p,\sigma)=m^{2j}\left[b_{v}\Gamma+\mathcal{D}(L(p))\mathcal{D}^{\dag}(L(p))\right].
\end{equation}
The second equality of the spin-sums are obtained using the general identity
\begin{equation}
\mathcal{D}(L(p))\Gamma\mathcal{D}(L(p))=\Gamma
\end{equation}
where
\begin{equation}
\Gamma=\left(\begin{matrix}
O & I \\
I & O \end{matrix}\right).
\end{equation}

The term $\mathcal{D}(L(p))\mathcal{D}^{\dag}(L(p))$ takes the form of a direct sum
\begin{equation}
\mathcal{D}(L(p))\mathcal{D}^{\dag}(L(p))=
\left(\begin{matrix}
\exp(2\J\cdot\bv) & O \\
O & \exp(-2\J\cdot\bv) \end{matrix}\right).
\end{equation}
In app.~\ref{AppendixE}, we reproduce the proof given by Weinberg in~\cite[app.~A]{Weinberg:1964cn} that there exists a traceless, symmetric rank $2j$ tensor $t^{\mu_{1}\mu_{2}\cdots\mu_{2j}}$ such that $\exp(-2\J\cdot\bv)$ can be expressed as a Lorentz-covariant polynomial
\begin{equation}
\exp(-2\J\cdot\bv)=\frac{(-1)^{2j}}{m^{2j}}t^{\mu_{1}\mu_{2}\cdots\mu_{2j}}p_{\mu_{1}}p_{\mu_{2}}\cdots p_{\mu_{2j}}.
\end{equation}
Since $\exp(2\J\cdot\bv)$ is related to $\exp(-2\J\cdot\bv)$ by a parity transformation, it can also be written in a covariant form
\begin{equation}
\exp(2\J\cdot\bv)=\frac{(-1)^{2j}}{m^{2j}}\bar{t}^{\mu_{1}\mu_{2}\cdots\mu_{2j}}p_{\mu_{1}}p_{\mu_{2}}\cdots p_{\mu_{2j}}
\end{equation}
where
\begin{equation}
\bar{t}^{\mu_{1}\mu_{2}\cdots\mu_{2j}}=\mathscr{P}^{\mu_{1}}_{\quad\nu_{1}}\mathscr{P}^{\mu_{2}}_{\quad\nu_{2}}\cdots\mathscr{P}^{\mu_{2j}}_{\quad\nu_{2j}}t^{\nu_{1}\nu_{2}\cdots\nu_{2j}}.
\end{equation}
Rewriting $\mathcal{D}(L(p))\mathcal{D}^{\dag}(L(p))$ by introducing the generalised $\gamma$-matrix
\begin{equation}
\mathcal{D}(L(p))\mathcal{D}^{\dag}(L(p))=\frac{1}{m^{2j}}\gamma^{\mu_{1}\mu_{2}\cdots\mu_{2j}}p_{\mu_{1}}p_{\mu_{2}}\cdots p_{\mu_{2j}}\Gamma
\end{equation}
where
\begin{equation}
\gamma^{\mu_{1}\mu_{2}\cdots\mu_{2j}}=(-1)^{2j}
\left(\begin{matrix}
O & \bar{t}^{\mu_{1}\mu_{2}\cdots\mu_{2j}}\\
t^{\mu_{1}\mu_{2}\cdots\mu_{2j}} & O 
\end{matrix}\right),
\end{equation}
the spin-sums become
\begin{equation}
N(\p)=(\gamma^{\mu_{1}\mu_{2}\cdots\mu_{2j}}p_{\mu_{1}}p_{\mu_{2}}\cdots p_{\mu_{2j}}+b_{u}m^{2j}I)\Gamma,
\end{equation}
\begin{equation}
M(\p)=(\gamma^{\mu_{1}\mu_{2}\cdots\mu_{2j}}p_{\mu_{1}}p_{\mu_{2}}\cdots p_{\mu_{2j}}+b_{v}m^{2j}I)\Gamma.
\end{equation}

When $j=\frac{1}{2}$, the above spin-sums coincide with eqs.~(\ref{eq:u_spin_sum}) and (\ref{eq:v_spin_sum}). Substituting them into eq.~(\ref{eq:general_locality}) yields
\begin{equation}
[\Psi(x),\Psi^{\dag}(y)]_{\pm}=\Big[(b_{u}\pm b_{v})m^{2j}I+[(-i)^{2j}\pm(i)^{2j}]\gamma^{\mu_{1}\mu_{2}\cdots\mu_{2j}}\partial_{\mu_{1}}\partial_{\mu_{2}}\cdots\partial_{\mu_{2j}}
\Big]\Delta_{+}(x-y)\Gamma.\label{eq:general_locality2}
\end{equation}
Equation (\ref{eq:general_locality2}) can be interpreted as a statement of the spin-statistics theorem. Quantum fields of integral and half-integral spin commute or anti-commute at space-like separation with their Hermitian conjugate respectively, provided that
\begin{equation}
b_{u}=1,\hspace{0.5cm}b_{v}=(-1)^{2j}.
\end{equation}
The expansion coefficients are then determined to be
\begin{equation}
u(\0,j)=m^{j}\left(\begin{matrix}
1 \\
0 \\
\vdots \\
0 \\
1 \\
0 \\
\vdots \\
0
\end{matrix}\right),\,
u(\0,j-1)=m^{j}\left(\begin{matrix}
0 \\
1 \\
\vdots \\
0 \\
0 \\
1 \\
\vdots \\
0
\end{matrix}\right),\,\cdots\,,
u(\0,-j)=m^{j}\left(\begin{matrix}
0 \\
\vdots\\
0 \\
1 \\
0 \\
\vdots\\
0 \\
1 \\
\end{matrix}\right)
\end{equation}
\begin{equation}
v(\0,j)=m^{j}\left(\begin{matrix}
0 \\
\vdots\\
0\\
(-1)^{2j+1}\\
0 \\
\vdots\\
0\\
-1
\end{matrix}\right)\,,
v(\0,j-1)=m^{j}\left(\begin{matrix}
0 \\
\vdots\\
(-1)^{2j}\\
0 \\
0 \\
\vdots\\
1\\
0 \end{matrix}\right),\,\cdots ,
v(\0,-j)=m^{j}\left(\begin{matrix}
-1\\
0\\
\vdots\\
0\\
(-1)^{2j+1}\\
0\\
\vdots\\
0\end{matrix}\right).
\end{equation}

\subsubsection{Parity}

The parity-transformed field can now be written as
\begin{equation}
\mathsf{P}\Psi(x)\mathsf{P}^{-1}=(2\pi)^{-3/2}\int\frac{d^{3}p}{\sqrt{2p^{0}}}\sum_{\sigma}\left[e^{-ip\cdot(\mathscr{P}x)}\eta^{*}u(\p,\sigma)a(\p,\sigma)+e^{ip\cdot(\mathscr{P}x)}(-1)^{2j}\bar{\eta}v(\p,\sigma)b^{\dag}(\p,\sigma)\right]\label{eq:general_parity}
\end{equation}
where we have used the identities
\begin{equation}
u(\p,\sigma)=\Gamma u(-\p,\sigma),\hspace{0.5cm}
v(\p,\sigma)=(-1)^{2j}\Gamma v(-\p,\sigma).
\end{equation}
Parity is conserved when the intrinsic parity phases in eq.~(\ref{eq:general_parity}) satisfy the relation
\begin{equation}
\eta^{*}=(-1)^{2j}\bar{\eta}.
\end{equation}
Therefore, fermions and anti-fermions have opposite intrinsic parity phases. Bosons and anti-bosons have even intrinsic parity phases. Both the fermionic and bosonic quantum fields satisfy
\begin{equation}
\mathsf{P}\Psi(x)\mathsf{P}^{-1}=\eta^{*}\Gamma\Psi(\mathscr{P}x).
\end{equation}

\subsubsection{Charge-conjugation}

Applying charge-conjugation operator to the quantum field gives us
\begin{equation}
\mathsf{C}\Psi(x)\mathsf{C}^{-1}=(2\pi)^{-3/2}\int\frac{d^{3}p}{\sqrt{2p^{0}}}\sum_{\sigma}\left[e^{-ip\cdot x}u(\p,\sigma)\varsigma^{*}b(\p,\sigma)+
e^{ip\cdot x}v(\p,\sigma)\bar{\varsigma}a^{\dag}(\p,\sigma)\right].
\end{equation}
Charge-conjugation is a symmetry if $\mathsf{C}\Psi(x)\mathsf{C}^{-1}$ is proportional to $\Psi^{*}(x)$. Rewriting eq.~(\ref{eq:js_j}) in terms of matrices
\begin{equation}
-\J^{*}=C\J C^{-1}
\end{equation}
with
\begin{equation}
C_{\ell\bar{\ell}}=-(-1)^{-j-\ell}\delta_{(-\ell)\bar{\ell}},
\end{equation}
the boost matrix $\mathcal{D}(L(p))$ is related to $\mathcal{D}^{*}(L(p))$ by
\begin{equation}
\mathcal{C}\mathcal{D}(L(p))\mathcal{C}^{-1}=\mathcal{D}^{*}(L(p))
\end{equation}
where 
\begin{equation}
\mathcal{C}=\left(\begin{matrix}
O & (-1)^{2j}C \\
C & O \end{matrix}\right).
\end{equation}
Since the rest coefficients are real, we get
\begin{equation}
u(\p,\sigma)=\mathcal{C}v^{*}(\p,\sigma),\hspace{0.5cm}
v(\p,\sigma)=\mathcal{C}u^{*}(\p,\sigma). \label{eq:cuv}
\end{equation}
Substituting eq.~(\ref{eq:cuv}) into $\mathsf{C}\Psi(x)\mathsf{C}^{-1}$, we see that charge-conjugation is a symmetry when particles and anti-particles have the same charge-conjugation phases
\begin{equation}
\varsigma^{*}=\bar{\varsigma}
\end{equation}
thus giving us
\begin{equation}
\mathsf{C}\Psi(x)\mathsf{C}^{-1}=\varsigma^{*}\mathcal{C}\Psi^{*}(x).
\end{equation}

\subsubsection{Time-reversal}

Applying time-reversal operator to the quantum field gives us
\begin{eqnarray}
\hspace{-1cm}\mathsf{T}\Psi(x)\mathsf{T}^{-1}&=&(2\pi)^{-3/2}\int\frac{d^{3}p}{\sqrt{2p^{0}}}\sum_{\sigma}(-1)^{j-\sigma}
\Big[e^{ip\cdot x}u^{*}(\p,\sigma)\varrho^{*}a(-\p,-\sigma)\nonumber \\
&&\hspace{5cm}+e^{-ip\cdot x}v^{*}(\p,\sigma)\bar{\varrho}b^{\dag}(-\p,-\sigma)\Big].
\end{eqnarray}
Time-reversal is a symmetry if $\mathsf{T}\Psi(x)\mathsf{T}^{-1}$ is proportional to $\Psi(\mathscr{T}x)$. Using the identity
\begin{equation}
\mathcal{T}\mathcal{D}(L(\mathscr{P}p))\mathcal{T}^{-1}=\mathcal{D}^{*}(L(p))
\end{equation}
where
\begin{equation}
\mathcal{T}=(-1)^{2j}
\left(\begin{matrix}
C & O \\
O & C \end{matrix}\right),
\end{equation}
the expansion coefficients satisfy
\begin{equation}
\mathcal{T}u(-\p,-\sigma)=(-1)^{j-\sigma}u^{*}(\p,\sigma),\hspace{0.5cm}
\mathcal{T}v(-\p,-\sigma)=(-1)^{j-\sigma}v^{*}(\p,\sigma).
\end{equation}
Substituting these into $\mathsf{T}\Psi(x)\mathsf{T}^{-1}$, we see that time-reversal is a symmetry when particles and anti-particles have identical time-reversal phases
\begin{equation}
\varrho^{*}=\bar{\varrho}
\end{equation}
thus giving us
\begin{equation}
\mathsf{T}\Psi(x)\mathsf{T}^{-1}=\varrho^{*}\mathcal{T}\Psi(\mathscr{T}x).
\end{equation}

We are now ready to derive the field equation for the massive spin-$j$ quantum field. Following the same procedure for the Dirac field, we define the field adjoint as
\begin{equation}
\overline{\Psi}(x)=\Psi^{\dag}(x)\Gamma
\end{equation}
so that the dual coefficients are given by
\begin{equation}
\bar{u}(\p,\sigma)=u^{\dag}(\p,\sigma)\Gamma,\hspace{0.5cm}
\bar{v}(\p,\sigma)=v^{\dag}(\p,\sigma)\Gamma.
\end{equation}
The norms of the coefficients are orthonormal, satisfying
\begin{equation}
\bar{u}(\p,\sigma)u(\p,\sigma')=2m^{2j}\delta_{\sigma\sigma'},\hspace{0.5cm}
\bar{v}(\p,\sigma)v(\p,\sigma')=2m^{2j}(-1)^{2j}\delta_{\sigma\sigma'}.
\end{equation}
The spin-sums are
\begin{equation}
\sum_{\sigma}u(\p,\sigma)\bar{u}(\p,\sigma)=(\gamma^{\mu_{1}\mu_{2}\cdots\mu_{2j}}p_{\mu_{1}}p_{\mu_{2}}\cdots p_{\mu_{2j}}+m^{2j}I)
\end{equation}
\begin{equation}
\sum_{\sigma}v(\p,\sigma)\bar{v}(\p,\sigma)=(\gamma^{\mu_{1}\mu_{2}\cdots\mu_{2j}}p_{\mu_{1}}p_{\mu_{2}}\cdots p_{\mu_{2j}}+(-1)^{2j}m^{2j}I)
\end{equation}
thus giving us the following equations in the momentum space
\begin{equation}
(\gamma^{\mu_{1}\mu_{2}\cdots\mu_{2j}}p_{\mu_{1}}p_{\mu_{2}}\cdots p_{\mu_{2j}}-m^{2j}I)u(\p,\sigma)=0,
\end{equation}
\begin{equation}
(\gamma^{\mu_{1}\mu_{2}\cdots\mu_{2j}}p_{\mu_{1}}p_{\mu_{2}}\cdots p_{\mu_{2j}}-(-1)^{2j}m^{2j}I)v(\p,\sigma)=0.
\end{equation}
The field equation in the configuration space is therefore
\begin{equation}
\left[\gamma^{\mu_{1}\mu_{2}\cdots\mu_{2j}}\partial_{\mu_{1}}\partial_{\mu_{2}}\cdots\partial_{\mu_{2j}}-(im)^{2j}I\right]\Psi(x)=0
\label{eq:higher_spin_field_equation}
\end{equation}
When $j=\frac{1}{2}$, it gives us the Dirac equation. For $j=1$, in the massless limit, the resulting field equation is equivalent to the Maxwell equations.

In the literature, the representation of the constructed massive spin-$j$ quantum field is known as the $(j,0)\oplus(0,j)$ representation of the Lorentz group where $(j,0)$ and $(0,j)$ are the irreducible spin-$j$ representations. From eq.~(\ref{eq:rb_generators}), we see that $(j,0)$ and $(0,j)$ representations have the same rotation generator $\J$ while the boost generators are $\K=-i\J$ and $\K=i\J$ respectively. As a result, the parity operator $\mathsf{P}$ maps the quantum fields of the two representations to each other. Therefore, parity is not a symmetry of the irreducible spin-$j$ representation. The quantum field of the $(j,0)\oplus(0,j)$ representation furnishes a reducible representation under continuous Lorentz transformation but becomes irreducible when parity is included.

\section{Massless quantum fields}\label{Massless QF}

The massless quantum fields can be constructed following the same procedure as their massive counterpart. For massive quantum fields of the $(j,0)$, $(0,j)$ and $(j,0)\oplus(0,j)$ representations, their massless limit is coincide with the massless quantum fields of those representations. Therefore, we will only construct the massless spin-half field. Their higher spin generalisations do not reveal further insight beyond what we have gained from the previous section. The construction of these general massless fields are given~\cite{Weinberg:1964ev}. On the other hand, the correspondence between massive and massless fields fails for the tensor-product representation $(j,j')=(j,0)\otimes(0,j')$~\cite{Weinberg:1964ev}. In fact, the massless quantum fields of the $(j,j)$ representation with the exception of scalar fields, are not Lorentz-covariant. This difficulty and the subsequent resolution as we will show, is the origin of gauge invariance~\cite{Weinberg:1965rz}. This issue will be discussed in the next chapter. For now, we focus on the massless spin-half field.

\subsection{Massless spin-half field}

We divide this section to construct the massless spin-half fields of  $(\frac{1}{2},0)$ and $(0,\frac{1}{2})$ representations. They are also known as the right-handed and left-handed fields respectively. We will show that for the right-handed field, particles and anti-particles have only $\frac{1}{2}$ and $-\frac{1}{2}$ helicities. For the left-handed field, particles and anti-particles have only $-\frac{1}{2}$ and $\frac{1}{2}$ helicities.

\subsubsection{Right-handed field: $\mathbf{(\frac{1}{2},0)}$}
We use the same constraints given by eqs.~(\ref{eq:con3}) and (\ref{eq:con4}) to determine the coefficients
\begin{equation}
\sum_{\bar{\sigma}}(u_{R})_{\bar{\ell}}(\L\p,\bar{\sigma})D^{(j)}_{\bar{\sigma}\sigma}(W(\Lambda,p))=\sum_{\ell}\mathcal{D}_{\bar{\ell}\ell}(\Lambda)(u_{R})_{\ell}(\mathbf{p},\sigma), 
\end{equation}
\begin{equation}
\sum_{\bar{\sigma}}(v_{R})_{\bar{\ell}}(\L\p,\bar{\sigma})D^{(j)*}_{\bar{\sigma}\sigma}(W(\Lambda,p))=\sum_{\ell}\mathcal{D}_{\bar{\ell}\ell}(\Lambda)(v_{R})_{\ell}(\mathbf{p},\sigma) 
\end{equation}
where the subscript $R$ tells us the coefficients are right-handed. The main difference between the massive and massless particle is the little group. The little group for massless particle is the  Euclidean group $ISO(2)$, its unitary irreducible representation is given by eq.~(\ref{eq:DW_massless}) to be
\begin{equation}
D_{\sigma'\sigma}(W(\Lambda,p))=\exp[i\phi(\Lambda,p)\sigma]\delta_{\sigma'\sigma}.
\end{equation}
Therefore,
\begin{equation}
u_{R}(\L\p,\sigma)e^{i\phi(\Lambda,p)\sigma}=\mathcal{D}(\Lambda)u_{R}(\p,\sigma),
\end{equation}
\begin{equation}
v_{R}(\L\p,\sigma)e^{-i\phi(\Lambda,p)\sigma}=\mathcal{D}(\Lambda)v_{R}(\p,\sigma).
\end{equation}

In the $(\frac{1}{2},0)$ representation, $\J=\s/2$ and $\K=-i\s/2$. We solve the above constraints starting with the Lorentz boost $\Lambda=L(p)$. Let $p=k=(\kappa,0,0,\kappa)$ be the standard vector, then $W(L(p),p)=I$ so that $\phi(L(p),p)=0$ thus giving us 
\begin{equation}
u_{R}(\p,\sigma)=\mathcal{D}(L(p))u_{R}(\k,\sigma),\label{eq:massless_boost_u_spinor}
\end{equation}
\begin{equation}
v_{R}(\p,\sigma)=\mathcal{D}(L(p))v_{R}(\k,\sigma).\label{eq:massless_boost_v_spinor}
\end{equation}
Now take $\Lambda=W(\alpha,\beta,\phi)=S(\alpha,\beta)R(\phi)$ to be an element of the little group where $S(\alpha,\beta)$ and $R(\phi)$ are given by eqs.~(\ref{eq:sab}) and (\ref{eq:rphi}) respectively.
Setting $\alpha=\beta=0$, the little group becomes a rotation about the $3$-axis, $W(0,0,\phi)=R(\phi)$. Therefore, the constraints are
\begin{equation}
u_{R}(\k,\sigma)e^{i\sigma\phi}=\mathcal{D}(R)u_{R}(\k,\sigma),\label{eq:massless_rot}
\end{equation}
\begin{equation}
v_{R}(\k,\sigma)e^{-i\sigma\phi}=\mathcal{D}(R)v_{R}(\k,\sigma)
\end{equation}
where $\phi$ is the angle of rotation. Finally, we take $\phi=0$, so that $W(\alpha,\beta,0)=S(\alpha,\beta)$ to obtain
\begin{equation}
u_{R}(\k,\sigma)=\mathcal{D}(S(\alpha,\beta))u_{R}(\k,\sigma), \label{eq:massless_sabx}
\end{equation}
\begin{equation}
v_{R}(\k,\sigma)=\mathcal{D}(S(\alpha,\beta))v_{R}(\k,\sigma). \label{eq:massless_sab}
\end{equation}

We now solve eqs.~(\ref{eq:massless_rot}-\ref{eq:massless_sab}) to determine the coefficients. Under rotation, $\mathcal{D}(R)$ is
\begin{equation}
\mathcal{D}(R)=\exp\left(\frac{1}{2}i\sigma^{3}\phi\right)=
\left(\begin{matrix}
e^{\frac{1}{2}i\phi} & O \\
O & e^{-\frac{1}{2}i\phi}\end{matrix}\right).
\end{equation}
Substituting the matrix into the rotation constraints for $\sigma=\pm\frac{1}{2}$, we obtain
\begin{equation}
u_{R}(\k,\textstyle{\frac{1}{2}})=\left(\begin{matrix}
u_{R1}(\k,\textstyle{\frac{1}{2}}) \\
0 \end{matrix}\right),\hspace{0.5cm}
u_{R}(\k,-\textstyle{\frac{1}{2}})=\left(\begin{matrix}
0 \\
u_{R2}(\k,-\textstyle{\frac{1}{2}})\end{matrix}\right),
\end{equation}
\begin{equation}
v_{R}(\k,\textstyle{\frac{1}{2}})=\left(\begin{matrix}
0 \\
v_{R2}(\k,\textstyle{\frac{1}{2}})\end{matrix}\right),\hspace{0.5cm}
v_{R}(\k,-\textstyle{\frac{1}{2}})=\left(\begin{matrix}
v_{R1}(\k,-\textstyle{\frac{1}{2}}) \\
0 \end{matrix}\right).
\end{equation}
In the $(\frac{1}{2},0)$ representation, the matrix $\mathcal{D}(S(\alpha,\beta))$ is given by
\begin{equation}
\mathcal{D}(S(\alpha,\beta))=e^{i\alpha A}e^{i\beta B}
\end{equation}
where
\begin{equation}
A=K^{1}+J^{2}=\left(\begin{matrix}
0 &-i \\
0 & 0 \end{matrix}\right),
\end{equation}
\begin{equation}
B=K^{2}-J^{1}=\left(\begin{matrix}
0 &-1 \\
0 & 0 \end{matrix}\right).
\end{equation}
Therefore, the explicit expression of $\mathcal{D}(S(\alpha,\beta))$ reads
\begin{equation}
\mathcal{D}(S(\alpha,\beta))=
\left(\begin{matrix}
1 & \alpha-i\beta\\
0 & 1 \end{matrix}\right).
\end{equation}
Substituting $\mathcal{D}(S(\alpha,\beta))$ into the eqs.~(\ref{eq:massless_sabx}) and (\ref{eq:massless_sab}), we find it does not further constrain $u_{R}(\k,\textstyle{\frac{1}{2}})$ and $v_{R}(\k,-\textstyle{\frac{1}{2}})$, but for the remaining coefficients, we get
\begin{equation}
u_{R}(\k,-\textstyle{\frac{1}{2}})=\left(\begin{matrix}
0 \\
u_{R2}(\k,-\textstyle{\frac{1}{2}}) \end{matrix}\right)=
\left(\begin{matrix}
(\alpha-i\beta)u_{R2}(\k,-\textstyle{\frac{1}{2}})\\
u_{R2}(\k,-\textstyle{\frac{1}{2}})\end{matrix}\right),
\end{equation}
\begin{equation}
v_{R}(\k,\textstyle{\frac{1}{2}})=\left(\begin{matrix}
0 \\
v_{R2}(\k,\textstyle{\frac{1}{2}}) \end{matrix}\right)=
\left(\begin{matrix}
(\alpha-i\beta)v_{R2}(\k,\textstyle{\frac{1}{2}})\\
v_{R2}(\k,\textstyle{\frac{1}{2}})\end{matrix}\right).
\end{equation}
Generally $\alpha$ and $\beta$ are non-zero, so the coefficients $u_{R}(\k,-\textstyle{\frac{1}{2}})$ and $v_{R}(\k,\textstyle{\frac{1}{2}})$ must identically vanish
\begin{equation}
u_{R}(\k,-\textstyle{\frac{1}{2}})=v_{R}(\k,\textstyle{\frac{1}{2}})=0.
\end{equation}
Therefore, the right-handed massless spin-half field of the $(\frac{1}{2},0)$ representation is
\begin{equation}
\Psi_{R}(x)=(2\pi)^{-3/2}\int\frac{d^{3}p}{\sqrt{2p^{0}}}\left[e^{-ip\cdot x}u_{R}(\p,\textstyle{\frac{1}{2}})a(\p,\textstyle{\frac{1}{2}})+e^{ip\cdot x}v_{R}(\p,-\textstyle{\frac{1}{2}})b^{\dag}(\p,-\textstyle{\frac{1}{2}})\right].
\end{equation}

To determine the coefficients further we compute the following commutator/anti-commutator at space-like separation
\begin{equation}
[\Psi_{R}(x),\Psi_{R}^{\dag}(y)]_{\pm}=(2\pi)^{-3}\int\frac{d^{3}p}{2p^{0}}\left[e^{-ip\cdot(x-y)}|u_{R}(\p,\textstyle{\frac{1}{2}})|^{2}
\pm e^{ip\cdot(x-y)}|v_{R}(\p,-\textstyle{\frac{1}{2}})|^{2}\right]
\end{equation}
The coefficients of arbitrary momentum are obtained by eqs.~(\ref{eq:massless_boost_u_spinor}) and (\ref{eq:massless_boost_v_spinor}) using the same method described in sec.~\ref{Massless_particle_state}. First, boost $k^{\mu}$ along the $3$-axis, then rotate it to the direction $\hat{\p}$. The boost matrix $\mathcal{D}(L(p))$ is then given by
\begin{eqnarray}
\mathcal{D}(L(p))&=&\mathcal{D}(R(\hat{\p}))\mathcal{D}(B(|\p|))\label{eq:massless_boost_coefficients}\\
&=&\exp\left(-i\frac{\sigma^{3}}{2}\phi\right)\exp\left(-i\frac{\sigma^{2}}{2}\theta\right)\exp\left(\frac{\sigma^{3}}{2}\varphi\right)\nonumber\\
&=&\left(\begin{matrix}
e^{-i\phi/2} e^{\varphi/2} \cos\frac{\theta}{2} & -e^{-i\phi/2}e^{-\varphi/2} \sin\frac{\theta}{2}\nonumber \\
e^{i\phi/2}e^{\varphi/2} \sin\frac{\theta}{2}& e^{i\phi/2}e^{-\varphi/2} \cos\frac{\theta}{2}\end{matrix}\right)
\end{eqnarray}
where $\varphi=\ln(p^{0}/k^{0})$ is the rapidity parameter. Explicit computation of $u_{R}(\p,\textstyle{\frac{1}{2}})u_{R}^{\dag}(\p,\textstyle{\frac{1}{2}})$ and $v_{R}(\p,\textstyle{\frac{1}{2}})v_{R}^{\dag}(\p,\textstyle{\frac{1}{2}})$ yields
\begin{equation}
u_{R}(\p,\textstyle{\frac{1}{2}})u_{R}^{\dag}(\p,\textstyle{\frac{1}{2}})=
\left(\begin{matrix}
\cos^{2}\frac{\theta}{2} & \frac{1}{2}e^{-i\phi} \sin\theta \\
\frac{1}{2}e^{i\phi}\sin\theta & \sin^{2}\frac{\theta}{2}\end{matrix}\right)e^{\varphi}|u_{R1}(\k,\textstyle{\frac{1}{2}})|^{2},\label{eq:uudag}
\end{equation}
\begin{equation}
v_{R}(\p,-\textstyle{\frac{1}{2}})v_{R}^{\dag}(\p,-\textstyle{\frac{1}{2}})=
\left(\begin{matrix}
\cos^{2}\frac{\theta}{2} & \frac{1}{2}e^{-i\phi} \sin\theta \\
\frac{1}{2}e^{i\phi}\sin\theta & \sin^{2}\frac{\theta}{2}\end{matrix}\right)e^{\varphi}|v_{R1}(-\k,-\textstyle{\frac{1}{2}})|^{2}.\label{eq:vvdag}
\end{equation}
Upon choosing
\begin{equation}
e^{\varphi}|u_{R1}(\k,\textstyle{\frac{1}{2}})|^{2}=e^{\varphi}|v_{R1}(-\k,-\textstyle{\frac{1}{2}})|^{2}=2p^{0}
\end{equation}
we get
\begin{equation}
u_{R1}(\k,\textstyle{\frac{1}{2}})=\pm\sqrt{2k^{0}},
\end{equation}
\begin{equation}
v_{R1}(\k,-\textstyle{\frac{1}{2}})=\pm\sqrt{2k^{0}}.
\end{equation}
Equations (\ref{eq:uudag}) and (\ref{eq:vvdag}) can now be written in a covariant form
\begin{equation}
u_{R}(\p,\textstyle{\frac{1}{2}})u_{R}^{\dag}(\p,\textstyle{\frac{1}{2}})=
v_{R}(\p,-\textstyle{\frac{1}{2}})v_{R}^{\dag}(\p,-\textstyle{\frac{1}{2}})
=\frac{1}{2}(I p^{0}+\s\cdot\p).
\end{equation}
The commutator/anti-commutator becomes
\begin{eqnarray}
[\Psi_{R}(x),\Psi_{R}^{\dag}(y)]_{\pm}&=&(2\pi)^{-3}\int\frac{d^{3}p}{4p^{0}}(p^{0}I+\s\cdot\p)\left[e^{-ip\cdot(x-y)}
\pm e^{ip\cdot(x-y)}\right]\nonumber\\
&=&\frac{1}{2}i\left[\left(I\frac{\partial}{\partial t}-\sigma^{i}\partial_{i}\right)\mp\left(I\frac{\partial}{\partial t}-\sigma^{i}\partial_{i}\right)\right]\Delta_{+}(x-y).
\end{eqnarray}
The right-hand side vanishes under anti-commutator. Therefore, by the demand of locality, the massless right-handed field furnishes fermionic-statistics. Choosing both expansion coefficients of momentum $\k$ to be positive, we obtain
\begin{equation}
u_{R}(\k,\textstyle{\frac{1}{2}})=v_{R}(\k,-\textstyle{\frac{1}{2}})=\sqrt{k^{0}}
\left(\begin{matrix}
1 \\
0 \end{matrix}\right).
\end{equation}
At arbitrary momentum, they become
\begin{equation}
u_{R}(\p,\textstyle{\frac{1}{2}})=v_{R}(\p,\textstyle{-\frac{1}{2}})=
\sqrt{p^{0}}
\left(\begin{matrix}
e^{-i\phi/2}\cos\frac{\theta}{2}\\
e^{i\phi/2}\sin\frac{\theta}{2}\end{matrix}\right).
\end{equation}

The field equation for $\Psi_{R}(x)$ is derived by noting that the expansion coefficients are eigenvectors of the helicity operator
$\s\cdot\hat{\p}$
\begin{equation}
\s\cdot\hat{\p}\,u_{R}(\p,\textstyle{\frac{1}{2}})=u_{R}(\p,\textstyle{\frac{1}{2}}),\hspace{0.5cm}
\s\cdot\hat{\p}\,v_{R}(\p,\textstyle{-\frac{1}{2}})=v_{R}(\p,\textstyle{-\frac{1}{2}}).
\end{equation}
In the configuration space, the field equation for $\Psi_{R}(x)$ is
\begin{equation}
\left(I\frac{\partial}{\partial t}+\sigma^{i}\partial_{i}\right)\Psi_{R}(x)=0.
\end{equation}

Before we study the discrete symmetry of $\Psi_{R}(x)$, it is more instructive to construct the left-handed field $\Psi_{L}(x)$ and investigate their discrete transformations simultaneously. We note, the parity transformation on a massless particle $|p,\sigma\rangle$ given by eq.~(\ref{eq:massless_particle_parity})
\begin{equation}
\mathsf{P}|p,\sigma\rangle=(-1)^{j+\sigma}\eta_{\sigma}e^{\mp i\sigma\pi}|\mathscr{P}p,-\sigma\rangle
\end{equation}
reflects both the momentum and the helicity. Since the right-handed particles and anti-particles only have $\frac{1}{2}$ and $-\frac{1}{2}$ helicities, we see that $\Psi_{R}(x)$ violates parity. For similar reason, $\Psi_{L}(x)$ also violates parity. After constructing the left-handed field $\Psi_{L}(x)$, we show that parity maps the right-handed and left-handed field to each other.

\subsubsection{Left-handed field: $\mathbf{(0,\frac{1}{2})}$}
In this representation, the rotation and boost generators are $\J=\s/2$ and $\K=i\s/2$ respectively. The rotation constraints on the right-handed and left-handed coefficients are identical, since their rotation generators are the same. The difference in boost generators means that the constraints given by $\mathcal{D}(S(\alpha,\beta))$ are different. Here, the generator $A$ and $B$ are
\begin{equation}
A=\left(\begin{matrix}
 0 & 0\\
 i & 0\end{matrix}\right),\hspace{0.5cm}
B=\left(\begin{matrix}
 0 & 0\\
-1 & 0 \end{matrix}\right)
\end{equation}
so that
\begin{equation}
\mathcal{D}(S(\alpha,\beta))=
\left(\begin{matrix}
1 & 0 \\
-\alpha-i\beta & 1\end{matrix}\right).
\end{equation}
Substituting $\mathcal{D}(S(\alpha,\beta))$ into eqs.~(\ref{eq:massless_sabx}) and (\ref{eq:massless_sab}), the coefficients are
\begin{equation}
u_{L}(\k,\textstyle{\frac{1}{2}})=0,\hspace{0.5cm}
u_{L}(\k,-\textstyle{\frac{1}{2}})=\left(\begin{matrix}
0 \\
u_{L2}(\k,-\textstyle{\frac{1}{2}})\end{matrix}\right),
\end{equation}
\begin{equation}
v_{L}(\k,\textstyle{\frac{1}{2}})=\left(\begin{matrix}
0 \\
v_{L2}(\k,\textstyle{\frac{1}{2}}) \end{matrix}\right),\hspace{0.5cm}
v_{L}(\k,-\textstyle{\frac{1}{2}})=0.
\end{equation}
Therefore, the massless left-handed quantum field is given by
\begin{equation}
\Psi_{L}(x)=(2\pi)^{-3/2}\int\frac{d^{3}p}{\sqrt{2p^{0}}}\left[e^{-ip\cdot x}u_{L}(\p,-\textstyle{\frac{1}{2}})a(\p,-\textstyle{\frac{1}{2}})+
e^{ip\cdot x}v_{L}(\p,\textstyle{\frac{1}{2}})b^{\dag}(\p,\textstyle{\frac{1}{2}})\right].
\end{equation}
The boost for the left-handed coefficients can be obtained simply by mapping $\varphi\rightarrow-\varphi$ in eq.~(\ref{eq:massless_boost_coefficients})
\begin{equation}
\mathcal{D}(L(p))=
\left(\begin{matrix}
e^{-i\phi/2} e^{-\varphi/2} \cos\frac{\theta}{2} & -e^{-i\phi/2}e^{\varphi/2} \sin\frac{\theta}{2}\nonumber \\
e^{i\phi/2}e^{-\varphi/2} \sin\frac{\theta}{2}& e^{i\phi/2}e^{\varphi/2} \cos\frac{\theta}{2}\end{matrix}\right).
\end{equation}
Following the same calculation as we did for the right-handed field, the coefficients of arbitrary momentum for a local left-handed field takes the form
\begin{equation}
u_{L}(\p,-\textstyle{\frac{1}{2}})=v_{L}(\p,\textstyle{\frac{1}{2}})=\sqrt{p^{0}}
\left(\begin{matrix}
-e^{-i\phi/2}\sin\frac{\theta}{2}\\
e^{i\phi/2}\cos\frac{\theta}{2}\end{matrix}\right).
\end{equation}
They are eigenvectors of the helicity operator with eigenvalue $-\frac{1}{2}$
\begin{equation}
\s\cdot\hat{\p}\, u_{L}(\p,-\textstyle{\frac{1}{2}})=-u_{L}(\p,-\textstyle{\frac{1}{2}}),\hspace{0.5cm}
\s\cdot\hat{\p}\, v_{L}(\p,\textstyle{\frac{1}{2}})=-v_{L}(\p,\textstyle{\frac{1}{2}}).
\end{equation}
Consequently, the field equation for $\Psi_{L}(x)$ is
\begin{equation}
\left(I\frac{\partial}{\partial t}-\sigma^{i}\partial_{i}\right)\Psi_{L}(x)=0.
\end{equation}
Having determined the coefficients and field equations for $\Psi_{R}(x)$ and $\Psi_{L}(x)$, we now study their discrete symmetries.

\subsubsection{Parity}

Due to the discontinuity in the action of parity on the massless states with respect to their momenta, we write the field as a sum of two terms integrating over positive and negative momentum
\begin{eqnarray}
\Psi_{R}(x)&=&(2\pi)^{-3/2}\int_{0}^{\infty}\frac{d^{3}p}{\sqrt{2p^{0}}}u_{R}(\p,\textstyle{\frac{1}{2}})\left[e^{-ip\cdot x}a(\p,\textstyle{\frac{1}{2}})+e^{ip\cdot x}b^{\dag}(\p,\textstyle{-\frac{1}{2}})\right]\nonumber\\
&&+(2\pi)^{-3/2}\int_{-\infty}^{0}\frac{d^{3}p}{\sqrt{2p^{0}}}u_{R}(\p,\textstyle{\frac{1}{2}})\left[e^{-ip\cdot x}a(\p,\textstyle{\frac{1}{2}})+e^{ip\cdot x}b^{\dag}(\p,\textstyle{-\frac{1}{2}})\right].
\end{eqnarray}
Acting parity on $\Psi_{R}(x)$, we get
\begin{eqnarray}
\mathsf{P}\Psi_{R}(x)\mathsf{P}^{-1}&=&(2\pi)^{-3/2}\int_{0}^{\infty}\frac{d^{3}p}{\sqrt{2p^{0}}}e^{\frac{1}{2}i\pi}u_{R}(\p,\textstyle{\frac{1}{2}})\left[e^{-ip\cdot x}(-\eta_{1/2}^{*})a(-\p,-\textstyle{\frac{1}{2}})+e^{ip\cdot x}\bar{\eta}_{-1/2}b^{\dag}(-\p,\textstyle{\frac{1}{2}})\right]\nonumber\\
&&\hspace{-2cm}+(2\pi)^{-3/2}\int_{-\infty}^{0}\frac{d^{3}p}{\sqrt{2p^{0}}}e^{-\frac{1}{2}i\pi}u_{R}(\p,\textstyle{\frac{1}{2}})\left[e^{-ip\cdot x}(-\eta_{1/2}^{*})a(-\p,-\textstyle{\frac{1}{2}})+e^{ip\cdot x}\bar{\eta}_{-1/2}b^{\dag}(-\p,\textstyle{\frac{1}{2}})\right].
\end{eqnarray}
For positive and negative momentum along the 2-axis, the phase is $e^{\frac{1}{2}i\pi}$ and $e^{-\frac{1}{2}i\pi}$ respectively. The parity-reflected particle and anti-particle states now have helicity 
$-\frac{1}{2}$ and $\frac{1}{2}$ respectively which coincide with the degrees of freedom of the left-handed field 
$\Psi_{L}(x)$. Therefore, we consider $\Psi_{L}(\mathscr{P}x)$ and rewrite it as
\begin{eqnarray}
\Psi_{L}(\mathscr{P}x)&=&(2\pi)^{-3/2}\int_{0}^{\infty}\frac{d^{3}p}{\sqrt{2p^{0}}}u_{L}(-\p,-\textstyle{\frac{1}{2}})\left[e^{-ip\cdot x}a(-\p,-\textstyle{\frac{1}{2}})+e^{ip\cdot x}b^{\dag}(-\p,\textstyle{\frac{1}{2}})\right]\nonumber\\
&&\hspace{-1.3cm}+(2\pi)^{-3/2}\int_{-\infty}^{0}\frac{d^{3}p}{\sqrt{2p^{0}}}u_{L}(-\p,-\textstyle{\frac{1}{2}})\left[e^{-ip\cdot x}a(-\p,-\textstyle{\frac{1}{2}})+e^{ip\cdot x}b^{\dag}(-\p,\textstyle{\frac{1}{2}})\right].
\end{eqnarray}
Using the identity
\begin{equation}
u_{L}(-\p,-\textstyle{\frac{1}{2}})=e^{\pm\frac{1}{2}i\pi}u_{R}(\p,\textstyle{\frac{1}{2}})
\end{equation}
where the respective top and bottom signs are for positive and negative momentum along the $2$-axis, we get
\begin{eqnarray}
\Psi_{L}(\mathscr{P}x)&=&(2\pi)^{-3/2}\int_{0}^{\infty}\frac{d^{3}p}{\sqrt{2p^{0}}}e^{\frac{1}{2}i\pi}u_{R}(\p,\textstyle{\frac{1}{2}})\left[e^{-ip\cdot x}a(-\p,-\textstyle{\frac{1}{2}})+e^{ip\cdot x}b^{\dag}(-\p,\textstyle{\frac{1}{2}})\right]\nonumber\\
&&+(2\pi)^{-3/2}\int_{-\infty}^{0}\frac{d^{3}p}{\sqrt{2p^{0}}}e^{-\frac{1}{2}i\pi}u_{R}(\p,\textstyle{\frac{1}{2}})\left[e^{-ip\cdot x}a(-\p,-\textstyle{\frac{1}{2}})+e^{ip\cdot x}b^{\dag}(-\p,\textstyle{\frac{1}{2}})\right].\nonumber
\end{eqnarray}
Comparing this with $\mathsf{P}\Psi_{R}(x)\mathsf{P}^{-1}$, we see that the two are related when 
\begin{equation}
\eta_{1/2}^{*}=-\bar{\eta}_{-1/2} \label{eq:R_intrinsic_parity}
\end{equation}
so that
\begin{equation}
\mathsf{P}\Psi_{R}(x)\mathsf{P}^{-1}=-\eta_{1/2}^{*}\Psi_{L}(\mathscr{P}x).\label{eq:massless_right_parity}
\end{equation}
Performing the same calculation for $\Psi_{L}(x)$, under the condition
\begin{equation}
\eta^{*}_{-1/2}=-\bar{\eta}_{1/2} \label{eq:L_intrinsic_parity}
\end{equation}
we get
\begin{equation}
\mathsf{P}\Psi_{L}(x)\mathsf{P}^{-1}=\eta_{-1/2}^{*}\Psi_{R}(\mathscr{P}x).\label{eq:massless_left_parity}
\end{equation}
Equations (\ref{eq:R_intrinsic_parity}) and (\ref{eq:L_intrinsic_parity}) have opposite helicity reflect the intrinsic properties of the right and left-handed fields. Both equations contain the same physics, that particle and anti-particle have opposite intrinsic parity.

Therefore, parity maps $\Psi_{R}(x)$ and $\Psi_{L}(x)$ to each other. Consequently, parity is violated for interacting theories containing massless right-handed or left-handed fields. A well-known example is the SM neutrinos where it is described by a left-handed massless spin-half field. This explains why the weak-interaction violates parity. However, the preference of left-handed over right-handed fields in the SM remains an open question.

\subsubsection{Charge-conjugation}

Acting the charge-conjugation operator on $\Psi_{R}(x)$ gives us
\begin{equation}
\mathsf{C}\Psi_{R}(x)\mathsf{C}^{-1}=(2\pi)^{-3/2}\int\frac{d^{3}p}{\sqrt{2p^{0}}}u_{R}(\p,\textstyle{\frac{1}{2}})\left[e^{-ip\cdot x}\varsigma^{*}b(\p,\textstyle{\frac{1}{2}})
+e^{ip\cdot x}\bar{\varsigma}a^{\dag}(\p,-\textstyle{\frac{1}{2}})\right].
\end{equation}
Using the identity
\begin{equation}
i\sigma^{2}u_{L}^{*}(\p,-\textstyle{\frac{1}{2}})=u_{R}(\p,\textstyle{\frac{1}{2}})
\end{equation}
and taking the charge-conjugation phases to be the same for particle and anti-particle
\begin{equation}
\varsigma^{*}=\bar{\varsigma}
\end{equation}
we get
\begin{equation}
\mathsf{C}\Psi_{R}(x)\mathsf{C}^{-1}=i\varsigma^{*}\sigma^{2}\Psi_{L}^{*}(x).
\end{equation}
Repeat the same calculation for $\Psi_{L}(x)$,
\begin{equation}
\mathsf{C}\Psi_{L}(x)\mathsf{C}^{-1}=-i\varsigma^{*}\sigma^{2}\Psi_{R}^{*}(x).
\end{equation}
The charge-conjugation operator like parity, maps the $\Psi_{R}(x)$ and $\Psi_{L}(x)$ to each other so it is not a conserved symmetry. Since charge-conjugation and parity interchanges the representation of the field, the combination of $\mathsf{CP}$ is therefore conserved. Applying $\mathsf{CP}$ on $\Psi_{R}(x)$ and $\Psi_{L}(x)$ yields
\begin{equation}
(\mathsf{CP})\Psi_{R}(x)(\mathsf{CP})^{-1}=-i(\eta_{1/2}\varsigma)^{*}\sigma^{2}\Psi_{R}^{*}(\mathscr{P}x),
\end{equation}
\begin{equation}
(\mathsf{CP})\Psi_{L}(x)(\mathsf{CP})^{-1}=-i(\eta_{-1/2}\varsigma)^{*}\sigma^{2}\Psi_{L}^{*}(\mathscr{P}x).
\end{equation}

\subsubsection{Time-reversal}

The time-reversal operator $\mathsf{T}$ acts on the massless annihilation and creation operator in a similar way to the parity
\begin{equation}
\mathsf{T}a(\p,\sigma)\mathsf{T}^{-1}=\varrho^{*}_{\sigma}e^{\mp\frac{1}{2}i\pi}a(-\p,\sigma),
\end{equation}
\begin{equation}
\mathsf{T}b^{\dag}(\p,\sigma)\mathsf{T}^{-1}=\bar{\varrho}_{\sigma}e^{\pm\frac{1}{2}i\pi}b^{\dag}(-\p,\sigma)
\end{equation}
where the top and bottom signs apply when the momentum along the 2-axis are positive and negative respectively. Performing a similar calculation as parity by writing the field as a sum of two terms integrating over positive and negative momentum, taking into account of the anti-unitarity of $\mathsf{T}$ and equating the phases
\begin{equation}
\varrho_{\pm1/2}^{*}=\bar{\varrho}_{\mp1/2}
\end{equation}
we get 
\begin{equation}
\mathsf{T}\Psi_{R}(x)\mathsf{T}^{-1}=-i\varrho_{1/2}^{*}\sigma^{2}\Psi_{R}(\mathscr{T}x),
\end{equation}
\begin{equation}
\mathsf{T}\Psi_{L}(x)\mathsf{T}^{-1}=i\varrho_{-1/2}^{*}\sigma^{2}\Psi_{L}(\mathscr{T}x).
\end{equation}
Therefore, time-reversal is conserved. We note, since $\mathsf{CP}$ is a symmetry for both $\Psi_{R}(x)$ and $\Psi_{L}(x)$, time-reversal must be a symmetry by the $\mathsf{CPT}$ theorem.

\section{Summary}
The harmony between theories and experiments in particle physics have established that in the low energy limit (with respect to the Planck scale), on scale where gravitational effects are negligible, the space-time symmetry is determined by the Poincar\'{e} group. The universality of the Poincar\'{e} symmetry suggests all the known physical laws of nature (except gravity) can be derived from first principle by studying its symmetries and representations. In this chapter, we review the seminal work of Wigner~\cite{Wigner:1939cj} and Weinberg~\cite{Weinberg:1964cn,Weinberg:1964ev,Weinberg:1995mt} in which this task was completed. 

The main purpose of this chapter is to lay the foundation for the remaining thesis. In the subsequent chapters, we apply and in some cases extend the formalism to our research. The objective is to conduct these investigations in a systematic manner with emphasis on the relevant symmetries.

The properties of the free particle states are uniquely determined by the unitary irreducible representations of the Poincar\'{e} algebra while their interactions are described by the $S$-matrix. Using the $S$-matrix, we can compute the the relevant observables for the theory of interest and compare them with the experiments. In order to satisfy the Poincar\'{e} symmetry and the cluster decomposition principle, a unification of quantum mechanics and special relativity is required. As far as we know, in the low energy limit, quantum field theory is the only formalism that offers a consistent unification.

Unlike most textbooks, where the quantum fields are derived as solutions to the field equations,  using the formalism presented in sec.~\ref{chap:quantum_fields}, they are uniquely determined purely through the demand of Poincar\'{e} symmetry. The corresponding field equations and Lagrangian can then be determined through the properties of the quantum fields.

To demonstrate the formalism, we construct the massive and massless Dirac field and its massive higher spin generalisations. In the process, we reproduced well-known general properties of quantum field theory, namely the spin-statistics theorem and that fermions and bosons have opposite and even intrinsic parities respectively. For the massless spin-half fields, right (left)-handed particles and anti-particles have only $\frac{1}{2}$ ($-\frac{1}{2}$) and $-\frac{1}{2}$ ($\frac{1}{2}$) helicities respectively and both violate $\mathsf{C}$ and $\mathsf{P}$ but conserves $\mathsf{CP}$ and $\mathsf{T}$.

However, the formalism also has its limitations. The particle states of definite mass and spin are inadequate to describe experimentally observed phenomena such as neutrino oscillation~\cite{Cleveland:1998nv}. To describe neutrino oscillation, one must abandon the notion that all particle states are mass-eigenstates and consider neutrino states as linear combination of mass eigenstates~\cite{Pontecorvo:1957cp,Bilenky:1978nj}. Extending this idea further, one is lead to the hypothesis that the most general quantum states is a linear combination of particle states with different mass and spin~\cite{Ahluwalia:1997hc}
\begin{equation}
|\rho,\lambda\rangle=\sum_{ij,k}A_{\rho i}B_{\lambda j}|m_{i},s_{j};\p_{k}\rangle \label{eq:most_general_particle}
\end{equation}
where $A_{\rho i}$ and $B_{\lambda j}$ are the unitary mass and spin mixing matrix respectively. The summation over the mass and spin is equivalent to abandoning the demand of irreducibility. Contrary to conventional wisdom, reducible representations do lead to new physics as we have seen in neutrino oscillation. A general investigation on the state $|\rho,\lambda\rangle$ and its quantum field is beyond the scope of the thesis. Instead, in the next chapter, we study massive vector particle states with indefinite spin containing both scalar and spin-one degrees of freedom.

Another possible limitation of the formalism is one we have discussed at the beginning of this chapter, that the Poincar\'{e} symmetry is a reflection of the properties of the SM particles. Indeed, this statement by itself is meaningless and circular, since the SM predictions are in agreement with the experiments. On the other hand, the statement becomes meaningful if we consider the existence of particles that do not satisfy Poincar\'{e} symmetry. Given the increasing evidence for dark matter, it is conceivable that dark matter do not satisfy Poincar\'{e} symmetry. Depending on the properties of its symmetry group, it may provide a first-principle explanation on their limited interactions with the SM particles. That is, the allowed interactions between the two sectors must simultaneously satisfy Poincar\'{e} symmetry and the symmetry of dark matter. This possibility is discussed in chapter~\ref{Chapter4} when we study Elko and VSR. 


\chapter{Vector fields} 
\label{Chapter3}
\lhead{Chapter 3. \emph{Vector fields}} 




Traditionally, the massive vector fields are derived from the irreducible spin-one representations of the Lorentz group. The massive vector fields were initially used in the intermediate vector boson model~\cite{Glashow:1961tr} to describe the weak interaction. However, the model violates unitarity and are non-renormalisable. A serious difficulty of the theory is the propagator
\begin{equation}
S_{\mu\nu}^{(j=1)}(y,x)=i\int\frac{d^{4}p}{(2\pi)^{4}}e^{-ip\cdot(x-y)}\frac{-\eta_{\mu\nu}+\frac{p^{\mu}p^{\nu}}{m^{2}}}{p^{\lambda}p_{\lambda}-m^{2}+i\epsilon}.
\end{equation}
For higher-order perturbation, the momentum dependence in the numerator of the propagator yields divergent integrals. Additionally, when the vector bosons are external particles, at tree-level, the resulting cross-sections violate unitarity at about 1TeV~\cite[chap.~21]{Aitchison:2004cs}. These difficulties are solved in the SM using non-Abelian gauge theory, but it requires the introduction of the Higgs boson and spontaneous symmetry breaking for the vector bosons to gain masses~\cite{Higgs:1964ia,Higgs:1964pj,Guralnik:1964eu,Kibble:1967sv}.

Here, we propose an alternative solution to these problems. Our solution originates from the observation that since
the vector representation of the Lorentz group contains scalar and spin-one degrees of freedom, the most general massive vector fields should contain both degrees of freedom. This observation is not new, as far as we know, it was first noted by Corben and Schwinger~\cite{Corben:1940zz}. Also, the view that one should include all the dynamical physical degrees of freedom was expressed by Schwinger in~~\cite{Schwinger:1962wd}

We construct the new massive vector field that defines its primitive state vectors to span the eigenvalues $- 0\times m^2$ {and} $- 2 m^2$ of the second Casimir operator $C_{2}$ so that it has both scalar and spin-one degrees of freedom.\footnote{Recall that the second Casimir operator is defined as $C_{2}= W^\mu W_\mu$, where  $W_{\mu}=\frac{1}{2}\epsilon_{\mu\nu\rho\sigma}J^{\nu\rho}P^{\sigma}$. The eigenvalues of $W^{\mu}W_{\mu}$ are $-m^{2}j(j+1)$ where $j=0,\frac{1}{2},1,\cdots$.} In doing so, we have abandoned the irreducibility demand with respect to $C_2$. Contrary to naive expectations, such a construction leads to new physics. We explain why the naive expectations fail and how certain freedom of phases in defining the dual of the expansion coefficients for the field adjoint result in the unexpected theoretical discovery. This field has the property that
\begin{enumerate}
\item For certain choice of phases it leads to the Veltman propagator~\cite{Veltman:2000xv}.
\begin{equation}
S_{\mu\nu}^{(j=0,1)}(y,x)=i\int\frac{d^{4}p}{(2\pi)^{4}}e^{-ip\cdot(x-y)}\frac{-\eta_{\mu\nu}}{p^{\lambda}p_{\lambda}-m^{2}+i\epsilon}.
\label{eq:Velt_prop}
\end{equation}
\item The state space does not require an indefinite metric.
\item The theory is unitary preserving, with better convergent behaviour than the intermediate vector boson model and Fermi theory.
\item The existence of the $- 0\times m^2$ degree of freedom slightly alters the decay rate of $W^{-}\rightarrow\bar{\nu}_{\ell}+\ell$ while the decay rate of $Z\rightarrow\bar{\nu}_{\ell}+\nu_{\ell}$ remains essentially unchanged.
\end{enumerate}
We refer to eq.~(\ref{eq:Velt_prop}) as the Veltman propagator since to the best of our knowledge, it was Veltman who first raised the possibility of obtaining such a propagator through the introduction of particles with indefinite metric~\cite{Veltman:2000xv}. Here our derivation do not require these particles. The Veltman propagator can be obtained by exploiting the freedom of relative phases between the scalar and spin-one sector.~\footnote{We are grateful to the hospitality offered by Graeme Murray at Lake Tekapo where the initial part of the work was completed.}


After our study of the massive vector field, we review the theory of the massless vector field. Following the Weinberg formalism, only the derivative of a scalar field $\partial_{\mu}\phi(x)$ transforms as a four-vector. The field $a^{\mu}(x)$ with non-trivial expansion coefficients transforms as a four-vector up to a derivative term. In the words of Weinberg, the non-covariance of $a^{\mu}(x)$ is a \textit{blessing in disguise} for it provides a first-principle explanation on the origin of gauge-symmetry~\cite{Weinberg:1965rz}.
While this result is not new, its implication becomes important in chapter~\ref{Chapter4} when we study Elko.

The first section introduces the massive vector boson state space with scalar and spin-one degrees of freedom. This particle transforms under the reducible representation of the Lorentz group. Later we construct the massive vector fields and their adjoint. Choosing the appropriate phases for the dual coefficients of the adjoint, we are able to derive the Veltman propagator. The phenomenology is studied by replacing the spin-one vector fields with the new ones in the SM Lagrangian densities. 

In the second section, we review the massless vector fields of the $(\frac{1}{2},\frac{1}{2})$ representation. In this case, only the derivative of scalar field $\partial_{\mu}\phi(x)$ is Lorentz covariant and there is no consistent massless limit for the spin-one vector fields. We find, the massless spin-one vector fields transform as four-vector up to a derivative term. As a result, the interacting Lagrangian densities involving massless spin-one fields will transform as a scalar only if the transformation is independent of the derivative term. We show that this demand is equivalent to the principle of local gauge-invariance.

\section{Massive vector fields}\label{m_vector_fields}

We now introduce a new Hilbert space for the description of  massive vector particles which  superimposes eigenvalues $- 0 (\times m^2)$ and $-2m^{2}$ of $C_{2}$. This is accomplished by abandoning the demand of irreducibility of the Lorentz group with respect to $C_{2}$.  We then construct the new vector field. The demand of Lorentz symmetry is applied to determine the form of the expansion coefficients of the vector field and their transformation properties.

\subsection{One-particle states}
A reducible representation with respect to $C_{2}$ is obtained by constructing $D(W)$ as a direct sum of two irreducible representations of spin-$j$ and $j'$
\begin{equation}
D(W)=\left(\begin{array}{cc}
D^{(j)}(W) & O \\
O & D^{(j')}(W) \end{array}\right).\label{eq:direct_sum}
\end{equation}
The resulting particle one-states $|p,\sigma\rangle$ remain eigenstate of $C_{1}$ but its eigenvalues associated with $C_{2}$ are now $\sigma$-dependent,
\begin{equation}
C_{2}|p,\sigma\rangle=
\begin{cases}
-m^{2}j(j+1)|p,\sigma\rangle, & \sigma=-j,\cdots j\\
-m^{2}j'(j'+1)|p,\sigma\rangle, & \sigma=-j',\cdots j' \label{eq:c2_j1j2}
\end{cases}.
\end{equation}
The one-particle states of such a representation then has $2[(j+j')+1]$ degeneracy with $\sigma=-j,\cdots+j,-j',\cdots j'$. The Lorentz transformation of the state remains the same as eq.~(\ref{eq:massive_tran})
\begin{equation}
U(\Lambda)|p,\sigma\rangle=\sqrt{\frac{(\Lambda p)^{0}}{p^{0}}}\sum_{\sigma'}
D_{\sigma'\sigma}(W(\Lambda,p))|\Lambda p,\sigma'\rangle
\label{eq:state_trans2}
\end{equation}
but with $D(W(\Lambda,p))$ given by eq.~(\ref{eq:direct_sum}). Due to the direct sum of $D(W)$, states belonging to $\sigma=-j,\cdots j$ and $\sigma=-j'\cdots j'$ do not mix with each other under Lorentz transformation. Therefore, the inner-product between the states remains orthonormal and positive-definite
\begin{equation}
\langle p',\sigma'|p,\sigma\rangle=\delta_{\sigma\sigma'}\delta^{3}(\p-\p')\label{eq:inner_prod}
\end{equation}
so there are no states with negative norms.

Conventional wisdom states that since all representations can be decomposed into direct sums of irreducible representations, all the physics are encoded in the irreducible representations. Hence, while reducible representations do not violate any physical principles, no new physics can result from such constructions. Here, to the contrary, we will show that the massive vector field containing both spin-one and scalar degrees of freedom describing particle states constructed via  eq.~(\ref{eq:direct_sum}) does lead to new physics. The matrix $D(W)$ is taken to be a direct sum of scalar and spin-one representation
\begin{equation}
D(W)=\left(\begin{array}{cc}
D^{(\widetilde{0})}(W) & 0 \\
0 & D^{(1)}(W) \end{array}\right)=
\left(\begin{array}{cc}
1 & 0 \\
0 & D^{(1)}(W) \end{array}\right).\label{eq:D_0_1}
\end{equation}
The degeneracy label is $\sigma=\pm1,0$ and $\widetilde{0}$ where $\widetilde{0}$ denotes the scalar degree of freedom. The rotation generator $\J=(J^{1},J^{2},J^{3})$ for $D(W)$ is given by
the direct sum of the rotation generator of scalar and the $(1,0)$ or $(0,1)$ representation of the Lorentz group,
\begin{equation}
J^{1}=\frac{1}{\sqrt{2}}
\left(\begin{array}{cccc}
0 & 0 & 0 & 0 \\
0 & 0 & 1 & 0 \\
0 & 1 & 0 & 1 \\
0 & 0 & 1 & 0  \end{array}\right),\hspace{0.5cm}
J^{2}=\frac{1}{\sqrt{2}}
\left(\begin{array}{cccc}
0 & 0 & 0 & 0 \\
0 & 0 &-i & 0 \\
0 & i & 0 &-i \\
0 & 0 & i & 0 \end{array}\right),\hspace{0.5cm}
J^{3}=
\left(\begin{array}{cccc}
0 & 0 & 0 & 0\\
0 & 1 & 0 & 0 \\
0 & 0 & 0 & 0 \\
0 & 0 & 0 &-1 \end{array}\right).\label{eq:bJ}
\end{equation}
 
We now have a particle state with scalar and spin-one degrees freedom with well-defined Lorentz transformation. The states have orthonormal and positive-definite norm. The next step is to introduce the massive vector field describing the particle state.

\subsection{The new massive vector field}
Let $V(x)$ be a new charged massive vector field for the above particle state\footnote{From now on, unless otherwise specified, $\sum_{\sigma}=\sum_{\sigma=\pm1,0,\widetilde{0}}$.}
\begin{equation}
V(x)=(2\pi)^{-3/2}\int\frac{d^{3}p}{\sqrt{2p^{0}}}\sum_{\sigma=\pm1,0,\widetilde{0}}
\left[e^{-ip\cdot x}u(\p,\sigma)c(\p,\sigma)+e^{ip\cdot x}v(\p,\sigma)d^{\ddag}(\p,\sigma)\right]
\end{equation}
where $u(\p,\sigma)$ and $v(\p,\sigma)$ are the expansion coefficients. The operators $c(\p,\sigma)$ and $d^{\ddag}(\p,\sigma)$ satisfy the commutation relations,
\begin{equation}
[c(\p',\sigma'),c^{\ddag}(\p,\sigma)]=[d(\p',\sigma'),d^{\ddag}(\p,\sigma)]\label{eq:commutator}
=\delta_{\sigma\sigma'}\delta^{3}(\p-\p')
\end{equation}
with all other combinations identically vanish and
\begin{equation}
[c^{\ddag}(\p,\sigma)]^{\ddag}=c(\p,\sigma),\hspace{0.5cm}
[d^{\ddag}(\p,\sigma)]^{\ddag}=d(\p,\sigma).
\end{equation}
The symbol $\ddag$ represents the adjoint for the particle creation operator. The reason for not using the usual $\dag$ operation will become transparent when we construct the field adjoint. Their action on the vacuum are defined as
\begin{equation}
c(\p,\sigma)|\,\,\rangle=d(\p,\sigma)|\,\,\rangle=0, \nonumber
\end{equation}
\begin{equation}
c^{\ddag}(\p,\sigma)|\,\,\rangle=|p,\sigma\rangle,\hspace{0.5cm}
d^{\ddag}(\p,\sigma)|\,\,\rangle=|p^{c},\sigma\rangle \label{eq:creation_op_vacuum}
\end{equation}
where $|p,\sigma\rangle$ and $|p^{c},\sigma\rangle$ are the particle and anti-particle states respectively. The field contains both scalar and spin-one degrees of freedom so it sums over $\sigma=\pm 1,0$ and $\widetilde{0}$.

Apart from the symbolic differences between $\ddag$ and $\dag$, their action on particle states in the Hilbert space remains identical to $\dag$ as evident from eqs.~(\ref{eq:commutator}) and (\ref{eq:creation_op_vacuum}). The only qualitative between $\ddag$ and $\dag$ is their action on the expansion coefficients of the vector fields. This is discussed in sec.~\ref{vdual_coefficients}

The vector field given are already manifestly covariant under space-time translation. The demand of Lorentz covariance requires the field $V(x)$ to transform as
\begin{equation}
U(\Lambda)V(x)U^{-1}(\Lambda)=\mathcal{D}^{-1}(\Lambda)V(\Lambda x)
\label{eq:field_trans}
\end{equation}
where $\mathcal{D}(\Lambda)$ is a finite-dimensional vector representation of the Lorentz group. When the Lorentz transformation is expanded about the identity, $\Lambda=I+\omega$, its finite-dimensional presentation $\mathcal{D}(I+\omega)$ takes the form
\begin{equation}
\mathcal{D}(I+\omega)=I-\frac{1}{2}i\omega_{\mu\nu}\mathcal{J}^{\mu\nu}\label{eq:D_omega}
\end{equation}
where $\mathcal{J}^{\mu\nu}$ is the generator of the Lorentz group. The rotation and boost generators are  identified as the components of $\mathcal{J}^{\mu\nu}$
\begin{equation}
\mJ=(\mathcal{J}^{23},\mathcal{J}^{31},\mathcal{J}^{12})=
(\mathcal{J}^{1},\mathcal{J}^{2},\mathcal{J}^{3}),
\end{equation}
\begin{equation}
\mK=(\mathcal{J}^{01},\mathcal{J}^{02},\mathcal{J}^{03})=
(\mathcal{K}^{1},\mathcal{K}^{2},\mathcal{K}^{3}).
\end{equation}
 Here, we choose to work in a different but physically equivalent basis to the four-vectors. The finite-dimensional rotation and boost generators in this basis are~\cite{Ahluwalia:2000pj} 
\begin{equation}
\mathcal{J}^{1}=\frac{1}{2}
\left(\begin{array}{cccc}
0 & 1 & 1 & 0 \\
1 & 0 & 0 & 1 \\
1 & 0 & 0 & 1 \\
0 & 1 & 1 & 0 \end{array}\right),\hspace{0.5cm}
\mathcal{J}^{2}=\frac{1}{2}
\left(\begin{array}{cccc}
0 &-i &-i & 0 \\
i & 0 & 0 &-i \\
i & 0 & 0 &-i \\
0 & i & i & 0 \end{array}\right),\hspace{0.5cm}
\mathcal{J}^{3}=
\left(\begin{array}{cccc}
1 & 0 & 0 & 0 \\
0 & 0 & 0 & 0 \\
0 & 0 & 0 & 0 \\
0 & 0 & 0 &-1 \end{array}\right),\label{eq:cJ}
\end{equation}
\begin{equation}
\mathcal{K}^{1}=\frac{1}{2}
\left(\begin{array}{cccc}
 0 & i &-i & 0 \\
 i & 0 & 0 &-i \\
-i & 0 & 0 & i \\
 0 &-i & i & 0 \end{array}\right),\hspace{0.5cm}
\mathcal{K}^{2}=\frac{1}{2}
\left(\begin{array}{cccc}
 0 & 1 &-1 & 0 \\
-1 & 0 & 0 &-1 \\
 1 & 0 & 0 & 1 \\
 0 & 1 &-1 & 0 \end{array}\right),\hspace{0.5cm}
\mathcal{K}^{3}=
\left(\begin{array}{cccc}
0 & 0 & 0 & 0 \\
0 &-i & 0 & 0 \\
0 & 0 & i & 0 \\
0 & 0 & 0 & 0 \end{array}\right).\label{eq:cK}
\end{equation}
The mapping between the above generators and the generators of the four-vector representation are given by the following similarity transformation 
\begin{equation}
(S\mathcal{J}^{i}S^{-1})^{\mu}_{\,\,\,\nu}=(\mathscr{J}^{i})^{\mu}_{\,\,\,\nu},\hspace{0.5cm}
(S\mathcal{K}^{i}S^{-1})^{\mu}_{\,\,\,\nu}=(\mathscr{K}^{i})^{\mu}_{\,\,\,\nu}, \label{eq:similarity_trans}
\end{equation}
where $S$ is a unitary matrix~\cite{Ahluwalia:2000pj}
\begin{equation}
S=\frac{1}{\sqrt{2}}
\left(\begin{array}{cccc}
 0 & i &-i & 0 \\
-i & 0 & 0 & i \\
 1 & 0 & 0 & 1 \\
 0 & i & i & 0 \end{array}\right)\label{eq:S_matrix}
\end{equation}
and the generators on the right-hand side are in the four-vector basis given by eqs.~(\ref{eq:vector_j}) and (\ref{eq:vector_k}).

The reason for this choice of basis is that it allows us to construct the field $V(x)$ in the same manner as the Dirac field in the $(\frac{1}{2},0)\oplus(0,\frac{1}{2})$ representation. Therefore, we do not have to worry about the four-vector index of the field and the raising and lowering of indices when defining the field adjoint.


The demand of Poincar\'{e} symmetry as derived from sec.~\ref{chap:quantum_fields}, gives us the following constraints
\begin{equation}
\sum_{\bar{\sigma}}u(\L\p,\bar{\sigma})D_{\bar{\sigma}\sigma}(W)=\mathcal{D}(\Lambda)u(\p,\sigma),
\end{equation}
\begin{equation}
\sum_{\bar{\sigma}}v(\L\p,\bar{\sigma})D^{*}_{\bar{\sigma}\sigma}(W)=\mathcal{D}(\Lambda)v(\p,\sigma).
\end{equation}
Taking $\Lambda=L(p)$ and $\p=0$, the little group element becomes $W(L(p),p)=I$. The coefficients of arbitrary momentum are then given by
\begin{equation}
u(\p,\sigma)=\exp\left(i\mK\cdot\bv\right)u(\0,\sigma),\hspace{0.5cm} 
v(\p,\sigma)=\exp\left(i\mK\cdot\bv\right)v(\0,\sigma)
\end{equation}
where $\mK=(\mathcal{K}^{1},\mathcal{K}^{2},\mathcal{K}^{3})$ and $\bv=\varphi\hat{\p}$ is the rapidity parameter. Taking $\Lambda=\mathcal{R}$ as rotation and using the identity $W(\mathcal{R},p)=\mathcal{R}$, we obtain the constraints imposed by rotation symmetry 
\begin{equation}
\sum_{\bar{\sigma}}u_{\bar{\ell}}(\0,\bar{\sigma})\mathbf{J}_{\bar{\sigma}\sigma}=\mJ_{\bar{\ell}\ell} u_{\ell}(\0,\sigma),\label{eq:v_rot1}
\end{equation}
\begin{equation}
\sum_{\bar{\sigma}}v_{\bar{\ell}}(\0,\bar{\sigma})\mathbf{J}_{\bar{\sigma}\sigma}^{*}=-\mJ_{\bar{\ell}\ell}v_{\ell}(\0,\sigma)\label{eq:v_rot2}
\end{equation}
where $\mathbf{J}$ and $\mJ$ are given by eqs.~(\ref{eq:bJ}) and (\ref{eq:cJ}). Solving eqs.~(\ref{eq:v_rot1}) and (\ref{eq:v_rot2}) with the appropriate normalisation yields
\begin{equation}
u(\0,\widetilde{0})=\frac{1}{\sqrt{2}}
\left(\begin{array}{cccc}
 0 \\
 c_{1} \\
 -c_{1} \\
 0 \end{array}\right),
u(\0,1)=
\left(\begin{array}{cccc}
 c_{2}\\
 0 \\
 0 \\
 0 \end{array}\right),
 u(\0,0)=\frac{1}{\sqrt{2}}
\left(\begin{array}{cccc}
 0 \\
 c_{2} \\
 c_{2} \\
 0 \end{array}\right),
 u(\0,-1)=
\left(\begin{array}{cccc}
 0 \\
 0 \\
 0 \\
 c_{2} \end{array}\right), 
\end{equation}
\begin{equation}
v(\0,\widetilde{0})=\frac{1}{\sqrt{2}}
\left(\begin{array}{cccc}
 0 \\
 d_{1} \\
 -d_{1} \\
 0 \end{array}\right),
v(\0,1)=
\left(\begin{array}{cccc}
 0\\
 0 \\
 0 \\
 -d_{2} \end{array}\right),
 v(\0,0)=\frac{1}{\sqrt{2}}
\left(\begin{array}{cccc}
 0 \\
 d_{2} \\
 d_{2} \\
 0 \end{array}\right),
 v(\0,-1)=
\left(\begin{array}{cccc}
 -d_{2} \\
 0 \\
 0 \\
 0 \end{array}\right), 
\end{equation}
where $c_{i}$ and $d_{i}$ are the proportionality constants. They will be determined by the demand of a Lorentz covariant propagator once the field adjoint is defined. 

We note, in the study of massive vector fields, the coefficients $u(\mathbf{0},\widetilde{0})$ and $v(\mathbf{0},\widetilde{0})$ are usually projected out since they belong to the scalar sector. However, when one studies non-Abelian gauge theory, scalar degree of freedom appears as ghost fields~\cite{Faddeev:1967fc}. As we have shown, by relaxing the condition of irreducibility, the scalar degrees of freedom can be naturally incorporated in the particle state and the vector field.  In the subsequent sections we will show that by an appropriate definition of the field adjoint for $V(x)$, a renormalisable and unitarity-preserving theory of massive vector field can be obtained without the introduction of ghosts. 

\subsection{Field adjoint and the dual coefficients}
Here, we briefly review some of the results in secs.~(\ref{perturbation_theory}) and (\ref{chap:quantum_fields}) for the discussion here. The inevitability of quantum field theory is that it allows us to construct consistent Lorentz invariant $S$-matrix describing particle interactions. Perturbatively, the $S$-matrix is expanded in powers of the interaction density $\mathcal{V}(x)$ using the Dyson series
\begin{equation}
S=1+\sum_{n=1}^{\infty}\frac{(-i)^{n}}{n!}\int_{-\infty}^{\infty}d^{4}x_{1}\cdots d^{4}x_{n}T[\mathcal{V}(x_{1})\cdots\mathcal{V}(x_{n})]
\end{equation}
where $\mathcal{V}(x)$ is a scalar in the sense that
\begin{equation}
U(\Lambda,a)\mathcal{V}(x)U^{-1}(\Lambda,a)=\mathcal{V}(\Lambda x+a).
\end{equation}
Due to the time-ordered product, the $S$-matrix will only become Lorentz invariant if $\mathcal{V}(x)$ commutes at space-like separation
\begin{equation}
[\mathcal{V}(x),\mathcal{V}(y)]=0.
\end{equation}
The purpose of introducing the field adjoint $\widehat{V}(x)$ is to construct scalar interaction density in terms of the field and its adjoint.

When $V(x)$ and $\widehat{V}(x)$ are involved in an interaction, the corresponding interaction density usually contains terms of the form $\widehat{V}(x)V(x)$. Since $V(x)$ contains only bosonic degrees of freedom, it is sufficient to demand the field and its adjoint to commute at space-like separation
\begin{equation}
[V(x),\widehat{V}(y)]=0
\end{equation}
to preserve the locality of the interaction density. The scalar nature of the interaction density is realised by demanding the adjoint to transform as
\begin{equation}
U(\Lambda,a)\widehat{V}(x)U^{-1}(\Lambda,a)=\widehat{V}(\Lambda x+a)\mathcal{D}(\Lambda).\label{eq:field_adjoint_trans}
\end{equation}

We now define the adjoint $\widehat{V}(x)$ for the vector field.  After constructing the adjoint, the propagator is obtained by evaluating the vacuum expectation value of the time-ordered product $\langle\,\,|T[V(x)\widehat{V}(y)]|\,\,\rangle$. Let $\widehat{V}(x)$ be defined as
\begin{equation}
\widehat{V}(x)=(2\pi)^{-3/2}\int\frac{d^{3}p}{\sqrt{2p^{0}}}\sum_{\sigma}\left[e^{ip\cdot x}\widehat{u}(\p,\sigma)c^{\ddag}(\p,\sigma)+e^{-ip\cdot x}\widehat{v}(\p,\sigma)d(\p,\sigma)\right] \label{eq:a_dual}
\end{equation}
where $\widehat{u}(\p,\sigma)$ and $\widehat{v}(\p,\sigma)$ are the dual coefficients. From the space-time translation phases of $\widehat{V}(x)$, we see that the adjoint involves complex-conjugation. 

\subsubsection{The dual coefficients}\label{vdual_coefficients}

When defining the dual coefficients for a particle corresponding to an irreducible representation, there exists  a freedom of a global phase which is later fixed by the demand of a positive-definite free Hamiltonian.

The massive vector particle state we have constructed transforms according to a reducible representation given by a direct sum of scalar and spin-one irreducible representation so the quantum field sums over both degrees of freedom. As we have noted earlier, the two irreducible sectors do not mix with each other. In terms of kinematics, this means the resulting Hamiltonian must be a sum of the free-Hamiltonian of the scalar and spin-one representation. The dual coefficients for the field-adjoint will then have two phases, one for the scalar and another for the spin-one sector. They must be chosen in such a way to ensure a positive definite Hamiltonian. Therefore, we define the dual coefficients as
\begin{equation}
\widehat{u}(\p,\sigma)=u^{\ddag}(\p,\sigma)g=
	\begin{cases}
  	  \xi u^{\dag}(\p,\sigma)g, & \sigma=\widetilde{0} \\
	\zeta u^{\dag}(\p,\sigma)g, & \sigma=\pm1, 0 	 \label{eq:vector_dual1}
	\end{cases}
\end{equation}
\begin{equation}
\widehat{v}(\p,\sigma)=v^{\ddag}(\p,\sigma)g=
	\begin{cases}
   	  \xi v^{\dag}(\p,\sigma)g, & \sigma=\widetilde{0} \\
	\zeta v^{\dag}(\p,\sigma)g, & \sigma=\pm1, 0 \label{eq:vector_dual2}
	\end{cases}
\end{equation}
where $g$ is the metric to be determined. The operation $\ddag$ is defined to take Hermitian conjugate the coefficients and including the relative phases for the coefficients with $\sigma$ dependence. For complex numbers and functions, the action of $\ddag$ is the same as Hermitian conjugation. Apart from the phases, the defined adjoint is the same as Hermitian conjugate. Therefore, successive application of the adjoint is proportional to the identity. In sec.~\ref{Veltman_propagator} we show that in order to obtain a positive-definite free Hamiltonian and the Veltman propagator the phases must take the values $|\xi|^{2}=|\zeta|^{2}=1$ so that
\begin{equation}
[u^{\ddag}(\p,\sigma)]^{\ddag}= u(\p,\sigma),\hspace{0.5cm}
[v^{\ddag}(\p,\sigma)]^{\ddag}=v(\p,\sigma)
\end{equation}
which is the same as Hermitian conjugate.

%

\subsubsection{The field adjoint}

The metric $\eta$ can be determined using the transformation properties of the field and its adjoint. Since the phases $\xi$ and $\zeta$ are momentum and space-time independent, we can rewrite the field adjoint as
\begin{equation}
\widehat{V}(x)=V^{\ddag}(x)g.
\end{equation}
Apply the operation $\ddag$ to eq.~(\ref{eq:field_trans}) then multiply from the right by $\eta$ gives us
\begin{equation}
U(\Lambda)\widehat{V}(x)U^{-1}(\Lambda)=\widehat{V}(\Lambda x)g^{-1}[\mathcal{D}^{-1}(\Lambda)]^{\dag}g.
\end{equation}
The matrix $\mathcal{D}(\Lambda)$ has no $\sigma$ dependence so $\mathcal{D}^{\ddag}(\Lambda)=\mathcal{D}^{\dag}(\Lambda)$. Equating the above expression to eq.~(\ref{eq:field_adjoint_trans}) and expanding $\mathcal{D}(\Lambda)$ about the identity yields the following constraint on $\eta$
\begin{equation}
g\mathcal{J}^{\mu\nu}g^{-1}=\mathcal{J}^{\dag\mu\nu}.
\end{equation}
In the finite-dimensional representation of the Lorentz group, the rotation generators are Hermitian while the boost generators are anti-Hermitian. This allows us to rewrite the constraints for $\eta$ as
\begin{equation}
[g,\mJ]=\{g,\mK\}=0.\label{eq:eta_constraint}
\end{equation}
Substituting eqs.~(\ref{eq:cJ}) and (\ref{eq:cK}) into the above equation gives us the solution
\begin{equation}
g=\lambda\left(\begin{array}{cccc}
1 & 0 & 0 & 0 \\
0 & 0 & 1 & 0 \\
0 & 1 & 0 & 0 \\
0 & 0 & 0 & 1 \end{array}\right)\label{eta}
\end{equation}
where $\lambda$ is a proportionality constant to be determined. An alternative derivation of the metric is by demanding the Lorentz invariance of $\widehat{u}(\p,\sigma)u(\p,\sigma)$.
The norms for the coefficients are
\begin{equation}
\widehat{u}(\0,\sigma)u(\0,\sigma')=\begin{cases}
 0 & \sigma\neq\sigma' \\
 -\xi|c_{1}|^{2}\lambda & \sigma=\sigma'=\widetilde{0} \\
 \zeta|c_{2}|^{2}\lambda & \sigma=\sigma'=\pm 1,0
 \end{cases}
\end{equation}
\begin{equation}
\widehat{v}(\0,\sigma)v(\0,\sigma')=\begin{cases}
 0 & \sigma\neq\sigma' \\
 -\xi|d_{1}|^{2}\lambda & \sigma=\sigma'=\widetilde{0} \\
 \zeta|d_{2}|^{2}\lambda & \sigma=\sigma'=\pm 1,0.
 \end{cases}
\end{equation}
An important feature of the norm is the orthonormality between different coefficients which becomes important when evaluating the free Hamiltonian. For now, the determination of $\eta$ allows us to derive the propagator.

Before proceed to derive the propagator, we show that the definition of the dual coefficient does not violate Lorentz symmetry. To show this, we first recall the rotation constraints on the coefficients given by eqs.~(\ref{eq:rot1}) and (\ref{eq:rot2})
\begin{equation}
\sum_{\bar{\sigma}}u_{\bar{\ell}}(\mathbf{0},\bar{\sigma})\J_{\bar{\sigma}\sigma}=\sum_{\ell}\mJ_{\bar{\ell}\ell}u_{\ell}(\mathbf{0},\sigma)\nonumber
\end{equation}
\begin{equation}
\sum_{\bar{\sigma}}v_{\bar{\ell}}(\mathbf{0},\bar{\sigma})\J_{\bar{\sigma}\sigma}^{*}=-\sum_{\ell}\mJ_{\bar{\ell}\ell}v_{\ell}(\mathbf{0},\sigma)\nonumber
\end{equation}
Complex conjugate both sides and multiply from the right by matrix $g$, with some simplification, it yields the constraints on the dual coefficients
\begin{equation}
\sum_{\bar{\sigma}}\widehat{u}_{\ell}(\mathbf{0},\bar{\sigma})\J^{*}_{\bar{\sigma}\sigma}=\sum_{\ell}\widehat{u}_{\bar{\ell}}(\mathbf{0},\sigma)\mJ_{\bar{\ell}\ell},\label{eq:dual_rot_constraint1}
\end{equation}
\begin{equation}
\sum_{\bar{\sigma}}\widehat{v}_{\ell}(\mathbf{0},\bar{\sigma})\J_{\bar{\sigma}\sigma}=-\sum_{\ell}\widehat{v}_{\bar{\ell}}(\mathbf{0},\sigma)\mJ_{\bar{\ell}\ell}.\label{eq:dual_rot_constraint2}
\end{equation}
where we have used the Hermiticity of $\mJ$ and eq.~(\ref{eq:eta_constraint}). We see that when $\J$ is an irreducible representation where $\sigma=-j,\cdots,j$ Lorentz symmetry does not allow the existence of local phases for individual dual coefficients of a given $\sigma$. But since the rotation generator we have chosen in eq.~(\ref{eq:bJ}) is a direct sum of scalar and spin-one rotation generator, the dual coefficients defined in eqs.~(\ref{eq:vector_dual1}) and (\ref{eq:vector_dual1}) satisfy the above constraints. 

The derived constraints on the dual coefficients are very general. In most cases, the boost and rotation generators of the Lorentz group are Hermitian and anti-Hermitian respectively so the matrix $\eta$ must satisfy eq.~(\ref{eq:eta_constraint}) to construct Lorentz-invariant scalars.

\subsection{The propagator}\label{Veltman_propagator}
According to eq.~(\ref{eq:g_prop}), we compute the propagator by first evaluating the spin-sums. In the process we will determine $c_{i}$ and $d_{i}$ for the expansion coefficients.
The relevant spin-sums are
\begin{equation}
\Pi_{1}(p)=\sum_{\sigma}u(\p,\sigma)\widehat{u}(\p,\sigma),\label{eq:pi_1}
\end{equation}
\begin{equation}
\Pi_{2}(p)=\sum_{\sigma}v(\p,\sigma)\widehat{v}(\p,\sigma).
\end{equation}
Explicit computation of $\Pi_{1}(p)$ and $\Pi_{2}(p)$ gives the covariant spin-sums quadratic in momentum	
\begin{equation}
\Pi_{1}(p)=\lambda\left[\xi|c_{1}|^{2}C_{1}+\zeta|c_{2}|^{2}C_{2}+
(\xi |c_{1}|^{2}+\zeta |c_{2}|^{2})\frac{\alpha_{\mu\nu}p^{\mu}p^{\nu}}{2m^{2}}\right],
\end{equation}
\begin{equation}
\Pi_{2}(p)=\Pi_{1}(p)\big\vert_{c_{i}\rightarrow d_{i}}.
\end{equation}
The explicit expression of the matrices $C_{1}$, $C_{2}$ and $\alpha_{\mu\nu}$ are not given here as they are not important for our discussion here.

Since the spin-sums $\Pi_{1}(p)$ and $\Pi_{2}(p)$ are quadratic in momentum, so generally, it is not possible to obtain a covariant propagator unless $c_{i}$, $d_{i}$ and the phases take particular values. In this case, the only way to obtain a covariant propagator is to choose the proportionality constants and the phases of the dual coefficients such that the spin-sums are momentum independent. Without the loss of generality, this can be achieved by the following choice 
\begin{equation}
c_{1}=-d_{1}=-i,\hspace{0.5cm} c_{2}=d_{2}=1.
\end{equation}
On the other hand, the phases $\xi$ and $\zeta$ must be chosen as
\begin{equation}
\xi=-\zeta=1\label{eq:phase_value}
\end{equation}
in order to obtain the correct propagator and ensuring the probability for the physical processes to be positive-definite.
The coefficients now take the form
\begin{equation}
u(\0,\widetilde{0})=\frac{1}{\sqrt{2}}
\left(\begin{array}{cccc}
 0 \\
 -i \\
 i \\
 0 \end{array}\right),
u(\0,1)=
\left(\begin{array}{cccc}
 1\\
 0 \\
 0 \\
 0 \end{array}\right),
 u(\0,0)=\frac{1}{\sqrt{2}}
\left(\begin{array}{cccc}
 0 \\
 1 \\
 1 \\
 0 \end{array}\right),
 u(\0,-1)=
\left(\begin{array}{cccc}
 0 \\
 0 \\
 0 \\
 1 \end{array}\right), 
\end{equation}
\begin{equation}
v(\0,\widetilde{0})=\frac{1}{\sqrt{2}}
\left(\begin{array}{cccc}
 0 \\
 i \\
 -i \\
 0 \end{array}\right),
v(\0,1)=
\left(\begin{array}{cccc}
 0\\
 0 \\
 0 \\
 -1 \end{array}\right),
 v(\0,0)=\frac{1}{\sqrt{2}}
\left(\begin{array}{cccc}
 0 \\
 1 \\
 1 \\
 0 \end{array}\right),
 v(\0,-1)=
\left(\begin{array}{cccc}
 -1 \\
 0 \\
 0 \\
 0 \end{array}\right).
\end{equation}
The spin-sums become proportional to the identity matrix
\begin{equation}
\Pi_{1}(p)=\Pi_{2}(p)=-\lambda I.
\end{equation}
and the norms of the coefficients become
\begin{equation}
\widehat{u}(\0,\sigma)u(\0,\sigma')=\widehat{v}(\0,\sigma)v(\0,\sigma')=-\lambda\delta_{\sigma\sigma'}.
\end{equation}
Substituting the spin-sums into the general propagator given by eq.~(\ref{eq:g_prop}) gives us
\begin{equation}
S(y,x)=\langle\,\,|T[V(x)\widehat{V}(y)]|\,\,\rangle=i\int\frac{d^{4}p}{(2\pi)^{4}}e^{-ip\cdot(y-x)}\frac{-\lambda I}{p^{\mu}p_{\mu}-m^{2}+i\epsilon}. \label{eq:prop_psi}
\end{equation}
When we transform to the four-vector basis, the identity matrix becomes proportional to the metric $\eta^{\mu}_{\,\,\nu}$. 

Here it is appropriate for us to comment on the generic properties of propagators. Generally for higher-spin fields, while their spin-sums are Lorentz covariant, they are also non-linear. As a result, the final propagators are usually not Lorentz covariant. The non-covariant terms can usually be cancelled by introducing appropriate local interactions. Since they do not contribute to the interaction, we may introduce the $T^{*}$-product which has the same definition as the $T$-product but ignores the non-covariant terms.

\subsection{The four-vector basis}
We now transform the result to the standard four-vector basis. To this end, multiply eqs.~(\ref{eq:v_rot1}) and (\ref{eq:v_rot2}) from the left by the matrix $S$ given by eq.~(\ref{eq:S_matrix}), we get a set of four-vectors that satisfy the rotation symmetry
\begin{equation}
\sum_{\bar{\sigma}}(Su)^{\mu}(\0,\bar{\sigma})\mathbf{J}_{\bar{\sigma}\sigma}=\mathscr{J}^{\mu}_{\,\,\nu}(Su)^{\nu}(\0,\sigma),
\end{equation}
\begin{equation}
\sum_{\bar{\sigma}}(Sv)^{\mu}(\0,\bar{\sigma})\mathbf{J}^{*}_{\bar{\sigma}\sigma}=-\mathscr{J}^{\mu}_{\,\,\nu}(Su)^{\nu}(\0,\sigma).
\end{equation}
These vectors, up to some proportionality constants, are the expansion coefficients for the massive vector field $A^{\mu}(x)$ in~\cite[sec.~5.3]{Weinberg:1995mt}. However they were studied separately for $\sigma=\pm1,0$ and $\sigma=\widetilde{0}$ due to the demand of irreducible representation for the particle state. By relaxing this demand, the resulting vector field is
\begin{equation}
A^{\mu}(x)=(2\pi)^{-3/2}\int\frac{d^{3}p}{\sqrt{2p^{0}}}\sum_{\sigma}\left[e^{-ip\cdot x}e^{\mu}(\p,\sigma)c(\p,\sigma)+e^{ip\cdot x}e^{\mu*}(\p,\sigma)d^{\ddag}(\p,\sigma)\right].
\end{equation}
The four-vectors $e^{\mu}(\0,\sigma)$ are related to $u(\0,\sigma)$ by
\begin{equation}
u_{\ell}(\0,\sigma)=i(S^{-1})_{\ell\mu}e^{\mu}(\0,\sigma)\label{eq:u_to_w}
\end{equation}
so the vector field $A^{\mu}(x)$ is related to $V(x)$ by the unitary transformation
\begin{equation}
A^{\mu}(x)=-i[SV(x)]^{\mu}.\label{eq:amu_al}
\end{equation}
The four-vector coefficients are
\begin{equation}
e(\0,\widetilde{0})=
\left(\begin{array}{cccc}
-i \\
0 \\
0 \\
0 \end{array}\right),
e(\0,1)=\frac{1}{\sqrt{2}}
\left(\begin{array}{cccc}
 0 \\
-1 \\
-i \\
 0 \end{array}\right),
e(\0,0)=
\left(\begin{array}{cccc}
0 \\
0 \\
0 \\
1 \end{array}\right),
e(\0,-1)=\frac{1}{\sqrt{2}}
\left(\begin{array}{cccc}
 0 \\
 1 \\
-i \\
 0 \end{array}\right).
\end{equation}

We now compute the propagator for $A^{\mu}(x)$ by evaluating the spin-sum 
\begin{equation}
\Pi^{\mu}_{\,\,\nu}(p)=\sum_{\sigma}e^{\mu}(\p,\sigma)\widehat{e}_{\nu}(\p,\sigma). \label{eq:pi_spin_sum}
\end{equation}
where $\widehat{e}_{\nu}(\p,\sigma)$ is the dual four-vector which can be obtained using eq.~(\ref{eq:u_to_w}). In component form, $u^{\dag}(\0,\sigma)g$ is related to $e^{\mu}(\0,\sigma)$ by
\begin{eqnarray}
u_{\ell}^{*}(\0,\sigma)g_{\ell k}&=&-i(S^{-1})_{\ell\mu}^{*}e^{\mu*}(\0,\sigma)g_{\ell k} \nonumber \\
&=&-ie^{\mu*}(\0,\sigma)[S_{\mu\ell}g_{\ell m}(S^{-1})_{m\nu}]S_{\nu k} \nonumber \\
&=&i\lambda e^{\mu*}(\0,\sigma)\eta_{\mu\nu}S_{\nu k}. \label{eq:uw_dual}
\end{eqnarray}
The second line of the equation uses the unitarity of the matrix $(S^{-1})_{\mu k}=S^{*}_{k\mu}$ and the third line uses the identity 
\begin{equation}
S_{\mu\ell}g_{\ell m}(S^{-1})_{m\nu}=-\lambda\eta_{\mu\nu}.
\end{equation}
Removing the matrix $S$ from the right-hand side of equation (\ref{eq:uw_dual}), the dual for the four-vector $e^{\mu}(\0,\sigma)$ with the inclusion of relevant phases is given by
\begin{equation}
\widehat{e}_{\mu}(\0,\sigma)=\begin{cases}
\xi\lambda\eta_{\mu\nu}e^{\nu*}(\0,\sigma) & \sigma=\widetilde{0}\\
\zeta\lambda\eta_{\mu\nu}e^{\nu*}(\0,\sigma) & \sigma=\pm1,0.\end{cases}\label{eq:e_dual}
\end{equation}
The phases $\xi$ and $\zeta$ take the values given in eq.~(\ref{eq:phase_value}). The norm of the coefficients in the four-vector basis becomes
\begin{equation}
e^{\mu}(\p,\sigma')\widehat{e}_{\mu}(\p,\sigma)=\lambda\delta_{\sigma\sigma'}.\label{eq:norm_four_vector}
\end{equation}
The field adjoint $\widehat{A}_{\mu}(x)$ is given by
\begin{equation}
\widehat{A}_{\mu}(x)=(2\pi)^{-3/2}\int\frac{d^{3}p}{\sqrt{2p^{0}}}\sum_{\sigma}\left[e^{ip\cdot x}\widehat{e}_{\mu}(\p,\sigma)c^{\ddag}(\p,\sigma)+e^{ip\cdot x}\widehat{e}_{\mu}^{\,*}(\p,\sigma)d(\p,\sigma)\right].\label{eq:vector_field_adjoint}
\end{equation}
where we have used the demand $[d^{\ddag}(\p,\sigma)]^{\ddag}=d(\p,\sigma)$.
Computing the spin-sum of eq.~(\ref{eq:pi_spin_sum}) using eq.~(\ref{eq:e_dual}) yields
\begin{equation}
\Pi^{\mu}_{\,\,\,\nu}(p)=\sum_{\sigma}e^{\mu}(\p,\sigma)\widehat{e}_{\nu}(\p,\sigma)=-\lambda\zeta\eta^{\mu}_{\,\,\nu}+(\xi+\zeta)\frac{p^{\mu}p_{\nu}}{m^{2}}.\label{eq:vector_spin_sum}
\end{equation}
By writing the spin-sum with the phases explicitly included, we see how the choice of phases affect the spin-sum. Moreover, the masses between the scalar and spin-one sector cannot be different if one wishes to preserve the covariance of the spin-sum.
Substituting the phases given by equation (\ref{eq:phase_value}) into the spin-sum gives us
\begin{equation}
\Pi^{\mu}_{\,\,\nu}=\lambda\eta^{\mu}_{\,\,\nu}
\end{equation}
and the covariant Veltman propagator for $A^{\mu}(x)$
\begin{equation}
S^{\mu}_{\,\,\nu}(y,x)=\langle\,\,|T[A^{\mu}(x)\widehat{A}_{\nu}(y)]|\,\,\rangle=i\int\frac{d^{4}p}{(2\pi)^{4}}e^{-ip\cdot(y-x)}\frac{\lambda\eta^{\mu}_{\,\,\nu}}{p^{\alpha}p_{\alpha}-m^{2}+i\epsilon}.\label{eq:v_propagator}
\end{equation}
The propagator for $A^{\mu}(x)$ can also be obtained by a similarity transformation on the propagator for $V(x)$ using equation (\ref{eq:amu_al}).

\subsection{The kinematics}
The vector field $A^{\mu}(x)$ constructed in this chapter, does not have a definite spin. Consequently, its Lagrangian density differs from the standard massive spin-one vector field. 
In the four-vector basis, the coefficients for the spin-one sector at rest have vanishing time-components
\begin{equation}
e^{0}(\0,\sigma)=0,\hspace{0.5cm}\sigma=\pm1,0
\end{equation}
so the spin-one vector field is divergence-free. On the other hand, the scalar coefficient has a non-vanishing time-component
\begin{equation}
e^{0}(\0,\widetilde{0})=1
\end{equation}
so that the vector field, with the scalar degree of freedom, is not divergence-free~\cite{Ahluwalia:2000pj,Kirchbach:2002nu}
\begin{equation}
\partial_{\mu}A^{\mu}(x)\neq0\label{eq:dA}
\end{equation}
where the non-vanishing term on the right-hand side is proportional to the square of the mass of the particle. 

The propagator obtained in the previous section suggests the Lagrangian density to be Klein-Gordon
\begin{equation}
\mathscr{L}=-\partial_{\mu}\widehat{A}_{\nu}\partial^{\mu}A^{\nu}+m^{2}\widehat{A}_{\nu}A^{\nu}.\label{eq:vector_L}
\end{equation}
The resulting Hamiltonian is given by
\begin{equation}
H=\int d^{3}x\,\large[\Pi_{\mu}\frac{\partial{A}^{\mu}}{\partial t}+\widehat{\Pi}_{\mu}\frac{\partial\widehat{A}^{\mu}}{\partial t}-\mathscr{L}\large]
\end{equation}
where $\Pi_{\mu}(x)$ and $\widehat{\Pi}_{\mu}(x)$ are the conjugate momenta
\begin{equation}
\Pi_{\mu}(x)=\frac{\partial\mathscr{L}}{\partial(\partial A^{\mu}/\partial t)}=-\frac{\partial\widehat{A}_{\mu}}{\partial t},\hspace{0.5cm}
\widehat{\Pi}_{\mu}(x)=\frac{\partial\mathscr{L}}{\partial(\partial\widehat{A}^{\mu}/\partial t)}=-\frac{\partial A_{\mu}}{\partial t}.
\end{equation}
Substituting all the relevant terms into the Hamiltonian and normal-ordering the creation and annihilation operators, we get
\begin{eqnarray}
H&=&-\int d^{3}p\,(p^{0})\sum_{\sigma,\sigma'}\big[e^{\mu}(\p,\sigma')\widehat{e}_{\mu}(\p,\sigma)c^{\ddag}(\p,\sigma)c(\p,\sigma')\nonumber \\
&&\hspace{3.5cm}+e^{\mu*}(\p,\sigma')\widehat{e}_{\mu}^{\,*}(\p,\sigma)d^{\ddag}(\p,\sigma)d(\p,\sigma')\big].
\end{eqnarray}
The norm given in eq.~(\ref{eq:norm_four_vector}), with the demand of a positive-definite free Hamiltonian fixes the proportionality constant of the metric to be
\begin{equation}
\lambda=-1.
\end{equation}

The positivity of the Hamiltonian resulting from the Lagrangian density support our claim that the scalar degree of freedom is physical and the importance of the relative phases $\xi$ and $\zeta$. The propagator now reads
\begin{equation}
S_{\mu\nu}(y,x)=\langle\,\,|T[A_{\mu}(x)\widehat{A}_{\nu}(y)]|\,\,\rangle=i\int\frac{d^{4}p}{(2\pi)^{4}}e^{-ip\cdot(y-x)}\frac{-\eta_{\mu\nu}}{p^{\lambda}p_{\lambda}-m^{2}+i\epsilon}.
\end{equation}
Although the propagator and the Hamiltonian suggest the Lagrangian density of the vector field to be Klein-Gordon, these conditions are insufficient to guarantee that we have the correct Lagrangian density. This has to do with the renormalisability of the theory which is discussed in sec.~\ref{vector_phenomenologies}.

\subsection{Locality structure}
In this section, we analyse the locality structure of the vector field by computing the relevant equal-time commutators. So far, we have only considered the electrically charged vector field. However, it is also possible to have electrically neutral vector field analogous to the Majorana fermion for the Dirac field~\cite{Majorana:1937vz}. Here, the locality structure of both fields are analysed. We divide the analysis into two subsections, one for the charged and the other for the neutral vector field. 

\subsubsection{Charged vector field}
Since the particle and anti-particle are distinct for $A^{\mu}(x)$, the following commutators trivially vanish at equal-time
\begin{equation}
[A^{\mu}(\mathbf{x},t),A^{\nu}(\mathbf{y},t)]=[\Pi^{\mu}(\mathbf{x},t),\Pi^{\nu}(\mathbf{y},t)]=0.\label{eq:charged_v_commutator1}
\end{equation}
The equal-time commutator between $A^{\mu}(x)$ and $\Pi_{\mu}(x)$ computed using eq.~(\ref{eq:commutator}) yields
\begin{equation}
[A^{\mu}(\x,t),\Pi_{\nu}(\mathbf{y},t)]=-i(2\pi)^{-3}\int\frac{d^{3}p}{2}\sum_{\sigma}e^{-i\mathbf{p}(\mathbf{x-y})}\left[e^{\mu}(\p,\sigma)\widehat{e}_{\nu}(\p,\sigma)+e^{\mu*}(-\p,\sigma)\widehat{e}_{\nu}^{\,*}(-\p,\sigma)\right]
\end{equation}
Explicit computation of the spin-sum yields
\begin{equation}
\sum_{\sigma}\left[e^{\mu}(\p,\sigma)\widehat{e}_{\nu}(\p,\sigma)+e^{\mu*}(-\p,\sigma)\widehat{e}_{\nu}^{\,*}(-\p,\sigma)\right]=-2\eta^{\mu}_{\,\,\,\nu}
\end{equation}
giving us the standard commutator
\begin{equation}
[A^{\mu}(\x,t),\Pi_{\nu}(\mathbf{y},t)]=i\eta^{\mu}_{\,\,\nu}\delta^{3}(\mathbf{x-y}). \label{eq:charged_v_commutator2}
\end{equation}
Equations (\ref{eq:charged_v_commutator1}) and (\ref{eq:charged_v_commutator2}) show that the charged vector field is local.

\subsubsection{Neutral vector field}
Apart from the electrically charged scalar-vector field $A^{\mu}(x)$, one can also construct an electrically neutral scalar-vector field $a^{\mu}(x)$,
\begin{equation}
a^{\mu}(x)=(2\pi)^{-3/2}\int\frac{d^{3}p}{\sqrt{2p^{0}}}\sum_{\sigma}\left[e^{-ip\cdot x}e^{\mu}(\p,\sigma)c(\p,\sigma)+e^{ip\cdot x}e^{\mu*}(\p,\sigma)c^{\ddag}(\p,\sigma)\right]
\end{equation}
with the particle being identical to the anti-particle. The commutator between the neutral vector field and its conjugate-momentum remains the same as their charged counterpart
\begin{equation}
[a^{\mu}(\mathbf{x},t),\pi_{\nu}(\mathbf{y},t)]=i\eta^{\mu}_{\,\,\nu}\delta^{3}(\mathbf{x-y}).
\end{equation}
However, the commutators given by eq.~(\ref{eq:charged_v_commutator1}) is now non-trivial. Explicit computation of the commutator between the neutral vector field yields
\begin{eqnarray}
[a^{\mu}(\mathbf{x},t),a^{\nu}(\mathbf{y},t)]&=&(2\pi)^{-3}\int\frac{d^{3}p}{2p^{0}}\sum_{\sigma}\left[e^{i\mathbf{p}\cdot(\mathbf{x-y})}e^{\mu}(\p,\sigma)e^{\nu*}(\p,\sigma)-e^{-i\mathbf{p}\cdot\mathbf{(x-y)}} e^{\mu*}(\p,\sigma)e^{\nu}(\p,\sigma)\right] \nonumber \\
&=&(2\pi)^{-3}\int\frac{d^{3}p}{2p^{0}}\left[e^{i\mathbf{p}\cdot(\mathbf{x-y})}\left(-\eta^{\mu\nu}+2\frac{p^{\mu}p^{\nu}}{m^{2}}\right)-e^{-i\mathbf{p}\cdot(\mathbf{x-y})}\left(-\eta^{\mu\nu}+2\frac{p^{\mu}p^{\nu}}{m^{2}}\right)\right]\nonumber \\
&=&(2\pi)^{-3}\int\frac{d^{3}p}{m^{2}p^{0}}e^{i\mathbf{p}\cdot(\mathbf{x-y})}\left[p^{\mu}p^{\nu}-(\mathscr{P}p)^{\mu}(\mathscr{P}p)^{\nu}\right]\label{eq:aa_commutator}
\end{eqnarray}
where the second line makes use of the spin-sum
\begin{equation}
\sum_{\sigma}e^{\mu}(\p,\sigma)e^{\nu*}(\p,\sigma)=-\eta^{\mu\nu}+2\frac{p^{\mu}p^{\nu}}{m^{2}}.\label{eq:vector_spin_sums}
\end{equation}
When $\mu=\nu=0$ and $\mu=i$, $\nu=j$ the terms in the bracket cancel, giving us
\begin{equation}
[a^{0}(\mathbf{x},t),a^{0}(\mathbf{y},t)]=[a^{i}(\mathbf{x},t),a^{j}(\mathbf{y},t)]=0.
\end{equation}
As for $\mu=0$ and $\nu=i$, we get
\begin{eqnarray}
[a^{0}(\mathbf{x},t),a^{i}(\mathbf{y},t)]&=&\frac{2}{(2\pi)^{3}m}\int d^{3}p\, e^{i\mathbf{p}\cdot(\mathbf{x-y})}p^{i} \nonumber \\
&=&-\frac{2i}{m^{2}}\,\partial_{i}\delta^{3}(\mathbf{x-y})\nonumber \\
&=&\frac{2 i}{m^{2}(x^{i}-y^{i})}\delta^{3}(\mathbf{x-y})
\end{eqnarray}
where we have used the identity
\begin{equation}
\partial_{i}\delta^{3}(\mathbf{x-y})=-\frac{\delta^{3}(\mathbf{x-y})}{x^{i}-y^{i}}.\label{eq:d_delta}
\end{equation}
Since the right-hand side is proportional to a delta function, the equal-time commutator between the neutral vector field preserves locality.

Now we consider the commutator between the conjugate momentum. Following the same computation procedure as we have done to obtain eq.~(\ref{eq:aa_commutator}), we get
\begin{eqnarray}
[\pi^{\mu}(\mathbf{x},t),\pi^{\nu}(\mathbf{y},t)]&=&(2\pi)^{-3}\int\frac{d^{3}p}{2}(-p^{0})\sum_{\sigma}\Big[e^{-i\mathbf{p}\cdot(\mathbf{x-y})}\widehat{e}^{\mu}(\p,\sigma)\widehat{e}^{\,\nu*}(\p,\sigma)\nonumber\\
&&\hspace{5cm}-e^{i\mathbf{p}\cdot\mathbf{(x-y)}}\widehat{e}^{\,\mu*}(\p,\sigma)\widehat{e}^{\nu}(\p,\sigma)\Big] \nonumber \\
&=&(2\pi)^{-3}\int\frac{d^{3}p}{2}(-p^{0})\Big[e^{-i\mathbf{p}\cdot(\mathbf{x-y})}\left(-\eta^{\mu\nu}+2\frac{p^{\mu}p^{\nu}}{m^{2}}\right)\nonumber\\
&&\hspace{5cm}-e^{i\mathbf{p}\cdot(\mathbf{x-y})}\left(-\eta^{\mu\nu}+2\frac{p^{\mu}p^{\nu}}{m^{2}}\right)\Big]\nonumber \\
&=&(2\pi)^{-3}\int\frac{d^{3}p}{m^{2}}(-p^{0})e^{-i\mathbf{p}\cdot(\mathbf{x-y})}\left[p^{\mu}p^{\nu}-(\mathscr{P}p)^{\mu}(\mathscr{P}p)^{\nu}\right].
\end{eqnarray}
In the second line, we have used the identity
\begin{equation}
\sum_{\sigma}\widehat{e}^{\mu}(\p,\sigma)\widehat{e}^{\,\nu*}(\p,\sigma)=\sum_{\sigma}e^{\mu}(\p,\sigma)e^{\nu*}(\p,\sigma).
\end{equation}
The purely temporal and spatial component of the commutator vanishes
\begin{equation}
[\pi^{0}(\mathbf{x},t),\pi^{0}(\mathbf{y},t)]=[\pi^{i}(\mathbf{x},t),\pi^{j}(\mathbf{y},t)]=0.
\end{equation}
As for the temporal and spatial component of the conjugate-momentum, we get
\begin{equation}
[\pi^{0}(\mathbf{x},t),\pi^{i}(\mathbf{y},t)]=2i\frac{\delta^{3}(\mathbf{x-y})}{x^{i}-y^{i}}-\frac{2i}{m^{2}}\partial^{j}\partial_{j}\left[\frac{\delta^{3}(\mathbf{x-y})}{x^{i}-y^{i}}\right].
\end{equation}
Here we do not give the explicit expression of the second term involving the Laplacian of the delta function since it is quite complicated and does not provide additional insights. It is sufficient to see from eq.~(\ref{eq:d_delta}) that the second term is proportional to a delta function so the final result is
\begin{equation}
[\pi^{0}(\mathbf{x},t),\pi^{i}(\mathbf{y},t)]\sim\delta^{3}(\mathbf{x-y}).
\end{equation}

The equal-time commutators for the neutral vector field and its conjugate momentum either vanishes or are proportional to the delta function. Therefore, the neutral vector fields are local.

\subsection{Phenomenologies}\label{vector_phenomenologies}
In the SM, the $W^{\pm}$ and $Z$ bosons are described by the non-Abelian gauge theory with spontaneous symmetry breaking to give them masses. Here we explore the possibility that the $W^{\pm}$ and $Z$ bosons are described by the massive vector field with an additional scalar degree of freedom. We show that the decay rates of $W^{-}$ and $Z$ to leptons are in agreement with the SM predictions.

\subsubsection{The $W^{-}\rightarrow \ell+\bar{\nu}_{\ell}$ decay}

The charged-current interaction is described by the following Lagrangian density
\begin{equation}
\mathscr{L}_{CC}=\frac{ig}{\sqrt{2}}\bar{e}\gamma^{\mu}\frac{(I-\gamma^{5})}{2}\nu_{\ell}A_{\mu}
\end{equation}
where $e(x)$, $\nu_{\ell}(x)$ are the electron and the $\ell$-flavoured neutrino fields and $A_{\mu}(x)$ is the charged vector field. Here we calculate the decay process $W^{-}\rightarrow \ell+\bar{\nu}_{\ell}$ 
where the lepton $\ell$ is massive but anti-neutrino $\bar{\nu}_{\ell}$ is massless.
For completeness, we first compute the decay rate as predicted by SM using the spin-one $(j=1)$ vector field. Subsequently, we replace the spin-one vector field with the new vector field containing scalar and spin-one degrees of freedom $(j=0,1)$. 


At tree-level, the transition amplitude for $W^{-}\rightarrow \ell+\bar{\nu}_{\ell}$ is 
\begin{equation}
M_{(\bar{\nu}_{\ell}\ell)(W^{-})}=(2\pi)^{-3/2}\frac{ig}{\sqrt{8p^{0}_{\ell}p^{0}_{\bar{\nu}_{\ell}}p^{0}_{W}}}\bar{u}(\p_{\ell},\sigma_{\ell})\gamma^{\mu}
\left(\frac{I-\gamma^{5}}{2}\right)v(\p_{\bar{\nu}_{\ell}},\sigma_{\bar{\nu}_{\ell}})e_{\mu}(\p_{W},\sigma_{W}).
\end{equation}
This expression applies for both $j=1$ and $j=0,1$. In most experiments, the initial and final spin-projections are not measured, so we sum over all the degrees of freedom. The transition probability for $W^{-}$ when $j=1$ is given by
\begin{eqnarray}
\sum_{\sigma_{\ell}\sigma_{\nu_{\ell}},\sigma_{W}=\pm1,0}|M_{(\bar{\nu}_{\ell}e^{-})(W^{-})}^{(j=1)}|^{2} &=&\sum_{\sigma_{\ell}\sigma_{\nu_{\ell}},\sigma_{W}=\pm1,0}
\frac{g^{2}}{(8\pi)^{3} p^{0}_{\ell}p^{0}_{\bar{\nu}_{\ell}}p^{0}_{W}}
\left[\bar{u}(\p_{\ell},\sigma_{\ell})\gamma^{\mu}
(I-\gamma^{5})v(\p_{\bar{\nu}_{\ell}},\sigma_{\bar{\nu}_{\ell}})\right] \nonumber \\
&&\times\left[\bar{u}(\p_{\ell},\sigma_{\ell})\gamma^{\nu}
(I-\gamma^{5})v(\p_{\bar{\nu}_{\ell}},\sigma_{\bar{\nu}_{\ell}})\right]^{\dag}w_{\mu}(\p_{W},\sigma_{W})w^{\dag}_{\nu}(\p_{W},\sigma_{W})\nonumber\\
&=&\frac{g^{2}}{(8\pi)^{3} p^{0}_{\ell}p^{0}_{\bar{\nu}_{\ell}}p^{0}_{W} m_{W}^{2}}
\mbox{Tr}\left[(\slashed{p}_{\ell}+m_{\ell})\gamma^{\mu}(I-\gamma^{5})\slashed{p}_{\bar{\nu}_{\ell}}\gamma^{\nu}(I-\gamma^{5})\right]\nonumber \\
&&\hspace{1cm}\times\left[-\eta_{\mu\nu}+\frac{(p_{W})_{\mu}(p_{W})_{\nu}}{m_{W}^{2}}\right]
\end{eqnarray}
where the fourth line is obtained using the spin-sum
\begin{equation}
\sum_{\sigma=\pm1,0}w_{\mu}(\p,\sigma)w^{\dag}_{\nu}(\p,\sigma)=-\eta_{\mu\nu}+\frac{(p_{W})_{\mu}(p_{W})_{\nu}}{m_{W}^{2}}.\label{eq:vector_ss1}
\end{equation}
Taking the $W^{-}$ boson to be at rest $p^{\mu}_{W}=(m_{W},\mathbf{0})$, the traces contributing to the transition probability is given by
\begin{equation}
\mbox{Tr}\left[(\slashed{p}_{\ell}+m_{\ell})\gamma^{\mu}(I-\gamma^{5})\slashed{p}_{\bar{\nu}_{\ell}}\gamma^{\nu}(I-\gamma^{5})\right](-\eta_{\mu\nu})=
16p^{0}_{\bar{\nu}_{\ell}}(p^{0}_{\ell}+|\mathbf{p}_{\ell}|),
\end{equation}
\begin{equation}
\mbox{Tr}\left[\slashed{p}_{\bar{\nu}_{\ell}}\slashed{p}_{W}(I-\gamma^{5})\slashed{p}_{\ell}\slashed{p}_{W}(I-\gamma^{5})\right]
=8m_{W}^{2}p^{0}_{\bar{\nu}_{\ell}}(p^{0}_{\ell}-|\mathbf{p}_{\ell}|).
\end{equation}
In computing the traces, we have used the identity
\begin{equation}
\mbox{Tr}\left[\gamma^{5}\gamma^{\mu}\gamma^{\nu}\gamma^{\rho}\gamma^{\sigma}\right]=-4i\epsilon^{\mu\nu\rho\sigma}
\end{equation}
where $\epsilon^{\mu\nu\rho\sigma}$ is the Levi-Civita tensor, so it vanishes upon contracting with $(p_{W})_{\mu}(p_{W})_{\nu}$. Substituting the traces into the transition probability gives us
\begin{eqnarray}
\sum_{\sigma_{\ell}\sigma_{\nu_{\ell}},\sigma_{W}=\pm1,0}|M_{(\bar{\nu}_{\ell}e^{-})(W^{-})}^{(j=1)}|^{2}&=&\frac{g^{2}}{(8\pi)^{3} p^{0}_{\ell}p^{0}_{\bar{\nu}_{\ell}}p^{0}_{W}}
\left[16p^{0}_{\bar{\nu}_{\ell}}(p^{0}_{\ell}+|\mathbf{p}_{\ell}|)+8m_{W}^{2}p^{0}_{\bar{\nu}_{\ell}}(p^{0}_{\ell}-|\mathbf{p}_{\ell}|)\right]\nonumber \\
&=&\frac{g^{2}}{(8\pi)^{3} p^{0}_{\ell}p^{0}_{\bar{\nu}_{\ell}}p^{0}_{W}}
\left[16p^{0}_{\bar{\nu}_{\ell}}m_{W}+8m_{W}^{2}p^{0}_{\bar{\nu}_{\ell}}(p^{0}_{\ell}-|\mathbf{p}_{\ell}|)\right]\nonumber \\
&=&\left[\frac{16g^{2}}{(8\pi)^{3}p^{0}_{\ell}}+
\frac{8g^{2}(p^{0}_{\ell}-|\mathbf{p}_{\ell}|)}{(8\pi)^{3}p^{0}_{\ell} m_{W}}\right]
\end{eqnarray}
where the second line is obtained by conservation of energy. The main contribution to the probability comes from the first term. The second term makes a small but non-zero contribution since the lepton $\ell$ is massive $m_{\ell}>0$.

The average differential decay rate is obtained using eq.~(\ref{eq:decay_rate})
\begin{eqnarray}
d\Gamma^{(j=1)}_{\mbox{\footnotesize{avg}}}(W^{-}\rightarrow\bar{\nu}_{\ell}\ell)
&=&\frac{1}{3}\Gamma^{(j=1)}(W^{-}\rightarrow\bar{\nu}_{\ell}\ell)\nonumber\\
&=&\frac{2\pi |\k'|p^{0}_{\bar{\nu}_{\ell}}p^{0}_{\ell}}{3m_{W}}\left[\frac{16g^{2}}{(8\pi)^{3}p^{0}_{\ell}}+
\frac{8g^{2}(p^{0}_{\ell}-|\mathbf{p}_{\ell}|)}{(8\pi)^{3}p^{0}_{\ell} m_{W}}\right]d\Omega
\end{eqnarray}
where $d\Omega=\sin\theta d\theta d\phi$ is the solid angle and $|\k'|$ is 
\begin{equation}
|\k'|=\frac{m_{W}^{2}-m_{\ell}^{2}}{2m_{W}}.
\end{equation}
Integrating over the angles, and after some algebraic manipulation, the average decay rate of $W^{-}\rightarrow\bar{\nu}_{\ell}+\ell$ is
\begin{equation}
\Gamma^{(j=1)}_{\mbox{\footnotesize{total}}}(W^{-}\rightarrow\bar{\nu}_{\ell}\ell)=\frac{\sqrt{2}G_{F}(m_{W}^{2}-m_{\ell}^{2})p^{0}_{\bar{\nu}_{\ell}}}{6\pi}+\frac{\sqrt{2}G_{F}(m_{W}^{2}-m_{\ell}^{2})p^{0}_{\bar{\nu}_{\ell}}(p^{0}_{\ell}-|\mathbf{p}_{\ell}|)}{12\pi m_{W}}
\end{equation}
where $G_{F}=(\sqrt{2} g^{2})/(8m_{W}^{2})$ is the Fermi coupling constant. This expression can be further simplified using energy-momentum conservation,
\begin{equation}
p^{0}_{\ell}+p^{0}_{\bar{\nu}_{\ell}}=m_{W},\hspace{0.5cm}\mathbf{p}_{\bar{\nu}_{\ell}}=-\mathbf{p}_{\ell}.
\end{equation}
Since neutrinos are considered to be massless so that $p^{0}_{\bar{\nu}_{\ell}}=|\mathbf{p}_{\bar{\nu}_{\ell}}|=|\mathbf{p}_{\ell}|$, the energy and momentum of the lepton $\ell$ are found to be
\begin{equation}
p^{0}_{\ell}=\frac{m_{W}^{2}+m_{\ell}^{2}}{2m_{W}},\hspace{0.5cm} |\mathbf{p}_{\ell}|=\frac{m_{W}^{2}-m_{\ell}^{2}}{2m_{W}}.
\end{equation}
Substituting them into the average decay rates yields
\begin{equation}
\Gamma^{(j=1)}_{\mbox{\footnotesize{avg}}}(W^{-}\rightarrow\bar{\nu}_{\ell}\ell)=\frac{\sqrt{2}G_{F}(m_{W}^{2}-m_{\ell}^{2})^{2}}{12\pi m_{W}}\left[1+\frac{1}{2}\left(\frac{m_{\ell}}{m_{W}}\right)^{2}\right]. \label{eq:spin_one_decay}
\end{equation}

In our theory, the vector boson carries both scalar and spin-one degrees of freedom, which results in a modification of the decay rate. An important issue we have to address is whether we take average the decay rate by multiplying by a factor $1/3$ or $1/4$. To answer this question, we note that the decay rate of a scalar $W$-boson to leptons is suppressed by a factor $\mathcal{O}(m_{\ell}^{2}/m_{W}^{2})$. This situation is somewhat analogous to electrodynamics with massive photon where the extra longitudinal degree of freedom is \textit{sterile} so the effective number of degrees of freedom taking part in the relevant interactions remains two~\cite[sec. II.F]{Goldhaber:2008xy}.\footnote{I am grateful to Matt Visser for bringing this to my attention.} Since the scalar degree of freedom of $W$ is sterile, we propose that the average decay rate is obtained by multiplying the decay rate by 1/3. Therefore, the deviation from the SM prediction originates from the fact that the spin-sum is momentum-independent
\begin{equation}
\sum_{\sigma=\pm1,0,\widetilde{0}}w_{\mu}(\p,\sigma)w_{\nu}^{\ddag}(\p,\sigma)=-\eta_{\mu\nu}.\label{eq:vector_ss2}
\end{equation}
The two spin-sums given by eqs.~(\ref{eq:vector_ss1}) and (\ref{eq:vector_ss2}) differs by the $p^{\mu}p^{\nu}/m^{2}$ term. The term $p^{\mu}p^{\nu}/m^{2}$, upon contracting with the $\gamma$-matrices give the second term in eq.~(\ref{eq:spin_one_decay}). Therefore, the predicted average decay rate of a $W^{-}$ boson with scalar and spin-one degrees of freedom is
\begin{equation}
\Gamma^{(j=0,1)}_{\mbox{\footnotesize{avg}}}(W^{-}\rightarrow\bar{\nu}_{\ell}\ell)=\frac{\sqrt{2}G_{F}(m_{W}^{2}-m_{\ell}^{2})^{2}}{12\pi m_{W}}.
\end{equation}
The ratio of the two decay rate is
\begin{equation}
\frac{\Gamma^{(j=1)}_{\mbox{\footnotesize{avg}}}}{\Gamma^{(j=0,1)}_{\mbox{\footnotesize{avg}}}}(W^{-}\rightarrow\bar{\nu}_{\ell}\ell)=1+\frac{1}{2}\left(\frac{m_{\ell}}{m_{W}}\right)^{2}.
\end{equation}
Identifying the lepton to be electron, muon and tau, their mass ratio squared are
\begin{equation}
\left(\frac{m_{e}}{m_{W}}\right)^{2}\sim 10^{-11},\hspace{0.5cm}
\left(\frac{m_{\mu}}{m_{W}}\right)^{2}\sim 10^{-6},\hspace{0.5cm}
\left(\frac{m_{\tau}}{m_{W}}\right)^{2}\sim 10^{-4}.
\end{equation}
We see that the deviation from the SM prediction is extremely small. The difference ranges from one part in $10^{11}$ to one part in $10^{4}$. 

\subsubsection{The $Z\rightarrow\nu_{\ell}+\bar{\nu}_{\ell}$ decay}

The neutral current interaction is describing by the following Lagrangian density
\begin{equation}
\mathscr{L}_{NC}=\frac{i}{2}\sqrt{g^{2}+g'^{2}}\bar{\nu}_{\ell}\gamma^{\mu}\left(\frac{I-\gamma^{5}}{2}\right)\nu_{\ell}a_{\mu}.
\end{equation}
The tree-level amplitude of $Z\rightarrow\nu_{\ell}+\bar{\nu}_{\ell}$ is given by
\begin{equation}
M_{(\nu_{\ell}\bar{\nu}_{\ell})(Z)}=(2\pi)^{-3/2}\frac{i\sqrt{g^{2}+g'^{2}}}{2\sqrt{8p^{0}_{\nu_{\ell}}p^{0}_{\bar{\nu}_{\ell}}p^{0}_{W}}}\bar{v}(\p_{\bar{\nu}_{\ell}},\sigma_{\bar{\nu}_{\ell}})\gamma^{\mu}
\left(\frac{I-\gamma^{5}}{2}\right)u(\p_{\nu_{\ell}},\sigma_{\nu_{\ell}})e_{\mu}(\p_{Z},\sigma_{Z}).
\end{equation}
First, we compute the average decay rate by summing over all the spin degrees of freedom for the massive spin-one vector field. The transition probability is
\begin{equation}
\sum_{\sigma_{Z}=\pm1,0}|M_{(\nu_{\ell}\bar{\nu}_{\ell})(Z)}^{(j=1)}|^{2}=\frac{g^{2}+g'^{2}}{2(8\pi)^{3}p^{0}_{\nu_{\ell}}p^{0}_{\bar{\nu}_{\ell}}p^{0}_{Z}}
\mbox{Tr}\left[\slashed{p}_{\nu_{\ell}}\gamma^{\mu}(I-\gamma^{5})\slashed{p}_{\bar{\nu_{\ell}}}\gamma^{\nu}(I-\gamma^{5})\right]
\left[-\eta_{\mu\nu}+\frac{(p_{Z})_{\mu}(p_{Z})_{\nu}}{m_{Z}^{2}}\right].
\end{equation}
The neutrino mass is negligible so the traces with the momenta of $Z$-boson vanishes. Therefore, the transition probability for both the $j=1$ and $j=0,1$ are identical, thus giving us
\begin{equation}
\sum_{\sigma_{Z}=\pm1,0,\widetilde{0}}|M_{(\nu_{\ell}\bar{\nu}_{\ell})(Z)}^{(j=0,1)}|^{2}=\sum_{\sigma_{Z}=\pm1,0}|M_{(\nu_{\ell}\bar{\nu}_{\ell})(Z)}^{(j=1)}|^{2}=\frac{g^{2}+g'^{2}}{32\pi^{3}m_{Z}}.
\end{equation}
In the centre of mass frame, where the $Z$-boson is at rest, its differential and average decay rate to $\nu_{\ell}$ and $\bar{\nu_{\ell}}$, following the previous calculation, is  given by
\begin{equation}
\frac{d\Gamma^{(j=0,1)}_{\mbox{\footnotesize{avg}}}(Z\rightarrow\nu_{\ell}\bar{\nu}_{\ell})}{d\Omega}=
\frac{d\Gamma^{(j=1)}_{\mbox{\footnotesize{avg}}}(Z\rightarrow\nu_{\ell}\bar{\nu}_{\ell})}{d\Omega}=
\frac{1}{3}\left[\frac{(g^{2}+g'^{2})m_{Z}}{128\pi^{2}}\right]
\end{equation}
and
\begin{equation}
\Gamma^{(j=0,1)}_{\mbox{\footnotesize{avg}}}(Z\rightarrow\nu_{\ell}\bar{\nu}_{\ell})=
\Gamma^{(j=1)}_{\mbox{\footnotesize{avg}}}(Z\rightarrow\nu_{\ell}\bar{\nu}_{\ell})=\frac{(g^{2}+g'^{2})m_{Z}}{96\pi}.
\end{equation}
Theoretically, there is a difference in the decay rate since neutrinos are massive. But since $m_{\nu_{\ell}}\ll m_{\ell}\ll m_{Z}$, the resulting decay rate for the $Z\rightarrow\nu_{\ell}+\bar{\nu_{\ell}}$ for all practical purposes are the same as the SM prediction.

\subsection{Renormalisation and the Higgs mechanism}
\begin{figure}
\begin{center}
\includegraphics[scale=0.8]{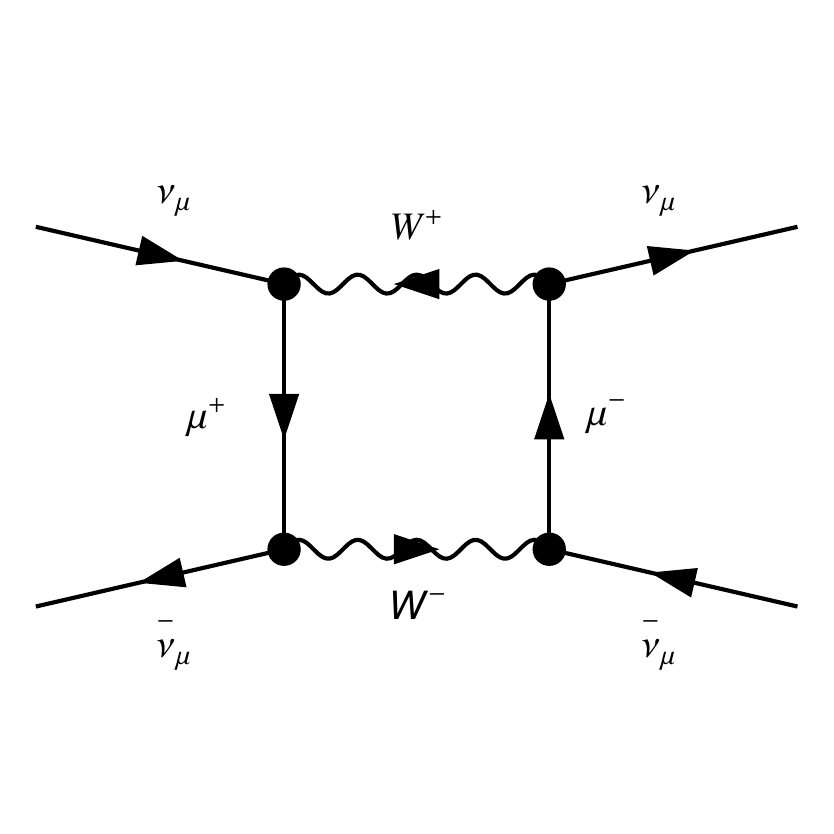}
\end{center}
\caption{Higher-order contribution to $\nu_{\mu}+\bar{\nu}_{\mu}\rightarrow \nu_{\mu}+\bar{\nu}_{\mu}$}\label{fig:box_diagram}
\end{figure}

The new vector field with scalar and spin-one degrees of freedom is an improvement of the intermediate vector boson model in terms of unitarity and renormalisability. This can be seen by considering the propagator of a pure spin-one massive vector field
\begin{equation}
S_{\mu\nu}^{(j=1)}(y,x)=i\int\frac{d^{4}p}{(2\pi)^{4}}e^{-ip\cdot(x-y)}\frac{-\eta_{\mu\nu}+\frac{p_{\mu}p_{\nu}}{m^{2}}}{p^{\lambda}p_{\lambda}-m^{2}+i\epsilon}.
\label{eq:spin_one_propagator}
\end{equation}
At tree-level, the intermediate vector boson theory solves the unitarity problem for processes such as $\bar{\nu}_{\mu}+\mu^{-}\rightarrow \bar{\nu}_{e}+e^{-}$ of Fermi theory~\cite[chap.~21]{Aitchison:2004cs}. However, the theory is still not immune to unitarity violation. In processes where the vector bosons are either the in or out states, the total cross-section would depend on its spin-sum which is
\begin{equation}
\sum_{\sigma=\pm1,0}e^{\mu}(\p,\sigma)e^{\nu*}(\p,\sigma)=-\eta^{\mu\nu}+\frac{p^{\mu}p^{\nu}}{m^{2}}. \label{eq:spin_one_spin_sum}
\end{equation}
Since spin-sum is quadratic in momentum, processes such as $\nu_{\mu}+\bar{\nu}_{\mu}\rightarrow W^{+}+W^{-}$ has a cross-section that is unbounded by energy $\sigma\sim(p^{0})^{2}$.
As a result, the unitarity bound is violated at about $1$TeV. Furthermore, the integrals involving $p^{\mu}p^{\nu}/m^{2}$ in the propagator is divergent in higher-order processes
such as fig.\ref{fig:box_diagram}~\cite[chap.~21]{Aitchison:2004cs}.

 Here, we have shown that the above mentioned problems can be resolved by including an additional scalar degree of freedom in the massive vector field with appropriately defined field adjoint. The momentum term in the numerator of the propagator and the spin-sum are cancelled between the scalar and spin-one sector. Therefore, the resulting propagator given in eq.~(\ref{eq:v_propagator}) does not contain momentum terms in the numerator so the integral for the higher-order to  $\nu_{\mu}+\bar{\nu}_{\mu}\rightarrow\nu_{\mu}+\bar{\nu}_{\mu}$ as given by fig.\ref{fig:box_diagram} is now convergent.  

Although our theory preserves unitarity at tree-level and the box-diagram is convergent, it is important to note that the theory has not been proved to be renormalisable to all orders of perturbation.
The absence of the $p^{\mu}p^{\nu}/m^{2}$ term in the spin-sum and propagator means the theory has a better behaviour in higher-order scattering processes comparing with the intermediate vector boson model and the Fermi theory. However, this is insufficient for renormalisability to all orders. The propagator and the Hamiltonian suggest that the vector field is described by a Klein-Gordon Lagrangian density, but until we can show that all the divergences associated with higher-order scattering amplitudes can be cancelled by the relevant counter-terms generated by the renormalised Lagrangian density, we cannot be sure whether our choice is correct.

The non-renormalisability and unitarity violation of the intermediate vector boson are the main motivation for a non-Abelian gauge theoretic description of the electroweak processes. As a matter of fact, 't Hooft and Veltman have shown that non-Abelian gauge theory and hence the SM plus the Higgs boson are renormalisable~\cite{'tHooft:1972fi}. Given the recent discovery of a \textit{Higgs-like} boson at the LHC~\cite{:2012gu,:2012gk}, the question of relevance to us is what are the similarities and differences between our construct and the Higgs mechanism, or are they equivalent? Presently, we do not have a definite answer to this question, but we can still make some useful observations. In our construct, the Veltman propagator is obtained by introducing an additional scalar degree of freedom and the appropriate field adjoint provided that the scalar and spin-one sector have the same mass. On the other hand, in the SM, the Higgs and vector bosons have \textit{different} masses and the Veltman propagator is obtained by choosing the Feynman gauge after spontaneous symmetry breaking~\cite[eq. (21.2.20)]{Weinberg:1996kr}. These observations suggest that the scalar degree of freedom in our construct cannot be identified as the Higgs boson. 

It should be noted that the idea of using the new massive vector fields to describe the $W^{\pm}$ and $Z$ bosons are purely phenomenological and is not based on a underlying gauge symmetry. Nevertheless, since the description of the weak force using the massive spin-one vector fields are unitary violating and non-renormalisable,\footnote{Quantum electrodynamics with massive photons is unitary and renormalisable~\cite{vanHees:2003dk}.} it is natural to investigate whether our construct does any better. As far as we can see, our construct passes the initial test as the computed decay rates in sec.~\ref{vector_phenomenologies} are in agreement with the SM predictions and the propagator has a better high energy behaviour than their spin-one counterpart. In this thesis, we did not explore the possible connexions between the new massive vector field and non-Abelian gauge theory. This is definitely an important issue to be studied in the future.

\section{Massless vector fields}\label{massless_VF}
The massless spin-half fields and their constraints have already been derived in sec.~\ref{Massless QF}, so all we need to do is to repeat the same calculation for the vector representation of the Lorentz group. Let us define the massless vector field as\footnote{The massless field $a^{\mu}(x)$ is electrically neutral, but it does not affect the discussion here.} 
\begin{equation}
a^{\mu}(x)=(2\pi)^{-3/2}\int\frac{d^{3}p}{\sqrt{2p^{0}}}\sum_{\sigma=\pm1,0}\left[e^{-ip\cdot x}u^{\mu}(\p,\sigma)c(\p,\sigma)+e^{ip\cdot x}v^{\mu}(\p,\sigma)c^{\dag}(\p,\sigma)\right].
\end{equation}
According to our arguments at the beginning of the chapter, the most general massless field  would sum over both the scalar and spin-one degrees of freedom. This possibility is considered at the end of the chapter. For now, we will be content with a massless vector field where $\sigma=\pm1,0$.

The constraints on the expansion coefficients which for momentum $k=(\kappa,0,0,\kappa)$ are
\begin{equation}
u^{\mu}(\k,\sigma)e^{i\sigma\phi(\Lambda,p)}=W^{\mu}_{\,\,\,\nu}(\Lambda,p)u^{\nu}(\k,\sigma), \label{eq:u_w_u}
\end{equation}
\begin{equation}
v^{\mu}(\k,\sigma)e^{-i\sigma\phi(\Lambda,p)}=W^{\mu}_{\,\,\,\nu}(\Lambda,p)v^{\nu}(\k,\sigma).
\end{equation}
Since the little group element for massless particle is separable $W(\phi,\alpha,\beta)=R(\phi)S(\alpha,\beta)$, we obtain the following constraints from $R(\phi)$ and $S(\alpha,\beta)$
\begin{equation}
u^{\mu}(\k,\sigma)e^{i\sigma\phi}=R^{\mu}_{\,\,\nu}(\phi)u^{\nu}(\k,\sigma),\hspace{0.5cm}
u^{\mu}(\k,\sigma)=S^{\mu}_{\,\,\nu}(\alpha,\beta)u^{\nu}(\k,\sigma),
\end{equation}
\begin{equation}
v^{\mu}(\k,\sigma)e^{-i\sigma\phi}=R^{\mu}_{\,\,\nu}(\phi)v^{\nu}(\k,\sigma),\hspace{0.5cm}
v^{\mu}(\k,\sigma)=S^{\mu}_{\,\,\nu}(\alpha,\beta)v^{\nu}(\k,\sigma).
\end{equation}
The matrices $R(\phi)$ and $S(\alpha,\beta)$ are the elements of the little group given by
\begin{equation}
R(\phi)=\left(\begin{matrix}
1 & 0 & 0 & 0 \\
0 &\cos\phi & \sin\phi & 0\\
0 &-\sin\phi & \cos\phi & 0\\
0 & 0 & 0 & 1\end{matrix}\right),\hspace{0.5cm}
S(\alpha,\beta)=\left(\begin{matrix}
\gamma+1 & \alpha & \beta & -\gamma \\
\alpha & 1 & 0 & -\alpha \\
\beta & 0 & 1 & -\beta\\
\gamma & \alpha & \beta & 1-\gamma\end{matrix}\right)
\end{equation}
with $\gamma=\frac{1}{2}(\alpha^{2}+\beta^{2})$. Since $R(\phi)$ and $S(\alpha,\beta)$ are both real matrices, without losing generality, we make the following choice of coefficient
\begin{equation}
u^{\mu}(\k,\sigma)=v^{\mu*}(\k,\sigma)=e^{\mu}(\k,\sigma)
\end{equation}
so the field becomes
\begin{equation}
a^{\mu}(x)=(2\pi)^{-3/2}\int\frac{d^{3}p}{\sqrt{2p^{0}}}\sum_{\sigma=\pm1,0}\left[e^{-ip\cdot x}e^{\mu}(\p,\sigma)c(\p,\sigma)+e^{ip\cdot x}e^{\mu*}(\p,\sigma)c^{\dag}(\p,\sigma)\right]
\end{equation}
and constraints we need to solve reduce to
\begin{equation}
e^{\mu}(\k,\sigma)e^{i\sigma\phi}=R^{\mu}_{\,\,\nu}(\phi)e^{\nu}(\k,\sigma),\label{eq:e_rot}
\end{equation}
\begin{equation}
e^{\mu}(\k,\sigma)=S^{\mu}_{\,\,\nu}(\alpha,\beta)e^{\nu}(\k,\sigma).\label{eq:e_sab}
\end{equation}
For eq.~(\ref{eq:e_rot}), the values of $\sigma$ are $\pm1,0$ are eigenvalues of $\mathscr{J}^{3}$. Solving eq.~(\ref{eq:e_rot}) for each $\sigma$ yields
\begin{equation}
e(\k,\pm1)=f_{\pm}\left(\begin{matrix}
0 \\
1 \\
\pm i\\
0 \end{matrix}\right),\hspace{0.5cm}
e(\k,0)=\left(\begin{matrix}
g_{+} \\
0 \\
0 \\
g_{-} \end{matrix}\right).
\end{equation}
where $f_{\pm}$ and $g_{\pm}$ are proportionality constants to be determined. Substituting the coefficients into eq.~(\ref{eq:e_sab}), for $\sigma=0$ we obtain one additional condition
\begin{equation}
g_{+}=g_{-}.
\end{equation}
However, for $\sigma=\pm1$, the result is
\begin{equation}
\left(\begin{matrix}
0 \\
1 \\
\pm i\\
0 \end{matrix}\right)=
\left(\begin{matrix}
\alpha\pm i\beta \\
1 \\
\pm i\\
\alpha\pm i\beta \end{matrix}\right).
\end{equation}
This equation is cannot be satisfied for all $\alpha$ and $\beta$, so $e^{\mu}(\k,\pm1)$ must identically vanish, leaving us with $e(\k,0)=g_{+}(1,0,0,1)$.
Boosting $e^{\mu}(\k,0)$ using $L(p)$ given in eq.~(\ref{eq:massless_boost}) gives us
\begin{equation}
e^{\mu}(\p,0)=\frac{g_{+}}{\kappa} p^{\mu}.
\end{equation}
Taking $g_{+}=-i\kappa$, we get $e^{\mu}(\p,\sigma)=-ip^{\mu}$. Therefore, the only massless vector field that satisfies Lorentz symmetry is the scalar field of the form $a_{\mu}^{(j=0)}(x)=\partial_{\mu}\phi(x)$.

\subsection{Gauge-invariance}\label{local_gauge_invariance}
The vanishing of $e^{\mu}(\k,\pm1)$ is in agreement with a general theorem proven by Weinberg in~\cite{Weinberg:1964ev}. Any massless field of the $(j,j')$ representation of the Lorentz group can only describe particle of spin $j-j'$. In this notation, the vector representation is the $(\frac{1}{2},\frac{1}{2})$ representation so the massless vector field is given by the derivative of scalar field. This result can also be anticipated from eq.~(\ref{eq:e_rot}) since for $\sigma=\pm1$, the coefficient $e^{\mu}(\k,\pm1)$ does not transform as a four-vector. Their transformation under the little group is
\begin{equation}
W^{\mu}_{\,\,\nu}(\phi,\alpha,\beta)e^{\nu}(\k,\pm1)=e^{\pm i\phi(\Lambda,p)} e^{\mu}(\k,\pm1)+(\alpha\pm i\beta)\frac{k^{\mu}}{\kappa}\label{eq:w_e}
\end{equation}
where we have used $W(\phi,\alpha,\beta)=S(\alpha,\beta)R(\phi)$. Comparing eq.~(\ref{eq:w_e}) with (\ref{eq:u_w_u}), the second term violates the required constraint.  In spite of this, the structure of the transformation shows the following coefficient $e^{\mu}(\k,\pm1)k^{\nu}-k^{\mu}e^{\nu}(\k,\pm1)$ is covariant under the little group
\begin{equation}
W^{\mu}_{\,\,\,\rho}(\phi,\alpha,\beta)W^{\nu}_{\,\,\,\lambda}(\phi,\alpha,\beta)[e^{\rho}(\k,\pm1)k^{\lambda}-k^{\rho}e^{\lambda}(\k,\pm1)]=
e^{\pm i\phi(\Lambda,p)}[e^{\mu}(\k,\pm1)k^{\nu}-k^{\mu}e^{\nu}(\k,\pm1)].\label{eq:w_de}
\end{equation}
Equation (\ref{eq:w_de}) is precisely the constraint for a second rank anti-symmetric tensor $f^{\mu\nu}(x)$ with expansion coefficients proportional to $e^{\mu}(\k,\pm1)k^{\nu}-k^{\mu}e^{\nu}(\k,\pm1)$. It can be derived following the same calculation performed in the previous chapter starting from the usual tensor transformation law
\begin{equation}
U(\Lambda,a)f^{\mu\nu}(x)U^{-1}(\Lambda,a)=\Lambda_{\rho}^{\,\,\mu}\Lambda_{\lambda}^{\,\,\nu}f^{\rho\lambda}(\Lambda x+a).
\end{equation}
Therefore, a massless particle of helicity $\pm1$ is described by the field strength tensor $f^{\mu\nu}(x)$. The field strength tensor, with the appropriate normalisation is defined as
\begin{equation}
f^{\mu\nu}(x)=(2\pi)^{-3/2}\int\frac{d^{3}p}{\sqrt{2p^{0}}}\sum_{\sigma=\pm 1}\left[e^{-ip\cdot x}e^{\mu\nu}(\p,\sigma)c(\p,\sigma)+e^{ip\cdot x}e^{\mu\nu*}(\p,\sigma)c^{\dag}(\p,\sigma)\right]
\end{equation}
where
\begin{equation}
e^{\mu\nu}(\p,\sigma)=i[e^{\mu}(\p,\sigma)p^{\nu}-p^{\mu}e^{\nu}(\p,\sigma)].
\end{equation}
The factor of $i$ is introduced so that the field strength tensor can be expressed as
\begin{equation}
f^{\mu\nu}(x)=\partial^{\mu}a^{\nu}(x)-\partial^{\nu}a^{\mu}(x)
\end{equation}
where
\begin{equation}
a^{\mu}(x)=(2\pi)^{-3/2}\int\frac{d^{3}p}{\sqrt{2p^{0}}}\sum_{\sigma=\pm1}\left[e^{-ip\cdot x}e^{\mu}(\p,\sigma)c(\p,\sigma)+e^{ip\cdot x}e^{\mu*}(\p,\sigma)c^{\dag}(\p,\sigma)\right].
\end{equation}
We choose the proportionality constant to be $f_{\pm}=1/\sqrt{2}$ so that
\begin{equation}
e(\k,\pm1)=\frac{1}{\sqrt{2}}\left(\begin{matrix}
0 \\
1 \\
\pm i\\
0 \end{matrix}\right).
\end{equation}

In hindsight, we could have started with $f^{\mu\nu}(x)$ instead of $a^{\mu}(x)$, side-stepping the problem of the non-covariance. However,  it is the field $a^{\mu}(x)$ and not $f^{\mu\nu}(x)$ that participates in the interactions with other fields. In 3+1 dimensions, interactions involving $f^{\mu\nu}(x)$ such as the Pauli term $\overline{\Psi}[\gamma_{\mu},\gamma_{\nu}]\Psi f^{\mu\nu}$ is non-renormalisable whereas $\overline{\Psi}\slashed{a}\Psi$ is renormalisable.  To determine the form of interactions, we need determine the Lorentz transformation of $a^{\mu}(x)$. To this end, we consider the inverse of eq.~(\ref{eq:w_e}),
\begin{equation}
(W^{-1})^{\mu}_{\,\,\nu}(\phi,\alpha,\beta)e^{\nu}(\k,\pm1)=e^{\mp i\phi(\Lambda,p)} e^{\mu}(\k,\pm1)-(\alpha\pm i\beta)\frac{k^{\mu}}{\kappa}.\label{eq:iw_e}
\end{equation}
Substituting the definition $W(\Lambda,p)=L^{-1}(\Lambda p)\Lambda L(p)$ into the above, we get
\begin{equation}
\Lambda_{\rho}^{\,\,\,\mu}e^{\rho}(\L\p,\pm1)=e^{\mp i\phi}e^{\mu}(\p,\pm1)-(\alpha\pm i\beta)\frac{p^{\mu}}{\kappa}.
\end{equation}
Taking $\mu=0$ and use $e^{0}(\p,\pm1)=0$, we obtain
\begin{equation}
\frac{\alpha\pm i\beta}{\kappa}=-\frac{\Lambda_{\rho}^{\,\,\,0}e^{\rho}(\Lambda p,\pm1)}{p^{0}}
\end{equation}
which gives us the desired result
\begin{eqnarray}
\Lambda_{\rho}^{\,\,\,\mu}e^{\rho}(\L\p,\pm1)&=&e^{\mp i\phi(\Lambda,p)}e^{\mu}(\p,\pm1)+\frac{\Lambda_{\rho}^{\,\,\,0}e^{\rho}(\Lambda p,\pm1)}{p^{0}}p^{\mu} \nonumber\\
&=&e^{\mp i\phi(\Lambda,p)}e^{\mu}(\p,\pm1)+p^{\mu}\Omega_{\pm}(\Lambda,p).\label{eq:L_eL}
\end{eqnarray}
We can now calculate $U(\Lambda)a^{\mu}(x)U^{-1}(\Lambda)$. Using eqs.~(\ref{eq:DW_massless}) and (\ref{eq:ann},\ref{eq:cre}),
\begin{eqnarray}
U(\Lambda)a^{\mu}(x)U^{-1}(\Lambda)&=&(2\pi)^{-3/2}\int\frac{d^{3}(\Lambda p)}{\sqrt{2(\Lambda p)^{0}}}\sum_{\sigma}
\Big[e^{-i\sigma\phi(\Lambda,p)}e^{-ip\cdot x}e^{\mu}(\p,\sigma)c(\L\p,\sigma)\nonumber \\
&&\hspace{3cm}+e^{i\sigma\phi(\Lambda,p)}e^{ip\cdot x}e^{\mu*}(\p,\sigma)c^{\dag}(\L\p,\sigma)\Big].\label{eq:massless_amu_trans}
\end{eqnarray}
Substituting eq.~(\ref{eq:L_eL}) into (\ref{eq:massless_amu_trans}), we obtain the transformation for $a^{\mu}(x)$
\begin{equation}
U(\Lambda)a^{\mu}(x)U^{-1}(\Lambda)=\Lambda_{\nu}^{\,\,\,\mu}a^{\nu}(\Lambda x)+\partial^{\mu}\Omega(\Lambda,x).
\end{equation}
The definition for $\Omega(\Lambda,x)$ can be obtained in terms of integrals of $\Omega_{\pm}(\Lambda,p)$ but the explicit form is not important for the discussion. The important point to note is that the term $\partial^{\mu}\Omega(\Lambda,x)$ is cancelled in the transformation of $f^{\mu\nu}(x)$.

Given an interacting field theory which satisfies Lorentz symmetry, its corresponding Lagrangian density $\mathscr{L}(x)$ must be invariant under a fixed point Lorentz transformation. However, the field $a^{\mu}(x)$ does not transform as a four-vector, so in addition to Lorentz invariance, we must also impose local gauge invariance so that the extra term $\partial^{\mu}\Omega(\Lambda,x)$ does not contribute to the Lagrangian density. Therefore, the interacting Lagrangian density containing $a^{\mu}(x)$ must also be invariant under a local gauge transformation
\begin{equation}
a^{\mu}(x)\rightarrow a^{\mu}(x)+\partial^{\mu}\epsilon(x).\label{eq:gauge_trans}
\end{equation}
where $\epsilon(x)$ is an arbitrary space-time dependent function. It follows that the corresponding action $I_{M}$ must be invariant under
\begin{equation}
\delta I_{M}=\int d^{4}x\frac{\delta I_{M}}{\delta a_{\mu}(x)}\partial_{\mu}\epsilon(x)=0.
\end{equation}
Integrating by part, this amounts to
\begin{equation}
\partial_{\mu}\frac{\delta I_{M}}{\delta a_{\mu}(x)}=\partial_{\mu}J^{\mu}(x)=0
\end{equation}
where $J^{\mu}(x)=\delta I_{M}/\delta a_{\mu}(x)$ is a conserved current independent of $a^{\mu}(x)$. By Noether's theorem, the existence of $J^{\mu}(x)$ is guaranteed if the Lagrangian density of the field $\Psi(x)$ involved in the interaction is invariant under the transformation
\begin{equation}
\delta\Psi(x)=iq\epsilon(x)\Psi(x)\label{eq:gauge_trans2}
\end{equation}
where $q$ is a coupling constant.

Equations (\ref{eq:gauge_trans}) and (\ref{eq:gauge_trans2}) are the familiar local $U(1)$ gauge transformations. Apart from Weinberg~\cite[chap.~8]{Weinberg:1995mt}, most textbooks take the principle of local gauge invariance a priori to determine the precise form of interactions. However, in doing so, the direct connexion between gauge symmetry and Lorentz symmetry is no longer apparent, whereas a systematic study on the symmetries of the massless vector fields reveals gauge symmetry as an artefact of Lorentz symmetry. 

Generally, for theories satisfying Lorentz symmetry, the principle of gauge invariance can always be used to determine the interaction. Nevertheless there are limitations that must be realised. Firstly, the above analysis only applies to Abelian gauge symmetries. As far as we know, a similar derivation has not been carried out for non-Abelian gauge fields. The case for non-Abelian gauge fields would be considerably more difficult as the field equation is non-linear and contains self-interaction. Secondly, since gauge symmetry is a consequence of Lorentz symmetry, there is no guarantee that it also applies to theories with other symmetries that are not Lorentz. However our result does suggest that for any field theories, regardless whether the underlying symmetry is Lorentz or not, the form of the interactions, at least for Abelian gauge fields, can in principle be determined by the underlying symmetry group.

\section{Summary}
Since this chapter is divided into two main sections, we separate the summary into two sections on massive and massless vector fields. The main results and outstanding questions are summarised.

\subsection{Massive vector field}

Our results show that contrary to conventional wisdom, the massive vector particles and their fields with both scalar and spin-one degrees of freedom which transform under a reducible representation do lead to new physics. In the SM, the massive vector fields we have constructed present an interesting alternative description of the $W^{\pm}$ and $Z$ boson in the electroweak theory.

Results obtained in this chapter shows that the difficulties faced by the intermediate vector boson model and the Fermi theory are connected to the demand of irreducibility, that the massive vector fields must transform under the irreducible spin-one representation. Once this demand is relaxed, these problems are circumvented by an appropriate choice of phases for the fields and their adjoint. The resulting fields satisfy Poincar\'{e} symmetry, are local and have positive-definite free Hamiltonian.

In physical processes where vector bosons act as mediators, we do not expect deviations from the SM since the propagator is effectively unchanged.\footnote{The propagator for the massive spin-one vector boson is $[i/(2\pi)^{4}](-\eta_{\mu\nu}+p_{\mu}p_{\nu}/m^{2})/(p^{\lambda}p_{\lambda}-m^{2}+i\epsilon)$ in the momentum space. In most experiments, the vector bosons are non-relativistic so the $p_{\mu} p_{\nu}/m^{2}$ term can be ignored.} As for processes where vector bosons appear as initial states, at the leading order, we have shown that the scalar degrees of freedom contributes very little to the decay rate to the lepton sector. Therefore, for the considered processes, our theory is in agreement with the SM predictions.

\subsection{Massless vector field}
The systematic study of symmetries of the massless vector fields revealed that the massless particle of helicity $\pm1$ is described by the field strength tensor $f^{\mu\nu}(x)$ while 
the field $a^{\mu}(x)$ transforms as a four-vector only up to a local gauge transformation. 

The Weinberg formalism revealed the deep connexion between local Abelian gauge symmetry and Lorentz symmetry. Unless one approaches quantum field theory with an emphasis on Lorentz symmetry, this connexion would be hidden. We would be misled into believing the universality of gauge invariance and would be unaware of the possibility that it may only be applicable to Lorentz-invariant field theories. For field theories with other symmetries that is not Lorentz-invariant there is no guarantee that local gauge invariance can be applied. In that case, if gauge symmetry is applicable, it must be derived and not postulated.

%

\subsection{Future works}

The issue of gauge invariance is potentially important if we want the massive vector field we have constructed to be fully renormalisable. Conventional wisdom tells us that gauge invariance and renormalisation are closely related. In quantum electrodynamics, the renormalisability of the theory not only depends on the 
mass-dimensionality of $ie\overline{\Psi}\slashed{a}\Psi$, it also depends on the fact that the action  $-\frac{1}{4}\int d^{4}x f^{\mu\nu}f_{\mu\nu}$ is gauge invariant and that it generates the appropriate counter-term to cancel the divergence.

We note, the mass-dimension requirement on the interaction is a necessary but not sufficient condition for renormalisability. In the $S$-matrix, the high-order amplitudes involving loop diagrams can generate divergent terms that are gauge-invariant. These terms can only be cancelled by the counter-terms generated by the correct gauge-invariant Lagrangian. Therefore, it is important for us to study the possible connexions between the new massive vector fields and gauge symmetry. 
A possible path to explore is to see whether the action for the massive vector field can be made gauge invariant using the St$\ddot{\mbox{u}}$ckelberg method. Alternatively, there may exist a different symmetry breaking mechanism which yields the massive vector fields we have constructed. 

Phenomenologically, we have only computed the decay rates for $W^{-}$ and $Z$. These calculations are expected to reproduce the SM predictions since the masses of  $W^{\pm}$ and $Z$ are much heavier than the leptons, so energy-momentum conservation guarantees that the term present in the SM $p^{\mu}p^{\nu}/m_{W,Z}^{2}\sim(m_{l}/m_{W,Z})^{2}$ where $m_{l}$ is the mass of the relevant lepton is negligible. A potential class of processes where substantial deviation from the SM may occur are the interactions involving $W^{\pm}$ and $Z$ bosons only where $p^{\mu}p^{\nu}/m_{W,Z}^{2}$ is now of order unity even in the low energy regime.

Our discussion of the massless vector field is a review on the relationship between local Abelian gauge symmetry and Lorentz symmetry. But upon closer inspection, there is potentially new physics. The massless field $a^{\mu}(x)$ and $f^{\mu\nu}(x)$ only contain $\pm1$ helicity degrees of freedom. But since the spectrum of $\mathscr{J}^{3}$ are $\pm1,0$ and $\widetilde{0}$. A possible new massless field is then
\begin{equation}
b^{\mu}(x)=a^{\mu}(x)+\partial^{\mu}\phi(x).
\end{equation}
Similarly, like $a^{\mu}(x)$, $b^{\mu}(x)$ does not transform as a four-vector, so its kinematics is also described by an anti-symmetric second-rank tensor. Since partial derivative commutes, we still get
\begin{equation}
h^{\mu\nu}(x)=\partial^{\mu}b^{\nu}(x)-\partial^{\nu}b^{\mu}(x)=f^{\mu\nu}(x).
\end{equation}
Nevertheless, the interactions for $b^{\mu}(x)$ with an extra scalar degree of freedom may now differ from $a^{\mu}(x)$. This construct can be generalised to higher spin tensor fields following the prescription given in~\cite{Weinberg:1965rz} and potentially modify their interactions.


\chapter{Elko dark matter} 
\label{Chapter4}
\lhead{Chapter 4. \emph{Elko dark matter}} 

%

Quantum field theory is successful in describing the known elementary particles of the SM. Given its success, it is natural to try and extend the existing structure to explore the particle nature of dark matter. This chapter is devoted to study the quantum field theoretic aspect of a spin-half dark matter particle called Elko proposed by Ahluwalia and Grumiller~\cite{Ahluwalia:2004ab,Ahluwalia:2004sz}. Elko is a German acronym for \textit{Eigenspinoren des Ladungskonjugationsoperators}. In English, it means eigenspinor of the charge conjugation operator. The precise definition of Elko is given in sec.~\ref{elko_spinors_sec}.

The origin of Elko was an unexpected theoretical discovery, while trying to understand the work of Majorana~\cite{Majorana:1937vz} and its relation to the Majorana spinors. The authors
had no intention of proposing another dark matter candidate. However, it was discovered that Elko has mass dimension one instead of three-half so it cannot enter the SM doublets. Moreover, it has a renormalisable self-interaction, a desirable property for dark matter~\cite{Spergel:1999mh,Wandelt:2000ad}. These properties make Elko a dark matter candidate. In sec.~\ref{Elko_DM_candidate}, we provide further evidence to support Elko as a dark matter candidate.

Elko has attracted interests from various fields of research. In cosmology, it was shown by various authors that Elko has the properties to generate inflation~\cite{Boehmer:2010ma,Boehmer:2010tv,Boehmer:2009aw,Boehmer:2008ah,Boehmer:2008rz,Boehmer:2007ut,
Boehmer:2007dh,Boehmer:2006qq,Chee:2010ju,Wei:2010ad,Shankaranarayanan:2010st,Shankaranarayanan:2009sz,Gredat:2008qf}. The mathematical properties of Elko spinors have been studied in detail by da Rocha et al.~\cite{HoffdaSilva:2009is,
daRocha:2008we,daRocha:2007pz,daRocha:2005ti,daRocha:2011yr,Bernardini:2012sc}. In quantum field theory, Fabbri has shown that Elko does not violate causality ~\cite{Fabbri:2010va,Fabbri:2009aj,Fabbri:2009ka}. Wunderle and Dick have used Elko to construct
supersymmetric Lagrangians for fermionic fields of mass dimension one~\cite{Wunderle:2010yw}.

The main results of sec.~\ref{Elko_QF} are taken from~\cite{Ahluwalia:2008xi,Ahluwalia:2009rh} where the locality structure of Elko was analysed and the field equation and propagator were derived. While the propagator, field equation and mass-dimension of Elko remain unchanged from the original results~\cite{Ahluwalia:2004ab,Ahluwalia:2004sz}, the Elko fields given in~\cite{Ahluwalia:2008xi,Ahluwalia:2009rh} have improved locality structure. However, they are local only when the momentum of the particle is aligned to a preferred axis which we called the axis of locality. The axis of locality for the Elko fields originates from the Elko spin-sums which intrinsically contain a preferred direction and are not Lorentz-covariant. Therefore, it was concluded that Elko violates Lorentz symmetry. Following the Weinberg formalism presented in sec.~\ref{chap:quantum_fields}, we show that the generalisation of Elko fields to arbitrary spin violate Lorentz symmetry. Wunderle and Dick have also shown that Elko violates Lorentz symmetry within a specific case of supersymmetry breaking~\cite{Wunderle:2012mj}.

Although Elko violates Lorentz symmetry, in our opinion, this fact along is insufficient to rule out such a construct. In fact, since it is a dark matter candidate, it cannot be easily dismissed on the grounds of Lorentz violation since there are no direct experimental evidence showing dark matter must satisfy Lorentz symmetry. Furthermore, if the history behind parity violation in the weak interaction has taught us anything, it is one of not taking seemingly obvious but untested principles for granted~\cite{Yang:2012zze}.\footnote{I learnt this analogy from D. V. Ahluwalia.}

Progress has been made by Ahluwalia and Horvath to resolve the problem of Lorentz violation~\cite{Ahluwalia:2010zn}. Their results suggest Elko satisfies the symmetry of very special relativity (VSR) proposed by Cohen and Glashow where the VSR groups are subgroups of the Poincar\'{e} group~\cite{Cohen:2006ky}. They showed that the Elko spin-sums are covariant under $ISIM(2)$ transformations of VSR. Additionally the VSR algebra has an intrinsic preferred direction which coincides with the axis of locality. Also, under VSR transformations, the speed of light is the maximum attainable velocity and is invariant in all inertial reference frames~\cite{Horvath}.

After reviewing the VSR transformations in the momentum space, we derive the particle states and their quantum fields from the unitary irreducible representations of VSR.
We show that apart from gravity, the SM and VSR particles can only interact through massive scalar particles thus making the VSR particles dark matter candidates.

In sec.~\ref{Elko_QF}, we review the results obtained in~\cite{Ahluwalia:2008xi,Ahluwalia:2009rh}. Section \ref{Elko_LV} shows that the Elko fields and their generalisation to arbitrary spin violate Lorentz symmetry. In sec.~\ref{Elko_VSR}, we study the symmetries of VSR and derive the corresponding particle states and quantum fields. Section \ref{Elko_DM_candidate} shows that the VSR particles are dark matter candidates since they can only interact with the SM particles through gravity and massive scalar particles. Therefore, if Elko satisfies the VSR symmetry, they automatically qualify as a dark matter candidate. Finally, we study the Elko scattering amplitudes and discuss the importance of the adjoint.

\section{The Elko quantum field}\label{Elko_QF}
Unlike the previous chapters where the expansion coefficients for quantum fields are derived from  Poincar\'{e} symmetry, we start by defining the Elko spinors and construct the quantum field. The reason behind the difference in our approach is due to the fact that the Lorentz group is not the symmetry group for Elko. In sec.~\ref{Elko_LV}, using the Weinberg formalism, we show that the Elko fields and their generalisation to arbitrary spin violate Lorentz symmetry. 

We begin this section by defining the Elko spinors and study their properties. In particular, since Elko originated from the study of Majorana spinors, we discuss the similarities and differences between the two.

Presently we do not have a rigorous way to derive them. In spite of this, by choosing the Elko spinors carefully with the appropriate phases, we show that the resulting fields and their adjoints are local along the axis-of-locality and have positive-definite free Hamiltonian. Although Elko violates Lorentz symmetry, the existing properties indicate the field describes a well-defined particle state with an underlying structure that remains to be fully uncovered. In sec.~\ref{Elko_VSR} we study the connexion between Elko and VSR and discuss its implications.


\subsection{A detour to the Dirac field}
Before we define the Elko spinors, we take a brief detour to present the Dirac field and its spinors from a slightly different perspective as a motivation for Elko. The charged Dirac field constructed in sec.~\ref{Dirac_field} is
\begin{equation}
\Psi(x)=(2\pi)^{-3/2}\int\frac{d^{3}p}{\sqrt{2p^{0}}}\sum_{\sigma}
\left[e^{-ip\cdot x}u(\p,\sigma)a(\p,\sigma)+e^{ip\cdot x}v(\p,\sigma)b^{\dag}(\p,\sigma)\right].
\end{equation}
Symbolically, the expansion coefficients can be written as
\begin{equation}
u(\0,\textstyle{\frac{1}{2}})=\left(\begin{matrix}
\uparrow \\
\uparrow \end{matrix}\right),\hspace{0.5cm}
u(\0,-\textstyle{\frac{1}{2}})=\left(\begin{matrix}
\downarrow \\
\downarrow \end{matrix}\right)\label{eq:symbolic_u_spinor}
\end{equation}
\begin{equation}
v(\0,\textstyle{\frac{1}{2}})=\left(\begin{matrix}
\downarrow \\
-\downarrow \end{matrix}\right),\hspace{0.5cm}
v(\0,-\textstyle{\frac{1}{2}})=\left(\begin{matrix}
-\uparrow \\
\uparrow \end{matrix}\right)\label{eq:symbolic_v_spinor}
\end{equation}
where
\begin{equation}
\uparrow=\sqrt{m}\left(\begin{matrix}
1 \\
0 \end{matrix}\right),\hspace{0.5cm}
\downarrow=\sqrt{m}\left(\begin{matrix}
0 \\
1 \end{matrix}\right)
\end{equation}
in the polarisation basis. The arrows $\uparrow$ and $\downarrow$ symbolically represent the positive and negative eigenvalues of the spinors with respect to $J^{3}=\sigma^{3}/2$. 

As we have shown in sec.~\ref{Dirac_field}, the numerical values of the coefficients and the relative phases between the spinors are derived without reference to the Dirac equation. The notation used here emphasises the importance of the relative phases and the eigenvalues of the spinors as they play an important role when we define the Elko spinors. For the Dirac field, had the phases (minus signs) been absent for the $v(\p,\sigma)$ coefficients, the Majorana field defined by eq.~(\ref{eq:Majorana_fermion}) would be non-local and violate Lorentz symmetry.\footnote{In our research group, this was first discovered by T. F. Watson in 2007}

Although the transformations of the Dirac spinors are derived from symmetry consideration, they can also be obtained by studying the finite-dimensional representations of the Lorentz group, independent of the quantum fields. The spin-half representation of the Lorentz group is given by
\begin{equation}
\mJ=\left(\begin{matrix}
\J & \mathbf{O} \\
\mathbf{O} & \J \end{matrix}\right),\hspace{0.5cm}
\mK=\left(\begin{matrix}
-i\J & \mathbf{O} \\
\mathbf{O} & i\J\end{matrix}\right)
\end{equation}
where $\J=\s/2$. Now let $\psi(\0)$ be an arbitrary four-component spinor at rest
\begin{equation}
\psi(\0)=\left(\begin{matrix}
\psi_{R}(\0)\\
\psi_{L}(\0) \end{matrix}\right)
\end{equation}
where $\psi_{R}(\0)$ and $\psi_{L}(\0)$ represent the right and left-handed Weyl spinors respectively. The spinors of arbitrary momentum are obtained by the following boost transformation
\begin{equation}
\psi(\p)=\mathcal{D}(L(p))\psi(\0)
\end{equation}
where
\begin{equation}
\mathcal{D}(L(p))=\left(\begin{matrix}
\mathcal{D}_{R}(L(p)) & O \\
O & \mathcal{D}_{L}(L(p)) \end{matrix}\right)=
\left(\begin{matrix}
\exp(\J\cdot\bv) & \mathbf{O} \\
\mathbf{O} & \exp(-\J\cdot\bv)\end{matrix}\right)
\end{equation}
The definitions and explicit forms of the boost are given by eqs.~(\ref{eq:spin_half_boost}-\ref{eq:left_right_boost}). 

Substituting the Dirac spinors for $\psi(\p)$, the Dirac equation can be derived by computing the spin-sums for the Dirac spinors and exploiting the orthonormality relation. This is precisely what we have done in sec.~\ref{Dirac_field}. However, this derivation is not entirely satisfactory. A rigorous derivation requires us to compute the Dirac propagator given by the vacuum expectation value $\langle\,\,|T[\Psi(x)\overline{\Psi}(y)]|\,\,\rangle$ and find the relevant operator for which the Dirac propagator is proportional to Green's function.
Using eq.~(\ref{eq:g_prop}) and the spin-sums, we get
\begin{equation}
\langle\,\,|T[\Psi(x)\overline{\Psi}(y)]|\,\,\rangle=i\int\frac{d^{4}q}{(2\pi)^{4}}e^{-iq\cdot(x-y)}\left[\frac{\gamma^{\mu}q_{\mu}+m I}{q^{\nu}q_{\nu}-m^{2}+i\epsilon}\right].\label{eq:Dirac_propagator}
\end{equation}
Applying the operator $(i\gamma^{\mu}\partial/\partial x^{\mu}-m I)$ from the left on eq.~(\ref{eq:Dirac_propagator}) yields
\begin{equation}
\left(i\gamma^{\mu}\frac{\partial}{\partial x^{\mu}}-m I\right)\langle\,\,|T[\Psi(x)\overline{\Psi}(y)]|\,\,\rangle=iI\delta^{4}(x-y).
\end{equation}
We see that for the Dirac field, the field equation can either be derived from the properties of the spinors and the propagator. But as we will demonstrate in sec.~\ref{elko_spinors_sec}, the field equation derived from the Elko spinors is not entirely satisfactory. A proper derivation of the Elko field equation requires us to compute its propagator.


\subsection{A critique on Majorana spinors}
We begin this section by defining the Majorana spinors. The Majorana spinors have the same Lorentz transformations as the Dirac spinors but with an extra condition. Taking $\phi_{L}(\0)$ to be a left-handed Weyl spinor at rest. Its boost is given by
\begin{equation}
\phi_{L}(\p)=\exp\left(-\frac{\s}{2}\cdot\bv\right)\phi_{L}(\0)
\end{equation}
it is straightforward to show that $\vartheta\Theta\phi_{L}^{*}$ where $\vartheta$ is an arbitrary phase and $\Theta=-i\sigma^{2}$ transforms as a right-handed spinor
\begin{equation}
\vartheta\Theta\phi_{L}^{*}(\p)=\exp\left(\frac{\s}{2}\cdot\bv\right)[\vartheta\Theta\phi_{L}^{*}(\0)].
\end{equation}
The Majorana spinor is defined as ($\vartheta=i$)~\cite[eq.~(1.4.52)]{Ramond:1981pw}\footnote{In~\cite{Ramond:1981pw}, Ramond adopts the convention
\begin{equation}
\Psi_{M}=\left(\begin{matrix}
\Psi_{L} \\
\Psi_{R}\end{matrix}\right)
=\left(\begin{matrix}
\Psi_{L} \\
i\Theta\Psi^{*}_{L}\end{matrix}\right)\nonumber
\end{equation} 
where $\Psi_{L}$ and $\Psi_{R}$ are the left and right-handed Weyl spinor.}
\begin{equation}
\psi_{M}(\p)=\left(\begin{matrix}
i\Theta\phi^{*}_{L}(\p) \\
\phi_{L}(\p) \end{matrix}\right).\label{eq:Majorana_spinors}
\end{equation}
The matrix $\Theta$ is generally known as the Wigner time-reversal operator. For arbitrary representation, it satisfies the general identity
\begin{equation}
\Theta\J\Theta^{-1}=-\J^{*}. \label{eq:Wigner_time_reversal}
\end{equation}

From the above definition, it is evident that the Majorana spinor is different from the Majorana field. The Majorana field defined in eq.~(\ref{eq:Majorana_fermion}) is just a spin-half quantum field with Dirac spinors as expansion coefficients and the additional conditions that particles are identical with the anti-particles. Here, the Majorana spinor defined in the momentum space is not a quantum field operator. At best, they may be taken as the expansion coefficients for a new quantum field which is what Ahluwalia and Grumiller have done in~\cite{Ahluwalia:2004ab,Ahluwalia:2004sz} and gave birth to Elko.

Historically, the Majorana field was first introduced by Majorana in 1937~\cite{Majorana:1937vz} but the origin of Majorana spinors is unclear, the earliest references we have found are papers by McLennan~\cite{McLennan:1957} and Case~\cite{Case:1957}.

It is interesting to note that in the literature, while the Dirac spinors are treated as commuting numbers, the Majorana spinors are taken to be Grassmann variables (variables that anti-commutes among themselves)~\cite{Ramond:1981pw,Weinberg:2000cr,MullerKirsten:1986cw}. We find this to be unsatisfactory. While there are areas such as the path-integral formulation of quantum field theory and supersymmetry where the Grassmann variables are an indispensable tool, for ordinary quantum fields in the operator formalism, there is no need to introduce Grassmann variables.\footnote{For a nice introduction on the use of Grassmann variables in path-integral, see~\cite[chap.~II.5]{Zee:2010}. The application of Grassmann variables in supersymmetry can be found in~\cite{Weinberg:2000cr,MullerKirsten:1986cw}.}  The fermionic statistics is intrinsically encoded in the anti-commutation relations of the creation and annihilation operators.

%

A reasonable task to undertake is therefore to construct quantum fields with the Majorana spinors as expansion coefficients. But before we can proceed, we note that there are at most only two Majorana spinors. For massive particles, this is inconsistent with Lorentz symmetry. A massive spin-half field, by Lorentz symmetry, must have four degrees of freedom equally shared between particles and anti-particles distinguished by the spin-projection~\cite{Ahluwalia:1994uy}. Constructing a field theory with only Majorana spinors would be akin to projecting out the anti-particle spinors of the Dirac field. In the next section, we show that the use of Elko spinors solves this problem by including the Majorana spinors with two additional spinors. Together, it gives us four spinors, allowing us to construct the Elko quantum field with Elko spinors as expansion coefficients. 

\subsection{Elko spinors}\label{elko_spinors_sec}
Currently the best way to construct the Elko spinors is by analogy with respect to the Dirac spinors. In the $(\frac{1}{2},0)\oplus(0,\frac{1}{2})$ representation, parity operation on the Dirac spinors is implemented by the $\gamma^{0}$ matrix
\begin{equation}
\gamma^{0}u(\p,\sigma)=u(-\p,\sigma),\hspace{0.5cm}
\gamma^{0}v(\p,\sigma)=-v(-\p,\sigma).
\end{equation} 
We define the following matrix operator
\begin{equation}
\mathcal{S}(\mathscr{P})=\gamma^{0}\mathcal{R}
\end{equation}
where for $\p=p(\sin\theta\cos\phi,\sin\theta\sin\phi,\cos\theta)$ 
\begin{equation}
\mathcal{R}:(\phi,\theta,p)\rightarrow(\phi\pm\pi,-\theta+\pi,p).
\end{equation}
The plus and minus sign apply when the momentum along the 2-axis is positive or negative respectively. Therefore, we get
\begin{equation}
\mathcal{S}(\mathscr{P})u(\p,\sigma)=u(\p,\sigma),\hspace{0.5cm}
\mathcal{S}(\mathscr{P})v(\p,\sigma)=-v(\p,\sigma).
\end{equation}
with $u(\p,\sigma)$ and $v(\p,\sigma)$ having eigenvalues $+1$ and $-1$ respectively. 

For charge-conjugation, we have the following identities
\begin{equation}
u(\p,\sigma)=i\gamma^{2}v^{*}(\p,\sigma),\hspace{0.5cm}
v(\p,\sigma)=i\gamma^{2}u^{*}(\p,\sigma).
\end{equation}
The operator
\begin{equation}
\mathcal{C}=-\gamma^{2}K=
\left(\begin{matrix}
O & i\Theta \\
-i\Theta & O \end{matrix}\right)K
\end{equation}
where $K$ is the complex-conjugation operator maps the $u(\p,\sigma)$ and $v(\p,\sigma)$ spinors to each other
\begin{equation}
\mathcal{C}u(\p,\sigma)=iv(\p,\sigma),\hspace{0.5cm}
\mathcal{C}v(\p,\sigma)=iu(\p,\sigma).
\end{equation}

The above results show that the Dirac spinors are eignespinors of the parity operator $\mathcal{S}(\mathscr{P})$ with eigenvalues +1 and $-1$. In analogy to this observation, the Elko spinors are defined to be eigenspinors of $\mathcal{C}$ with eigenvalues +1 and $-1$. To construct the Elko spinors, let $\phi_{L}(\p,\sigma)$ be a left-handed Weyl spinor so that under Lorentz boost, it transforms as
\begin{equation}
\phi_{L}(\p,\sigma)=\exp\left(-\frac{\s}{2}\cdot\bv\right)\phi_{L}(\e)
\end{equation}
where $\e$ is defined as $\p\vert_{p\rightarrow0}$ instead of $\p\vert_{p=0}$. Here $\phi_{L}(\p,\sigma)$ is taken to be eigenspinors of the helicity operator with eigenvalues $\sigma=\pm\frac{1}{2}$ 
\begin{equation}
\frac{1}{2}\s\cdot\hat{\p}\,\phi_{L}(\p,\sigma)=\sigma\phi_{L}(\p,\sigma).
\end{equation}
In the previous section, it is shown that $\vartheta\Theta\phi_{L}^{*}(\p,\sigma)$ where $\vartheta$ is a phase to be determined, transforms as a right-handed Weyl spinor. Explicit computation shows that $\vartheta\Theta\phi_{L}^{*}(\p,\sigma)$ has opposite helicity with respect to $\phi_{L}(\p,\sigma)$~\cite[sec.~3.1]{Ahluwalia:2004sz}
\begin{equation}
\frac{1}{2}\s\cdot\hat{\p}[\vartheta\Theta\phi_{L}^{*}(\p,\sigma)]=-\sigma[\vartheta\Theta\phi_{L}^{*}(\p,\sigma)].
\end{equation}
We define a four-component spinor $\chi(\p,\alpha)$ as
\begin{equation}
\chi(\p,\alpha)=
\left(\begin{matrix}
\vartheta\Theta\phi_{L}^{*}(\p,\sigma)\\
\phi_{L}(\p,\sigma)\end{matrix}\right)
\end{equation}
where $\alpha=\mp\sigma$ represents the dual-helicity nature of the spinor with top and bottom sign denoting the helicity of the right and left-handed Weyl spinors respectively. The spinor $\chi(\p,\alpha)$ becomes the eigenspinors of the charge-conjugation operator $\mathcal{C}$ (and hence Elko) with the following choice of phases
\begin{equation}
\mathcal{C}\chi(\p,\alpha)\vert_{\vartheta=\pm i}=\pm\chi(\p,\alpha)\vert_{\vartheta=\pm i}
\end{equation}
thus giving us four Elko spinors. The Majorana spinors defined in eq.~(\ref{eq:Majorana_spinors}) now become a subset of Elko spinors with eigenvalue $+1$ with respect to $\mathcal{C}$.

In the helicity basis, we take the left-handed Weyl spinors at rest to be
\begin{equation}
\phi_{L}(\e,\textstyle{\frac{1}{2}})=\sqrt{m}
\left(\begin{matrix}
\cos(\theta/2)e^{-i\phi/2}\\
\sin(\theta/2)e^{i\phi/2} \end{matrix}\right),
\end{equation}
\begin{equation}
\phi_{L}(\e,-\textstyle{\frac{1}{2}})=\sqrt{m}
\left(\begin{matrix}
-\sin(\theta/2)e^{-i\phi/2} \\
\cos(\theta/2)e^{i\phi/2}\end{matrix}\right).
\end{equation}
The Elko spinors at rest, are divided into the self-conjugate spinors ($\vartheta=i$) and the anti-self-conjugate spinors ($\vartheta=-i$)
\begin{equation}
\xi(\p,\mp\textstyle{\frac{1}{2}})(\e)=+\chi(\e,\mp\textstyle{\frac{1}{2}})\vert_{\vartheta=+i},
\end{equation}
\begin{equation}
\xi(\p,\pm\textstyle{\frac{1}{2}})(\e)=+\chi(\e,\pm\textstyle{\frac{1}{2}})\vert_{\vartheta=+i},
\end{equation}
\begin{equation}
\zeta(\p,\mp\textstyle{\frac{1}{2}})(\e)=+\chi(\e,\pm\textstyle{\frac{1}{2}})\vert_{\vartheta=-i},
\end{equation}
\begin{equation}
\zeta(\p,\pm\textstyle{\frac{1}{2}})(\e)=-\chi(\e,\mp\textstyle{\frac{1}{2}})\vert_{\vartheta=-i}.
\end{equation}
Section \ref{elko_locality_structure} shows that the above definitions of the four Elko spinors with the specific phases are necessary for the Elko fields to be local.

Comparing to the Dirac spinors of eqs.~(\ref{eq:symbolic_u_spinor}) and (\ref{eq:symbolic_v_spinor}), in the polarisation basis, the Elko spinors takes symbolic form~\cite{Ahluwalia:2009rh}
\begin{equation}
\xi(\e,\mp\textstyle{\frac{1}{2}})=\left(\begin{matrix}
i \Downarrow\\
\Uparrow \end{matrix}\right),\hspace{0.5cm}
\xi(\e,\pm\textstyle{\frac{1}{2}})=\left(\begin{matrix}
-i\Uparrow \\
\Downarrow \end{matrix}\right)\label{eq:elko_polarisation1}
\end{equation}
\begin{equation}
\zeta(\e,\mp\textstyle{\frac{1}{2}})=\left(\begin{matrix}
i\Uparrow\\
\Downarrow\end{matrix}\right),\hspace{0.5cm}
\zeta(\e,\pm\textstyle{\frac{1}{2}})=-\left(\begin{matrix}
-i\Downarrow\\
\Uparrow
\end{matrix}\right)\label{eq:elko_polarisation2}
\end{equation}
where
\begin{equation}
\Uparrow=\sqrt{m}\left(\begin{matrix}
e^{-i\phi/2} \\
0 \end{matrix}\right),\hspace{0.5cm}
\Downarrow=\sqrt{m}\left(\begin{matrix}
0 \\
e^{i\phi/2} \end{matrix}\right).
\end{equation}
The arrows $\Uparrow$ and $\Downarrow$ differ from $\uparrow$ and $\downarrow$ by the phase $e^{\pm i\phi/2}$. In the polarisation basis, it is essential to keep the phases to preserve the locality structure of the quantum field. 

The angle $\phi$ in the definitions of the Elko rest spinors intrinsically contain a preferred plane as required by the demand of locality. The existence of a preferred plane inevitably leads to Lorentz violation. Our task is to investigate the effects of Lorentz violation.

\subsubsection{The dual coefficients}

The explicit computation of the Lorentz-invariant norms of the Elko spinors under the Dirac dual identically vanish
\begin{equation}
\bar{\xi}(\p,\alpha)\xi(\p,\alpha)=0,\hspace{0.5cm}
\bar{\zeta}(\p,\alpha)\zeta(\p,\alpha)=0. \label{eq:Elko_vanishing_norm}
\end{equation}
The above identities suggest it may not be possible to construct non-vanishing Lorentz-invariant scalars for Elko spinors. However, it was found in~\cite{Ahluwalia:1994uy} that Elko spinors are bi-orthonormal under the Dirac dual
\begin{equation}
\bar{\xi}(\p,\mp\textstyle{\frac{1}{2}})\xi(\p,\pm\textstyle{\frac{1}{2}})=-2im,\hspace{0.5cm}
\bar{\xi}(\p,\pm\textstyle{\frac{1}{2}})\xi(\p,\mp\textstyle{\frac{1}{2}})=2im,
\end{equation}
\begin{equation}
\bar{\zeta}(\p,\mp\textstyle{\frac{1}{2}})\zeta(\p,\pm\textstyle{\frac{1}{2}})=-2im,\hspace{0.5cm}
\bar{\zeta}(\p,\pm\textstyle{\frac{1}{2}})\zeta(\p,\mp\textstyle{\frac{1}{2}})=2im
\end{equation}
with all other combinations being identically zero. Demanding the Elko spinors to have real orthonormal norms, we define the Elko dual coefficients as
\begin{equation}
\gdual{\xi}(\p,\mp\textstyle{\frac{1}{2}})=-i\xi^{\dag}(\p,\pm\textstyle{\frac{1}{2}})\gamma^{0},\hspace{0.5cm}
\gdual{\xi}(\p,\pm\textstyle{\frac{1}{2}})=i\xi^{\dag}(\p,\mp\textstyle{\frac{1}{2}})\gamma^{0},\label{eq:xi_dual}
\end{equation}
\begin{equation}
\gdual{\zeta}(\p,\mp\textstyle{\frac{1}{2}})=-i\zeta^{\dag}(\p,\pm\textstyle{\frac{1}{2}})\gamma^{0},\hspace{0.5cm}
\gdual{\zeta}(\p,\pm\textstyle{\frac{1}{2}})=i\zeta^{\dag}(\p,\mp\textstyle{\frac{1}{2}})\gamma^{0}.\label{eq:zeta_dual}
\end{equation}
Under the new dual, we obtain the following orthonormality relation
\begin{equation}
\gdual{\xi}(\p,\alpha)\xi(\p,\alpha')=2m\delta_{\alpha\alpha'}
\end{equation}
\begin{equation}
\gdual{\zeta}(\p,\alpha)\zeta(\p,\alpha')=-2m\delta_{\alpha\alpha'}.
\end{equation}

The Elko dual also gives us the completeness relation
\begin{equation}
\frac{1}{2m}\sum_{\alpha}\left[\xi(\p,\alpha)\gdual{\xi}(\p,\alpha)-\zeta(\p,\alpha)\gdual{\zeta}(\p,\alpha)\right]=I.
\end{equation}
This establishes the fact that both self-conjugate and anti-self-conjugate spinors are needed to include all the degrees of freedom of Elko. 

\subsubsection{Elko spinor equation}
We now derive the field equation for Elko spinors in the momentum space. A more rigorous derivation requires us to compute the propagator which is carried out in sec.~\ref{Elko_kinematics} after constructing the Elko fields. 

In the spin-half representation we have chosen, the $\gamma$-matrices are given in the chiral-representation
\begin{equation}
\gamma^{0}=\left(\begin{matrix}
O & I \\
I & O \end{matrix}\right),\hspace{0.5cm}
\gamma^{i}=\left(\begin{matrix}
O &-\sigma^{i} \\
\sigma^{i} & O \end{matrix}\right)
\end{equation}
where the Dirac spinors are eigenspinors of the Dirac operator $\gamma^{\mu}p_{\mu}$ with eigenvalues $\pm m$. Acting the operator $\gamma^{\mu}p_{\mu}$ on the Elko spinors yields the following result
\begin{equation}
\gamma^{\mu}p_{\mu}\xi(\p,\mp\textstyle{\frac{1}{2}})=im\xi(\p,\pm\textstyle{\frac{1}{2}}),
\end{equation}
\begin{equation}
\gamma^{\mu}p_{\mu}\xi(\p,\pm\textstyle{\frac{1}{2}})=-im\xi(\p,\mp\textstyle{\frac{1}{2}}),
\end{equation}
\begin{equation}
\gamma^{\mu}p_{\mu}\zeta(\p,\mp\textstyle{\frac{1}{2}})=-im\zeta(\p,\pm\textstyle{\frac{1}{2}}),
\end{equation}
\begin{equation}
\gamma^{\mu}p_{\mu}\zeta(\p,\pm\textstyle{\frac{1}{2}})=im\zeta(\p,\mp\textstyle{\frac{1}{2}}).
\end{equation}
Therefore, Elko spinors do not satisfy the Dirac equation. 

The Dirac operator maps the Elko spinors to each other. Applying $\gamma^{\nu}p_{\nu}$ from the left on $\gamma^{\mu}p_{\mu}\xi(\p,\alpha)$ and $\gamma^{\mu}p_{\mu}\zeta(\p,\alpha)$ then gives us
\begin{equation}
(p^{\mu}p_{\mu}-m^{2})\xi(\p,\alpha)=0,\label{eq:elko_spinor_equation1}
\end{equation}
\begin{equation}
(p^{\mu}p_{\mu}-m^{2})\zeta(\p,\alpha)=0.\label{eq:elko_spinor_equation2}
\end{equation}                                                                                                                                                                                                                                                                                                                                                                                                                                                                                                                                                                                                                                                                                                                                                                                                                                                                                                                                                                                                                                                                                                                                                                                                                                                                                                                                                                                                                                                                                                                                                                                                                                                             
Therefore, Elko spinors satisfy the Klein-Gordon but not the Dirac equation. However, given that the Klein-Gordon equation is a statement of the dispersion relation, it is trivially satisfied for all spinors. Therefore, to establish the proper field equation for Elko, we need to compute its propagator.

\subsubsection{Spin-sums and the preferred axis}
Direct evaluation of the Elko spin-sums yields
\begin{equation}
\sum_{\alpha}\xi(\p,\alpha)\gdual{\xi}(\p,\alpha)=m[\mathcal{G}(\phi)+I],\label{eq:elko_spinsum1}
\end{equation}
\begin{equation}
\sum_{\alpha}\zeta(\p,\alpha)\gdual{\zeta}(\p,\alpha)=m[\mathcal{G}(\phi)-I]\label{eq:elko_spinsum2}
\end{equation}
where $\mathcal{G}(\phi)$ is an off-diagonal matrix
\begin{equation}
\mathcal{G}(\phi)=i\left(\begin{matrix}
0 & 0 & 0 & -e^{-i\phi} \\
0 & 0 & e^{i\phi} & 0 \\
0 & -e^{-i\phi} & 0 & 0\\
e^{i\phi} & 0 & 0 & 0 \end{matrix}\right).
\end{equation}
For later reference we note that $\mathcal{G}(\phi)$ is an odd function of $\p$
\begin{equation}
\mathcal{G}(\phi)=-\mathcal{G}(\phi\pm\pi).
\end{equation}

Multiply eqs.~(\ref{eq:elko_spinsum1}) and (\ref{eq:elko_spinsum2}) from the left by $\xi(\p,\alpha')$ and $\zeta(\p,\alpha')$ respectively and using the orthonormality relation, we obtain the identity
\begin{equation}
\left[\mathcal{G}(\phi)-I\right]\xi(\p,\alpha)=0,
\end{equation}
\begin{equation}
\left[\mathcal{G}(\phi)+I\right]\zeta(\p,\alpha)=0.
\end{equation}
Although $\left[\mathcal{G}(\phi)\pm I\right]$ annihilates the Elko spinors, the equations contain no time dependence. Consequently, they cannot be considered as field equations for Elko.

Unlike the Dirac spin-sums, the Elko spin-sums contain a preferred plane characterised by $\mathcal{G}(\phi)$ which is independent of the angle $\theta$ and is not Lorentz-covariant.  In sec.~\ref{elko_locality_structure}, from the equal-time anti-commutator, we will see that the direction perpendicular to the plane defined by angle $\phi$ defines a preferred axis. The properties of the Elko spinors and the uniqueness of the Dirac field in the $(\frac{1}{2},0)\oplus(0,\frac{1}{2})$ representation suggests that the Elko fields violate Lorentz symmetry. In sec.~\ref{Elko_LV}, using the Weinberg formalism, we show that the Elko fields and their generalisation to higher spin violate Lorentz symmetry.  

We note, the Lorentz symmetry is violated in a specific manner. This led AH to show that Elko satisfies the $ISIM(2)$ symmetry of VSR~\cite{Ahluwalia:2010zn}. For now, we confine ourselves to construct the Elko field under the Lorentz group and study their properties before moving onto VSR. This will help us understand the physical significance of the preferred axis and why Elko satisfies the VSR symmetry.

\subsection{Elko fields}

We construct two Elko fields $\Lambda(x)$ and $\lambda(x)$. The former is defined as 
\begin{equation}
\Lambda(x)=(2\pi)^{-3/2}\int\frac{d^{3}p}{\sqrt{2mp^{0}}}\sum_{\alpha}\left[e^{-ip\cdot x}\xi(\p,\alpha)a(\p,\alpha)+e^{ip\cdot x}\zeta(\p,\alpha)b^{\ddag}(\p,\alpha)\right]
\end{equation}
where particles are distinguishable from the anti-particle and the later is 
\begin{equation}
\lambda(x)=\Lambda(x)\vert_{b^{\ddag}\rightarrow a^{\ddag}}
\end{equation}
with particles being identical to the anti-particles. The $\ddag$ symbol is used to allow for the possibility that the underlying state space maybe different from the Dirac particle. We assume the creation and annihilation operator satisfy the standard anti-commutation relations
\begin{equation}
\{a(\p,\alpha),a^{\ddag}(\p',\alpha')\}=\{b(\p,\alpha),b^{\ddag}(\p',\alpha')\}=
\delta^{3}(\p'-\p)\delta_{\alpha'\alpha}
\end{equation}
while all other anti-commutators identically vanish. The field adjoint $\gdual{\Lambda}(x)$ is defined as
\begin{equation}
\gdual{\Lambda}(x)=(2\pi)^{-3/2}\int\frac{d^{3}p}{\sqrt{2mp^{0}}}\sum_{\alpha}\left[e^{ip\cdot x}\gdual{\xi}(\p,\alpha)a^{\ddag}(\p,\alpha)+e^{-ip\cdot x}\gdual{\zeta}(\p,\alpha)b(\p,\alpha)\right]
\end{equation}
where we demand that $[b^{\ddag}(\p,\alpha)]^{\ddag}=b(\p,\alpha)$.

\subsubsection{The kinematics of Elko}\label{Elko_kinematics}
Equations (\ref{eq:elko_spinor_equation1}) and (\ref{eq:elko_spinor_equation2}) suggest that the Elko fields satisfy the Klein-Gordon equation. To verify this, we compute the propagator by substituting the Elko spin-sums into eqs.~(\ref{eq:g_prop}) and take into account the $1/\sqrt{m}$ normalisation factor for the fields. Using the fact that $\mathcal{G}(\phi)$ is an odd function of momentum and Elko has fermionic statistics, the propagator is
\begin{equation}
\langle\,\,|T[\Lambda(x)\gdual{\Lambda}(y)]|\,\,\rangle=i\int\frac{d^{4}q}{(2\pi)^{4}}e^{-iq\cdot(x-y)}\frac{I+\mathcal{G}(\phi)}{q^{\mu}q_{\mu}-m^{2}+i\epsilon}.\label{eq:elko_propagator}
\end{equation}
The Elko propagator is identical to the Klein-Gordon propagator up to a $\mathcal{G}(\phi)$ term. Therefore, Elko has mass-dimension one instead of three-half. 

Aligning $\x-\y$ to the 3-axis, a simple computation in the spherical coordinate then shows the integral over $\mathcal{G}(\phi)$  of eq.~(\ref{eq:elko_propagator}) identically vanishes.\footnote{de Oliveira and Rodrigues have shown explicitly that the integral over $\mathcal{G}(\phi)$ vanishes for arbitrary direction~\cite{deOliveira:2012wp}.} However, since Elko violates Lorentz symmetry and the spin-sums contain a preferred plane, it is possible that the space-time in which Elko resides has a non-trivial topology thus rendering the integral over $\mathcal{G}(\phi)$ non-vanishing unless $\x-\y$ is aligned along the 3-axis. If this is indeed the case, the propagator is not a Green's function of the Klein-Gordon operator $\partial^{\mu}\partial_{\mu}+m^{2}$ where $\partial_{\mu}=\partial/\partial x^{\mu}$. In particular, 
\begin{equation}
(\partial^{\mu}\partial_{\mu}+m^{2})\langle\,\,|T[\Lambda(x)\gdual{\Lambda}(y)]|\,\,\rangle=-iI\delta^{4}(x-y)
-i\int\frac{d^{4}q}{(2\pi)^{4}}e^{-iq\cdot(x-y)}\frac{\mathcal{G}(\phi)}{q^{\mu}q_{\mu}-m^{2}+i\epsilon}.
\end{equation}
Guided by the eqs.~(\ref{eq:elko_spinor_equation1}) and (\ref{eq:elko_spinor_equation2}) and the propagator, the simplest possible local Lagrangian density with renormalisable self-interaction for $\Lambda(x)$ and $\lambda(x)$ are
\begin{equation}
\mathscr{L}_{\Lambda}=\partial^{\mu}\gdual{\Lambda}\partial_{\mu}\Lambda-m^{2}\gdual{\Lambda}\Lambda-\frac{g_{0}}{4}(\gdual{\Lambda}\Lambda)^{2}
\end{equation}
\begin{equation}
\mathscr{L}_{\lambda}=\mathscr{L}_{\Lambda}\vert_{\Lambda\rightarrow\lambda}.
\end{equation}
where $g_{0}$ is the coupling constant.

By taking the Lagrangian densities to be Klein-Gordon, we see that whenever the Elko propagator is involved in the interaction, the non-covariant term would make an additional non-local contribution to the interacting Hamiltonian. It is non-local in the sense that it depends on both $x^{0}$ and $y^{0}$. Apart from the non-local interaction, it is shown that the free normal-ordered Hamiltonian obtained from the Klein-Gordon Lagrangian is positive-definite~\cite[sec.~7]{Ahluwalia:2004ab}. As a result, we can be confident that the kinematics of Elko is described by the Klein-Gordon Lagrangian.


\subsection{Locality structure of Elko}\label{elko_locality_structure}
Similar to the preceding chapter on vector fields, we divide the locality structure analysis into two sections for $\Lambda(x)$ and $\lambda(x)$. 

\subsubsection{Charged Elko field: $\L\mathbf{(x)}$}
The conjugate momentum is
\begin{equation}
\Pi(x)=\frac{\partial\mathscr{L}_{\Lambda}}{\partial(\partial\Lambda/\partial t)}=\frac{\partial\gdual{\Lambda}}{\partial t}(x).
\end{equation}
Since particles and anti-particles are distinguishable, the anti-commutators involving the creation and annihilation operators trivially vanish thus giving us
\begin{equation}
\{\Lambda(\x,t),\Lambda(\y,t)\}=0,\hspace{0.5cm}
\{\Pi(\x,t),\Pi(\y,t)\}=0. \label{eq:elko_anticommutator1}
\end{equation}
As for the equal-time anti-commutator between field and conjugate momentum, we have
\begin{equation}
\{\Lambda(\x,t),\Pi(\y,t)\}=i\int\frac{d^{3}p}{(2\pi)^{3}}\frac{1}{2m}e^{i\mathbf{p\cdot(x-y)}}
\sum_{\alpha}\left[\xi(\p,\alpha)\gdual{\xi}(\p,\alpha)-\zeta_{\alpha}(-\p,\alpha)\gdual{\zeta}(-\p,\alpha)\right].
\end{equation}
Substituting the Elko spin-sums into the anti-commutator yields
\begin{equation}
\{\Lambda(\x,t),\Pi(\y,t)\}=i\delta^{3}(\x-\y)I+i\int\frac{d^{3}p}{(2\pi)^{3}}e^{i\mathbf{p\cdot(x-y)}}\mathcal{G}(\phi).\label{eq:elko_anticommutator2}
\end{equation}
Similar to the non-covariant term in the propagator, second term also vanish along the 3-axis yielding the standard result. Therefore, we refer to the 3-axis as the axis of locality. Here, it is important to recall that physics is independent of the choice of basis. For Elko,   the identification of the preferred axis to be the 3-axis is basis-dependent but its existence is an intrinsic property of Elko and is thus basis-independent.

The anti-commutator between field and field-adjoint is
\begin{equation}
\{\Lambda(\x,t),\gdual{\Lambda}(\y,t)\}=(2\pi)^{-3}\int\frac{d^{3}p}{2mp^{0}}\sum_{\alpha}
\left[e^{i\mathbf{p\cdot(x-y)}}\xi(\p,\alpha)\gdual{\xi}(\p,\alpha)+e^{-i\mathbf{p\cdot(x-y)}}\zeta(\p,\alpha)\gdual{\zeta}(\p,\alpha)\right].
\end{equation}
Substituting the spin-sums into the anti-commutator, we get
\begin{equation}
\{\Lambda(\x,t),\gdual{\Lambda}(\y,t)\}=0.
\end{equation}

\subsubsection{Neutral Elko field: $\l\mathbf{(x)}$}

The equal-time anti-commutators $\{\lambda(\x,t),\pi(\y,t)\}$ and $\{\lambda(\x,t),\gdual{\lambda}(\y,t)\}$ remain unchanged. However, the equal-time anti-commutators  $\{\lambda(\x,t),\lambda(\y,t)\}$ and $\{\pi(\x,t),\pi(\y,t)\}$ require us to compute non-trivial spin-sums. Firstly,
\begin{equation}
\{\lambda(\x,t),\lambda(\y,t)\}=\int\frac{d^{3}p}{2mp^{0}}e^{i\mathbf{p\cdot(x-y)}}
\sum_{\alpha}\left[\xi(\p,\alpha)\zeta^{T}(\p,\alpha)+\zeta(-\p,\alpha)\xi^{T}(-\p,\alpha)\right].
\end{equation}
The spin-sum in the bracket upon direct computation vanishes, thus giving us
\begin{equation}
\{\lambda(\x,t),\lambda(\y,t)\}=0.\label{eq:elko_anticommutator3}
\end{equation}
Secondly we have,
\begin{equation}
\{\pi(\x,t),\pi(\y,t)\}=\int\frac{d^{3}p}{2mp^{0}}e^{-i\mathbf{p\cdot(x-y)}}
\sum_{\alpha}\left[\gdual{\xi}^{T}(\p,\alpha)\gdual{\zeta}(\p,\alpha)
+\gdual{\zeta}^{T}(-\p,\alpha)\gdual{\xi}(-\p,\alpha)\right].
\end{equation}
The spin-sum also vanishes yielding
\begin{equation}
\{\pi(\x,t),\pi(\y,t)\}=0. \label{eq:elko_anticommutator4}
\end{equation}
Equations (\ref{eq:elko_anticommutator1},\ref{eq:elko_anticommutator2}) and (\ref{eq:elko_anticommutator3},\ref{eq:elko_anticommutator4}) show that $\Lambda(x)$ and $\lambda(x)$ are local along the axis of locality. These results can be obtained in both the helicity and polarisation basis. In the polarisation basis, it is important to keep the phase $e^{\pm i\phi/2}$ for the Elko spinors at rest to preserve the locality structure.

\begin{figure}
\begin{center}
\includegraphics[scale=0.5]{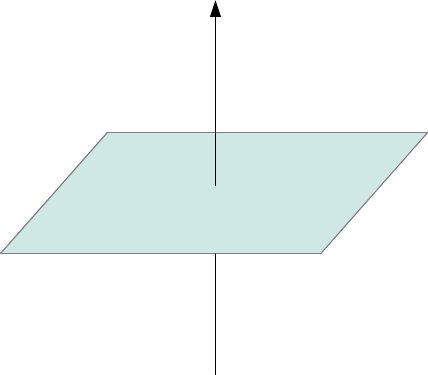}
\end{center}
\caption{The equivalence between the preferred plane and axis}\label{fig:preferred_plane_axis}
\end{figure}

The existence of a preferred axis is equivalent to the existence of a preferred plane. In this case, axis of locality conveys more information regarding the locality structure of the field. For scattering processes, it is more informative to discuss the direction of incoming and outgoing particles in terms of the preferred plane.
Figure \ref{fig:preferred_plane_axis} shows the two descriptions are equivalent, the preferred plane is defined by the angle $\phi$ for which the axis of locality is perpendicular to.

\subsection{Lorentz violation}\label{Elko_LV}

The fact that the Elko spinors transform correctly under the $(\frac{1}{2},0)\oplus(0,\frac{1}{2})$ representation of the Lorentz group does not imply the fields $\Lambda(x)$ and $\lambda(x)$ satisfy Lorentz symmetry. Indeed, from the non-covariance of the spin-sums and the propagator, it is sufficient to conclude that Elko is Lorentz violating. However, to compute the spin-sums and propagator, we have assumed the Elko spinors to take particular forms and phases. Although these choices guarantees locality and positivity of the free Hamiltonian, they are not derived from first principle. As a result, we cannot completely eliminate the possibility that the Lorentz violation is a result of incorrect choices of the Elko spinors. 

To prove Elko violates Lorentz symmetry without making any assumptions on the forms of the spinors, we turn to Weinberg formalism where the expansion coefficients of a quantum field are determined by the demand of Poincar\'{e} symmetry. If the Elko fields satisfy Lorentz symmetry, then we should be able to determine their expansions coefficients $\xi(\p,\alpha)$ and $\zeta(\p,\alpha)$. However, this is not possible. We now prove this for Elko and its generalisation to arbitrary spin.  

Let $\Psi(x)$ be a general massive spin-$j$ in the $(j,0)\oplus(0,j)$ representation
\begin{equation}
\Psi(x)=(2\pi)^{-3/2}\int\frac{d^{3}p}{\sqrt{2mp^{0}}}\sum_{\alpha}\big[e^{-ip\cdot x}u(\p,\alpha)a(\p,\alpha)
+e^{ip\cdot x}v(\p,\alpha)b^{\ddag}(\p,\alpha)\big].
\end{equation}
We take $\Psi(x)$ to be a generalisation of the spin-half Elko field by taking the expansion coefficients to have the form
\begin{equation}
u(\0,\alpha)=\left(\begin{array}{cc}
\vartheta\Theta\phi^{*}(\0,\sigma) \\
\phi(\0,\sigma)
\end{array}\right),\hspace{0.5cm}
v(\0,\alpha)=\left(\begin{array}{cc}
\wp\Theta\phi^{*}(\0,-\sigma) \\
\phi(\0,-\sigma)
\end{array}\right)
\label{eq:generalised_elko_spinors}
\end{equation}
where $\vartheta$ and $\wp$ are arbitrary phases and $\Theta$ is the Wigner time-reversal operator of dimension $(2j+1)\times(2j+1)$ satisfying eq.~(\ref{eq:Wigner_time_reversal}).

Taking $\J$ to be the rotation generator of the $(j,0)$ and $(0,j)$ representation spaces, rotation symmetry requires the coefficients at rest to satisfy the following equations
\begin{equation}
\sum_{\bar{\sigma}}u_{\ell}(\0,\bar{\sigma})\J_{\bar{\sigma}\sigma}=\sum_{\bar{\ell}}\mJ_{\ell\bar{\ell}}\,u_{\bar{\ell}}(\0,\sigma),
\end{equation}
\begin{equation}
\sum_{\bar{\sigma}}v_{\ell}(\0,\bar{\sigma})\J^{*}_{\bar{\sigma}\sigma}=-\sum_{\bar{\ell}}\mJ_{\ell\bar{\ell}}\,v_{\bar{\ell}}(\0,\sigma)
\end{equation}
where $\ell=1,\cdots,2\times(2j+1)$ denote the components of the coefficients. Here, without the loss of generality, we choose the rotation and boost generators to be
\begin{equation}
\mJ=\frac{1}{2}
\left(\begin{array}{cc}
\J & \mathbf{O}  \\
 \mathbf{O} & \J \end{array}\right),\hspace{0.5cm}
\mK=\frac{1}{2}
\left(\begin{array}{cc}
-i\J & \mathbf{O}  \\
 \mathbf{O} & i\J \end{array}\right)
\end{equation}
where $\J$ is the rotation generator of dimension $(2j+1)\times(2j+1)$.
Substituting $u(\0,\alpha)$ and $v(\0,\alpha)$ into eqs.~(\ref{eq:rot1}) and (\ref{eq:rot2}), we find the equalities do not hold. 
It follows that Elko and its higher spin generalisations violates Lorentz symmetry.

The Lorentz violations for Elko are expected, since it was noted that its spin-sums contain a preferred direction and the Elko spinors do not satisfy the Dirac equation in momentum space~\cite{Ahluwalia:2004ab,Ahluwalia:2004sz,Ahluwalia:2008xi,Ahluwalia:2009rh}. 
The above calculation generalises this result. It shows that the massive quantum fields of the $(j,0)\oplus(0,j)$ representation with expansion coefficients of the form given by eq.~(\ref{eq:generalised_elko_spinors}) also violate Lorentz symmetry.

\section{Elko and very special relativity}\label{Elko_VSR}


In 2010, the results obtained by AH suggested that Elko satisfies the $ISIM(2)$ symmetry of VSR~\cite{Ahluwalia:2010zn} proposed by Cohen and Glashow~\cite{Cohen:2006ky}.\footnote{The $ISIM(2)$ group is an extension of $SIM(2)$ including space-time translation generators.} The VSR groups are generated from the VSR algebras summarised in 
tab.4.1 where\footnote{The table
and caption are taken from~\cite{Ahluwalia:2010zn}
with permissions from AH. The generators $T^{1}$ and $T^{2}$ are identical to the generators $A$ and $B$ introduced in sec.~\ref{Massless_particle_state}. Here we choose to follow the notation adopted by AH.}
\begin{equation}
T^{1}=K^{1}+J^{2},\hspace{0.5cm} T^{2}=K^{2}-J^{1}.\label{eq:T1T2}
\end{equation}
The VSR groups are proper Lorentz subgroups. The VSR transformations obtained through its generators satisfy the postulates of special relativity, namely the existence of a maximal velocity invariant in all inertial frames~\cite{Horvath}.

Their claim is supported by the fact that
\begin{enumerate}
\item  The $SIM(2)$ group contains the spinor representation of Elko.
\item  The VSR algebras contain a preferred direction coinciding with the axis of locality.
\item  The Elko spin-sums are covariant under VSR transformations~\cite{Ahluwalia:2010zn,Horvath}.
\item  Previously obtained results such as the mass-dimensionality and locality structure of Elko remain unchanged.
\end{enumerate}

While these are strong evidence indicating that Elko satisfies VSR symmetry, at this stage, we are unable to provide an ab-initio derivation of the Elko spinors as expansion coefficients of a quantum field with VSR symmetry. 

\begin{table}[!hbt]
\centering
\begin{tabular}{lll}
\toprule
Designation & Generators & Algebra \\
\midrule
$\mathfrak{t}(2)$ & $T^1,T^2$ & $[T^1,T^2]=0$ \\
$\mathfrak{e}(2)$ & $T^1,T^2,J^3$ & $[T^1,T^2]=0,\  [T^1,J^3] = - i T^2,\ [T^2,J^3] = i T^1$ \\
$\mathfrak{hom}(2)$ & $T^1,T^2,K^3$ & $[T^1,T^2]= 0,\ [T^1,K^3] = i T^1,\  [T^2,K^3] = i T^2$ \\
$\mathfrak{sim}(2)$ & $T^1,T^2,J^3,K^3$ & $[T^1,T^2]= 0,\ [T^1,K^3] = i T^1,\  [T^2,K^3] = i T^2$ \\
& & $[T^1,J^3] = - i T^2,\  [T^2,J^3] = i T^1,\  [J^3,K^3] = 0$ \\ 
\bottomrule
\end{tabular}
\caption{The four VSR algebras.}
\label{table:1}
\end{table}

There are two properties of VSR that are important for us. Firstly, the inclusion of  any of the following 
discrete symmetries $\mathsf{P}$, $\mathsf{T}$, $\mathsf{CP}$ or $\mathsf{CT}$ 
will yield the full Lorentz group. This suggest that the quantum fields with VSR symmetry may violate
the mentioned discrete symmetries. Nevertheless, we still expect $\mathsf{CPT}$ to be conserved. Secondly, all the VSR algebras have a preferred axis. Here, we chose the axis to be the 3-axis coinciding with the axis of locality. While it is possible to choose alternative axis, this would require us to transform the existing Elko spinors to a different basis. The reason for choosing the 3-axis is that the resulting $SIM(2)$ transformations on the existing Elko spinors preserves all their existing properties.

In VSR, the main results for Elko remain identical to what we have presented in secs.~\ref{elko_spinors_sec}-\ref{Elko_LV} so we will not reproduce them here. Detailed calculations can be found in~\cite{Ahluwalia:2010zn,Horvath}. Motivated by the work of AH, after reviewing the $SIM(2)$ transformations in the momentum space, we derive the particle states and quantum fields with VSR symmetry.\footnote{Unless otherwise stated, from now on when we mention VSR, depending on the context, it refers to either $SIM(2)$ or $ISIM(2)$.} Later, we show that the SM and VSR particles can only interact via massive scalar particles and gravity, thus making them dark matter candidates. 

Our proposal of VSR as the symmetry group for dark matter deviates from the original goals of reproducing the SM predictions and  explaining the phenomena of $\mathsf{CP}$ violation as originally envisaged by Cohen and Glashow~\cite{Cohen:2006ky}. In sec.~\ref{VSR_particle_state}, by studying the representations of the $SIM(2)$, we show that the corresponding particle states are physically distinct from their SM counterpart. Since no experiments thus far have shown that dark matter must satisfy Poincar\'{e} symmetry, our proposal is not ruled out.

\subsection{The VSR transformations in momentum space}\label{Sec:VSR-Review}
 
The VSR transformations on momentum vectors are generated by $\mathfrak{sim}(2)$.
Here, the finite-dimensional $\mathfrak{sim}(2)$ generators are obtained using eq.~(\ref{eq:T1T2}) and the generators of the vector representation of the Lorentz group given by eqs.~(\ref{eq:vector_j}) and (\ref{eq:vector_k})
\begin{equation}
\mathcal{J}^{3}=\mathscr{J}^{3}
 ,\hspace{0.5cm}
\mathcal{K}^{3}=\mathscr{K}^{3},
\end{equation}
\begin{equation}
\mathcal{T}^{1}=\mathscr{K}^{1}+\mathscr{J}^{2}
,\hspace{0.5cm}
\mathcal{T}^{2}=\mathscr{K}^{2}-\mathscr{J}^{1}.
 \end{equation}
The group elements of $SIM(2)$ are obtained by exponentiating the generators. The VSR rotation is identical to the rotation in the Lorentz group along the 3-axis, so we have
\begin{equation}
\mathcal{R}(\phi)=\exp(i\mathscr{J}^{3}\phi).\label{eq:VSR_rot}
\end{equation}
On the other hand, the VSR boost for momentum is defined as a product of all three group elements
\begin{equation}
\mathcal{L}(p)=e^{i\beta_{1}\mathcal{T}^{1}}e^{i\beta_{2}\mathcal{T}^{2}}e^{i\mathcal{K}^{3}\varsigma}\label{eq:VSR_boost}
\end{equation}
where the parameters are defined as
\begin{equation}
\beta_{1}=\frac{p^{1}}{p^{0}-p^{3}},\nonumber
\end{equation}
\begin{equation}
\beta_{2}=\frac{p^{2}}{p^{0}-p^{3}}, \nonumber
\end{equation}
\begin{equation}
\hspace{1.3cm}\varsigma=-\ln\left(\frac{p^{0}-p^{3}}{m}\right).\label{eq:vsr_parameters}
\end{equation}
The explicit expression of $\mathcal{L}(p)$ is
\begin{equation}
\mathcal{L}(p)=\left(\begin{array}{cccc}
\frac{p^{0}}{m} & \frac{p^{1}}{p^{0}-p^{3}} & \frac{p^{2}}{p^{0}-p^{3}} & \frac{m^{2}-p^{0}(p^{0}-p^{3})}{m(p^{0}-p^{3})} \\
\frac{p^{1}}{m} & 1 & 0 & -\frac{p^{1}}{m} \\
\frac{p^{2}}{m} & 0 & 1 & -\frac{p^{2}}{m} \\
\frac{p^{3}}{m} & \frac{p^{1}}{p^{0}-p^{3}} & \frac{p^{2}}{p^{0}-p^{3}} & \frac{m^{2}-p^{3}(p^{0}-p^{3})}{m(p^{0}-p^{3})}\label{eq:vsr_boost}
\end{array}\right).
\end{equation}
The boost takes the momentum $k=(m,\0)$ to arbitrary momentum $p=(p^{0},\p)$ where $p^{0}=\sqrt{|\p|^{2}+m^{2}}$ so the dispersion relation remains unchanged.

Similar to the VSR algebras,  the $SIM(2)$ parameters $\beta_{1}$, 
$\beta_{2}$ and $\varsigma$ also contain a preferred direction. Whenever $p^{1}$ or $p^{2}$ are non-zero, the parameter $\varsigma$ is non-zero even when $p^{3}=0$. For this purpose, it is instructive to write down the boost along each axis explicitly
\begin{equation}
\mbox{1-axis: } \mathcal{L}(p)\vert_{p^{2}=p^{3}=0}=e^{i\beta_{1}\mathcal{T}^{1}}e^{i\mathcal{K}^{3}\varsigma}\nonumber
\end{equation}
\begin{equation}
\mbox{2-axis: }\mathcal{L}(p)\vert_{p^{1}=p^{3}=0}=e^{i\beta_{2}\mathcal{T}^{2}}e^{i\mathcal{K}^{3}\varsigma}\nonumber
\end{equation}
\begin{equation}
\mbox{3-axis: }\mathcal{L}(p)\vert_{p^{1}=p^{2}=0}=e^{i\mathcal{K}^{3}\varsigma}.
\end{equation}
Therefore, when considering a VSR boost $\mathcal{L}(p)$, one must always include the factor $e^{i\mathcal{K}^{3}\varsigma}$ whenever $p^{1}$ or $p^{2}$ are non-zero. We may also consider the transformation $e^{i\beta_{i}\mathcal{T}^{i}}$ on the momentum vectors by itself, but it is no longer a VSR boost.

\subsection{Casimir invariants of $\mathfrak{isim}(2)$}\label{Casimir_invariants_VSR}
One of the success of the Poincar\'{e} group is that the eigenvalues of the Casimir operators have the physical interpretation of mass and spin and are invariant in all inertial frames.
Therefore, we would like emulate the success of the Poincar\'{e} group by defining the simplest particle states of VSR as simultaneous eigenstates of the Casimir invariants of $\mathfrak{isim}(2)$. The $\mathfrak{isim}(2)$ algebra has two Casimir invariants and they are given by~\cite[tab.VII, row 7]{Patera:1975bw}
\begin{equation}
C_{1}=P^{\mu}P_{\mu},\hspace{0.5cm}
C_{2}=J^{3}-\frac{P^{2}}{P^{0}-P^{3}}T^{1}+\frac{P^{1}}{P^{0}-P^{3}}T^{2} \label{eq:Casimir_VSR}
\end{equation}
The first Casimir invariant gives the mass of the particle states. However, the physical interpretation and eigenvalues of the second Casimir invariant are currently not known to us. What we can extract from eq.~(\ref{eq:Casimir_VSR}) is that mass remains a valid description of VSR particles but the notion of spin in the context of the Poincar\'{e} group is no longer appropriate. Fortunately, not knowing the physical meaning of $C_{2}$ does not stop us from deriving the particle states and their transformations since the eigenvalues of Casimir invariants are invariant in all inertial frames.
%

\subsection{The one-particle state}\label{VSR_particle_state}
The VSR one-particle state can be constructed by carrying out the exact same analysis as given in sec.~\ref{one_particle_state}. For this reason, we simply write down the relevant results omitting the intermediate details which can be found in sec.~\ref{one_particle_state}. Let $k^{\mu}$ be the standard momentum given in tab.\ref{table:2}, we define the state $|k,\sigma\rangle$ of momentum $k^{\mu}$ with the inner-product
\begin{equation}
\langle k',\sigma'|k,\sigma\rangle=\delta_{\sigma'\sigma}\delta^{3}(\k'-\k).
\end{equation}
The one-particle state of arbitrary momentum is then given by applying the VSR boost in the Hilbert space $U(\mathcal{L}(p))$
\begin{equation}
|p,\sigma\rangle=\sqrt{\frac{k^{0}}{p^{0}}}U(\mathcal{L}(p))|k,\sigma\rangle.
\end{equation}
The normalisation factor is chosen so that the inner-product remains orthonormal for arbitrary momentum
\begin{equation}
\langle p',\sigma'|p,\sigma\rangle=\delta_{\sigma\sigma'}\delta^{3}(\mathbf{p}-\mathbf{p}').
\end{equation}
It follows that the VSR transformation of the one-particle state is given by
\begin{equation}
U(\Lambda)|p,\sigma\rangle=\sqrt{\frac{(\Lambda p)^{0}}{p^{0}}}\sum_{\sigma'}
D_{\sigma'\sigma}(\mathcal{W}(\Lambda,p))|\Lambda p,\sigma'\rangle.\label{eq:trans_particle}
\end{equation}
Equation (\ref{eq:trans_particle}) takes the same form as eq.~(\ref{eq:Lorentz_particle_tran}), except this time the little group element $\mathcal{W}(\Lambda,p)$ is defined in terms of the VSR boost
\begin{equation}
\mathcal{W}(\Lambda,p)=\mathcal{L}^{-1}(\Lambda p)\Lambda\mathcal{L}(p)
\end{equation}
where $\Lambda\in SIM(2)$ is an arbitrary VSR transformation given by eqs.~(\ref{eq:VSR_rot}) and (\ref{eq:VSR_boost})
%
%
Intuitively, the little group for massive particle is $SO(2)$, the rotation about the 3-axis. Explicit calculation of the little group element under rotation $\mathcal{R}(\phi)$ and arbitrary boost $\mathcal{L}(q)$ gives us\footnote{The computation was carried out using Mathematica 7.0.}
\begin{equation}
\mathcal{W}(\mathcal{R}(\phi),p)=\mathcal{R}(\phi)
\end{equation}
\begin{equation}
\mathcal{W}(\mathcal{L}(q),p)=\mathcal{W}(e^{i\beta_{i}(q)\mathcal{T}^{i}},p)=I \label{eq:VSR_little_group_I}
\end{equation}
where $p\neq q$ thus confirming our intuition. As for massless particles, the little group is the
Euclidean group $E(2)$ generated by $\mathfrak{e}(2)$. This can be verified
by checking that $k=(\kappa,0,0,\kappa)$ is invariant
under the action of $\exp(i\beta_{i}\mathcal{T}^{i})$ and $\exp(i\mathcal{J}^{3}\phi)$.
\begin{table}[!hbt]
\centering
\begin{tabular}{lll}
\toprule
 & Standard $k^{\mu}$ & Little group \\
\midrule
$p^{\mu}p_{\mu}=m^{2}$, $p^{0}>0$ & $(m,0,0,0)$ & $SO(2)$ \\
$p^{\mu}p_{\mu}=0$, $p^{0}>0$ & $(\kappa,0,0,\kappa)$ & $E(2)$\\

\bottomrule
\end{tabular}
\caption{The little groups}
\label{table:2}
\end{table}



\subsubsection{Massive particle state}
The group $SO(2)$ is infinitely connected $SO(2)\sim R/Z_{\infty}$ and its universal covering group is $R$, the group of real numbers under addition~\cite[sec.~16.24]{Wybourne}. Therefore, all the multi-valued unitary irreducible representations of $SO(2)$ are~\cite{Binegar:1981gv}
\begin{equation}
D_{n'n}(R(\phi))
=e^{in\phi}\delta_{n'n},\hspace{0.5cm}n\in R
\end{equation}
where $n$ is a real number that labels the representation. The irreducible representations are one-dimensional since $SO(2)$ is an Abelian group.


The transformation of the massive one-particle state is
\begin{equation}
U(\Lambda)|p,n\rangle=\sqrt{\frac{(\Lambda p)^{0}}{p^{0}}}e^{in\phi(\Lambda,p)}|\Lambda p,n\rangle.\label{eq:vsr_massive_trans_particle}
\end{equation}
When $\Lambda=\mathcal{L}(p)$, $\phi(L(p),p)=0$ and for $\Lambda=R(\phi)$, $\phi(R,p)=\phi$ is the angle of rotation. The label $n$, at this stage is unrestricted and is invariant under VSR transformations.

%

\subsubsection{Massless particle state}

The little group for massless VSR particles is the Euclidean group $E(2)$ generated by~$\mathfrak{e}(2)$
\begin{equation}
[T^{1},J^{3}]=-iT^{2},\hspace{0.5cm}
[T^{2},J^{3}]=iT^{1},\hspace{0.5cm}
[T^{1},T^{2}]=0
\end{equation}
which is also the little group of massless particles in the Lorentz group. In sec.~\ref{Massless_particle_state}, it is shown that massless particle states of the Lorentz group must be annihilated by $T^{i}$ since they do not have continuous degrees of freedom. In VSR, there are no reasons to impose such constraints. Therefore, there are two types of massless particles in VSR
\begin{enumerate}
\item Discrete degrees of freedom: $T^{i}|k,\sigma\rangle=0$, $J^{3}|k,\sigma\rangle=\sigma|p,\sigma\rangle$, $\sigma=-j,\cdots j$.
\item Continuous degrees of freedom: $T^{i}|k,t^{1},t^{2}\rangle=t^{i}|k,t^{1},t^{2}\rangle$, $t^{i}\in R$.
\end{enumerate}
In the first case, the particle states are equivalent to their Lorentz counterpart so no further analysis is needed. As for the second possibility, it is a topic we have not investigated. For the interested reader, details on massless particles in the Lorentz group with continuous degrees of freedom can be found in~\cite{Brink:2002zx}.

%

Here, a problem of immediate concern, is the kinematics of the massless particles. The transformation $\exp(i\beta{i}\mathcal{T}^{i})$ in the vector representation, 
leaves $k=(\kappa,0,0,\kappa)$ invariant, the VSR boost only changes $k^{\mu}$ along the $3$-axis,
\begin{eqnarray}
\mathcal{L}^{\mu}_{\,\,\,\nu}(p)k^{\nu}&=&(e^{i\mathcal{K}^{3}\varsigma})^{\mu}_{\,\,\,\nu}k^{\nu} \nonumber \\
&=&(p,0,0,p).
\end{eqnarray}
The second line is obtained using the parameter $\ln\varsigma=(p/\kappa)$. Therefore, the motion of massless particles in VSR are constrained 
to 1+1 dimensions. The spatial direction of motion coincides with the preferred direction of the algebra.

Generally, due to the preferred direction in VSR, the choice of the standard vector $k^{\mu}$ for massless particle requires care. Here,
we have chosen $k=(\kappa,0,0,\kappa)$ where the motion of the particle is along the direction specified by the VSR algebra. Under this choice, the derivation of the transformations for the massless particle states is the same as the derivation in the Lorentz group. If we choose the standard momentum to be $k'=(\kappa,\kappa,0,0)$ the situation would be different. The null vector $k'$ is now invariant under $\exp(i\beta_{i}S\mathcal{T}^{i}S^{-1})$ where $k'^{\mu}=S^{\mu}_{\,\,\nu}k^{\nu}$. While the little group remains $E(2)$, the resulting massless particle states are no longer described by the original generators where the 3-axis is the preferred direction. Instead, they are now described by a new set of $SIM(2)$ generators where the 1-axis is the preferred direction.

Our analysis seem to have revealed a possible limitation of VSR, that the motion of massless particles are restricted to the preferred direction. Equivalently, this seems to imply that in a given frame, the VSR algebra can only describe massless particles moving along the preferred direction. Currently we do not have a solution to this problem, but we can think of two possible outcomes:
\begin{enumerate}
\item The VSR groups do not admit massless particles.
\item A deeper understanding on the physics of the preferred direction is required.
\end{enumerate}
In the Lorentz group, there is no such restriction. The boost that takes a massless
particle of momentum $k=(\kappa,0,0,\kappa)$ to arbitrary momentum is defined as 
a product of rotation and boost $R(\hat{\mathbf{p}})B(|\p|)$ where $B(|\p|)$ 
takes $k$ to $(p,0,0,p)$ and $R(\hat{\mathbf{p}})$ rotates this four-vector
to the direction $\hat{\mathbf{p}}$. This is not possible for VSR, since 
$R(\hat{\mathbf{p}})$ of the Lorentz group is composed of products of two rotations about two different 
axis and $ISIM(2)$, the largest VSR group only has one rotation generator. 
Moreover, different choices
of standard null vector do not affect the description of massless particles in the
Lorentz group, since the group is isotropic containing all the rotation generators.

\subsection{Quantum fields with VSR symmetry}

Due to the unresolved problem with massless particles in VSR,
in this section we will only focus on the construction of
massive quantum fields with VSR symmetry. 

Before we proceed with our task, it is instructive to recall why quantum field operators are needed. As we have mentioned before, this answer was provided by Weinberg~\cite{Weinberg:1995mt,Weinberg:1964cn}. Weinberg showed that quantum field theory is the only known way to unify quantum mechanics and special relativity, it is the inevitable consequence due to the demand of Poincar\'{e}-invariant $S$-matrix and the cluster decomposition principle. From an ab-initio perspective, it is therefore necessary to examine whether quantum fields are needed to construct VSR-invariant $S$-matrix. 

We approach this problem by recalling that the Dyson series
\begin{equation}
S=1+\sum_{n=1}^{\infty}\frac{(-i)^{n}}{n!}\int_{-\infty}^{\infty}d^{4}x_{1}\cdots d^{4}x_{n}T[\mathcal{V}(x_{1})\cdots\mathcal{V}(x_{n})]
\end{equation}
is Poincar\'{e}-invariant if the interaction densities commute at space-like separation
\begin{equation}
[\mathcal{V}(x),\mathcal{V}(y)]=0. \label{eq:local_interacting_density}
\end{equation}
In sec.~\ref{chap:quantum_fields} it is shown that such interaction densities can be constructed using local Poincar\'{e}-covariant quantum field operators. Following the same line of reasoning, if the Dyson series is manifestly VSR-invariant, the introduction of VSR-covariant quantum field operators pertaining the properties given in sec.~\ref{chap:quantum_fields} would be necessary. Their introduction would allow us to construct interaction densities satisfying eq.~(\ref{eq:local_interacting_density}) and hence VSR-invariant $S$-matrices.


\subsubsection{Causal structure of VSR} \label{causal_VSR}
For the Dyson series to be manifestly VSR-invariant, the causal structure of VSR must be the same as special relativity. That is, the the order of time-like events must be absolute while the order of space-like events are reversible. One would assume that the VSR coordinate transformation obtained by making the following substitution
\begin{equation}
p^{0}=m\gamma,\hspace{0.5cm}
\p=m\gamma\mathbf{u}
\end{equation}
for eq.~(\ref{eq:vsr_boost}) where $\mathbf{u}=(u^{1},u^{2},u^{3})$ is the velocity and $\gamma=(\sqrt{1-|\mathbf{u}|^{2}})^{-1/2}$ would have the same causal structure as special relativity. However, it is shown in app.~\ref{AppendixC} that this is not the case. For certain values of $\mathbf{u}$, the order of some particular time-like separated event becomes reversible. 

Here it is important to emphasise that this result does not imply the Dyson series violates VSR-symmetry. After all, the VSR groups are subgroups of the Poincar\'{e} group, so a Poincar\'{e}-invariant $S$-matrix must also be VSR-invariant. Moreover the VSR boost given by eq.~(\ref{eq:vsr_boost}) correctly takes the massive particles at rest to arbitrary momentum. Therefore, we shall explore the possibility that there may exist an alternative set of parameters for eq.~(\ref{eq:vsr_boost}) than the ones given in eq.~(\ref{eq:vsr_parameters}) specific to the coordinate space that preserve the causal structure of special relativity. This is reminiscent of the situation for the massless particles of the Lorentz group where the parameters of the coordinate transformation differs from the parameters used to boost the particle from $k=(\kappa,0,0,\kappa)$ to arbitrary momentum.

Presently, we do not have the precise parameters which yield the transformations that preserve the causal structure of special relativity, but we can put constraint on the parameters. Let $x=(t,\x)$ be an arbitrary time-like vector $x^{\mu}x_{\mu}>0$, a general VSR transformation on its temporal component gives us
\begin{eqnarray}
t'&=&(e^{i\alpha_{i}\mathcal{T}^{i}}e^{i\mathcal{K}^{3}\rho})^{0}_{\,\,\mu}t^{\mu}\nonumber\\
&=&\left[\frac{1}{2}e^{\rho}+\frac{1}{2}e^{-\rho}(1+\alpha_{1}^{2}+\alpha_{2}^{2})\right]t+\alpha_{1}x^{1}+\alpha_{2}x^{2}\nonumber\\
&&\hspace{4cm}+\left[\frac{1}{2}e^{\rho}-\frac{1}{2}e^{-\rho}(1+\alpha_{1}^{2}+\alpha_{2}^{2})\right]x^{3}
\end{eqnarray}
where $\alpha_{1}$, $\alpha_{2}$ and $\rho$ are parameters to be determined. The condition which preserve the order of time-like event is simply $t'>0$. After some simplification, we obtain the following inequality
\begin{equation}
t>\frac{1}{\cosh\rho+\frac{1}{2}e^{-\rho}(\alpha_{1}^{2}+\alpha_{2}^{2})}
\left[\left(\frac{1}{2}e^{-\rho}(\alpha_{1}^{2}+\alpha_{2}^{2})-\sinh\rho\right)x^{3}-\alpha_{1}x^{1}-\alpha_{2}x^{2}\right].
\end{equation}
Admittedly this inequality is insufficient to determine the parameters and may not be useful for this purpose. Nevertheless, any parameters must satisfy the above inequality. For now, we assume that such a set of parameters exist and proceed to construct massive quantum fields with VSR symmetry.

\subsection{Massive quantum fields}

Let $\psi^{(n)}(x)$ be a quantum field describing a
massive particle state labelled by $n$
\begin{equation}
\psi^{(n)}(x)=(2\pi)^{-3/2}\int \frac{d^{3}p}{\sqrt{2p^{0}}}\left[e^{-ip\cdot x}u(\p,n)a(\p,n)+
e^{ip\cdot x}v(\p,n)b^{\dag}(\p,n)\right]\label{eq:qf}
\end{equation}
where $u(\p,n)$ and $v(\p,n)$ are the expansion coefficients. The
operators $a^{\dag}(\p,n)$ and $b^{\dag}(\p,n)$ creates 
massive particle and anti-particle states when acted on the
vacuum state $|\,\,\rangle$\footnote{The charge-conjugation
symmetry is allowed in VSR since its 
inclusion does not take the VSR groups to the full Lorentz group.}
\begin{equation}
a^{\dag}(\p,n)|\,\,\rangle=|p,n\rangle,\hspace{0.5cm}
b^{\dag}(\p,n)|\,\,\rangle=|p^{c},n\rangle.
\end{equation}
We assume they satisfy the standard algebraic identities
\begin{equation}
[a(\p,n),a^{\dag}(\p',n')]_{\pm}=
[b(\p,n),b^{\dag}(\p',n')]_{\pm}=\delta_{nn'}\delta^{3}(\p-\p')
\end{equation}
with all other combinations identically vanish.

The field defined in eq.~(\ref{eq:qf}) is manifestly covariant
under space-time translations, the demand of VSR covariance
means that it must transform as
\begin{equation}
U(\Lambda)\psi^{(n)}_{\ell}(x)U^{-1}(\Lambda)= \mathcal{D}_{\ell\bar{\ell}}
(\Lambda^{-1})\psi^{(n)}_{\bar{\ell}}(\Lambda x)\label{eq:vsr_qf_transform}
\end{equation}
where $\mathcal{D}(\Lambda)$ is the finite-dimensional 
representation of $SIM(2)$. Equation (\ref{eq:vsr_qf_transform})
with the transformations of the massive particle states given by
eq.~(\ref{eq:vsr_massive_trans_particle}) is
sufficient for us to determine the coefficients and how they
transform. Following the same derivation performed in sec.~\ref{chap:quantum_fields},
the coefficients of arbitrary momentum are given by
\begin{equation}
u(\p,n)=\mathcal{D}(\mathcal{L}(p))u(\0,n)\label{eq:vsr_u_boost}
\end{equation}
\begin{equation}
v(\p,n)=\mathcal{D}(\mathcal{L}(p))v(\0,n)\label{eq:vsr_v_boost}
\end{equation}
and the coefficients at rest $u(\0,n)$ and $v(\0,n)$ are determined
by the demand of rotation symmetry
\begin{equation}
\sum_{\bar{n}}D_{n\bar{n}}(R)u_{\ell}(\0,\bar{n})=
\sum_{\bar{\ell}}\mathcal{D}_{\ell\bar{\ell}}(R)
u_{\bar{\ell}}(\0,n),
\end{equation}
\begin{equation}
\sum_{\bar{n}}D^{*}_{n\bar{n}}(R)v_{\ell}(\0,\bar{n})=
\sum_{\bar{\ell}}\mathcal{D}_{\ell\bar{\ell}}(R)
v_{\bar{\ell}}(\0,n).
\end{equation}
Expanding the matrices of the above equations about the identities with
$\mathcal{D}(R)=1+i\mathcal{J}^{3}\phi$ and $D_{n\bar{n}}(R)=\delta_{n\bar{n}}(1+in\phi)$,
we find that the coefficients at rest are eigenvectors of $\mathcal{J}^{3}$
\begin{equation}
(\mathcal{J}^{3})_{\ell\bar{\ell}}\,u_{\bar{\ell}}(\0,n)=nu_{\ell}(\0,n), \label{eq:eigenvectors1}
\end{equation}
\begin{equation}
(\mathcal{J}^{3})_{\ell\bar{\ell}}\,v_{\bar{\ell}}(\0,n)=-nv_{\ell}(\0,n). \label{eq:eigenvectors2}
\end{equation}
The label $n$ instead of being continuous, is now restricted to the eigenvalues of $\mathcal{J}^{3}$ 
and is therefore discrete.

Since the VSR generators are linear combinations of the Lorentz
generators, the finite-dimensional generators of 
$\mathcal{D}(\Lambda)$ can be obtained from the
finite-dimensional generators of the Lorentz group. 
Consequently, the rotation generator $\mathcal{J}^{3}$ in 
eqs.~(\ref{eq:eigenvectors1}) and (\ref{eq:eigenvectors2})
are identical to the one in the Lorentz sector along the $3$-axis.

As we have discussed in sec.~\ref{chap:quantum_fields}, a sufficient condition to construct local interaction densities is for the quantum fields to commute or anti-commute with its adjoint at space-like separation. The quantum field $\psi^{(n)}(x)$ given by eq.~(\ref{eq:qf}) does not in general satisfy this criteria since the expansion coefficients are non-zero. To solve this problem, we note, the rotation generator $\mathcal{J}^{3}$ has a spectrum
of $-n,\cdots,n$.  Therefore, there exists $2n+1$ fields $\psi^{(-n)}(x),\cdots,\psi^{(n)}(x)$ 
that transform according to eq.~(\ref{eq:vsr_qf_transform}) with the same finite-dimensional
representation $\mathcal{D}(\Lambda)$. This observation suggests that
different values of $n$ do not necessarily imply different species of 
particles. Here, it is possible that the species of massive VSR particles
are not completely determined by the unitary irreducible representations
of $SO(2)$. They are fully determined only after the finite-dimensional
representation is chosen. The eigenvalues of the  rotation generator
then corresponds to a $2n+1$ degeneracy of particle state. 

Interpreting $n$ as a degeneracy index suggests there must exist a 
unique quantum field with VSR symmetry for a given
finite-dimensional representation $\mathcal{D}(\Lambda)$ of dimension
$2n+1$. Therefore, without the loss of generality, we define such a quantum
field $\psi(x)$ as the sum of all fields from $\psi^{(-n)}(x)$ to
$\psi^{(n)}(x)$
\begin{eqnarray}
\hspace{-1cm}\psi(x)&=&\sum_{\alpha=-n}^{n}\psi^{(\alpha)}(x) \nonumber \\
&=&(2\pi)^{-3/2}\int \frac{d^{3}p}{\sqrt{2p^{0}}} \sum_{\alpha}
\left[e^{-ip\cdot x}u(\p,\alpha)a(\p,\alpha)+e^{ip\cdot x}v(\p,\alpha)b^{\dag}(\p,\alpha)\right].\label{eq:vsr_psi}
\end{eqnarray}
This field corresponds to a massive particle state $|p,\alpha\rangle$ with
degeneracy $\alpha=-n,\cdots,n$. 
In principle, the most general possibility is
$\sum_{\alpha=-n}^{n} f_{\alpha}\psi^{(\alpha)}$, where $f_{\alpha}$ are some
coefficients. But since we can always absorb $f_{\alpha}$ into the expansion coefficients,
this choice is physically equivalent to eq.~(\ref{eq:vsr_psi}).

By construction, the field
$\psi(x)$ is similar to a quantum field of spin-$n$ representation
of the Lorentz group with the same rotation generator along the $3$-axis. 
They have the same number of degrees of freedom and their coefficients
at rest take the same form by virtue of eqs.~(\ref{eq:eigenvectors1}) and (\ref{eq:eigenvectors2}).
However, their coefficients at arbitrary momentum are different since the VSR
boost differs from the Lorentz boost. Moreover, quantum fields with VSR and Lorentz
symmetry cannot be physically equivalent since the little group for massive particle states
of VSR is $SO(2)$ and not $SU(2)$.

\subsubsection{Spin-half field}\label{vsr_spin_field}

As we have discussed in sec.~\ref{Casimir_invariants_VSR}, the concept of spin associated with particles of the Poincar\'{e} group does not apply to VSR particles. Nevertheless, it remains a helpful label which helps us to categorise the particles and fields of belonging to different representations. In the context of VSR, we define the spin of a quantum field by the eigenvalues of the rotation generator $\mathcal{J}^{3}$.

In the Lorentz group,
given the rotation generators $\J$, we can always find two 
solutions to the boost generators given by $\mathbf{K}_{\pm}=\pm i\J$
such that the Lorentz algebra is satisfied. The spin-half generators of the Lorentz group are
\begin{equation}
\J=\frac{\s}{2},\hspace{0.5cm}\mathbf{K}_{\pm}=\pm i\frac{\s}{2}
\end{equation}
where $\s=(\sigma^{1},\sigma^{2},\sigma^{3})$ are the Pauli matrices.

Since $SIM(2)$ is a proper subgroup of the Lorentz group, we can construct
two set of generators that satisfy the $SIM(2)$ algebra. They
are given by
\begin{equation}
K^{3}_{\pm}=\pm iJ^{3},\hspace{0.5cm}
T^{1}_{\pm}=K^{1}_{\pm}+J^{2},\hspace{0.5cm}
T^{2}_{\pm}=K^{2}_{\pm}-J^{1}. \label{eq:vsr_spin_half_rep}
\end{equation}
%
Fields that transform under the above irreducible representation of $SIM(2)$ has two components. To compare the massive spin-half field with VSR symmetry with the Dirac field, we take the following direct sum to obtain
\begin{equation}
\mathcal{J}^{3}=
\left(\begin{array}{cc}
J^{3} & O \\
O & J^{3}
\end{array}\right),\hspace{0.5cm}
\mathcal{K}^{3}=
\left(\begin{array}{cc}
K^{3}_{-} & O \\
O & K^{3}_{+}
\end{array}\right)
\end{equation}
\begin{equation}
\mathcal{T}^{1}=
\left(\begin{array}{cc}
T^{1}_{-} & O \\
O & T^{1}_{+}
\end{array}\right),\hspace{0.5cm}
\mathcal{T}^{2}=
\left(\begin{array}{cc}
T^{2}_{-} & O \\
O & T^{2}_{+}
\end{array}\right). 
\end{equation}
We define the four-component field $\psi(x)$ as
\begin{equation}
\psi(x)=(2\pi)^{-3/2}\int\frac{d^{3}p}{\sqrt{2p^{0}}}\sum_{\alpha=\pm1/2}\left[e^{-ip\cdot x}u(\p,\alpha)a(\p,\alpha)+e^{ip\cdot x}v(\p,\alpha)
b^{\dag}(\p,\alpha)\right].
\end{equation}
According to eqs.~(\ref{eq:eigenvectors1}) and (\ref{eq:eigenvectors2}) the coefficients at rest are eigenvectors of $\mathcal{J}^{3}$
\begin{equation}
\mathcal{J}^{3}u(\0,\alpha)=\alpha u(\0,\alpha),
\end{equation}
\begin{equation}
\mathcal{J}^{3}v(\0,\alpha)=-\alpha v(\0,\alpha).
\end{equation}
Their general solutions, with the appropriate normalisation, are given by
\begin{equation}
u(\0,\textstyle{\frac{1}{2}})=\sqrt{m}\left(\begin{array}{cccc}
c_{+} \\
0\\
c_{-} \\
0 \end{array}\right),\hspace{0.5cm}
u(\0,-\textstyle{\frac{1}{2}})=\sqrt{m}\left(\begin{array}{cccc}
0 \\
c_{+} \\
0\\
c_{-} \end{array}\right),\label{eq:c_coefficients}
\end{equation}
\begin{equation}
v(\0,\textstyle{\frac{1}{2}})=\sqrt{m}\left(\begin{array}{cccc}
0 \\
d_{+}\\
0\\
d_{-} \end{array}\right),\hspace{0.5cm}
v(\0,-\textstyle{\frac{1}{2}})=\sqrt{m}\left(\begin{array}{cccc}
d_{+}\\
0\\
d_{-}\\
0 \end{array}\right)\label{eq:d_coefficients}.
\end{equation}

The proportionality constants $c_{\pm}$ and $d_{\pm}$
are determined by the demand of locality
\begin{equation}
[\psi(x),\psi^{\dag}(y)]_{\pm}=0.
\end{equation}
where $x$ and $y$ are separated by space-like interval.
Explicit computation yields
\begin{equation}
[\psi(x),\psi^{\dag}(y)]_{\pm}=\int\frac{d^{3}p}{2p^{0}}
\left[e^{-ip\cdot(x-y)}N(p)\pm e^{ip\cdot(x-y)}M(p)\right]
\end{equation}
where $N(p)$ and $M(p)$ are the spin-sums
\begin{equation}
N(p)=\sum_{\alpha}u(\p,\alpha)u^{\dag}(\p,\alpha)=\mathcal{D}(\mathcal{L}(p))N(0)\mathcal{D}^{\dag}(\mathcal{L}(p)),
\end{equation}
\begin{equation}
M(p)=\sum_{\alpha}v(\p,\alpha)v^{\dag}(\p,\alpha)=\mathcal{D}(\mathcal{L}(p))M(0)\mathcal{D}^{\dag}(\mathcal{L}(p)).
\end{equation}
We note, the coefficients at rest given in eqs.~(\ref{eq:c_coefficients}) and (\ref{eq:d_coefficients})
take the same form as those in the $(\frac{1}{2},0)\oplus(0,\frac{1}{2})$ representation of
the Lorentz group, so their spin-sums at rest are also of the same form. The 
possible point of departure is the spin-sums at arbitrary momentum since
$\mathcal{D}(\mathcal{L}(p))\neq\mathcal{D}(L(p))$ where $L(p)$ is the
standard Lorentz boost and $\mathcal{D}(L(p))$ its finite dimensional
representation.

Generally, spin-sums of the $(j,0)\oplus(0,j)$ representation of the
the Lorentz group at arbitrary momentum are
determined by $\mathcal{D}(L(p)\mathcal{D}^{\dag}(L(p))$. For spin-half, we perform the same calculation for VSR and find them
to be identical
\begin{equation}
\mathcal{D}(\mathcal{L}(p))\mathcal{D}^{\dag}(\mathcal{L}(p))=
\mathcal{D}(L(p)\mathcal{D}^{\dag}(L(p))=\frac{\gamma^{\mu}p_{\mu}}{m}\gamma^{0}
\label{eq:VSR_Lorentz_equiv}
\end{equation}
where 
\begin{equation}
\gamma^{0}=\left(\begin{matrix}
O & I\\
I & O\end{matrix}\right),\hspace{0.5cm}
\gamma^{i}=\left(\begin{matrix}
O & -\sigma^{i}\\
\sigma^{i} & O\end{matrix}\right).
\end{equation}

Since the Dirac field is local, anti-commuting with its adjoint at space-like separation, by taking
the coefficients in eqs.~(\ref{eq:c_coefficients}) and (\ref{eq:d_coefficients})
to be identical to Dirac spinors at rest, we obtain a local field theory
\begin{equation}
u(\0,\textstyle{\frac{1}{2}})=\sqrt{m}\left(\begin{array}{cccc}
1 \\
0 \\
1 \\
0 \end{array}\right),\hspace{0.5cm}
u(\0,-\textstyle{\frac{1}{2}})=\sqrt{m}\left(\begin{array}{cccc}
0 \\
1 \\
0 \\
1 \end{array}\right),
\end{equation}
\begin{equation}
v(\0,\textstyle{\frac{1}{2}})=\sqrt{m}\left(\begin{array}{cccc}
 0 \\
 1 \\
 0 \\
-1 \end{array}\right),\hspace{0.5cm}
v(\0,-\textstyle{\frac{1}{2}})=\sqrt{m}\left(\begin{array}{cccc}
-1 \\
 0 \\
 1 \\
 0 \end{array}\right).
\end{equation}
The resulting spin-sums are
\begin{equation}
N(p)=(\gamma^{\mu}p_{\mu}-mI)\gamma^{0},
\end{equation}
\begin{equation}
M(p)=(\gamma^{\mu}p_{\mu}+mI)\gamma^{0}.
\end{equation}
Therefore, the massive spin-half VSR field furnishes fermionic statistics
\begin{equation}
\{\psi(x),\psi^{\dag}(y)\}=0.
\end{equation}

In order to derive the field equation we need to define the
dual for the coefficients. Here we may take the Dirac dual
\begin{equation}
\bar{u}(\p,\alpha)=u^{\dag}(\p,\alpha)\gamma^{0},\hspace{0.5cm}
\bar{v}(\p,\alpha)=v^{\dag}(\p,\alpha)\gamma^{0}
\end{equation}
since the resulting norms are orthonormal and VSR invariant
\begin{equation}
\bar{u}(\p,\alpha)u(\p,\alpha')=-\bar{v}(\p,\alpha)v(\p,\alpha')=2m\delta_{\alpha\alpha'}.
\end{equation}
It follows that the field $\psi(x)$ we have constructed satisfies the Dirac equation
\begin{equation}
(i\gamma^{\mu}\partial_{\mu}-mI)\psi(x)=0.
\end{equation}

Although the VSR spin-half fields are physically different from the Dirac field, their field equations remain identical. Repeating the same construction for higher spin fields, we have shown that the first equality of eq.~(\ref{eq:VSR_Lorentz_equiv}) remains true up to the spin-two representation. This structure suggests that eq.~(\ref{eq:VSR_Lorentz_equiv}) holds for arbitrary spin. Consequently the coefficients of a massive spin-$j$ quantum field with VSR symmetry can always be chosen such that it satisfies the same field equation as their Lorentz counterpart of the $(j,0)\oplus(0,j)$ representation given by eq.~(\ref{eq:higher_spin_field_equation}). 

In hindsight, this result is expected. Examining Weinberg's proof for the existence of $t^{\mu_{1}\mu_{2}\cdots\mu_{2j}}$ in app.~\ref{AppendixE}, we see that since $SIM(2)$ transformations are also Lorentz transformations, it follows that there must exists a symmetric traceless rank $2j$ tensor in which $\mathcal{D}(\mathcal{L}(p))\mathcal{D}^{\dag}(\mathcal{L}(p))$ may be expressed in for arbitrary spin. However, we did not prove the VSR tensor must be identical to $t^{\mu_{1}\mu_{2}\cdots\mu_{2j}}$.


When the coordinate is aligned along the preferred direction (the 3-axis)
$x=(t,0,0,x^{3})$, we have $\p=(0,0,p^{3})$ and the boost parameter $\varsigma$
reduces to the rapidity parameter of special relativity
\begin{equation}
\cosh\varsigma=\frac{\sqrt{(p^{3})^{2}+m^{2}}}{m},\hspace{0.5cm}
\sinh\varsigma=\frac{p^{3}}{m}.
\end{equation}
In this case, the VSR and Lorentz boost coincides
\begin{equation}
\mathcal{D}(\mathcal{L}(p^{3}))=\mathcal{D}(L(p^{3})).
\end{equation}
Therefore, massive quantum fields with VSR symmetry constructed according to the above prescription are physically equivalent to their Lorentz counterpart when their coordinates are aligned to the preferred direction specified by the algebra.


\subsection{Discrete symmetries}

The differences between VSR and Poincar\'{e}-covariant quantum fields are best demonstrated by 
examining their discrete symmetries. The VSR algebras cannot accommodate discrete symmetry
operators $\mathsf{P}$, $\mathsf{T}$, $\mathsf{CP}$ and $\mathsf{CT}$. Including any one of these operators
in the VSR algebra would yield the Poincar\'{e} algebra. Therefore, it is expected that the
VSR quantum fields would violate the said discrete symmetries but conserves charge-conjugation.

Here we study the discrete symmetries of the massive spin-half field constructed in the 
previous section as an example. Although VSR algebra does not admit $\mathsf{P}$ and $\mathsf{T}$
generators, it does not prevent us from studying the discrete transformation of spin-half VSR fields by
considering $\mathsf{P}\psi(x)\mathsf{P}^{-1}$ and $\mathsf{T}\psi(x)\mathsf{T}^{-1}$.  

Assuming the action of $\mathsf{P}$ and $\mathsf{T}$ on the creation and annihilation operators for the VSR particles to be identical to their Lorentz counterpart, we get
\begin{equation}
\mathsf{P}a(\p,\alpha)\mathsf{P}^{-1}=\eta^{*} a(-\p,\alpha),\hspace{0.5cm}
\mathsf{T}a(\p,\alpha)\mathsf{T}^{-1}=\varrho^{*}(-1)^{1/2-\alpha}a(-\p,-\alpha)
\end{equation}
\begin{equation}
\mathsf{P}b(\p,\alpha)\mathsf{P}^{-1}=\bar{\eta}^{*} b(-\p,\alpha),\hspace{0.5cm}
\mathsf{T}b(\p,\alpha)\mathsf{T}^{-1}=\bar{\varrho}^{*}(-1)^{1/2-\alpha}b(-\p,-\alpha)
\end{equation}
where $\eta$ and $\varrho$ are the parity and time-reversal phases for particles and
$\bar{\eta}$ and $\bar{\varrho}$ are the parity and time-reversal phases for
anti-particles. Additionally, we also consider charge-conjugation
\begin{equation}
\mathsf{C}a(\p,\alpha)\mathsf{C}^{-1}=\varsigma^{*}b(\p,\alpha),\hspace{0.5cm}
\mathsf{C}b(\p,\alpha)\mathsf{C}^{-1}=\bar{\varsigma}^{*}a(\p,\alpha)
\end{equation}
where $\varsigma$ and $\bar{\varsigma}$ are the charge-conjugation phases.

\subsubsection{Parity}

Acting the parity operator on $\psi(x)$, we get
\begin{equation}
\mathsf{P}\psi(x)\mathsf{P}^{-1}=
(2\pi)^{-3/2}\int\frac{d^{3}p}{\sqrt{2p^{0}}}\sum_{\alpha}
\left[e^{-ip\cdot x}\eta^{*}a(-\p,\alpha)+
e^{ip\cdot x}\bar{\eta}\,b^{\dag}(-\p,\alpha)\right].
\end{equation}
Parity conservation requires $\mathsf{P}\psi(x)\mathsf{P}^{-1}$
to be proportional to $\psi(\mathscr{P}x)$ which is equivalent to the conditions
\begin{equation}
Pu(\p,\alpha)=\eta^{*}u(-\p,\alpha),\hspace{0.5cm} Pv(\p,\alpha)=\bar{\eta}v(-\p,\alpha)\label{eq:finite_parity}
\end{equation}
where $P$ is a momentum-independent matrix. We find, eq.~(\ref{eq:finite_parity}) cannot be 
satisfied for all $\p$, they are only satisfied when $\p=(0,0,p^{3})$ with $P=\gamma^{0}$
\begin{equation}
\gamma^{0}u(p^{3},\alpha)=u(-p^{3},\alpha),\label{eq:gamma0_vsr1}
\end{equation}
\begin{equation}
\gamma^{0}v(p^{3},\alpha)=-v(-p^{3},\alpha).\label{eq:gamma0_vsr2}
\end{equation}
The two identities are identical to the ones for the Dirac spinors except for the restrictions
on the momentum. 

Evidently, since eq.~(\ref{eq:finite_parity}) is not satisfied for all momentum, parity is violated. This
is not surprising since the VSR algebra does not include the parity operator. However,
according to eqs.~(\ref{eq:gamma0_vsr1}) and (\ref{eq:gamma0_vsr2}), when $\eta^{*}=-\bar{\eta}$, 
parity is conserved when the momentum is aligned along the preferred axis (3-axis)
\begin{equation}
\mathsf{P}\psi(t,x^{3})\mathsf{P}^{-1}=\eta^{*}\gamma^{0}\psi(t,-x^{3}).
\end{equation}
This can be explained by the above observation that the field $\psi(t,x^{3})$  along the preferred direction is physically equivalent to the Dirac field.

\subsubsection{Time-reversal}

Acting the time-reversal operator on $\psi(x)$ gives us
\begin{eqnarray}
\hspace{-1cm}\mathsf{T}\psi(x)\mathsf{T}^{-1}&=&(2\pi)^{-3/2}\int\frac{d^{3}p}{\sqrt{2p^{0}}}\sum_{\alpha}(-1)^{1/2-\alpha}\Big[e^{ip\cdot x}u^{*}(\p,\alpha)\varrho^{*}a(-\p,-\alpha)\nonumber\\
&&\hspace{5.5cm}+e^{-ip\cdot x}v^{*}(\p,\alpha)\bar{\varrho}\,b^{\dag}(-\p,-\alpha)\Big].\label{eq:time_reversal_dfield}
\end{eqnarray}
For time-reversal to be a symmetry, $\mathsf{T}\psi(x)\mathsf{T}^{-1}$ must be proportional to $\psi(\mathscr{T}x)$. This requires the following identities to hold
\begin{equation}
u^{*}(\p,\alpha)=(-1)^{1/2-\alpha}\varrho^{*}Tu(-\p,-\alpha),
\end{equation}
\begin{equation}
v^{*}(\p,\alpha)=(-1)^{1/2-\alpha}\bar{\varrho}\,Tv(-\p,-\alpha)
\end{equation}
where $T$ is a momentum-independent matrix. Similar to parity, the above identities are only satisfied when $\p=(0,0,p^{3})$ with
$T=i\gamma^{0}\gamma^{2}\gamma^{5}$
\begin{equation}
u^{*}(p^{3},\alpha)=(-1)^{1/2-\alpha}i\gamma^{0}\gamma^{2}\gamma^{5}u(-p^{3},-\alpha),
\end{equation}
\begin{equation}
v^{*}(p^{3},\alpha)=(-1)^{1/2-\alpha}i\gamma^{0}\gamma^{2}\gamma^{5}v(-p^{3},-\alpha).
\end{equation}
Therefore, when $\varrho^{*}=\bar{\varrho}$, time-reversal symmetry is conserved along the preferred-axis
\begin{equation}
\mathsf{T}\psi(t,x^{3})\mathsf{T}^{-1}=\varrho^{*}i\gamma^{0}\gamma^{2}\gamma^{5}\psi(-t,x^{3})
\end{equation}
and is violated for general momentum whenever $p^{1}$ and $p^{2}$ are non-zero.

\subsubsection{Charge-conjugation}

Acting the charge-conjugation operator on the field $\psi(x)$ gives us
\begin{equation}
\mathsf{C}\psi(x)\mathsf{C}^{-1}=(2\pi)^{-3/2}\int\frac{d^{3}p}{\sqrt{2p^{0}}}\sum_{\alpha}\left[e^{-ip\cdot x}u(\p,\alpha)\varsigma^{*}b(\p,\alpha)+e^{ip\cdot x}v(\p,\alpha)\bar{\varsigma}a^{\dag}(\p,\alpha)\right].
\end{equation}
We find the coefficients $u(\p,\sigma)$ and $v(\p,\sigma)$ are related to their complex-conjugates by
\begin{equation}
u(\p,\alpha)=i\gamma^{2}v^{*}(\p,\alpha),
\end{equation}
\begin{equation}
v(\p,\alpha)=i\gamma^{2}u^{*}(\p,\alpha).
\end{equation}
Therefore, when 
\begin{equation}
\varsigma^{*}=\bar{\varsigma}
\end{equation}
the charge-conjugation symmetry is conserved for all momentum
\begin{equation}
\mathsf{C}\psi(x)\mathsf{C}^{-1}=i\varsigma\gamma^{2}\psi^{*}(x).
\end{equation}

The above calculations show that the field violates parity and time-reversal
but charge-conjugation is conserved. This is in agreement with the original
observation of Cohen and Glashow that VSR violates parity and time-reversal
but allows charge-conjugation. Combining the above results, we see that $\mathsf{CPT}$ is conserved
\begin{equation}
(\mathsf{CPT})\psi(x)(\mathsf{CPT})^{-1}=-(\eta\varrho\varsigma)^{*}\gamma^{5}\psi^{*}(-x)
\end{equation}
which is expected since the field is local.

Our analysis reveals a clear distinction between VSR and Poincar\'{e}-covariant quantum fields. It highlighted the fact that the VSR coefficients for the spin-half field at arbitrary momentum are different from the Dirac spinors since $\mathcal{D}(\mathcal{L}(p))\neq\mathcal{D}(L(p))$. Together, the uniqueness of the Dirac spinors and the effect of the preferred direction in VSR are sufficient for us to conclude that the VSR fields violate Lorentz symmetry.



\section{Interactions between the SM and VSR particles}\label{Elko_DM_candidate}



\begin{figure}
\begin{center}
\includegraphics[scale=0.8]{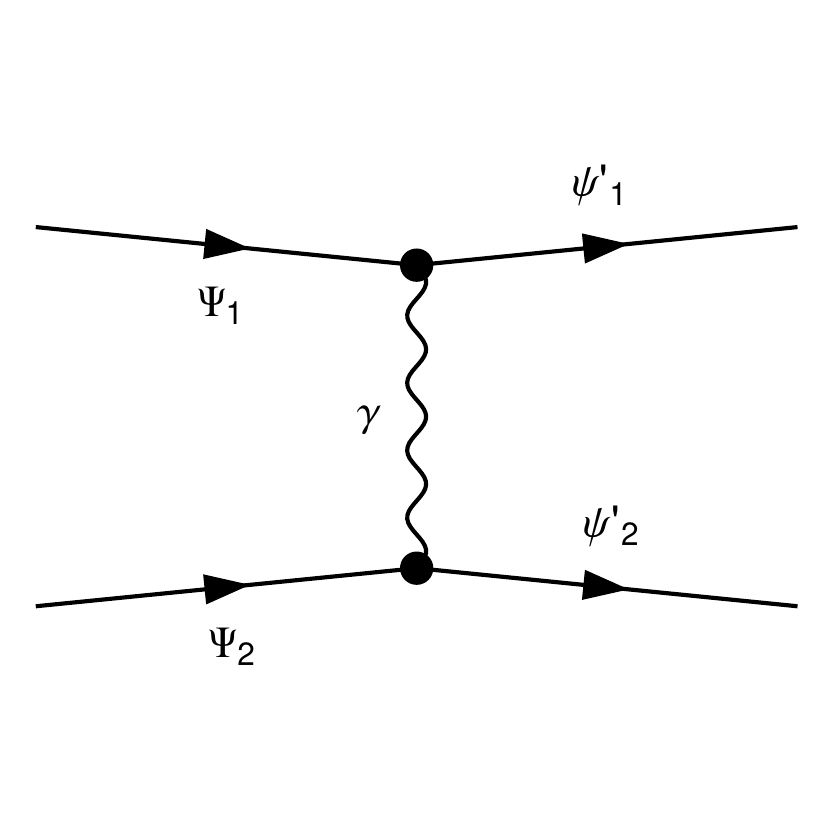}
\end{center}
\caption{A Lorentz-violating process where $\Psi_{1}$, $\Psi_{2}$ and $\gamma$ are the SM fermions and photon while $\psi'_{1}$ and $\psi'_{2}$ are the VSR fermions}\label{fig:vsr_qed}
\end{figure}

Cosmological observations and particles physics experiments have established that the visible universe (with respect to us) are comprised of the SM particles satisfying the Poincar\'{e} and not VSR symmetry. Therefore, the VSR particles have to be interpreted as a new set of particles yet to be discovered. To put constraints on these particles, we have to determine their interactions with the SM sector and make testable predictions. In the context of Elko, our analysis here will determine whether it qualifies as a dark mater candidate.

Since VSR groups are subgroups of the Poincar\'{e} group, any fields  with Poincar\'{e} symmetry automatically satisfy VSR symmetry. It follows that the interactions between the two sectors will also satisfy VSR symmetry and is thus unrestricted and unsuppressed as long as they are renormalisable. However, this expectation is incorrect. We show that in order to preserve Poincar\'{e} symmetry for all observable processes (with respect to us), the interactions between the two sectors must be limited.

Let us consider the massless field $a^{\mu}(x)$ of the $(\frac{1}{2},\frac{1}{2})$ representation of the Lorentz group. In sec.~\ref{massless_VF}, we have shown that $a^{\mu}$ transforms as
\begin{equation}
U(\Lambda)a^{\mu}(x)U^{-1}(\Lambda)=\Lambda_{\nu}^{\,\,\mu}a^{\nu}(\Lambda x)+\partial^{\mu}\Omega(\Lambda,x)
\end{equation}
where $\Lambda$ is an element of the Lorentz group. Since $SIM(2)$ is a subgroup of the Lorentz group, we can take $\Lambda$ to be an element of $SIM(2)$ and the transformation of $a^{\mu}(x)$ will remain covariant. This suggests its interaction with the massive spin-half VSR field $\psi(x)$ must be VSR and gauge-invariant and is therefore described by
\begin{equation}
\mathscr{L}=-\frac{1}{4}f^{\mu\nu}f_{\mu\nu}+\overline{\psi}(i\slashed{D}-m)\psi
\end{equation}
where $D_{\mu}=\partial_{\mu}-iqa_{\mu}(x)$ is the covariant derivative and $q$ the charge of $\psi(x)$. The Dirac field $\Psi(x)$ also satisfies VSR symmetry, so a more general Lagrangian density is
\begin{equation}
\mathscr{L}=-\frac{1}{4}f^{\mu\nu}f_{\mu\nu}+\overline{\psi}(i\slashed{D}-m)\psi+\overline{\Psi}(i\slashed{D}-m)\Psi \label{eq:vsr_QED}
\end{equation}
which describes the electromagnetic interactions between the SM and VSR fermions.

The Lagrangian density of eq.~(\ref{eq:vsr_QED}) gives us non-zero scattering amplitudes such as the one depicted by fig.\ref{fig:vsr_qed}
\begin{equation}
\Psi_{1}+\Psi_{2}\rightarrow \psi_{1}'+\psi_{2}'.
\end{equation}
This process satisfies VSR symmetry but violates Lorentz symmetry.\footnote{To be specific, the violation would be manifest only for the polarised cross-section. When the spin-projections of the fermions are not measured, the cross-section would be identical to prediction of the SM electrodynamics since the VSR fermionic spin-sums are identical to the SM counterpart.} Moreover, since both the SM and VSR fermions are external particles, the effect of Lorentz violation is observable in the SM sector. Demanding both VSR and Poincar\'{e} symmetry to be satisfied for all observable processes, eq.~(\ref{eq:vsr_QED}) is not a viable Lagrangian density. In fact, the two sectors cannot have direct interactions. The allowed processes, must then only contain the SM particles or VSR particles and not a mixture of both. 

A realistic scenario is for the two sectors to interact indirectly, mediated by particles that furnish representations of both the Poincar\'{e} and VSR group. We call these particles the \textit{mediators}. Due to the kinematics problem for the massless particles in the VSR sector, for now, we shall confine ourselves to determine the possible non-gravitational interactions between the SM and VSR sectors through the exchange of massive mediators. Of course, for the VSR particles to be dark matter candidates, they must have gravitational interactions. Since $C_{1}=P^{\mu}P_{\mu}$ remains a Casimir invariant in VSR, we expect them to have interact gravitationally. Determining the precise gravitational interactions is beyond the scope of this thesis but it is undoubtedly an important project for future investigation. 

To determine the massive non-gravitational mediator(s), we consider the transformations of massive particles, 
\begin{equation}
\mbox{SM: }U(\Lambda)|p,\sigma\rangle=\sqrt{\frac{(\Lambda p)^{0}}{p^{0}}}\sum_{\sigma'}D_{\sigma'\sigma}(W(\Lambda,p))|\Lambda p,\sigma\rangle,
\end{equation}
\begin{equation}
\mbox{VSR: }U(\widetilde{\Lambda})|p,n\rangle=\sqrt{\frac{(\widetilde{\Lambda} p)^{0}}{p^{0}}}e^{in\phi(\widetilde{\Lambda},p)}|\widetilde{\Lambda} p,n\rangle\label{eq:vsr_vsr_trans}
\end{equation}
where $\Lambda\in SO(3,1)$, $D(W(\Lambda,p))\in SU(2)$ and $\widetilde{\Lambda}\in SIM(2)$. Since the mediators furnish representation of both groups, they must have the same transformations when
\begin{equation}
\Lambda=\widetilde{\Lambda}\in SIM(2)\subset SO(3,1).
\end{equation}
Since $SIM(2)$ is a subgroup of the Lorentz group, a $SIM(2)$ transformation is also a Lorentz transformation so we get
\begin{equation}
U(\widetilde{\Lambda})|p,\sigma\rangle=\sqrt{\frac{(\widetilde{\Lambda} p)^{0}}{p^{0}}}\sum_{\sigma'}D_{\sigma'\sigma}(W(\widetilde{\Lambda},p))|\Lambda p,\sigma\rangle. \label{eq:vsr_lorentz_trans}
\end{equation}
We now show that only the massive scalar particles ($j=n=0$) have the same transformation under both eqs.~(\ref{eq:vsr_vsr_trans}) and (\ref{eq:vsr_lorentz_trans}) thus making them the mediator. To see this, it is sufficient to show that the transformation of particle states of non-zero spin ($j=n\neq0$) under eqs.~(\ref{eq:vsr_vsr_trans}) and (\ref{eq:vsr_lorentz_trans}) are different. Towards this end, we note that the VSR and Lorentz boost along the 3-axis are identical
\begin{equation}
\mathcal{L}(q)\vert_{q^{1}=q^{2}=0}=L(q)\vert_{q^{1}=q^{2}=0}
\end{equation}
but from eqs.~(\ref{eq:Lorentz_little_group_B}) and (\ref{eq:VSR_little_group_I}) we see that their corresponding little group elements are different
\begin{equation}
\mathcal{W}(\mathcal{L}(q)\vert_{q^{1}=q^{2}=0},p)=I,
\end{equation}
\begin{equation}
W(L(q)\vert_{q^{1}=q^{2}=0},p)=\exp\left(\frac{i}{2}\epsilon^{ij3}J^{i}\varphi^{j}_{p}\varphi^{3}_{q}\right).
\end{equation}
Therefore, massive particle states in Lorentz and VSR sector of non-zero spin cannot act mediators. By considering the particle state transformation of both sectors, we see that the only particle state that has the same transformation under VSR and Lorentz group is the massive scalar particle with $j=n=0$. The interactions between VSR and the SM sectors are therefore mediated by massive scalar particles.

In the SM, let us take the massive scalar particle to be the Higgs boson. 
Since the Higgs is neutral, the interacting Lagrangian density for Elko and Higgs with renormalisable interactions is
\begin{equation}
\mathscr{L}_{\Lambda\phi}=
-g_{1}\gdual{\Lambda}{\Lambda}\phi-g_{2}\gdual{\Lambda}{\Lambda}\phi^{2}.\label{eq:Elko_scalar1}
\end{equation}
where $\phi(x)$ is the Higgs field. Similarly for $\psi(x)$ constructed in sec.~\ref{vsr_spin_field} we have
\begin{equation}
\mathscr{L}_{\psi\phi}=
-h_{1}\overline{\psi}\psi\phi-h_{2}\overline{\psi}\psi\phi^{2}.\label{eq:Elko_scalar2}
\end{equation}

The interactions between the VSR and SM sector is now mediated by massive scalar particles. The general non-gravitational interacting mechanism is
\begin{equation}
\mbox{VSR interactions}\longleftrightarrow\mbox{Massive scalar particles}\longleftrightarrow\mbox{SM interactions}.\label{eq:mediator}
\end{equation}
Reading the equation from left to right, the VSR interactions produce the mediator which subsequently interact with the SM particles. This process is also reversible. Since the two sectors cannot interact directly, the probability of such a process must be calculated in two steps, namely
\begin{equation}
P(\mbox{VSR}\rightarrow\mbox{SM})=P(\mbox{VSR}\rightarrow\mbox{scalar})\times P(\mbox{scalar}\rightarrow\mbox{SM}).
\end{equation}

\subsection{Elko scattering amplitudes and adjoint}\label{Elko_interaction}

Here we present the relevant scattering amplitudes and cross-sections computed in app.~\ref{AppendixD}. The main focus of this section is on the definition of the Elko adjoint in which we have used to calculate the cross-sections. 

In sec.~\ref{elko_spinors_sec}, it is shown that the Elko dual 
\begin{equation}
\gdual{\xi}(\p,\mp1/2)=-i\xi^{\dag}(\p,\pm1/2)\gamma^{0},\hspace{0.5cm}
\gdual{\xi}(\p,\pm1/2)=i\xi^{\dag}(\p,\mp1/2)\gamma^{0},\nonumber
\end{equation}
\begin{equation}
\gdual{\zeta}(\p,\mp1/2)=-i\zeta^{\dag}(\p,\pm1/2)\gamma^{0},\hspace{0.5cm}
\gdual{\zeta}(\p,\pm1/2)=i\zeta^{\dag}(\p,\mp1/2)\gamma^{0}.\nonumber
\end{equation}
gives us a set of orthonormal and complete set of Elko spinors. We rewrite the dual as
\begin{equation}
\gdual{\xi}(\p,\alpha)=\xi^{\ddag}(\p,\alpha)\gamma^{0},\hspace{0.5cm}
\gdual{\zeta}(\p,\alpha)=\zeta^{\ddag}(\p,\alpha)\gamma^{0}.
\label{eq:ddag}
\end{equation}
The Elko adjoint has played an important role in uncovering the properties of Elko spinors and fields. Its introduction ensured the locality of the Elko fields and that the free Hamiltonian is positive-definite~\cite[sec.~7]{Ahluwalia:2004ab}.\footnote{The Elko fields would be non-local and the Hamiltonian non-physical had we used $\bar{\Lambda}(x)=\Lambda^{\dag}(x)\gamma^{0}$ instead of $\gdual{\Lambda}(x)$ for the Lagrangian density.} These results suggest, we should apply the Elko adjoint to the Elko fields when computing the transition probability.

For Elko scattering amplitudes, we take the standard definition of the $S$-matrix element for the process $\alpha\rightarrow\beta$ to be
\begin{equation}
S_{\beta\alpha}=-2\pi iM_{\beta\alpha}\delta^{4}(p_{\beta}-p_{\alpha}).
\end{equation}
Taking into account of the Elko adjoint, we propose that the transition probability to be proportional to
\begin{equation}
P(\alpha\rightarrow\beta)\propto |M_{\beta\alpha}|^{2}=(M_{\beta\alpha}^{\ddag})^{T}M_{\beta\alpha}
\label{eq:Elko_prob}
\end{equation}
instead of $M_{\beta\alpha}^{*}M_{\beta\alpha}$ where $*$ represents complex conjugation. The operator $T$ transposes the matrix and the operation $\ddag$ is  defined by eq.~(\ref{eq:ddag}).

The standard transition probability $M_{\beta\alpha}^{*}M_{\beta\alpha}$ is always positive definite. Under the new adjoint, this property may no longer be true. In the case of the $\Lambda_{1}+\Lambda_{2}\rightarrow\Lambda'_{1}+\Lambda'_{2}$ Elko self-interaction, eq.~(\ref{eq:avg_prob_self_interaction}) yields
\begin{eqnarray}
\frac{1}{4}\sum_{\alpha_{1},\alpha_{2},\alpha'_{1},\alpha'_{2}}|M_{(\Lambda'_{1}\Lambda'_{2})(\Lambda_{1}\Lambda_{2})}|^{2} &=&
\frac{1}{4}\left[\frac{g_{0}}{128\pi^{3}m^{2}\sqrt{p^{0}_{1}p^{0}_{2}p'^{0}_{1}p'^{0}_{2}}}\right]^{2}\nonumber\\
&&\times\Big\{32m^{4}\Big[-\cos(\phi_{1}-\phi_{2})+\cos(\phi_{1}-\phi'_{1})+\cos(\phi_{2}-\phi'_{1})\nonumber\\
&&\hspace{1.8cm}+\cos(\phi_{1}-\phi'_{2})+\cos(\phi_{2}-\phi'_{2})-\cos(\phi'_{1}-\phi'_{2})\nonumber\\
&&\hspace{1.8cm}+\cos(\phi_{1}+\phi_{2}-\phi'_{1}-\phi'_{2})\nonumber\\
&&\hspace{1.8cm}+\cos(\phi_{1}-\phi_{2}+\phi'_{1}-\phi'_{2})\nonumber\\
&&\hspace{1.8cm}+\cos(\phi_{1}-\phi_{2}-\phi'_{1}+\phi'_{2})+3\Big]\Big\}.
\end{eqnarray}
One can convince oneself that the right-hand side is positive definite by fixing $\phi_{1}$ to a particular value and plotting the three-dimensional graph with respect to $\phi'_{1}$ and $\phi'_{2}$ while varying $\phi_{2}$.
In the centre of mass frame, the cross-section is given by eq.~(\ref{eq:elko_self_cm})
\begin{equation}
\frac{d\sigma_{\mbox{\tiny{CM}}}}{d\Omega}(\Lambda_{1}\Lambda_{2}\rightarrow\Lambda'_{1}\Lambda'_{2})=\frac{g_{0}^{2}}{8\pi^{2}(p^{0}_{1})^{2}}\left[1+\cos^{2}(\phi_{1}-\phi'_{1})\right]\nonumber
\end{equation}
which is positive definite. Therefore, there are no difficulties at the tree-level for Elko self-interaction. However, for the Elko-Higgs interaction $\Lambda+\gdual{\Lambda}\rightarrow\phi$, we find the average probability to be
\begin{equation}
\frac{1}{2}\sum_{\alpha_{1},\alpha_{2}}|M_{(\phi'_{1})(\Lambda_{1}\gdual{\Lambda}_{2})}|^{2}
=\frac{g_{1}^{2}}{32\pi^{3}p^{0}_{1}p^{0}_{2}p'^{0}_{1}}
[\cos(\phi_{1}-\phi_{2})-1]\leq 0
\end{equation}
which is not physical. The transition probability should always be positive definite. Had we computed $M_{\beta\alpha}^{*} M_{\beta\alpha}$ instead of $(M^{\ddag})^{T}M$ the probability would indeed be positive definite. But this seems to be at odds with the use of $\ddag$ for Elko which until now has ensured that it has the desired physical properties.

In our opinion, the above results suggest that the definition of the adjoint for scattering amplitudes involving Elko must be revised and that it is premature to study the phenomenologies without first understanding the underlying symmetry. In sec.~\ref{causal_VSR}, we have shown that the Dyson series is not manifestly VSR-invariant since the  VSR coordinate transformations obtained from momentum transformations do not preserve the order of time-like events. Therefore, we must first determine the symmetries of Elko and the corresponding $S$-matrix before we can start computing the scattering amplitudes.

Generally, independent of the Elko coefficients, for a quantum field of the $(j,0)\oplus(0,j)$ representation, the definition of the Elko dual violates Lorentz symmetry. Assuming that the $u(\0,\alpha)$ and $v(\0,\alpha)$ coefficients satisfy the rotation constraints given by eqs.~(\ref{eq:rot1}) and (\ref{eq:rot2})
\begin{equation}
\sum_{\bar{\sigma}}u_{\ell}(\0,\bar{\sigma})\J_{\bar{\sigma}\sigma}=\sum_{\bar{\ell}}\mJ_{\ell\bar{\ell}}\,u_{\bar{\ell}}(\0,\sigma),
\end{equation}
\begin{equation}
\sum_{\bar{\sigma}}v_{\ell}(\0,\bar{\sigma})\J^{*}_{\bar{\sigma}\sigma}=-\sum_{\bar{\ell}}\mJ_{\ell\bar{\ell}}\,v_{\bar{\ell}}(\0,\sigma).
\end{equation}
It follows that their dual coefficients must satisfy eqs.~(\ref{eq:dual_rot_constraint1}) and (\ref{eq:dual_rot_constraint2}) which in this case become
\begin{equation}
\sum_{\bar{\alpha}}\bar{u}_{\ell}(\mathbf{0},\bar{\alpha})\J^{*}_{\bar{\alpha}\alpha}=\sum_{\ell}\bar{u}_{\bar{\ell}}(\mathbf{0},\alpha)\mJ_{\bar{\ell}\ell},\label{eq:Elko_dual_LV1}
\end{equation}
\begin{equation}
\sum_{\bar{\alpha}}\bar{v}_{\ell}(\mathbf{0},\bar{\alpha})\J_{\bar{\alpha}\alpha}=-\sum_{\ell}\bar{v}_{\bar{\ell}}(\mathbf{0},\alpha)\mJ_{\bar{\ell}\ell}\label{eq:Elko_dual_LV2}
\end{equation}
where $\bar{u}(\0,\alpha)=u^{\dag}(\0,\alpha)\Gamma$ is the Dirac dual with
\begin{equation}
\Gamma=\left(\begin{matrix}
O & I \\
I & O \end{matrix}\right)
\end{equation}
of dimension $2(2j+1)\times 2(2j+1)$. Since the constraints for the dual coefficients are derived from the constraints on the coefficients and these equations sum over the components of $\J$, there can be no different local phases for each dual coefficients. Since the Elko dual assigns a local phase for each $\alpha=\mp1/2,\pm1/2$, it violates Lorentz symmetry.
Incidentally, the above analysis can be seen as a further evidence that Elko satisfies VSR symmetry. The dual coefficients for the VSR fields are allowed to have local phases since the constraints on the coefficients only demand them to be eigenvectors of $\mathcal{J}^{3}$.

\section{Summary}

%

The Elko fields have positive-definite free Hamiltonians and the spinors form an orthonormal and complete set. Apart from the existence of the preferred direction and Lorentz violation, these properties suggest Elko to be a well-defined particle. Results obtained by AH suggest that Elko satisfies the VSR symmetry~\cite{Ahluwalia:2010zn}. Although we have not derived the Elko spinors from first principle using VSR symmetry, we are encouraged by their results that the axis of locality coincides with the preferred direction of the VSR algebra and the properties of Elko remain unchanged in VSR.

Based on the mass-dimension argument, Elko has a renormalisable self-interaction, so it is suggested that Elko is a dark matter candidate. But strictly speaking, these properties are desirable but not sufficient. For Elko to be a dark matter candidate, we must show that it has limited gauge interactions with the SM particles. To this end, we show that the VSR particles are dark matter candidates since they can only interact with the SM particles through massive scalar particles and gravity. So if Elko satisfies VSR symmetry, it automatically qualifies as a dark matter candidate.
\subsection{Future works}



The result obtained by AH is an important step towards understanding the symmetries of Elko. Although we are unable to derive Elko from the VSR group, we have shown that the VSR particles are dark matter candidate. 

The VSR particles and fields we have constructed in this chapter are derived from the irreducible representations of $ISIM(2)$, so they have a better mathematical foundation than Elko. In conjunction with our research on Elko, we hope to investigate their interactions both within the VSR sector and with the SM particles.

Before we can study the VSR phenomenologies, we need to solve the causality problem. In app.~\ref{AppendixC}, it is shown that the VSR coordinate transformation obtained from the transformation in the momentum space violates the causal structure of special relativity. Therefore, in order to use the Dyson series, we must find a different set of $SIM(2)$ transformation parameters that preserve the causality structure of special relativity so that the $S$-matrix is manifestly VSR-invariant.



While the Elko adjoint is needed to preserve locality and positivity of the free Hamiltonian, in app.~\ref{AppendixD}, we find the transition probability for the process $\gdual{\Lambda}+\Lambda\rightarrow \phi$ computed using the Elko adjoint is less than or equal to zero which is not physical. However, such a computation is not reliable for we have no complete knowledge of the underlying symmetry of Elko. But nevertheless, we expect this result to remain valuable for future investgation.

An important issue we have not discussed in detail is the physical interpretation of the preferred direction. Identifying the preferred direction for Elko was the first indication that it violates Lorentz symmetry and consequently led to the AH result. Currently we do not have a complete understanding of the physics of the preferred direction. Nevertheless, we suspect that the preferred direction to have non-trivial effects on the Elko scattering amplitudes since the Elko spin-sums contains a preferred plane. 

%

Ultimately, these problems converge to the need for a deeper understanding on the symmetry of Elko. It is conceivable that once this is achieved, we will be able to derive the Elko spinors and explain the origin of the Elko adjoint thus allowing us to compute positive-definite transition probability.

Finding the correct symmetry group would not only put Elko on a firm mathematical foundation, it would also allow us to study its gravitational interaction using the gauge-theoretic formalism~\cite{Kibble:1961ba,Hehl:1976kj,Blagojevic:2002du}. Such a task is necessary to ensure Elko has the correct gravitational interaction since Elko violates Lorentz symmetry.

\chapter{Conclusions} 
\label{Chapter5}
\lhead{Chapter 5. \emph{Conclusions}} 

The central theme of the thesis is space-time symmetries in quantum field theory and their applications in particle physics. 
Wigner~\cite{Wigner:1939cj} and Weinberg~\cite{Weinberg:1964cn,Weinberg:1964ev,Weinberg:1995mt,Weinberg:1965rz} showed that the particle states and the observables measured by experiments can be derived from first-principle by studying the representations and symmetries of the Poincar\'{e} group.

The formalism shows that, in the low energy limit, quantum field theory is the only known way that unifies quantum mechanics and special relativity and allows us to construct Poincar\'{e}-invariant $S$-matrix that satisfies the cluster decomposition principle. The fact that the SM particles are described by the particle states of this formalism reveals the deep connexion between space-time symmetries and quantum field theory.

The observation of neutrino oscillation and evidence for dark matter suggest that the SM, based on quantum field theory with Poincar\'{e} symmetry may be incomplete. Our work on the massive vector fields and Elko are motivated by these problems.

\section{The massive vector fields}



Generally, the axioms of quantum mechanics allow for the existence of states that are  linear combination of the most primitive states that are eigenstates with respect to some Hermitian operators. The phenomena of neutrino oscillation shows that the principle of superposition applies to mass eigenstates whose eigenvalues are determined by the first Casimir operator $C_{1}$ of the Poincar\'{e} algebra. Extending this observation, it is logical to investigate the possibility of states that are linear combination of eigenstates of $C_{2}$.

In this thesis, we have realised this possibility by suggesting that the most general massive vector fields should have both scalar and spin-one degrees of freedom. We showed that the additional scalar degree of freedom of the massive vector fields does not violate Poincar\'{e} symmetry. In fact, by choosing the appropriate dual coefficients, the resulting vector fields have the Veltman propagator and unitarity is preserved in scattering processes for the $W^{\pm}$ and $Z$ bosons.

On a broader picture, the important question is what are the fundamental properties of an elementary particle? The work of Wigner suggested that elementary particles can be uniquely defined by their mass and spin in terms of the Casimir operators of the Poincar\'{e} group. However, the discovery of neutrino oscillation forces us to revise this definition. In particular, it poses the possibility that the kinematical properties of the elementary particles may be intrinsically quantum mechanical in the sense that they may not have definite mass nor spin~\cite{Ahluwalia:1997hc}. 

In order to construct particle states with indefinite spin, one must say a few words about the topology of the Poincar\'{e} group.\footnote{See~\cite[sec.2.7]{Weinberg:1995mt} for a nice dicussion.} The Poincar\'{e} group is topologically the same as $R^{4}\times SL(2,C)/Z_{2}$ where $R^{4}$ is the real four-dimensional flat space, $Z_{2}=\{+1,-1\}$ and
\begin{equation}
SL(2,C)=\{M\in GL(2,C)|\mbox{det}M=1\}.
\end{equation}
This group is not simply connected, in fact it is doubly connected. As a result, the product rule is intrinsically projective
\begin{equation}
U(\bar{\Lambda},\bar{a})U(\Lambda,a)=\pm U(\bar{\Lambda}\Lambda,\bar{\Lambda}a+\bar{a})
\end{equation}
where the top and bottom sign correspond to bosonic and fermionic representations. It follows that the inner product of states that are linear combination of fermionic and bosonic states would not be Poincar\'{e}-invariant thus imposing a superselection rule forbidding the existence of states with indefinite spin. 

The superselection rule originates from the fact that the topology of the Lorentz group is $SL(2,C)/Z_{2}$ which is doubly connected. However, one can choose to work with $SL(2,C)$ instead of $SL(2,C)/Z_{2}$. This extension does not violate Lorentz symmetry and preserves the existing physics. The only new physics is that it now allows for superposition of fermionic and bosonic states since $SL(2,C)$ is simply connected.

Applying the principle of superposition, we propose that the most general elementary particle state is a linear combination of states of different mass and spin. Once we have acquired a better understanding of neutrinos and the massive vector fields, the next task is to explore the physics of $|\rho,\lambda\rangle$ given by eq.~(\ref{eq:most_general_particle})
\begin{equation}
|\rho,\lambda\rangle=\sum_{ij,k}A_{\rho i}B_{\lambda j}|m_{i},s_{j};\p_{k}\rangle.
\end{equation}

\section{The dark matter problem}

The accumulation of data from astrophysical and cosmological observations since Oort and Zwicky have made the existence of dark matter a well established fact. The agreement between observation and $\Lambda$CDM model with respect to the cosmic microwave background is one of the strongest indication for dark matter~\cite{Komatsu:2010fb}. Even in models of inhomogeneous cosmology where dark energy is absent, the amount of dark matter still dominates over baryonic matter~\cite{Wiltshire:2007jk,Wiltshire:2007fg}.

Despite having limited knowledge of dark matter, the fact that they have limited interactions with the SM particles is sufficient for theorists to propose various dark matter candidates. As the first test, these theories must be able to explain the limited interactions between the dark and the SM sector. Invariably, the majority explanations appeal to the principle of local gauge invariance, that these particles do not participate in gauge interactions due to the neutrality of charges.

The gauge theoretic approach is theoretically attractive. In addition to solve the dark matter problem, they also provide extensions to the SM. But this is not the only possibility. Our research on Elko and subsequent investigation on VSR have led us to consider the possibility that dark matter need not satisfy the Poincar\'{e} symmetry. Furthermore, no present experiments have shown that dark matter must satisfy Poincar\'{e} symmetry.

Effectively, we are proposing a new formalism to study dark matter based on the hypothesis that the space-time symmetries of dark matter may not be Poincar\'{e}. Since local gauge symmetry is a consequence of Lorentz symmetry, depending on the properties of the symmetry group of dark matter, they may no longer admit gauge symmetry of the SM thus limiting the interactions between the two sectors.

Our results on particles and quantum fields with VSR symmetry is a realisation of the our hypothesis. Although we have not derived the Elko spinors from first-principle, we have shown that the SM and VSR particles can only interact through massive scalar particles and gravity thus making the VSR particles dark matter candidates.

%
%

\section{Outlook}

Although the massive vector fields and Elko are not directly related, both theories extend the existing knowledge of quantum field theory and the SM. For the massive vector field, we realised the possibility of particles having indefinite spin. For Elko, given that it violates Lorentz symmetry and is a dark matter candidate suggests that dark matter may not satisfy Lorentz symmetry. For the last part of the thesis, it is later possibility that we which to elaborate on. This can potentially not only open a new field of research for dark matter but at the same time provide a deeper understanding on the connexion between space-time symmetry and quantum field theory.

The real power of the formalism of Wigner and Weinberg lies in their generality. In principle, given an appropriate symmetry group, we can derive the properties of the particle states and quantum fields by studying its representations and symmetries. Therefore, it provides us a systematic way to construct theories that are Lorentz violating, but are protected by different symmetries. 

Here, we are proposing the possibility that the universe is populated by particles satisfying different space-time symmetries. In this scenario, since particles of different symmetry can only interact through mediators, it provides a natural explanation for dark matter.\footnote{The precise meaning of mediator is defined in sec.~\ref{Elko_DM_candidate}} 
More importantly, if our hypothesis is true, it would have a much wider implication in physics. This would allow us to address the fundamental question whether our perception of space-time is observer dependent. Specifically in the low energy limit, where the effects of gravity are negligible, is Poincar\'{e} symmetry relative? Currently, we do not have a definite answer to this question. But given that we are able to derive physical particle states from $ISIM(2)$, a subgroup of the Poincar\'{e} group that are distinct from the SM particles, the initial answer seems to be affirmative.

%
%



Generally, this research program can be seen as an extension to the view of Brown~\cite{Brown:2005} that the space-time symmetry we perceive is a reflection of the symmetries of the SM particles since our experimental apparatus are built from the SM particles. One may find such a view to be circular since the theory must be self-consistent, which means that the symmetries of the SM particles and space-time must coincide.

In our opinion, this criticism is not completely true and that the symmetries of the particles has precedence over the symmetries of classical space-time. We note, our perception of space and time are purely based on classical physics. According to classical physics, at the macroscopic scale (relative to us), space-time coordinates and the energy-momentum vectors are observables. However, as we probe shorter distance at higher energy, at some point there is a transition from classical physics to quantum mechanics and subsequently to quantum field theory. At each stage of the transition, there is a reduction in the number of observables. 

From classical physics to quantum mechanics, the spatial-coordinate remains an observable but time becomes a parameter. Preceding to quantum field theory, both the spatial and temporal coordinates become parameters. Therefore, at the scale where quantum field theory is required, space and time are no longer observables. On the other hand, energy and momentum remain to be observables.

The classical picture of a deterministic space-time can be recovered from quantum field theory by appealing to the Ehrenfest theorem of quantum mechanics (after taking the non-relativistic limit of the Dirac equation), where the classical equation of motion emerges in terms of expectation values
\begin{equation}
m\frac{d^{2}\langle \x\rangle}{dt^{2}}=-\langle\nabla V(\x,t)\rangle\label{eq:Ehrenfest}
\end{equation}
where $m$ is the mass of the particle and $V(x,t)$ is the interaction potential. Rigorously speaking, the classical equation of motion is only an approximation of eq.~(\ref{eq:Ehrenfest}).\footnote{See~\cite[chap.~3.D.1.d and chap.~3 complement $\mbox{G}_{\mbox{\tiny{III}}}$]{CohenTannoudji} for discussions on the Ehrenfest theorem.} The important thing to note is that the quantity $\langle\x\rangle$ on left-hand side of eq.~(\ref{eq:Ehrenfest}) is what we associate as the trajectory of a particle in classical mechanics. The Ehrenfest theorem therefore shows that the spatial-coordinate in classical mechanics is an emergent phenomena from quantum mechanics.

The fact that space and time play a secondary role in quantum field theory can be anticipated from the Poincar\'{e} algebra since the algebra does not include an explicit space-time operator $X^{\mu}$. The only explicit observable is given by the energy-momentum  operator $P^{\mu}$ and angular momentum operator $\J$. Together, these operators with the Casimir operators of the Poincar\'{e} algebra provide a natural framework to define physical particle states. Observables such as cross-sections and decay rates can then be computed using the $S$-matrix. However, it is important to note that the $S$-matrix formalism is only applicable to theories with local interactions which allow us to define asymptotic free particle states long before  and after the interactions. In the presence of external gravitational fields and background with finite-temperature, we can no longer define the asymptotic free states~\cite{Weinberg:1996kw}. 

It is also worthwhile to note that in the research area of non-commutative space-time, the \textit{space-time} operator $X^{\mu}$ is often introduced as an additional operator to the Poincar\'{e} algebra with a generalised Heisenberg algebra~\cite{Snyder:1946qz,Yang:1947ud,VilelaMendes:1999xv}. In these theories, the resulting algebras become unstable so one needs to introduce deformations to find a stable algebra.\footnote{A unstable Lie algebra is one where small perturbation of the structure constant yields a different Lie algebra. Conversely a stable Lie algebra is invariant under such perturbation.} Such a stable algebra has been found by Chryssomalakos and Okon and is called the stabilised Poincar\'{e} Heisenberg algebra (SPHA)~\cite{Chryssomalakos:2004gk}. However, in the SPHA, the operator $X^{\mu}$ loses its original interpretation as a space-time operator.

This thesis focuses on symmetry groups where particle states are well-defined. We did not study the physics in the presence non-trivial backgrounds and the possibility of non-commutative space-time. While these issues are important, there remains open problems in this thesis and new possibility to explore within the present scenario. In particular, if classical space-time is indeed an emergent phenomena, it is not clear whether one should introduce a space-time operator. Using existing physics, our arguments suggest that the momentum space which defines the kinematics of the particle states is more fundamental than the classical space-time although they obey the same symmetry. 

Given that the classical space-time is an emergent phenomena, it is less objectionable to hypothesise that space-time symmetry is observer-dependent and propose particle states whose symmetries are not dictated by the Poincar\'{e} group. Depending on the properties of the symmetry groups, the macroscopic space-time that emerges from the resulting particle states is likely to differ from space-time described by special relativity. This does not contradict with observation so long as these particles have limited interactions with the SM particles. However, it may be difficult to justify the choice of symmetry groups. Therefore, it is instructive to start from the Poincar\'{e} group or other well-known symmetry groups such as the (anti) de Sitter group and study the symmetries and representations of their subgroups.


\appendix


\chapter{The existence of $t^{\mu_{1}\mu_{2}\cdots\mu_{2j}}$}
\label{AppendixE}
\lhead{Appendix A. \emph{The existence of $t^{\mu_{1}\mu_{2}\cdots\mu_{2j}}$}}

We reproduce two theorems proven by Weinberg in~\cite[app.~A]{Weinberg:1964cn}.

\textit{Theorem 1.} There exists a rank $2j$ tensor $t^{\mu_{1}\mu_{2}\cdots\mu_{2j}}$ in the $(0,j)$ representation of the Lorentz group such that
\begin{enumerate}
\item $t$ is symmetric in all $\mu$'s
\item $t$ is traceless in all $\mu$'s, that is $\eta_{\mu_{1}\mu_{2}}t^{\mu_{1}\mu_{2}\cdots\mu_{2j}}=0$
\item $t$ is a tensor in the sense that
\begin{equation}
\mathcal{D}^{(j)}(\Lambda)t^{\mu_{1}\mu_{2}\cdots\mu_{2j}}\mathcal{D}^{(j)\dag}(\Lambda)=
\Lambda_{\nu_{1}}^{\,\,\mu_{1}}\Lambda_{\nu_{2}}^{\,\,\mu_{2}}\cdots\Lambda_{\nu_{2j}}^{\,\,\mu_{2j}}
t^{\nu_{1}\nu_{2}\cdots\nu_{2j}}.
\end{equation}
\end{enumerate}
\textit{Proof.} Let $u_{a}$ be a $2j+1$ dimensional basis of the $(0,j)$ representation that transforms as
\begin{equation}
u'_{a}=\sum_{\bar{a}}\mathcal{D}^{(j)}_{\bar{a}a}(\Lambda)u_{\bar{a}}.
\end{equation}
Therefore, the product $u_{a}u_{b}^{*}$ is a basis of dimension $(2j+1)\times(2j+1)$ and transforms under the $(j,0)\otimes(0,j)=(j,j)$ representation. This representation consists of all the symmetric and traceless tensors of rank $2j$. 

The tensor $T^{\mu_{1}\mu_{2}\cdots\mu_{2j}}$ of interest can be constructed as a linear combination $u_{a}u_{b}^{*}$
\begin{equation}
T^{\mu_{1}\mu_{2}\cdots\mu_{2j}}=\sum_{ab}t^{\mu_{1}\mu_{2}\cdots\mu_{2j}}_{ab}u_{a}u_{b}^{*}.\label{eq:tensorT}
\end{equation}
By definition, $T^{\mu_{1}\mu_{2}\cdots\mu_{2j}}$ must transform as
\begin{equation}
T'^{\mu_{1}\mu_{2}\cdots\mu_{2j}}=\Lambda_{\nu_{1}}^{\,\,\mu_{1}}\Lambda_{\nu_{2}}^{\,\,\mu_{2}}\cdots\Lambda_{\nu_{2j}}^{\,\,\mu_{2j}}
T^{\nu_{1}\nu_{2}\cdots\nu_{2j}}.\label{eq:tensorT_transform}
\end{equation}
Substituting eq.~(\ref{eq:tensorT_transform}) into eq.~(\ref{eq:tensorT}), we obtain
\begin{equation}
\mathcal{D}^{(j)}(\Lambda)t^{\mu_{1}\mu_{2}\cdots\mu_{2j}}\mathcal{D}^{(j)\dag}(\Lambda)=
\Lambda_{\nu_{1}}^{\,\,\mu_{1}}\Lambda_{\nu_{2}}^{\,\,\mu_{2}}\cdots\Lambda_{\nu_{2j}}^{\,\,\mu_{2j}}
t^{\nu_{1}\nu_{2}\cdots\nu_{2j}}.
\end{equation}
Since $T^{\mu_{1}\mu_{2}\cdots\mu_{2j}}$ is symmetric and traceless, it follows that $t^{\mu_{1}\mu_{2}\cdots\mu_{2j}}$ must also be symmetric and traceless. This completes the proof.

\textit{Theorem 2.} Let $\Pi^{(j)}(p)=(-1)^{2j}t^{\mu_{1}\mu_{2}\cdots\mu_{2j}}p_{\mu_{1}}p_{\mu_{2}}\cdots p_{\mu_{2j}}$ be matrix. When $p^{\mu}p_{\mu}=m^{2}$ and $p^{0}>0$, we have the following identity
\begin{equation}
\mathcal{D}^{(j)}(L(p))\mathcal{D}^{(j)\dag}(L(p))=\exp(-2\J\cdot\bv)=\frac{\Pi^{(j)}(p)}{m^{2j}}.\label{eq:Pi_t}
\end{equation}

\textit{Proof.} The transformation of $\Pi(p)$ is
\begin{equation}
\mathcal{D}^{(j)}(\Lambda)\Pi(p)\mathcal{D}^{(j)\dag}(\Lambda)=\Pi(\Lambda p).
\end{equation}
Let us take $p=k=(m,\0)$ to be at rest and $\Lambda=R=\exp(i\J\cdot\th)$ a rotation. Since $k$ is rotation invariant, we get
\begin{equation}
\mathcal{D}^{(j)}(R)\Pi(k)\mathcal{D}^{(j)\dag}(R)=\Pi(k)
\end{equation}
which amounts to
\begin{equation}
[\J,t^{00\cdots0}]=0.
\end{equation}
Since $\J$ furnishes an irreducible representation, by Schur's Lemma, the matrix must be proportional to the identity matrix. We choose $t^{00\cdots0}=(-1)^{2j}I$ and $\Pi^{(j)}(k)=m^{2j}I$.

Now let us take $\Lambda=L(p)$ to be a boost such that $L(p)k=p$. Therefore, we get
\begin{equation}
\mathcal{D}^{(j)}(L(p))\mathcal{D}^{(j)\dag}(L(p))=\frac{\Pi^{(j)}(p)}{m^{2j}}.
\end{equation}
This completes the proof.

\textit{Some examples.} We provide explicit expression of the first two $\gamma$-matrices.
\begin{equation}
\hspace{-1cm}j=\frac{1}{2}:\hspace{0.5cm}t_{0}=I,\hspace{0.5cm} t_{i}=-\sigma_{i}
\end{equation}
\begin{eqnarray}
j=1:&& t^{00}=I, \hspace{0.5cm}
t^{0i}=t^{i0}=J^{i} \nonumber \\
&& t^{ij}=t^{ji}=\eta^{ij}I+\left\{J^{i},J^{j}\right\}
\end{eqnarray}
where $\sigma_{i}$ are the Pauli-matirces and $J_{i}$ are given by
\begin{equation}
J^{1}=\frac{1}{\sqrt{2}}
\left(\begin{array}{cccc}
 0 & 1 & 0 \\
 1 & 0 & 1 \\
 0 & 1 & 0  \end{array}\right),\hspace{0.5cm}
J^{2}=\frac{1}{\sqrt{2}}
\left(\begin{array}{cccc}
 0 &-i & 0 \\
 i & 0 &-i \\
 0 & i & 0 \end{array}\right),\hspace{0.5cm}
J^{3}=
\left(\begin{array}{cccc}
 1 & 0 & 0 \\
 0 & 0 & 0 \\
 0 & 0 &-1 \end{array}\right).
\end{equation}
Generally, the matrices $t^{\mu_{1}\cdots\mu_{2j}}$ can be determined by explicitly computing $\exp(-2\J\cdot\bv)$,
\begin{equation}
\exp(-2\J\cdot\bv)=\cosh(2\J\cdot\bv)-\sinh(2\J\cdot\bv).
\end{equation}
Let $\eta=2\J\cdot\bv$, we get the following expansions~\cite[eqs. (A.406-A.409)]{Ahluwalia:1991gs}

\textit{Integer spin: $j=1,2,\cdots$}
\begin{equation}
\cosh(\eta)=I+\sum_{n=0}^{j-1}\frac{(\eta)^{2}(\eta^{2}-2^{2}I)(\eta^{2}-4^{2}I)\cdots (\eta^{2}-(2n)^{2}I)}{(2n+2)!}\sinh^{2n+2}\varphi
\end{equation}
\begin{equation}
\sinh(\eta)=\eta\cosh\varphi\sum_{n=0}^{j-1}\frac{(\eta^{2}-2^{2}I)(\eta^{2}-4^{2}I)\cdots (\eta^{2}-(2n)^{2}I)}{(2n+1)!}\sinh^{2n+1}\varphi
\end{equation}

\textit{Half-integer spin: $j=\frac{1}{2},\frac{3}{2},\cdots$}
\begin{equation}
\cosh(\eta)=\cosh\varphi\left[I+\sum_{n=0}^{j-1/2}\frac{(\eta-I^{2})(\eta^{2}-3^{2}I)\cdots (\eta^{2}-(2n-1)^{2}I)}{(2n)!}\sinh^{2n}\varphi\right]
\end{equation}
\begin{equation}
\sinh(\eta)=\eta\sinh\varphi\left[I+\sum_{n=0}^{j-1/2}\frac{(\eta-I^{2})(\eta^{2}-3^{2}I)\cdots (\eta^{2}-(2n-1)^{2}I)}{(2n+1)!}\sinh^{2n}\varphi\right]
\end{equation}
Using the above expansions and the definition of the rapidity parameter, we see that $\exp(-2\J\cdot\bv)$ is a polynomial in $p^{\mu}$ of order $2j$ with matrices $t^{\mu_{1}\cdots\mu_{2j}}$ as coefficients.

\chapter{General free field propagator}
\label{AppendixB}
\lhead{Appendix B. \emph{General free field propagator}}

We compute the most general free-field propagator. Let operator $S(y,x)$ where $x=(t,\x)$ and $y=(t',\y)$ be the propagator for a general field $\psi(x)$
\begin{equation}
\psi(x)=(2\pi)^{-3/2}\int\frac{d^{3}p}{\sqrt{2p^{0}}}\sum_{\sigma}\left[e^{-ip\cdot x}u(\p,\sigma)a(\p,\sigma)+e^{ip\cdot x}v(\p,\sigma)b^{\dag}(\p,\sigma)\right]
\end{equation}
and its adjoint $\gdual{\psi}(x)$
\begin{equation}
\gdual{\psi}(x)=(2\pi)^{-3/2}\int\frac{d^{3}p}{\sqrt{2p^{0}}}\sum_{\sigma}\left[e^{ip\cdot x}\dual{u}(\p,\sigma)a^{\dag}(\p,\sigma)+e^{-ip\cdot x}\dual{v}(\p,\sigma)b(\p,\sigma)\right].
\end{equation}
The propagator $S(y,x)$ is defined as
\begin{eqnarray}
S(y,x)&=&\langle\,\,|T[\psi(x)\gdual{\psi}(y)]|\,\,\rangle \nonumber \\
&=&\theta(t-t')\langle\,\,|\psi(x)\gdual{\psi}(y)|\,\,\rangle\pm\theta(t'-t)\langle\,\,|\gdual{\psi}(y)\psi(x)|\,\,\rangle
\end{eqnarray}
where $T$ denotes the time-ordered product and $\theta(t)$ is the step function
\begin{equation}
\theta(t)=\begin{cases}
1 & t\geq 0 \\
0 & t<0
\end{cases}.
\end{equation}
The top and bottom signs are for the bosonic and fermionic fields respectively. 

Computing the vacuum expectation values, the non-vanishing matrix elements are
\begin{equation}
\langle\,\,|\psi(x)\gdual{\psi}(y)|\,\,\rangle=(2\pi)^{-3}\int\frac{d^{3}p}{2p^{0}}e^{-ip\cdot(y-x)}N(\p),
\end{equation}
\begin{equation}
\langle\,\,|\gdual{\psi}(y)\psi(x)|\,\,\rangle=(2\pi)^{-3}\int\frac{d^{3}p}{2p^{0}}e^{ip\cdot(y-x)}M(\p)
\end{equation}
where
\begin{equation}
N(\p)=\sum_{\sigma}u(\p,\sigma)\dual{u}(\p,\sigma),
\end{equation}
\begin{equation}
M(\p)=\sum_{\sigma}v(\p,\sigma)\dual{v}(\p,\sigma).
\end{equation}
Using the integral representation of the step function
\begin{equation}
\theta(t)=\lim_{\epsilon\rightarrow0^{+}}\int\frac{d\omega}{2\pi i}
\frac{e^{i\omega t}}{\omega-i\epsilon}
\end{equation}
the propagator can be written as
\begin{eqnarray}
S(y,x)&=&\int\frac{d\omega d^{3}p}{2\pi i}\frac{1}{2p^{0}}
\frac{e^{i(\omega-p^{0})(t'-t)}e^{-i\mathbf{p}\cdot(\mathbf{x-y})}}{\omega-i\epsilon}N(p) \nonumber \\
&&\pm\int\frac{d\omega d^{3}p}{2\pi i}\frac{1}{2p^{0}}
\frac{e^{i(\omega-p^{0})(t-t')}e^{-i\mathbf{p}\cdot(\mathbf{y-x})}}{\omega-i\epsilon}M(p).
\end{eqnarray}
Perform a change of integration variable, $\omega=p^{0}-q^{0}$ and $\p=\mathbf{q}$,
\begin{eqnarray}
S(y,x)&=&\int\frac{d^{4}q}{(2\pi)^{4}i}\frac{1}{(2\sqrt{|\mathbf{q}|^{2}+m^{2}})}
\frac{e^{-iq\cdot(y-x)}N(\mathbf{q})}{\sqrt{|\mathbf{q}|^{2}+m^{2}}-q^{0}-i\epsilon}\nonumber\\
&&\pm\int\frac{d^{4}q}{(2\pi)^{4}i}\frac{1}{(2\sqrt{|\mathbf{q}|^{2}+m^{2}})}
\frac{e^{-iq\cdot(x-y)}M(\mathbf{q})}{\sqrt{|\mathbf{q}|^{2}+m^{2}}-q^{0}-i\epsilon}.
\end{eqnarray}
It is important to note the difference between $p^{0}$ and $q^{0}$. The energy $p^{0}$ is
on the mass-shell satisfying the standard dispersion relation
\begin{equation}
p^{0}=\sqrt{|\q|^{2}+m^{2}}
\end{equation}
whereas $q^{0}$ is off the mass-shell, not constrained by any dispersion relations.
The integration for $q^{0}$ is unconstrained and integration measure is defined as
$d^{4}q=dq^{0}d^{3}q$. 

The two terms in $S(y,x)$ can be combined by changing $q^{0}\rightarrow-q^{0}$ and
$\mathbf{q}\rightarrow-\mathbf{q}$ in the second term thus giving us the general 
free-field propagator
\begin{equation}
S(y,x)=i\int\frac{d^{4}q}{(2\pi)^{4}}\frac{e^{-iq\cdot(y-x)}}{(2\sqrt{{|\mathbf{q}|^{2}+m^{2}}})}
\left[\frac{\sqrt{{|\mathbf{q}|^{2}+m^{2}}}[N(\mathbf{q})\pm M(-\mathbf{q})]+q^{0}[N(\mathbf{q})\mp M(-\mathbf{q})]}{q^{\mu}q_{\mu}-m^{2}+i\epsilon}\right].\label{eq:g_prop}
\end{equation}

\chapter{VSR and causality violation}
\label{AppendixC}
\lhead{Appendix C. \emph{VSR and causality violation}}

The VSR boost that takes $k=(m,\mathbf{0})$ to $p=(p^{0},\p)$ is given by
\begin{equation}
\mathcal{L}(p)=\left(\begin{array}{cccc}
\frac{p^{0}}{m} & \frac{p^{1}}{p^{0}-p^{3}} & \frac{p^{2}}{p^{0}-p^{3}} & \frac{m^{2}-p^{0}(p^{0}-p^{3})}{m(p^{0}-p^{3})} \\
\frac{p^{1}}{m} & 1 & 0 & -\frac{p^{1}}{m} \\
\frac{p^{2}}{m} & 0 & 1 & -\frac{p^{2}}{m} \\
\frac{p^{3}}{m} & \frac{p^{1}}{p^{0}-p^{3}} & \frac{p^{2}}{p^{0}-p^{3}} & \frac{m^{2}-p^{3}(p^{0}-p^{3})}{m(p^{0}-p^{3})}
\end{array}\right).
\end{equation}
One would assume that the following substitution
\begin{equation}
p^{0}=m\gamma,\hspace{0.5cm}
\p=m\gamma\mathbf{u}
\end{equation}
where $\mathbf{u}=(u^{1},u^{2},u^{3})$ is the velocity and $\gamma=(\sqrt{1-|\mathbf{u}|^{2}})^{-1/2}$ gives the corresponding coordinate transformation
\begin{equation}
\mathcal{L}(\mathbf{u})=\left(\begin{matrix}
\gamma & \frac{u^{1}}{1-u^{3}} & \frac{u^{2}}{1-u^{3}} & \gamma\left(\frac{u^{3}-|\mathbf{u}|^{2}}{1-u^{3}}\right)\\
\gamma u^{1} & 1 & 0 & -\gamma u^{1} \\
\gamma u^{2} & 0 & 1 & -\gamma u^{2} \\
\gamma u^{3} &\frac{u^{1}}{1-u^{3}} & \frac{u^{2}}{1-u^{3}} & \gamma\left(\frac{1-|\mathbf{u}|^{2}}{1-u^{3}}+u^{3}\right)\end{matrix}\right).
\label{eq:vsr_coord_trans}
\end{equation}

However, under such a transformation the VSR and special relativity would have different causal structure. That is, for certain cases, the order of events separated by time-like interval are no longer absolute under eq.~(\ref{eq:vsr_coord_trans}).

%
Consider a time-like interval where $0<|\x-\y|<y^{0}-x^{0}$. Under a general VSR boost, we get
\begin{equation}
(y')^{0}-(x')^{0}=\gamma (y^{0}-x^{0})+\frac{u^{1}}{1-u^{3}}(y^{1}-x^{1})+
\frac{u^{2}}{1-u^{3}}(y^{2}-x^{2})+\gamma\frac{u^{3}-|\mathbf{u}|^{2}}{1-u^{3}}(y^{3}-x^{3}).
\end{equation}
We parametrise the velocity using spherical polar coordinate
\begin{equation}
u^{1}=|\mathbf{u}|\cos\phi\sin\theta,\hspace{0.5cm}
u^{2}=|\mathbf{u}|\sin\phi\sin\theta,\hspace{0.5cm}
u^{3}=|\mathbf{u}|\cos\theta.
\end{equation}
We make the following choice $y^{1}-x^{1}=y^{2}-x^{2}>0$, $y^{3}=x^{3}=0$ so we get
\begin{equation}
(y')^{0}-(x')^{0}=\gamma(y^{0}-x^{0})+\left[\frac{|\mathbf{u}|\sin\theta(\cos\phi+\sin\phi)}{1-|\mathbf{u}|\cos\theta}\right](y^{1}-x^{1}).
\end{equation}
Let $|y^{1}-x^{1}|=\alpha (y^{0}-x^{0})$ where $0<\alpha<1$. This gives us
\begin{equation}
y^{1}>x^{1}:\,(y')^{0}-(x')^{0}=
\left[\gamma+\alpha\frac{|\mathbf{u}|\sin\theta(\cos\phi+\sin\phi)}{1-|\mathbf{u}|\cos\theta}\right](y^{0}-x^{0}),
\end{equation}
\begin{equation}
y^{1}<x^{1}:\,(y')^{0}-(x')^{0}=
\left[\gamma-\alpha\frac{|\mathbf{u}|\sin\theta(\cos\phi+\sin\phi)}{1-|\mathbf{u}|\cos\theta}\right](y^{0}-x^{0}).
\end{equation}
Since $y^{0}>x^{0}$, the ordering of events depend on the sign of the term in the bracket. Plotting this term against the angle $\theta$ and $\phi$ with 
specific values of $\alpha$ and $|\mathbf{u}|$ allow us to see whether ordering of time-like events are preserved. 

\begin{flushleft}
\begin{figure}[!htp]
\begin{tabular}{c}
\subfigure[ $y^{1}>x^{1}$ ]
{
\begin{minipage}{\textwidth}
\begin{center}
\includegraphics[scale=1]{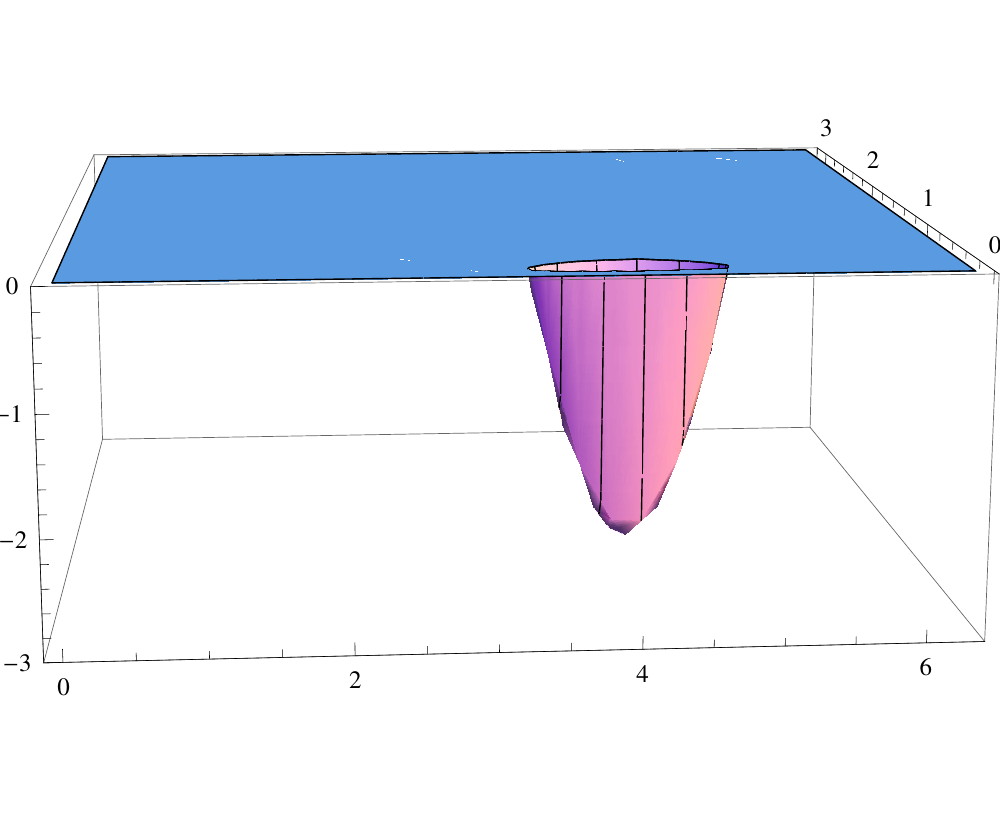}\label{fig:causal1}
\vspace{1cm}
\end{center}
\end{minipage}
}%
\\
\subfigure[ $y^{1}<x^{1}$ ]
{
\begin{minipage}{\textwidth}
\begin{center}
\includegraphics[scale=1]{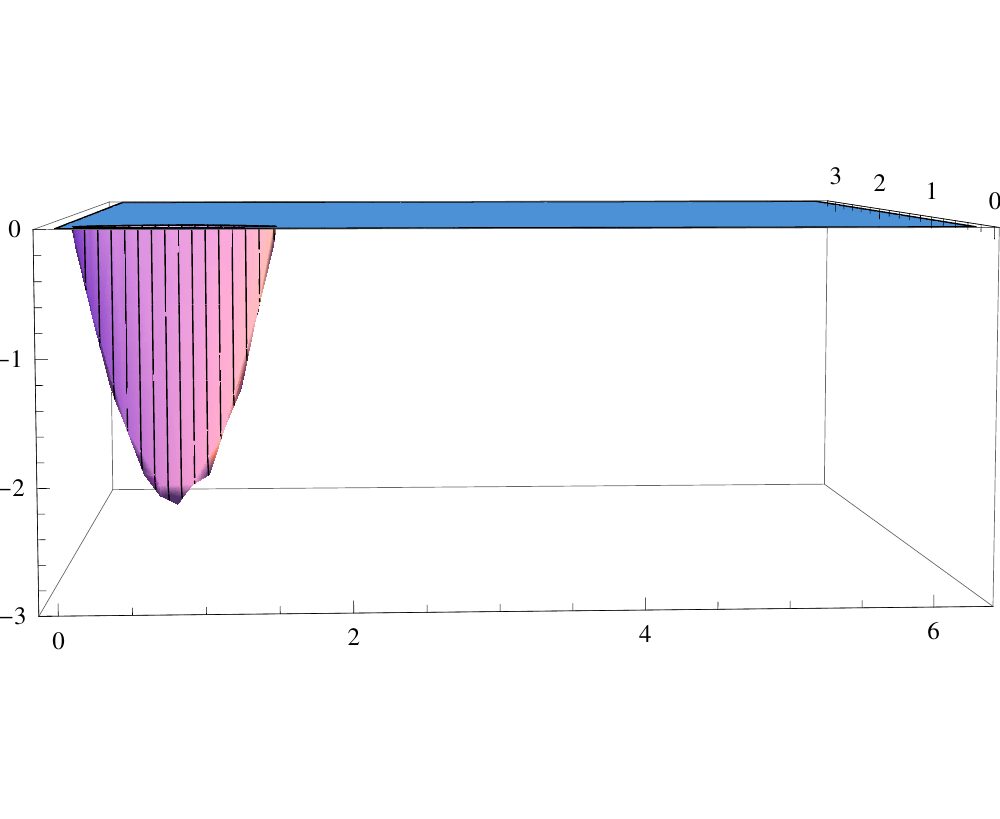}\label{fig:causal2}
\vspace{1cm}
\end{center}
\end{minipage}
}%
\\
\end{tabular}
\caption{$\alpha=0.95$, $|\mathbf{u}|=0.99$. In both cases, there are regions where the order of events are reversed. To make the violations apparent, the plot range for the vertical axis ranges from -3 to 0.}
\end{figure}
\end{flushleft}

In figs.\ref{fig:causal1} and \ref{fig:causal2}, we took $\alpha=0.95$ and $|\mathbf{u}|=0.99$.  In both cases, for certain domains of $\theta$ and $\phi$ the function is negative, so order of events are reversed. Therefore, if we want to use the Dyson series to compute scattering amplitudes in VSR, we must find an alternative set of parameters for VSR coordinate transformation such that it has identical causal structure as special relativity.


\chapter{Elko interactions}
\label{AppendixD}
\lhead{Appendix D. \emph{Elko phenomenologies}}

We divide the Elko phenomenologies into two parts. The first part is on the the self-interaction and the second part is on the Elko interaction with the Higgs boson.

\section{Elko self-interaction}

The self-interaction for Elko has a similar form to the $\phi^{4}$-theory. The point of departure between the two theory is that Elko is a four-component field with non-trivial expansion coefficients so the differential cross-section is expected to have angular dependence.

\begin{figure}
\begin{center}
  \includegraphics[scale=0.7]{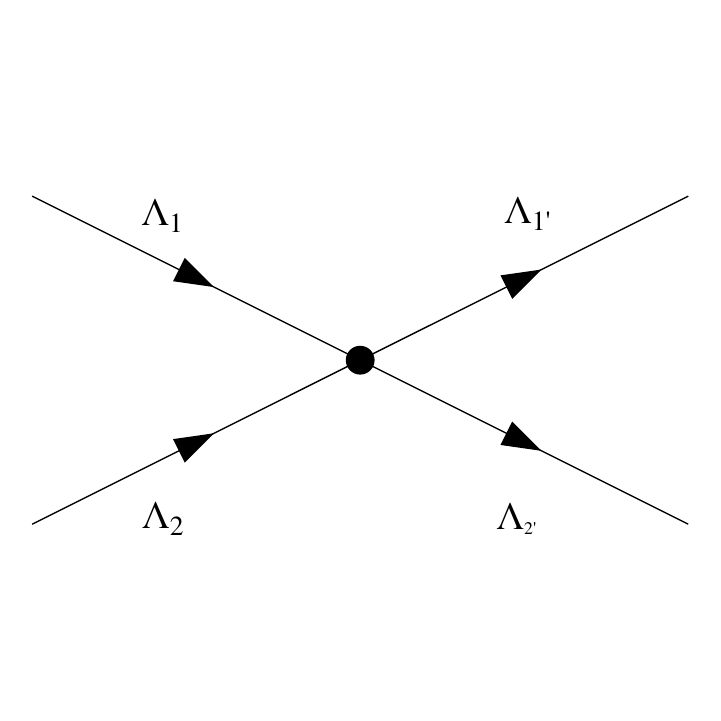}
  \includegraphics[scale=0.7]{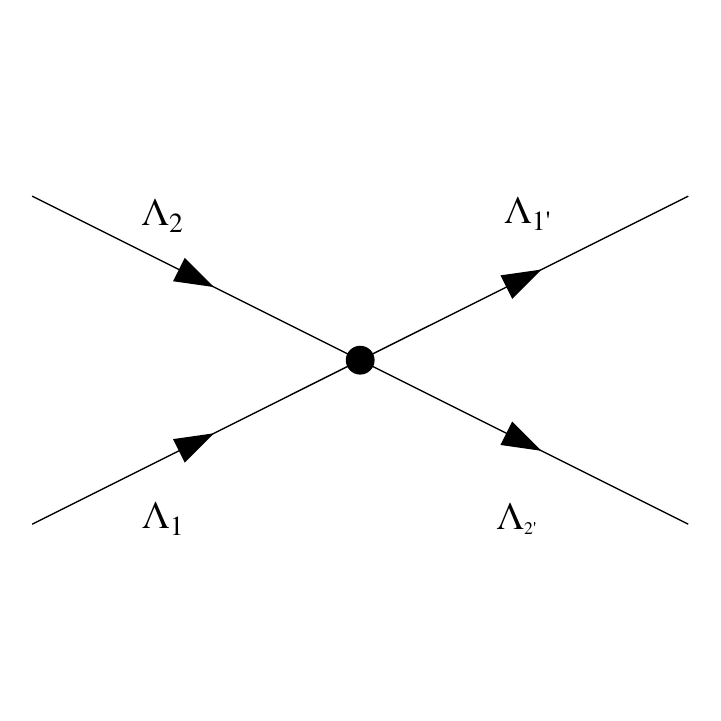}\\
  \includegraphics[scale=0.7]{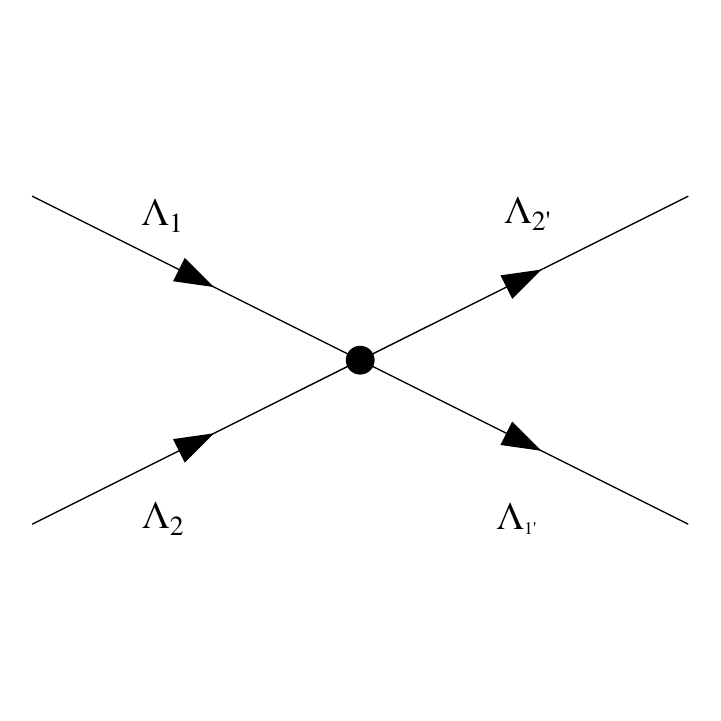}
  \includegraphics[scale=0.7]{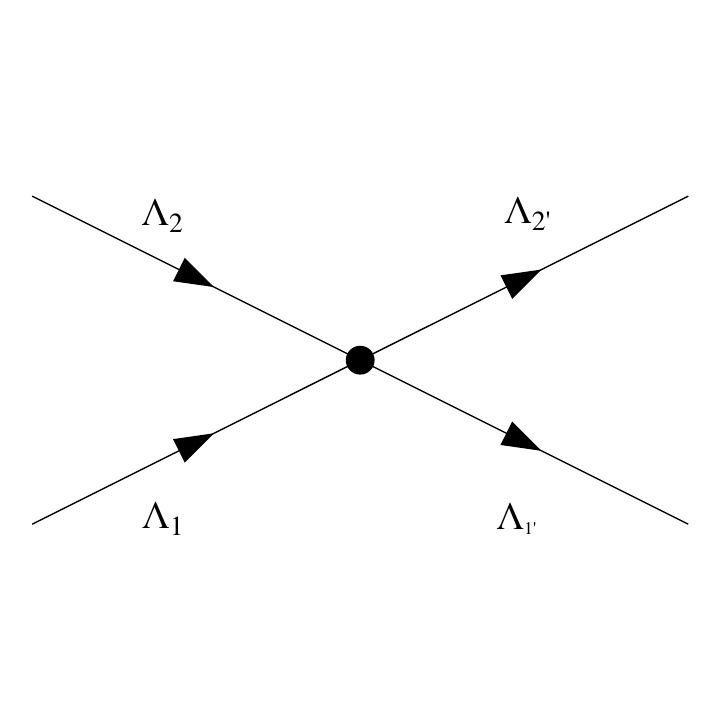}
  \caption{The leading order contribution to the $1+2\rightarrow 1'+2'$ Elko self-interaction.}\label{fig:self}
 \end{center}
\end{figure}

We consider the two-body scattering $1+2\rightarrow 1'+2'$ process at the lowest order in perturbation. The $S$-matrix for the diagrams of fig.\ref{fig:self} is
\begin{equation}
S_{(\Lambda'_{1}\Lambda'_{2})(\Lambda_{1}\Lambda_{2})}=\frac{-ig_{0}}{4}\int d^{4}x\langle \Lambda'_{1}\Lambda'_{2}|T[\gdual{\Lambda}_{\ell}(x)\Lambda_{\ell}(x)\gdual{\Lambda}_{k}(x)\Lambda_{k}(x)]|\Lambda_{1}\Lambda_{2}\rangle.
\end{equation}
Using the Wick's theorem by summing over all possible contractions between the fields and the particle states, we obtain
\begin{eqnarray}
S_{(\Lambda'_{1}\Lambda'_{2})(\Lambda_{1}\Lambda_{2})}&=&\frac{-ig^{0}\delta^{4}(p'_{1}+p'_{2}-p_{1}-p_{2})}{64\pi^{2}m^{2}\sqrt{p^{0}_{1}p^{0}_{2}p'^{0}_{1}p'^{0}_{2}}}\Big[
\gdual{\xi}_{\ell}(\p'_{2},\alpha'_{2})\gdual{\xi}_{k}(\p'_{1},\alpha'_{1})\xi_{\ell}(\p_{2},\alpha_{2})\xi_{k}(\p_{1},\alpha_{1})\nonumber\\
&&\hspace{3.5cm}-(1\leftrightarrow 2)-(1'\leftrightarrow 2')+(1\leftrightarrow 2,1'\leftrightarrow 2')\Big]
\end{eqnarray}
where the plus and minus signs of the last three terms are determined by fermionic statistics.
Using $S_{(\Lambda'_{1}\Lambda'_{2})(\Lambda_{1}\Lambda_{2})}=-2\pi iM_{(\Lambda'_{1}\Lambda'_{2})(\Lambda_{1}\Lambda_{2})}\delta^{4}(p'_{1}+p'_{2}-p_{1}-p_{2})$, we get
\begin{eqnarray}
M_{(\Lambda'_{1}\Lambda'_{2})(\Lambda_{1}\Lambda_{2})}&=&\frac{g_{0}}{128\pi^{3}m^{2}\sqrt{p^{0}_{1}p^{0}_{2}p'^{0}_{1}p'^{0}_{2}}}\Big[
\gdual{\xi}_{\ell}(\p'_{2},\alpha'_{2})\gdual{\xi}_{k}(\p'_{1},\alpha'_{1})\xi_{\ell}(\p_{2},\alpha_{2})\xi_{k}(\p_{1},\alpha_{1})\nonumber\\
&&\hspace{3.3cm}-(1\leftrightarrow 2)-(1'\leftrightarrow 2')+(1\leftrightarrow 2,1'\leftrightarrow 2')\Big].
\end{eqnarray}
To compute the transition probability $P(\alpha\rightarrow\beta)$ for $\alpha\rightarrow\beta$ involving Elko, we introduce the Elko adjoint for the amplitude $M_{\beta\alpha}$
\begin{equation}
P(\alpha\rightarrow\beta)\propto |M_{\beta\alpha}|^{2}=(M_{\beta\alpha}^{\ddag})^{T}M_{\beta\alpha}
\end{equation}
where $T$ transposes the matrix and $\ddag$ represents the modified conjugate for the Elko spinors. In terms of the Elko dual, it is defined as
\begin{equation}
\xi^{\ddag}(\p,\alpha)=\gdual{\xi}(\p,\alpha)\gamma^{0},\hspace{0.5cm}
\zeta^{\ddag}(\p,\alpha)=\gdual{\zeta}(\p,\alpha)\gamma^{0}.
\end{equation}

Summing over all $\alpha_{i}$ and $\alpha'_{i}$, the average probability is
\begin{equation}
\frac{1}{4}\sum_{\alpha_{1},\alpha_{2},\alpha'_{1},\alpha'_{2}}|M_{(\Lambda'_{1}\Lambda'_{2})(\Lambda_{1}\Lambda_{2})}|^{2}=\frac{1}{4}\sum_{\alpha_{1},\alpha_{2},\alpha'_{1},\alpha'_{2}}
(M^{\ddag}_{(\Lambda'_{1}\Lambda'_{2})(\Lambda_{1}\Lambda_{2})})^{T}
M_{(\Lambda'_{1}\Lambda'_{2})(\Lambda_{1}\Lambda_{2})}.
\end{equation}
The explicit form of the average probability now reads
\begin{eqnarray}
\frac{1}{4}\sum_{\alpha_{1},\alpha_{2},\alpha'_{1},\alpha'_{2}}|M_{(\Lambda'_{1}\Lambda'_{2})(\Lambda_{1}\Lambda_{2})}|^{2}&=&\frac{1}{4}\left[\frac{g_{0}}{128\pi^{3}m^{2}\sqrt{p^{0}_{1}p^{0}_{2}p'^{0}_{1}p'^{0}_{2}}}\right]^{2}\nonumber\\
&&\times\sum_{\alpha_{1},\alpha_{2},\alpha'_{1},\alpha'_{2}}\Big[
\gdual{\xi}_{\ell}(\p'_{2},\alpha'_{2})\gdual{\xi}_{k}(\p'_{1},\alpha'_{1})\xi_{\ell}(\p_{2},\alpha_{2})\xi_{k}(\p_{1},\alpha_{1})\nonumber\\
&&\hspace{1.5cm}-(1\leftrightarrow 2)-(1'\leftrightarrow 2')+(1\leftrightarrow 2,1'\leftrightarrow 2')\Big]\nonumber\\
&&\times\Big[
\xi_{\ell}(\p'_{2},\alpha'_{2})\xi_{k}(\p'_{1},\alpha'_{1})\gdual{\xi}_{\ell}(\p_{2},\alpha_{2})\gdual{\xi}_{k}(\p_{1},\alpha_{1})\nonumber\\
&&\hspace{1cm}-(1\leftrightarrow 2)-(1'\leftrightarrow 2')+(1\leftrightarrow 2,1'\leftrightarrow 2')\Big].\label{eq:cross_section_spin_sum}
\end{eqnarray}
Evaluating the spin-sums, we obtain\footnote{The computation was carried out using Mathematica 7.0}
\begin{eqnarray}
\frac{1}{4}\sum_{\alpha_{1},\alpha_{2},\alpha'_{1},\alpha'_{2}}|M_{(\Lambda'_{1}\Lambda'_{2})(\Lambda_{1}\Lambda_{2})}|^{2} &=&
\frac{1}{4}\left[\frac{g_{0}}{128\pi^{3}m^{2}\sqrt{p^{0}_{1}p^{0}_{2}p'^{0}_{1}p'^{0}_{2}}}\right]^{2}\nonumber\\
&&\times\Big\{32m^{4}\Big[-\cos(\phi_{1}-\phi_{2})+\cos(\phi_{1}-\phi'_{1})+\cos(\phi_{2}-\phi'_{1})\nonumber\\
&&\hspace{1.8cm}+\cos(\phi_{1}-\phi'_{2})+\cos(\phi_{2}-\phi'_{2})-\cos(\phi'_{1}-\phi'_{2})\nonumber\\
&&\hspace{1.8cm}+\cos(\phi_{1}+\phi_{2}-\phi'_{1}-\phi'_{2})\nonumber\\
&&\hspace{1.8cm}+\cos(\phi_{1}-\phi_{2}+\phi'_{1}-\phi'_{2})\nonumber\\
&&\hspace{1.8cm}+\cos(\phi_{1}-\phi_{2}-\phi'_{1}+\phi'_{2})+3\Big]\Big\}.
\label{eq:avg_prob_self_interaction}
\end{eqnarray}
In the centre of mass frame, the average differential cross-section is given by eq.~(\ref{eq:diff_cross_section}) and the spin-sum can be simplified using the following relations
\begin{equation}
\phi_{1}-\phi_{2}=\pm\pi,\hspace{0.5cm}
\phi'_{1}-\phi'_{2}=\pm\pi.\label{eq:phi_relations}
\end{equation}
After some algebraic simplification, the average differential cross-section becomes
\begin{equation}
\frac{d\sigma_{\mbox{\tiny{CM}}}}{d\Omega}(\Lambda_{1}\Lambda_{2}\rightarrow\Lambda'_{1}\Lambda'_{2})=\frac{g_{0}^{2}}{16\pi^{2}(p^{0}_{1})^{2}}\left[3+\cos(\phi_{1}-\phi'_{1}+\phi_{2}-\phi'_{2})\right].
\end{equation}
To simplify this expression further, let us rewrite eq.~(\ref{eq:phi_relations}) as
\begin{equation}
\phi_{1}=\phi_{2}+a\pi,\hspace{0.5cm}
\phi_{1}'=\phi_{2}'+b\pi,
\end{equation}
where $|a|=|b|=1$ but are otherwise independent of each other. Substituting the angles into $\cos(\phi_{1}-\phi'_{1}+\phi_{2}-\phi'_{2})$, we get
\begin{eqnarray}
\cos(\phi_{1}-\phi'_{1}+\phi_{2}-\phi'_{2})&=&\cos[2(\phi_{1}-\phi'_{1})]\cos[(b-a)\pi]\nonumber\\
&=&\cos[2(\phi_{1}-\phi'_{1})]=-1+2\cos^{2}(\phi_{1}-\phi'_{1}).
\end{eqnarray}
The second line uses the fact that $b-a=\{\pm2,0\}$ so that $\cos[(b-a)\pi]=1$. Therefore, the final form of the average differential cross-section reads
\begin{equation}
\frac{d\sigma_{\mbox{\tiny{CM}}}}{d\Omega}(\Lambda_{1}\Lambda_{2}\rightarrow\Lambda'_{1}\Lambda'_{2})=\frac{g_{0}^{2}}{8\pi^{2}(p^{0}_{1})^{2}}\left[1+\cos^{2}(\phi_{1}-\phi'_{1})\right].\label{eq:elko_self_cm}
\end{equation}

This result directly reflects the physical effect of the preferred plane. Although the spin-sum computed in eq.~(\ref{eq:cross_section_spin_sum}) is more complicated than the original Elko spin-sums of eqs.~(\ref{eq:elko_spinsum1}) and (\ref{eq:elko_spinsum2}), they only have dependence on the angle in the 12-plane. As a result, the above cross-section in arbitrary frames (not just the centre of mass frame) are independent on the direction of motion of the particles along the $3$-axis.

\section{Elko-Higgs interaction}
The lowest order $S$-matrix for the process $\Lambda+\gdual{\Lambda}\rightarrow\phi$ is given by
\begin{eqnarray}
S_{(\phi'_{1})(\Lambda_{1}\gdual{\Lambda}_{2})}&=&-ig_{1}\int d^{4}x\langle \phi'_{1}|T[\gdual{\Lambda}\Lambda\phi]|\Lambda_{1}\gdual{\Lambda}_{2}\rangle\nonumber\\
&=&\frac{-ig_{1}}{\sqrt{16\pi p^{0}_{1}p^{0}_{2}p'^{0}_{1}m^{2}}}\xi_{\ell}(\p_{1},\alpha_{1})\gdual{\zeta}_{\ell}(\p_{2},\alpha_{2}).
\end{eqnarray}
The amplitude $M_{(\phi'_{1})(\Lambda_{1}\gdual{\Lambda}_{2})}$ is given by
\begin{equation}
M_{(\phi'_{1})(\Lambda_{1}\gdual{\Lambda}_{2})}=\frac{-ig_{1}}{\sqrt{64\pi^{3} p^{0}_{1}p^{0}_{2}p'^{0}_{1}m^{2}}}\xi_{\ell}(\p_{1},\alpha_{1})\gdual{\zeta}_{\ell}(\p_{2},\alpha_{2}).
\end{equation}
Summing over $\alpha_{1}$ and $\alpha_{2}$, the average probability is proportional to
\begin{eqnarray}
\frac{1}{2}\sum_{\alpha_{1}\alpha_{2}}|M_{(\phi'_{1})(\Lambda_{1}\gdual{\Lambda}_{2})}|^{2}&=&\frac{g_{1}^{2}}{128\pi^{3}p^{0}_{1}p^{0}_{2}p'^{0}_{1}m^{2}}\sum_{\alpha_{1}\alpha_{2}}\xi_{\ell}(\p_{1},\alpha_{1})\gdual{\xi}_{k}(\p_{1},\alpha_{1})\zeta_{k}(\p_{2},\alpha_{2})\gdual{\zeta}_{\ell}(\p_{2},\alpha_{2})\nonumber\\
&=&\frac{g_{1}^{2}}{128\pi^{3}p^{0}_{1}p^{0}_{2}p'^{0}_{1}}\mbox{Tr}\left[(I+\mathcal{G}(\phi_{1}))(-I+\mathcal{G}(\phi_{2}))\right].
\end{eqnarray}
In the second line, we have used the Elko spin-sums. Evaluating the trace, we obtain
\begin{equation}
\frac{1}{2}\sum_{\alpha_{1}\alpha_{2}}|M_{(\phi'_{1})(\Lambda_{1}\gdual{\Lambda}_{2})}|^{2}=
\frac{g_{1}^{2}}{64\pi^{3}p^{0}_{1}p^{0}_{2}p'^{0}_{1}}[\cos(\phi_{1}-\phi_{2})-1]\leq 0.
\end{equation}
Since the transition probability is less than or equal to zero, this result is not physical. In our opinion, the problem is not with the process itself, but originates from our incomplete understanding of the particle state space of Elko and the adjoint.

\chapter{Addendum} 

Equation~(\ref{eq:Elko_prob}) that defines the transition probability
involving Elko is incorrect. The correct definition for transition probability should remain to be
\begin{equation}
P(\alpha\rightarrow\beta)\propto |M_{\beta\alpha}|^{2}=M^{\dag}_{\beta\alpha}M_{\beta\alpha}
\end{equation}
where $\dag$ represents Hermitian conjugation so that $P(\alpha\rightarrow\beta)$ is positive-definite. Both $\dag$ and $\ddag$ play an important role for Elko. 

In light of the recent work of Speran\c{c}a~\cite{Speranca:2013hqa} and Ahluwalia~\cite{Ahluwalia:2013uxa}, the following Elko spin-sums which is of relevance when computing scattering amplitudes can now be written in an elegant form
\begin{equation}
\sum_{\alpha}\xi(\p,\alpha)\bar{\xi}(\p,\alpha)=\gamma^{\mu}p_{\mu}\left[I+\mathcal{G}(\phi)\right]
\end{equation}
\begin{equation}
\sum_{\alpha}\zeta(\p,\alpha)\bar{\zeta}(\p,\alpha)=\gamma^{\mu}p_{\mu}\left[I-\mathcal{G}(\phi)\right].
\end{equation}
The dual coefficients for Elko can now be written as
\begin{equation}
\dual{\xi}(\p,\alpha)=\left[\Xi(\p)\xi(\p,\alpha)\right]^{\dag}\gamma^{0}
\end{equation}
\begin{equation}
\dual{\zeta}(\p,\alpha)=\left[\Xi(\p)\zeta(\p,\alpha)\right]^{\dag}\gamma^{0}
\end{equation}
where $\Xi(\p)$ is defined as
\begin{equation}
\Xi(\p)=\frac{1}{2m} \sum_{\alpha}\left[\xi(\p,\alpha)\bar{\xi}(\p,\alpha)-\zeta(\p,\alpha)\bar{\zeta}(\p,\alpha)\right].
\end{equation}
The above definition can be seen as a generalisation of the Dirac dual, since by replacing $\xi(\p,\alpha)$ and $\zeta(\p,\alpha)$ with the Dirac spinors $u(\p,\sigma)$ and $v(\p,\sigma)$ in $\Xi(\p)$ respectively, the matrix becomes the identity. This matrix is related to $\gamma^{\mu}p_{\mu}$ by 
\begin{equation}
\Xi(\p)=\frac{1}{m}\mathcal{G}(\phi)\gamma^{\mu}p_{\mu}
\end{equation}
so that the Elko spin-sums given in eqs.~(\ref{eq:elko_spinsum1}) and(\ref{eq:elko_spinsum2}) can be rewritten as
\begin{equation}
\sum_{\alpha}\xi(\p,\alpha)\dual{\xi}(\p,\alpha)=\left[\Xi(\p)\gamma^{\mu}p_{\mu}+m I \right]
\end{equation}
\begin{equation}
\sum_{\alpha}\zeta(\p,\alpha)\dual{\zeta}(\p,\alpha)=\left[\Xi(\p)\gamma^{\mu}p_{\mu}-m I \right]
\end{equation}
and we get
\begin{equation}
\left[\gamma^{\mu}p_{\mu}-m \Xi(\p)\right]\xi(\p,\alpha)=0
\end{equation}
\begin{equation}
\left[\gamma^{\mu}p_{\mu}+m \Xi(\p)\right]\zeta(\p,\alpha)=0.
\end{equation}
In this form, we see that the Lorentz violation of Elko is encoded in the matrix $\Xi(\p)$. For the latest development on Elko, please refer to ref.~\cite{Ahluwalia:2013uxa}.



\end{document}